\documentclass[12pt]{article}
\usepackage{graphicx}
\usepackage{newtxtext,xfrac,xcolor}
\usepackage{newtxmath}
\usepackage{natbib,tikz}
\usepackage{hyperref}

\usepackage{fancyhdr}
\allowdisplaybreaks
\hypersetup{
    colorlinks = true,
    urlcolor   = blue,
    citecolor  = black,
}

\newcommand{\RomanNumeralCaps}[1]
\linenumbers

\newcommand{\bseq}{\begin{subequations}}
\newcommand{\eseq}{\end{subequations}}
\newcommand{\beq}{\begin{equation}}
\newcommand{\eeq}{\end{equation}}
\newcommand{\bef}{\begin{figure}}
\newcommand{\eef}{\end{figure}}

\title{Spatiotemporal linear stability of viscoelastic Saffman-Taylor flows}

\author{ {\bf D. Bansal$\dagger$}, {\bf T. Chauhan$\dagger$} and {\bf S. Sircar$\dagger, \ddagger$}
\\{$\dagger$ \small Department of Mathematics, IIIT Delhi, India 110020} \\{\small $\ddagger$ Corresponding Author (email: sarthok@iiitd.ac.in)}}

\begin{document}
\maketitle
\newcommand{\blackline}{\raisebox{2pt}{\tikz{\draw[-,black,solid,line width = 1.5pt](0,0) -- (5mm,0);}}}
\newcommand{\blueline}{\raisebox{2pt}{\tikz{\draw[-,blue,solid,line width = 1.5pt](0,0) -- (5mm,0);}}}
\newcommand{\greenline}{\raisebox{2pt}{\tikz{\draw[-,green,solid,line width = 1.5pt](0,0) -- (5mm,0);}}}
\newcommand{\cyanline}{\raisebox{2pt}{\tikz{\draw[-,cyan,solid,line width = 1.5pt](0,0) -- (5mm,0);}}}

\begin{abstract}
A comprehensive, temporal and spatiotemporal linear stability analyses of a (driven) Oldroyd-B fluid in a horizontal, square, Hele-Shaw cell is reported to identify the viable regions of topological transition of the advancing interface. The base flow profile of the driving, Newtonian fluid of negligibly low viscosity, is the Poiseuille profile. The dimensionless groups that govern the stability are the Reynolds number, $Re = \frac{b^2 \rho \mathcal{U}_0}{12 \eta_0 L}$, the elasticity number, $E = \frac{12 \lambda (1-\nu)\eta_0}{\rho b^2}$ and the ratio of solvent to polymer solution viscosity, $\nu = \frac{\eta_s}{\eta_0}$; here $b$ is the cell gap, $L$ is the length/width of the cell, $\mathcal{U}_0$ is the maximum velocity of the mean flow, $\rho$ is the density of the driven fluid and $\lambda $ is the microstructural relaxation time. Under the assumption that the relative finger width of the interface is the wavelength of the most unstable (temporal) mode, excellent agreement between our model and the experiments in the Stokes and the inertial, Newtonian regime. In the asymptotic limit $E(1-\nu) \ll 1$, the critical Reynolds number, $Re_c$ (defined as the largest Reynolds number beyond which all wavenumbers are temporally unstable) diverges as per the scaling law $Re_c \sim \left[E(1-\nu)\right]^{-5/3}$ and the critical wavenumber increases as $\alpha_c \sim \left[E(1-\nu)\right]^{-2/3}$. The temporal stability analysis stipulate that (a) the inertial forces (proportional to $Re$) have a universally destabilizing influence on the advancing flow front, (b) the finite boundary has a destabilizing impact near the wall, and (c) farther away from the wall, elasticity combined with low (high) fluid inertia has a stabilizing (destabilizing) impact. The Briggs idea of analytic continuation is deployed to classify regions of absolute and convective instabilities, as well as the evanescent modes, and the results are compared with previously conducted experiments for Newtonian as well as viscoelastic flows with surface instabilities, exhibiting topological transitions. The spatiotemporal phase diagram reveals the presence of absolutely unstable region at high values of Reynolds and elasticity number, confirming the role of fluid inertia in triggering a pinch-off of the moving interface.
\end{abstract}


{\bf MSC Codes }  {\it(Optional)} Please enter your MSC Codes here

\section{Introduction}\label{sec:intro}
Injecting a fluid into a more viscous fluid in a thin linear channel (or the Hele-Shaw cell) triggers a two-dimensional viscous fingering pattern which is characterized by increasingly long fingers undergoing tip splitting and branching events, also known as the Saffman-Taylor instability (STI). These complex structures are considered to be a paradigm for interfacial pattern formation and have continued to receive prolonged interest in theoretical and experimental studies~\citep{Bensimon1986,Couder2000} as well as due to its practical applicability in crude oil recovery~\citep{Homsy1987}, surface coating~\citep{Grillet1999} and electrodeposition~\citep{Schroder2002}. In the classical (Newtonian) case, the viscosity difference drives the formation of fingers, since emerging fingers enhance the pressure gradients (treated as a Laplacian field for flow velocity governed by Darcy's law) acting in the tip region~\citep{Homsy1987}. Since the interface moves with a velocity proportional to the pressure gradient the feedback loop is closed, any minute initial displacement will turn unstable, as it increases the pressure gradient and further accelerates a protruding finger, which in turn becomes steeper and steeper as it continues to develop. This instability is damped by surface tension, which acts to minimize the interface area and opposes the formation of curved fingers. Thus, the parameter determining the stability of the fingers is the capillary number, $Ca$, which is the ratio between the viscous effects (given by the product of the velocity scale, $\mathcal{U}_0$, and the viscosity difference, $\triangle \mu$) against the capillary effects (represented by the surface tension, $\gamma$). Alternatively, the instability has also been studied for non-Newtonian fluids, for which strikingly different fingering patterns, such as fracture-like or fractal patterns are found~\citep{Lemaire1991}. The main mathematical challenge the `viscoelastic fingering' instability poses is that pressure may no longer be a Laplacian field, rendering the numerical prediction of the finger width somewhat complicated~\citep{Kondic1998}. Detailed theory on the viscoelastic fingering instability has attempted to account for the shear thinning behaviour~\citep{Kondic1996,Kondic1998,Poire1998,Fast2001, Fast2004, BenAmar2005}, as well as the effect of normal stresses~\citep{Lindner2002}. More recently, \citet{Shokri2017} reported the linear finger instability of Oldroyd-B fluids using a modified Darcy's law but a comprehensive theoretical explanation for the viscoelastic fingering pattern selection mechanism from first principles, in the spatiotemporal framework, remains elusive until now.

Instabilities detected via the spatiotemporal analysis are considered to be the precursor to the topological reconfiguration of fluid interfaces in Hele-Shaw flows~\citep{Goldstein1993}. The method of spatiotemporal characterization in confined~\citep{Nakamura1997} as well as in open flows~\citep{Huerre1985} by progressive moving of the isocontours in the complex frequency and wavenumber plane was first proposed by~\citet{Briggs1964} in the context of plasma physics. In this method, we search for the absolute instabilities (perturbations which grow exponentially in time at the point of excitation), the convective instabilities (disturbances which are swept downstream from the source and decay at any fixed position in space) and the evanescent modes (the non-propagating modes or false modes)~\citep{Patne2017}. While convectively unstable flows behave as spatial amplifiers of the incoming perturbations on the interface, absolutely unstable flows display intrinsic self-sustained dynamics or global modes~\citep{Huerre1990}. The transition between these two classes of flows has been experimentally evidenced in numerous situations for Newtonian flows. Absolute instability was recently experimentally verified by~\citet{Shoji2020} in liquid jets. There is also an extensive literature on Newtonian wakes and mixing layers, including the blunt body experiments listed by~\citet{Oertel1990}, the linear analysis by Chomaz~\citep{Delbende1998} and~\citet{Healy2009} and the Direct Numerical Simulation (DNS) studies by~\citet{Pier2008}. Overall, these studies found a destabilizing effect of finite boundaries and an existence as well as a transition to absolute instability in the near wake region. The spatiotemporal analyses of viscoelastic flows are more recent and scarce. An early experimental study by~\citet{Vihinen1997} reported absolute instability in viscoelastic liquid jets. \citet{Pipe2005} reported a stabilizing effect of polymer addition in his experiments on viscoelastic cylindrical wakes, which is counteracted by shear thinning and a transition from convective to absolute instability at higher polymer concentrations. In contrast, the linear analysis of dilute mixing layers~\citep{Ray2014} and dilute jets~\citep{Ray2015, Alhushaybari2019, Alhushaybari2020} relay a significant range of parameters where viscoelasticity was found to be destabilizing. A recent DNS study of jets necessitates the use of an extra (convective) timescale to characterize the memory fading property of viscoelastic fluids~\citep{Guimaraes2020}. In the context of STI, past research has shown that the temporal stability analysis yields a length scale that, in combination with the geometry of the system, determines the nonlinear growth of the viscous~\citep{Chuoke1959} as well as the viscoelastic fingers~\citep{Shokri2017}. However, from the above brief discussion, it is clear that a comprehensive spatiotemporal characterization of the STI, predicting the onset of topological transition~\citep{Lee2002}, is still largely an open problem in the instability and transition community that needs to be addressed.

Despite their consistent occurrence in vitro experiments, the topological reconfigurations of fluid interfaces in Hele-Shaw flows are poorly understood, since the interface restructuring entails intrinsically singular behavior~\citep{Lee2002}. An additional obstacle has been the lack of systems in which these topological transitions can be controlled to produce satisfactory experiments~\citep{Shelley1993}. Hence, the study of the transition from convective to absolute instability is crucial to understand how the perturbations on the interface change from a broad-band distribution to a sharp peak oscillatory behavior, in the latter case, resulting in interface detachment~\citep{Goldstein1993}. The study is complicated due to the presence of evanescent modes (which are stable modes) arising out of the direct resonance of two coalescing modes, originating from waves propagating in the same direction~\citep{Koch1986}. In the remainder of the introduction, we review the prior work relevant along the lines of the motivating questions raised above, under the following headings: (i) Saffman Taylor instability in Newtonian fluids, (ii) role of anisotropy in the modification of the finger width in non-Newtonian fluids, (iii) DNS and computational bifurcation studies, (iv) recent experimental findings, and (v) the temporal and the spatiotemporal stability analyses of viscous and viscoelastic Hele-Shaw flows. Finally, the specific objectives of the present work are presented in the context of the existing paradigm with regards to the spatiotemporal transition in viscoelastic Hele-Shaw flow.

\subsection{Saffman Taylor instability in Newtonian fluids}\label{subsec:L1}
The classical STI was outlined by~\citet{Saffman1958}, however, the finger selection mechanism in their experiments remained an enigma for a very long time. Omitting surface tension, they found a continuous family of solutions with the shape of the interface given by the expression $x = \frac{L(1-\Lambda)}{2\pi} \ln \left[ \frac{1}{2} (1 + \cos \frac{2\pi y}{\Lambda L})\right]$ (refer figure~\ref{fig1} for details). Although their analytical expression for the shape of the finger matched well with their experimental observations, the treatment did not explain the specific selection of the relative finger width or $\Lambda=0.5$ (defined as the ratio of the finger width to the cell width, $L$). The significance of surface tension on the shape selection procedure was earlier highlighted (numerically) by~\citet{McLean1981} and then later (analytically) by~\citet{Dong1986,Shraiman1986} and~\citet{Combescot1986}, who showed that the surface tension represented a singular perturbation leading to a solvability condition at the finger tip, thereby designating a particular value from the continuum of solutions proposed by Saffman (which is $\Lambda=0.5$). \citet{Lindner2006} described the nonlinear growth process in a linear channel of width $L$ (refer figure~\ref{fig1}): small fingers grow and begin to compete with the more advanced fingers, screening the less advanced fingers ultimately leading to a single finger propagating through the cell. The relative width of the single finger is described in terms of the control parameter, $\sfrac{1}{B} = 12 Ca (\sfrac{L}{b})^2$, where $\sfrac{L}{b}$ is the aspect ratio. For large aspect ratio (or thin channels), increasing the velocity $U$ of the finger tip leads to the viscous forces become increasingly dominant compared with the capillary forces and the relative finger width decreases. At higher velocity, the finger width stabilizes near a plateau value at $\Lambda=0.5$.

The above description presents an idealized two-dimensional setup. However, there is a thin wetting film that remains between the advancing finger and the glass plates, whose thickness, $t$, according to the Bretherton's law, is $Ca^{2/3} t / R = 0.643 (3 Ca)^{2/3}$, where $R$ is the radius of curvature of the finger~\citep{Tabeling1985}. This thin film effect leads to a continuous modification of the pressure jump at the interface and thus a slight modification of the finger width observed across different experimental geometries. \citet{Reinelt1987POF} developed the interface conditions by considering the variations in the thickness of the thin film, along with both transverse curvature (across the cell gap) and lateral curvature (along the interface edge) at the leading edge of the interface. He solved the shape of the interface and determined the unique relative finger width by conformally mapping the problem to a circular domain and expanding the solution (satisfying the interface condition) in terms of analytic functions~\citep{Reinelt1987JFM}. A good agreement between the numerical and the experimental results at low Reynolds number ($Re$) was achieved.

In another Hele-Shaw cell setup with ridges on the top and the bottom plates,~\citet{Rabaud1988} reported the formation of viscous fingers of relative width smaller than 0.5. The observation of such `anomalously' narrow fingers were explained using the argument that the carvings induced a local perturbation at the tip of the advancing finger, thereby removing the classical selection of the discrete set of solutions. He further showed that for a given value of $Ca$, it is not the relative finger width that is selected but the dimensionless radius of curvature at the tip, $\rho / b$, where $\rho$ is directly proportional to the relative finger width. At high velocities, one observes a saturation of $\rho$, leading to an increase in finger widths which might be attributed to the inertial effects~\citep{Lindner2006}, especially for small aspect ratios of the Hele-Shaw cell and lower viscosities of the driven fluid.

\subsection{Anisotropy effects in non-Newtonian fluids}\label{subsec:L2}
Experiments of viscous fluids pushing dilute solution of xanthate gum (a shear thinning polymeric liquid) reveal a strong modification of the finger selection process, in particular fingers were found to be narrower than the classical limit~\citep{Lindner2000}. Numerical simulations of shear thinning fluids~\citep{Kondic1996, Kondic1998, Fast2001} divulge that the viscosity is not uniform throughout the cell; regions of high fluid velocity (thus high shear rate) have low viscosity, especially in front of the finger tip, which leads to an anisotropic system. For weakly shear thinning case (with the exponent $0.65 < n < 1$ in Carreau fluids~\citep{Carreau1979}), \citet{Lindner2000} showed that simply replacing the constant viscosity by a shear dependent viscosity, in the control parameter, allows for the rescaling of the data for the relative finger width onto the universal curve for Newtonian fluids. For stronger shear-thinning fluids ($n < 0.65$), this rescaling fails and deviations from the classical limit result towards smaller fingers. A mechanism similar to the one proposed by \citet{Rabaud1988} is argued to be responsible for the selection of the fingers in shear thinning fluids with anisotropy playing the role of the local perturbation at the finger tip~\citep{Lindner2002}. The relation between $\rho / b$ and $\Lambda$ is found to depend on the shear thinning character of the fluids and thus the shear thinning exponent, $n$ and the knowledge of this relation solves the selection problem. \citet{Poire1998} solved the problem for power law fluids and their results are in good agreement with
the experimental observations.

In experiments of dilute solution of polyethylene oxide (a shear thickening liquid), a finger widening phenomena (compared with the Newtonian case) was observed~\citep{Lindner2002}. The presence of normal stresses in the thin wetting layer is held accountable for the finger widening; one can attempt to account for this effect by adding a supplementary pressure to the system. In classical theory, the pressure jump at the interface is given by the radius of curvature times the surface tension. \citet{Tabeling1985} showed that one can assimilate the effect of a finite thickness of the wetting film by adding the supplementary pressure caused by the normal stresses to the surface tension term in the control parameter, $\gamma^* = \gamma + \frac{1}{2} N_1 b$. For moderate normal stresses, this allows the rescaling of the data onto the universal curve for Newtonian fluids and again resolves the finger selection problem.

\subsection{DNS and computational bifurcation studies}\label{subsec:L3}
The instability of driven fronts when the invading fluid is less viscous than the defending fluid is a classical problem in nonlinear pattern formation~\citep{Park1985}. The early, sharp-interface simulations of fingering in Hele-Shaw cells were presented by~\citet{Tryggvason1983}, with results which were qualitatively similar to the
experiments of~\citet{Maher1985}. Early simulations using boundary integral methods were by~\citet{Park1985, Degregoria1986, Meiburg1988, Meiburg1989}, and a review of these early simulation methods was documented by~\citet{Whitaker1994}. A crucial series of simulations based on boundary element methods was developed by~\citet{Hou1997, Hou2001, Li2007} and~\citet{Li2009}. For non-Newtonian flow, a similar approach is used by~\citet{Almgren1993, Kondic1996, Kondic1998, Fast2001, Fast2004} and~\citet{Sarkar1989}.

The diffuse-interface theories of multiphase flow next emerged as encouraging tools to understand and simulate complex processes involving the simultaneous flow of two or more immiscible fluid phases~\citep{Anderson1998}. These models (originating from the material science community) described the interface evolution between the phases through an order parameter or `phase field' that defines a smooth transition between two phases~\citep{Cahn1958}. The fundamental idea of tracking and properly capturing interface dynamics through phase-field theories has been subsequently generalized and adapted to multiphase fluid flow scenarios. While variational or thermodynamic theories are perhaps the most common theoretical framework to rigorously derive these models~\citep{Anderson1998, Lowengrub1998, Jacqmin1999, Boyer2002, Badalassi2003, Yue2004, Kim2005, Ding2007}, diffuse-interface theories have been obtained from either averaging microscale interactions~\citep{Sun2004} or rationalized  as a micro force balance~\citep{Gurtin1994}. The common goal in these approaches is to formulate thermodynamically consistent stress tensors and mesoscale balance laws, including the impact of surface tension on the momentum balance, as well as properly tracking interfacial dynamics~\citep{Lowengrub1998}.

Other preliminary in silico studies include two-dimensional nonlinear simulations of viscous, miscible fingers in porous media using a velocity-dependent diffusivity tensor~\citep{Tan1988}, permittivity dependent on the spatiotemporal coordinates~\citep{Tan1992}, velocity-dependent diffusivity tensor and concentration dependent viscosity coefficients~\citep{Zimmerman1991}, three-dimensional viscous fingering with randomly perturbed initial conditions~\citep{Zimmerman1992a} and a unified two-dimensional nonlinear simulation combining the effects of viscosity contrast, anisotropic permittivities and velocity dependence on the diffusivity~\citep{Zimmerman1992b}; all the cases simulated using a combination of discrete Fourier and Hartley transform. Some of the issues tackled by these researchers were the evolution of the prediction of the growth rate and the wavelength of the most unstable fingers, the spreading and the shielding effects caused by the span wise secondary instability, branching and tip splitting caused by the crossflow stretching, predicting the lateral scale of the finger evolution in the absence of tip splitting, anisotropic finger pairing, the temporal evolution of the solvent concentration profile or the temporal growth of the mixing layer, prediction of more permeable paths in a heterogeneous media and the mechanism governing the nonlinear interaction and the shape selection of the fingers. More recent developments in the area of multiphase, immersed interface Hele-Shaw flows, include the utilization of a Fourier-Galerkin method to simulate the vorticity-streamfunction formulation, exploring the coupling between the viscosity and gravity~\citep{Ruith2000}, a combination of compact difference and spectral methods to numerically capture the density-driven instabilities between miscible fluids in vertical~\citep{Fernandez2002} as well as non-vertical~\citep{Upchurch2008} Hele-Shaw cells as well as the subsequent bifurcation studies~\citep{Goyal2007}, a high-resolution finite-volume method with flux limiters to investigate the evolution of the structure and the leading edge of the mixing zone in miscible viscous fingering~\citep{Booth2010}, a fractional step projection-based method for the three-dimensional simulation of miscible flows~\citep{Oliveira2011}, a fifth-order weighted essentially non-oscillatory (WENO) scheme for the computation of the three-dimensional, variable density and variable viscosity, miscible horizontal~\citep{John2013} as well as vertical~\citep{Heussler2014} Hele-Shaw setups, a parallel algorithm involving the volume-of-fluid method for heavy oil extraction applications~\citep{Lagree2016} and level set methods to simulate non-standard (tapered as well as rotating) cells with variable injection rates~\citep{Morrow2019}.

\subsection{Recent experimental observations}\label{subsec:L4}
A considerable experimental effort to study the bulk fingering instabilities in viscous/viscoelastic liquids is geared towards implementing novel strategies to control these instabilities. Motivated by the necessity to study carbon sequestration techniques, \citet{Slim2013} conducted an experimental study of the dissolution-driven Newtonian convection in a Hele-Shaw cell, showing the successive flow evolution regimes starting from a one-dimensional diffusional profile, followed by a region of linear growth in which the fingers initiate and grow quasi-exponentially and independently of one-another, and then a `flux-growth regime' (with the flux growing to a local maximum), accompanied with a region where the fingers interact and merge, and leading to a domain of reinitiation, where new fingers are created between the primary existing ones. Homsy~\citep{Haudin2014} reported the presence of thin stripes perpendicular to the moving interface inside the mixing zone of two miscible, Newtonian fluids, and attributed this instability due to the buoyancy effects within the gap of the cell which arises out of an unstable density stratification. \citet{Huerre2015} examined the motion of viscous droplets in a confined, micrometric Hele-Shaw cell, by focusing on the lubrication film between the droplet and the side wall. They deduced that, at lower values of $Ca$, the film thickness is constant and set by the pressure difference across the cell, and above a critical value of $Ca$, the interface behaviour is described by the viscous dissipation between the meniscus and the cell wall. \citet{Ecke2016} studied the mass-transport in the water-propyleneglycol system enclosed in a Hele-Shaw cell with variable permeability, representing a scaled-down laboratory analogue of the multi-species porous media convection. They observed a rapid decrease (and an approach to a constant steady value) of the critical wavenumber representing the plume pattern and reasoned the variability of the plume velocities via the microscopic merger and the renucleation of the advancing interface. Recent experiments on viscous fingering in a cell with rectangular occlusions~\citep{Gomez2016} have shown to support symmetric, asymmetric and oscillatory propagation states and multiple stable fingers at critical occlusion heights. Similarly, experiments in a radial, elastic walled Hele-Shaw cell~\citep{Puzovic2018}, indicated a wide variety of novel interfacial patterns including a delayed onset of instability to much larger values of the flow rate, periodic sideways fingers, dendritic-like patterns and short, blunt viscous fingers; due to the presence of the wall elasticity. The experiments on the lifting Hele-Shaw cell problem indicated an inertia-induced mechanism for dendritic-like patterns and an intensification of the finger competition events with increasing lifting velocities~\citep{Anjos2017}. For a fully miscible system, the surface tension of the air-suspension (which is treated as a simple Newtonian fluid with enhanced viscosity) interface that suppresses the fine fingers characteristic, was posited as the stabilizing mechanism~\citep{Hooshanginejad2019}; while for partially miscible systems, thermodynamic instability such as the phase separation due to the spinodal decomposition and the Korteweg convection induced by the compositional gradient in such a phase separation, were identified as the rationale behind the neoclassical fingering instability~\citep{Suzuki2020}.

More unorthodox setups include the investigation of the inverse STI (a case in which a viscous fluid displaces air, otherwise considered a stable scenario) with partially wettable hydrophilic particles adhering on the cell walls, which revealed a fingering instability at low capillary number resulting from the minimization of the interfacial energy~\citep{Bihi2016}; the studies exploring the role of STI in the propagation of premixed gaseous flames in a combustion chamber~\citep{Lopez2019}; and the reaction infiltration instability of a reactive fluid front in a soluble porous medium~\citep{Xu2019}. 

Similarly, purely elastic fingering instability was demonstrated by Saintyves~\citep{Saintyves2013}, who suggested that the confinement of the elastomer and its adhesion to the plates of the cell as the driving mechanism of the instability. \citet{Petrolo2020} developed an experimentally guided model to investigate the onset of the Darcy-B\'{e}nard instability in a two-dimensional porous medium saturated with a heated, non-Newtonian fluid, in a uniform horizontal pressure gradient. Discrepancies between theory and experiments were ascribed to a combination of factors, including the nonlinear phenomena, possible subcritical bifurcations, approximations in the rheological model, wall slippage, ageing and degradation of the fluid properties. In another in vitro study on the impact of surfactant addition in a non-Newtonian fingering instability, it was surmised that  the surfactant concentration locally decreased the interfacial tension, leading to a reduction in viscosity, and an increased impact on the capillary number which led to the evolution of wider fingers (in striking contrast with Newtonian flow counterpart~\citep{Ahmadikhamsi2020}). However, the relative finger width of both the experiments, with and without the surfactant, converged asymptotically to the same value, leading to a conclusion that finger widening is caused by the decrease in surfactant concentration in the vicinity of the tip so that only the shear-thinning feature of polymer prevails at long times.

\subsection{Temporal and spatiotemporal stability analyses}\label{subsec:L5}
The competition between the viscous and the capillary forces on the advancing front leads to the emergence of a characteristic length scale which determines the relative finger width, $\Lambda$ and the same can be calculated using the linear stability analysis. This physical reasoning has been made rigorous by deriving a dispersion relation relating the growth-rate, $\omega$ of the instability to the wavenumber $\alpha$, assuming a normal mode expansion  of the disturbance, $\exp {\it i} (\alpha x - \omega t)$ and by choosing $\alpha \in \mathbb{R}$ real and allowing for a complex $\omega \in \mathbb{C}$ or the so called temporal stability analysis~\citep{Huerre1985}. \citet{Chuoke1959} derived the dispersion relation in a rectilinear channel for Newtonian fluids,
\beq
\omega = {\it i} \left[ U \alpha - \frac{\gamma \alpha^3}{\mu_2} \frac{b^2}{12} \right],
\label{eqn:DR_DarcyRec}
\eeq
under the assumption when the viscosity of the driving fluid is negligible (or $\mu_1 \ll \mu_2$, refer figure~\ref{fig1}). $U, \gamma, b$ are the mean velocity (assumed constant and long the `flow' direction), surface tension coefficient and the gap between the cell plates. From equation~\eqref{eqn:DR_DarcyRec}, we conclude the existence of the most unstable temporal mode (obtained by setting $\frac{d \omega}{d \alpha} = 0$, leading to a value $\alpha_{\text{Temp}} = \frac{2}{b}\sqrt{\frac{ \mu_2 U}{\gamma}}$), resulting from the competition of the destabilization by the viscosity difference and the stabilization by surface tension~\citep{Chuoke1959}. A similar treatment for the circular injection case (of a negligibly low viscosity fluid driving a higher viscosity fluid) leads to the wavenumber of the fastest growing perturbation, $\alpha_{\text{Temp}} = \sqrt{\frac{1}{3}\left( 12 \frac{\mu_2 U}{\gamma} \frac{R^2}{b^2} + 1\right)}$ (and the corresponding most unstable wavelength, $\lambda_{\text{Temp}} = \sfrac{2\pi}{\alpha_{\text{Temp}}}$), from the following dispersion relation~\citep{Wilson1975},
\beq
\omega = {\it i} \left[ \frac{U}{R} \left(\alpha - 1\right) - \frac{b^2}{12}\frac{\gamma}{R^3}\frac{\alpha(\alpha^2-1)}{\mu_2} \right],
\label{eqn:DR_DarcyCir}
\eeq
where $R$ is the radius of the expanding interface. \citet{Maxworthy1989} experimentally verified the result of~\citet{Wilson1975} and showed that this prediction works well for low capillary numbers but does not reproduce the behavior at elevated values (i.~e., $Ca > 10^{-2}$). The predictions of both \citet{Chuoke1959} and \citet{Wilson1975} are based on the Darcy equation, or the classical Hele-Shaw limit of the Stokes equation for large aspect ratio, in which only viscous terms in the thin direction are retained. The flow in the thin direction is approximated by a Poiseuille profile and the equations can be depth-averaged in the shallow direction. This leads to the Darcy equation, which is a two-dimensional potential flow, where the pressure represents the potential. One inconsistency of this model was pointed out by \citet{Dai1993} who reasoned that for zero surface tension the fingers would form infinitely sharp cusps. 

\citet{Paterson1985} assumed that for zero surface tension flows, for example with two miscible fluids, the interface is influenced by the full three-dimensional stress tensor. He used a potential flow for which he derived the three-dimensional stress tensor at the fluid-fluid interface and looked for the perturbation wavelength that minimizes the dissipated energy, thereby using the property that low $Re$ flow minimizes the dissipation. In the limit of zero surface tension (or the equivalent limit $Ca \rightarrow \infty$), Paterson deduced the wavelength of the most unstable temporal mode, of the order $\lambda_{\text{Temp}} \approx 4 b$, which is close to the Maxworthy's result: $\lambda_{\text{Temp}} \approx 5 b$~\citep{Maxworthy1989}. A similar dissipation minimization approach was followed by~\citet{BenAmar2005} for 2D Darcy, lifting Hele-Shaw flows, including the influence of surface tension but neglecting all stress contribution except in the thin direction. In the short wavelength limit ($\alpha \gg 1$) they obtained the Wilson's result for the wavenumber of the most unstable mode with a refactor of $\sqrt{\sfrac{3}{5}}$.

\citet{Gadelha2009} treated the linear and the nonlinear evolution of the radial fingering problem using the concept of the viscous potential flow (VPF), where the flow is governed by the pressure potential but the boundary conditions include viscous normal stress. This concept was applied earlier to the fingering problem in a rotating Hele-Shaw cell by~\citet{Lacalle2004}. \citet{Kim2009} determined the most unstable mode using VPF and revealed improved agreement with Maxworthy's experiment even at capillary numbers of order one. This approach has been generalized by \citet{Dias2013} by including a second and a third order perturbative correction (coined as the `perturbative-mode-coupling' method) to the dispersion relation obtained by Wilson~\citep{Wilson1975}. They further included a Young-Laplace law for a prescribed wetting angle and a non-uniform out-of-plane curvature due to the dynamic film formation. Their work compares two selection criteria for the most unstable mode, one based on the maximum growth rate, like we use in this work, and another based on a maximum amplitude criterion. A maximum amplitude approach integrates the dynamic growth of the finger amplitude and permits an excellent fit but brings in additional parameters like the contact angle and the initial conditions. \citet{Logvinov2010} used the Brinkman equation to describe STI for miscible fluids in rectangular channels with zero surface tension and demonstrated the influence of viscous effects in the flow plane. They found a wavelength dependence, $\lambda_{\text{Temp}} \approx 2.5 b$, which is a factor 2 too small compared to Maxworthy's experiment. More recently, \citet{Housseiny2013} probed the linear stability of the Hele-Shaw flow in rectilinear and radial geometries with depth gradient and deduced that the wavenumber of the most unstable temporal mode was determined by both the viscosity contrast of the fluids and the ratio of the depth gradient to the capillary number of the system. 

Temporal stability analyses of viscoelastic fluids are recent. \citet{Mora2010} examined the STI of an Upper Convected Maxwell's (UCM) fluid. They observed a divergence of the temporal growth rate at a critical value of a dimensionless time parameter, $\tilde{\lambda}$, and associated this observation to a fracture-like pattern instability of the interface. \citet{Shokri2017} reported the fingering instability of Oldroyd-B fluids using a modified Darcy's law and deduced an elasticity induced stabilization of the interface at higher a Weissenberg number ($We$). However, a lack of a deeper classification (in the spatiotemporal framework) reduces the efficacy of their analysis in accurately capturing the instability transition. Hence, an in-depth spatiotemporal study of the transition pathway in viscoelastic STI is expedient.

%
The spatiotemporal evolution of a localized disturbance (located at the origin of the $x$-$t$ plane) is illustrated by considering the response of a given base velocity profile, $u(x, t)$, to an impulse excitation~\citep{Huerre1990},
\beq
D\left( -{\it i} \frac{\partial }{\partial x}, {\it i} \frac{\partial }{\partial t}, {\bf M} \right) u(x, t) = \delta(x) \delta(t), 
\label{eqn:DRP}
\eeq
where $D(\alpha, \omega, {\bf M})=0$ is the dispersion relation and ${\bf M}$ is the vector of material and fluid parameters. The solution to equation~\eqref{eqn:DRP} is dictated by the Green's function, 
\beq
G(x, t) = {\displaystyle \frac{1}{4\pi^2} \int_L \int_F \frac{e^{{\it i}(\alpha x - \omega t)}}{D(\alpha, \omega, {\bf M})} d \alpha d\omega }
\label{eqn:Green}
\eeq
where $F$ and $L$ are the Fourier contour in $\alpha$ and the Laplace contour in the $\omega$ plane. The Fourier contour integral is placed parallel to the Real($\alpha$) axis (in the ensuing description we denote real/imaginary components with subscript {\it r}/{\it i}, respectively), while the Laplace contour integral is placed above all singularities of the dispersion relation in the $\omega$-plane so as to satisfy the causality condition (i. e., $G \equiv 0$ if $t < 0$). An important criterion for the understanding of instability entails the study of the flow behavior in the `long term' (i. e., $t \rightarrow \infty$) where an analytical solution of equation~\eqref{eqn:DRP} is possible. In the asymptotic limit of long time, the integration of equation~\eqref{eqn:Green} is analytically accomplished by using the method of stationary phase~\citep{Ablowitz2003}, leading to the following expression for the Green's function,
\beq
G(x, t) \sim -\frac{1}{\sqrt{2\pi}} \frac{e^{{\it i}[\pi/4 + \alpha_* x - \omega_* t]}}{\frac{\partial D}{\partial \omega}\left[ \frac{d^2\omega}{d \alpha^2} \right]^{1/2}},
\label{eqn:ApproxGreen}
\eeq
where $\alpha_*$ is the saddle point in the $\alpha$ plane (i.~e., the root of $\frac{\partial D}{\partial \alpha}=0$) and $\omega_*$ is the corresponding branch point in the $\omega$ plane satisfying the dispersion relation. From equation~\eqref{eqn:ApproxGreen}, we can surmise the condition for which the flow will be {\it absolutely unstable} (signifying the growth of disturbance in both upstream and downstream direction from the origin, otherwise known as the `resonance mode'~\citep{Lingwood1997}), i.~e.,
\beq
G(x ,t) \xrightarrow[]{t \rightarrow \infty} \infty, \qquad \text{along the ray} \,\, {\displaystyle \sfrac{x}{t}} = 0,
\eeq
versus when the flow will be {\it convectively unstable} (where disturbances are swept downstream from the source and given sufficient time these disturbances decay at any fixed position in space, also known as the `driven mode'), i.~e.,
\beq
G(x ,t) \xrightarrow[]{t \rightarrow \infty} 0, \qquad \text{along the ray} \,\, \sfrac{x}{t} = 0.
\eeq
Our analysis also reveals the presence of evanescent modes (or false modes) in the flow field, which is the non-propagating mode or the locally concentrated mode~\citep{Drazin1979}, described later in \S \ref{sec:stsa}.

\subsection{Objectives of the present study}\label{subsec:L7}
The detailed exploration of the existing literature serves as a clear motivation for the work reported here, which provides a comprehensive picture of the stability of the viscoelastic Hele-Shaw flow using the Oldroyd-B model. The present work significantly differs from existing studies in the sense that we analyse the linear stability of the Hele-Shaw flow of dilute polymer solutions through a combined temporal and spatiotemporal stability analysis (rather than only temporal stability analysis~\citep{Shokri2017}) and aim to address the following intriguing questions: What is the critical flow/polymer relaxation condition for the onset of instability? and more crucially, what is the linear spatiotemporal, time asymptotic response of the flow at the critical value of the material parameters, leading to the topological transition of the advancing interface?

The rest of the paper is organized as follows. In \S \ref{sec:math}, we delineate the model of the depth-averaged, planar flow in a square Hele-Shaw cell coupled with the Oldroyd-B constitutive relation for the extra elastic stress tensor, along with the interface conditions (\S \ref{subsec:GE}), the mean flow (\S \ref{subsec:MF}) and the governing linearized differential equations (\S \ref{subsec:LSA}), followed by a brief description of the numerical method employed (\S \ref{subsec:NM}). The model is validated in \ref{sec:MV}, with the previously published experimental literature which quantify the size of the relative finger width in the Stokes as well as the inertial regime for Newtonian flows. Neutral stability curves are presented in \S \ref{sec:NSC} show that, for sufficiently small $E$, there is a remarkable coincidence of these curves in a suitably rescaled $Re-\alpha$ plane; a further coincidence is obtained in the dual limit $E(1-\nu) \ll 1$ and $(1-\nu) \ll 1$. \S \ref{sec:tsa} showcases the results of the temporal stability analysis, including a discussion on the most unstable mode (\S \ref{subsec:tgr}) and the fluctuations in the corresponding eigenfunctions (\S \ref{subsec:EigenV}). The sensitivity of the boundaries of the regions of absolutely/convective instability on the strength of the inertial and the elastic forces as well as the role of finite boundary effects, are outlined in the spatiotemporal phase diagrams (\S \ref{sec:stsa}). Finally, a brief discussion on the implication of these results as well as the focus of our future direction is summarized in \S \ref{sec:conclusion}. \S \ref{appA} and \S \ref{appB} lists all the expressions utilized in the dispersion relation and the dispersion relation, respectively.

\section{Problem formulation and numerical method}\label{sec:math}
%
\subsection{Governing equations}\label{subsec:GE}
\bef
\centering
\includegraphics[width=0.99\linewidth, height=0.3\linewidth]{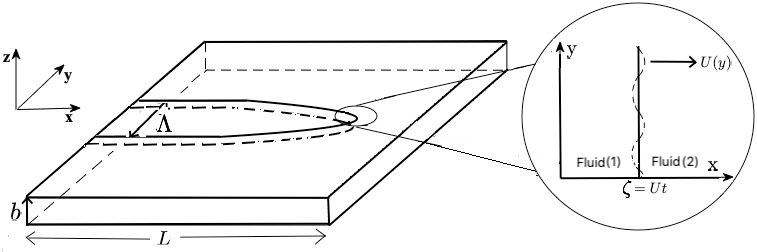}
\caption{Schematic diagram of the Hele-Shaw cell along with the coordinate system. $L$ and $b$ are the length and the cell gap-width, respectively. $\Lambda$ is the relative finger width. Inset: $U(y)$ is the velocity of the advancing interface, $\zeta$, separating the two fluids.}
\label{fig1}
\eef
We consider the linear stability of a steady fully developed flow of a low viscosity Newtonian fluid (referred as fluid 1, see figure~\ref{fig1}) displacing a high-viscosity viscoelastic fluid (i. e., fluid 2, see figure~\ref{fig1}) inside a square Hele-Shaw cell of length (or width), $L$, and a cell-gap, $b$. A rectilinear coordinate system is used with ${\bf x}, {\bf y}$ and ${\bf z}$ denoting the fluid flow direction, the transverse direction and the direction of the cell depth, respectively. Assuming that the length scale $b \ll L$ allows us to perform a normal mode perturbation expansion in the y-direction (refer \S \ref{subsec:LSA} for details). The following scales are used for non-dimensionalizing the governing equations: the length of the Hele-Shaw cell, $L$ for lengths, maximum base flow velocity, $\mathcal{U}_0$ for velocities, $\sfrac{L}{\mathcal{U}_0}$ for time and $\sfrac{(12 \eta_0 \mathcal{U}_0 L)}{b^2}$ for pressure and stresses, with the viscosity of the viscoelastic fluid (or fluid 2) being, $\eta_0 = \eta_s + \eta_p$ (where $\eta_s, \eta_p$ are the solvent and the polymeric contribution to the shear viscosity, respectively). The governing (non-dimensional) continuity and momentum equations for both fluids are given by,
\beq
\nabla \cdot {\bf v}_i = 0, \qquad Re_i^* \left[ \frac{\partial {\bf v}_i}{\partial t} + \frac{6}{5}({\bf v}_i \cdot \nabla){\bf v}_i \right] = -\nabla p_i - {\bf v}_i + \nabla \cdot \tau_i, \qquad \text{i=1, 2},
\label{eqn:Main}
\eeq
where the (non-dimensional) extra stress tensor, $\tau$, is given as follows,
\beq
\tau_i = \left\{
\begin{array}{ll}
{\bf D}_i, & \text{i=1} \\[2pt]
\nu {\bf D}_i + (1-\nu) {\bf A}, & \text{i=2},
\end{array} \right.
\label{eqn:TStress}
\eeq
where $Re_1^*=(\sfrac{\eta_0}{\eta_1})Re_2^*=(\sfrac{\eta_0}{\eta_1})Re^*$ ($\eta_1$ is the viscosity of fluid 1), ${\bf D}_i = \nabla {\bf v}_i + \nabla {\bf v}^T_i$ is the shear rate tensor and ${\bf A}$ is the elastic contribution to the stress tensor, satisfying the Oldroyd-B constitutive relation~\citep{Sircar2019},
\beq
\frac{\partial {\bf A}}{\partial t} + {\bf v}_2 \cdot \nabla {\bf A} -  \nabla{\bf v}_2^T\cdot {\bf A} - {\bf A} \cdot \nabla {\bf v}_2 = \frac{{\bf D} - {\bf A}}{We^*}.
\label{eqn:EStress}
\eeq
The variables ${\bf v}_i$, $p$ in equations~(\ref{eqn:Main}-\ref{eqn:EStress}) are the velocity and the pressure fields, respectively. The parameter, $\nu=\dfrac{\eta_s}{(\eta_s + \eta_p)}$, represents the viscous contribution to the total viscosity. $Re^*= \dfrac{1}{12}\frac{\kappa \rho \mathcal{U}_0 b}{\eta_0}$ and $We^*=\dfrac{\kappa \lambda \mathcal{U}_0}{b}$ are the depth-modified Reynolds and Weissenberg numbers, respectively ($\kappa = \sfrac{b}{L}$ being the inverse of the aspect ratio). In our subsequent discussion, we drop the superscript $(\cdot)^*$ on these two parameters.

Next, we outline the interface conditions. The kinematic boundary conditions at the interface are (a) the equality of the normal and the tangential velocities and the normal velocity equal to the interface velocity~\citep{Gallaire2017}, or 
\beq
\frac{\partial \zeta}{\partial t} + {\bf v} \cdot \nabla \zeta = {\bf v}_i\cdot{\bf n}
\label{eqn:Interface1}
\eeq
where $\zeta$ is the non-dimensional interface position and the vectors (${\bf n}, {\bf t}$) represent unit normal and unit tangential vectors on the interface, respectively. Since the interfacial stress component in the shallow direction is averaged out, only the normal and the in-plane tangential stresses are matched~\citep{Ro1995}. Hence, the non-dimensional dynamic boundary conditions are continuous tangential stress,
\beq
\left(\sfrac{\eta_1}{\eta_0}\right)({\bf t} \cdot \tau_1 \cdot {\bf n}) = {\bf t} \cdot \tau_2 \cdot {\bf n}, 
\label{eqn:Interface2}
\eeq
and jump in the normal stress according to the Laplace's law,
\beq
\left(\sfrac{\eta_1}{\eta_0}\right)({\bf n} \cdot \tau_1 \cdot {\bf n}) - {\bf n} \cdot \tau_2 \cdot {\bf n} = B K,
\label{eqn:Interface3}
\eeq
where the parameter $1/B = \dfrac{12 Ca}{\kappa^2}$ (where $Ca = \dfrac{\mathcal{U}_0 (\eta_0-\eta_1)}{\gamma}$) is the control parameter for the fingering problem and $K=\dfrac{\partial^2 \zeta / \partial y^2}{(1 + (\partial \zeta / \partial y)^2)^{3/2}}$ is the curvature of the interface~\citep{Lindner2002}. The factor $\left(\sfrac{\eta_1}{\eta_0}\right)$ appears in equations~(\ref{eqn:Interface2}, \ref{eqn:Interface3}) due to the choice of the dimensional scale for stresses.

Note that the second term on the right-hand side of equation~\eqref{eqn:Main}, ${\bf v}_i$, appears via averaging of the viscous stress in the direction of the plate spacing, i.~e., z-direction (Darcy's approximation), while the factor `$\sfrac{6}{5}$' on the second term on the left-hand side is obtained by averaging inertia in the third dimension~\citep{Dias2011}. We also note that the Oldroyd-B model describes the elastic stress in dilute polymer solutions in which the polymer chains are viewed as non-interacting Hookean dumbbells~\citep{Bird1987Vol1}. Consistent with the aforementioned microscopic picture, this constitutive equation assumes the relaxation time to be independent of both the shear rate and the polymer concentration. As the model predicts a shear-rate-independent viscosity, the elastic effects in this model arise from an effective tension along the streamlines (of the flow-aligned dumbbells), thereby revealing a shear-rate-independent first normal stress difference in viscometric flows. This model has been used extensively in earlier investigations of inertialess elastic instabilities in flows with curved streamlines~\citep{Grillet1999} as well as in the experimental realization of the effects of elasticity in Boger fluids~\citep{Boger1978}. Although shear thinning can play an important role especially in flow through microtubes~\citep{Chandra2020}, the Oldroyd-B model does have the necessary ingredients to qualitatively predict the instabilities observed in experiments related to viscoelastic fingering~\citep{Lindner2002}.

\subsection{Mean flow}\label{subsec:MF}
The non-dimensional mean flow velocity is the classical plane Poiseuille profile, given by  
\beq
{\bf V} = [U(y) \quad 0]^T,
\label{eqn:UMean}
\eeq
where $U(y) = 1 - (y/L)^2$. In our subsequent discussion, the mean flow variables will be denoted with capital letters. The Oldroyd-B base state stress tensor (denoted by the subscript `0') governed by equation~\eqref{eqn:EStress}, is given by,
\beq
\setlength{\arraycolsep}{3pt}
\renewcommand{\arraystretch}{1.3}
{\bf A}_0 = 
\left[
\begin{array}{c    c}
2 We U^2_y     &     U_y \\
U_y & 0 \\
\end{array} \right],
\label{eqn:EStressMean}
\eeq
where the subscript $(\cdot)_y$ denotes the derivative with respect to the spatial coordinate, $y$. The non-dimensional mean pressure of the respective fluids, satisfying equation~\eqref{eqn:Main}, is given by,
\beq
P_{0_i} = \left\{
\begin{array}{ll}
(U_{yy} - U)(x - Ut), & \text{i=1} \\[2pt]
\left[ \nu U_{yy} - U + (1-\nu) \dfrac{\partial A^{21}_0}{\partial y} \right](x - Ut), & \text{i=2}.
\end{array} \right.
\label{eqn:PMean}
\eeq

\subsection{Linearization via normal mode expansion}\label{subsec:LSA}
Due to the presence of the viscoelastic version of the Squire's theorem for plane parallel Oldroyd-B fluids~\citep{Bistagnino2007}, we restrict our stability analysis to the case when the disturbances are two-dimensional. Assuming an independent fate of each wavenumber, $\alpha$ (whose real part is chosen to be positive) and frequency, $\omega$, it is natural to consider disturbances in the form of a normal mode expansion, such that the total velocity, pressure and stress are expressed in terms of their mean values and perturbations (denoted by $\mathring{(\cdot)}$), as follows,
\bseq \label{eqn:NormalMode}
\begin{align}
&p_i = P_0 + \epsilon \mathring{p_i}, \label{eqn:pI} \\
&\zeta = \zeta_0 + \epsilon \mathring{\zeta}, \label{eqn:zeta} \\
&{\bf A} = {\bf A}_0 + \epsilon \mathring{{\bf a}}, \label{eqn:A} \\
&u_i = U(y) + \epsilon \mathring{u_i}, \label{eqn:uI} \\
&v_i = \epsilon \mathring{v_i}, \label{eqn:vI}
\end{align}
\eseq
where $\epsilon \ll 1$ and $\text{i = 1, 2}$. First, under the assumption that the pressure perturbations subside in the far-field (i.~e., $\mathring{p_1} \xrightarrow{x\rightarrow-L} 0$ and $\mathring{p_2} \xrightarrow{x\rightarrow L} 0$, where $L \gg b$), these disturbances are represented as follows,
\beq
(\mathring{p_1}, \mathring{p_2}) = (R_0 e^{\alpha x}, R_1 e^{-\alpha x}) e^{\{ i (\alpha y - \omega t) \}}. 
\eeq
A major assumption of our analysis is that the evolution of the interface occurs on a slow timescale compared with the timescale of the perturbation (or the so called `quasi stationary' approximation). Since the interface moves with the mean velocity, the mean position of the interface is given by $\zeta_0 = U(y) t$. The perturbations on the interface as well as on the elastic stresses are given by
\beq
\setlength{\arraycolsep}{3pt}
\renewcommand{\arraystretch}{1.3}
\mathring{\zeta} = R_2 e^{\{ i (\alpha y - \omega t) \}} \quad \text{and} 
\quad \mathring{{\bf a}} = 
\left[
\begin{array}{c    c}
R_3   & R_4 \\
R_5 & R_6 \\
\end{array} \right] e^{\{ i (\alpha y - \omega t) \}}.
\eeq
Finally, the velocity perturbations are given by,
\beq
(\mathring{u_i}, \mathring{v_i}) = (g_i, f_i)e^{\{ i (\alpha y - \omega t) \}}, 
\label{eqn:uvPerturbation}
\eeq
The functions $(g_i, f_i)$ are found by substituting the expressions~(\ref{eqn:uI}, \ref{eqn:vI}) in equations~(\ref{eqn:Main}, \ref{eqn:TStress}) leading to the following set of differential equations governing the $\mathcal{O}(\epsilon)$ perturbations,
\bseq \label{eqn:f_and_gODE}
\begin{align}
& g_{1_{xx}} - \left[ \left(\dfrac{6}{5} \dfrac{\eta_0}{\eta_1}\right)Re U\right] g_{1_{x}} + \left[ \left(\dfrac{\eta_0}{\eta_1}\right) i\omega Re - 1 - \alpha^2\right]g_1 = \alpha R_0 e^{\alpha x} + \left(\dfrac{6}{5} \dfrac{\eta_0}{\eta_1}\right)Re U_y f_1, \\
& \nu g_{2_{xx}} - \left[ \dfrac{6}{5} Re U\right] g_{2_{x}} + \left[ i\omega Re - 1 - \nu \alpha^2\right]g_2 = -\alpha R_1 e^{-\alpha x} -(1-\nu)i\alpha R_5 + \left(\dfrac{6}{5}\right) Re U_y f_2, \\
& f_{1_{xx}} - \left[ \left(\dfrac{6}{5} \dfrac{\eta_0}{\eta_1}\right)Re U\right] f_{1_{x}} + \left[ \left(\dfrac{\eta_0}{\eta_1}\right) i\omega Re - 1 - \alpha^2\right]f_1 = i \alpha R_0 e^{\alpha x}, \\
& \nu f_{2_{xx}} - \left[ \dfrac{6}{5} Re U\right] f_{2_{x}} + \left[ i\omega Re - 1 - \nu \alpha^2\right] f_2 = i \alpha R_1 e^{-\alpha x} -(1-\nu)i\alpha R_6,
\end{align}
\eseq
where the subscript $(\cdot)_x$ denotes the derivative with respect to the spatial coordinate, $x$. Equation~\eqref{eqn:f_and_gODE} is solved at the interface, $\zeta = \zeta_0$, subject to the far-field decaying disturbance conditions (i.e., $(g_i, f_i) \rightarrow 0$ as $(x,y) \rightarrow \pm L$, where $L \gg b$) and the kinematic boundary conditions (velocities are equal at the interface). The latter condition leads to the conclusion: $g_1 = g_2 = g$ and $f_1 = f_2 = f$ at the interface. The analytical expressions for $(g, f)$ are listed in \S \ref{appA} (equation~\eqref{appA:f_and_g}). Next, substituting the solution form~\eqref{eqn:NormalMode} in equations~(\ref{eqn:Main}-\ref{eqn:EStress}) and the interface conditions~(\ref{eqn:Interface1}-\ref{eqn:Interface3}) and assuming that the viscosity of the displacing fluid (or fluid 1) is negligible in comparison to the displaced fluid (or fluid 2), i.~e., $\sfrac{\eta_1}{\eta_0} \ll 1$, we choose the (shape dependent) normal and tangential vectors at the interface,
\beq
{\bf n} = [1 \,\, -\sfrac{\partial \zeta}{\partial y}], \qquad {\bf t} = [\sfrac{\partial \zeta}{\partial y} \,\, 1],
\label{eqn:UnitVectors}
\eeq
and retain the $\mathcal{O}(\epsilon)$ terms to arrive at the linearized equation governing the tangential stress at the interface,
\begin{align}
&2\nu U_y t \left[ g_{r_x} + i g_{i_x} + f_{i_x} - i f_{r_x} + \alpha (f_i - i f_r) \right] + \left[ (1-\nu)i\alpha (A^{11}_0 + A^{22}_0) -2 i\alpha U_y t \right. \nonumber \\
&\left. \left\{ \nu U_y + (1-\nu) A^{12}_0 \right\}\right] R_2 + t U_y (1-\nu)(R_3 - R_6) + \nu \left[ -g_{i_x} + ig_{r_x} + f_{r_x} + i f_{i_x} \right] \nonumber \\
& (1-\nu) R_5 - U^2_y t^2 \left[ \nu\left( -g_{i_x} + ig_{r_x} + f_{r_x} + i f_{i_x} \right) + (1-\nu) R_4\right]=0,
\label{eqn:linearized1}
\end{align}
the linearized equation describing the normal stress at the interface,
\begin{align}
&2\nu \left[g_{r_x} + i g_{i_x}\right] + 2\nu U^2_y t^2 \left[-f_{i_x} + f_{r_x} \right] -2\nu U_y t \left[ -g_{i_x} + i g_{r_x} + f_{r_x} + i f_{i_x} \right] - e^{-\alpha U t} \nonumber \\
& - R_2 \left[ \nu U_yy \!\!-\!\! U \!\!+\!\! (1\!\!-\!\!\nu)A^{21}_{0_y} \!\!+\!\!  i\alpha \left[ 2\nu U_y + (1\!\!-\!\!\nu)(A^{12}_0\!\!+\!\!A^{21}_0) \right] \!\!+\!\! B\alpha^2 \!\!-\!\! 2t(1\!\!-\!\!\nu)i\alpha U_yA^{22}_0 \right] \nonumber \\
& + (1-\nu) R_3 + t (1-\nu) U_y(R_4 + R_5) + t^2 (U_y)^2 (1-\nu) R_6 = 0,
\label{eqn:linearized2}
\end{align}
the interface condition,
\beq
i \omega R_2 - 2t U_y (f_r + i f_i) = 0,
\label{eqn:linearized3}
\eeq
the linearized equation governing the elastic stress component $A^{11}$,
\begin{align}
&i \omega R_3 \!\!-\!\! A^{11}_{0_y} \left[ f_r \!\!+\!\! i f_i \right] + 2A^{11}_{0}\left[ g_{r_x} \!\!+\!\! i g_{i_x} \right] + (A_0^{12} \!\!+\!\! A_0^{21})\left[ -g_{i_x} \!\!+\!\! i g_{r_x} \!\!+\!\! i\alpha (g_{r_x} + i g_{i_x}) \right] \nonumber \\
& + U_y (R_3 + R_5) + (\dfrac{1}{We})(g_{r_x} + i g_{i_x}) - \dfrac{R_3}{We} = 0,
\label{eqn:linearized4}
\end{align}
for the component $A^{12}$,
\begin{align}
&i \omega R_4 \!\!-\!\! A^{12}_{0_y} \left[ f_r \!\!+\!\! i f_i \right] \!\!+\!\! \left(A^{22}_{0}\!\!+\!\!\dfrac{1}{We}\right) \left[ -g_{i_x} \!\!+\!\! i g_{r_x} \!\!+\!\! i\alpha (g_{r_x} \!\!+\!\! i g_{i_x}) \right] \!\!+\!\! R_6 U_y \!\!+\!\! A^{11}_0 (f_{r_x} \!\!+\!\! i f_{i_x}) \nonumber \\
& + \left(\dfrac{1}{We}\right)(f_{r_x} + i f_{i_x}) - \dfrac{R_4}{We} = 0,
\label{eqn:linearized5}
\end{align}
and for the component $A^{22}$,
\beq
i \omega R_6 \!\!-\!\! A^{22}_{0_y} \left[ f_r \!\!+\!\! i f_i \right] \!\!+\!\! 2\!\left(A^{22}_{0}\!\!+\!\!\dfrac{1}{We}\right) \left[ -f_{i_x} \!\!+\!\! i f_{r_x} \!\!+\!\! i\alpha (f_{r_x} \!\!+\!\! i f_{i_x}) \right] \!\!+\!\! (A^{12}_0\!\!+\!\!A^{21}_0) (f_{r_x} \!\!+\!\! i f_{i_x}) \!\!-\!\! \dfrac{R_6}{We} = 0.
\label{eqn:linearized6}
\eeq
Equations~(\ref{eqn:linearized1}-\ref{eqn:linearized6}) may be written in a matrix-vector format as follows,
\beq
\setlength{\arraycolsep}{4pt}
\renewcommand{\arraystretch}{1.3}
\left[
\begin{array}{cccccc}
A + {\it i}P & {\it i}C & D & E & (1-\nu) & -D \\
H + {\it i}J & N + {\it i}M & -(1-\nu) & D & D & E\\
R + {\it i}S & -{\it i}\omega & 0 & 0 & 0 & 0\\
Q + {\it i}W & 0 & X - {\it i}\omega & 0 & -U_y & 0\\
Y + {\it i}Z & 0 & 0 & \sfrac{1}{We} - {\it i}\omega & 0 & -U_y\\
\delta + {\it i}\tau & 0 & 0 & 0 & 0 & \sfrac{1}{We} - {\it i}\omega\\
\end{array}  \right] 
\left[
\begin{array}{c}
R_1 \\
R_2 \\
R_3 \\
R_4 \\
R_5 \\
R_6 \\
\end{array} \right]
= 0,
\label{eqn:DRP_matrix}
\eeq
where the expressions in the coefficient matrix are listed in \S \ref{appA} (equation~\eqref{appA:DRP_exp}). A nontrivial solution for the system~\eqref{eqn:DRP_matrix}, imposes a zero determinant condition on the coefficient matrix which leads to the dispersion relation: $D(\alpha, \omega) = 0$ (equation~\eqref{appB:DRP}, \S \ref{appB}).

\subsection{Numerical Method}\label{subsec:NM}
The zeros of the dispersion relation ($D(\alpha, \omega)=0$, equation~\eqref{appB:DRP}) were explored within the complex $\alpha-\omega$ plane inside the region $-3.0 \le \omega_r \le 5.0, 0.0 \le \omega_i \le 10.0, \alpha_r \le 5.0 $ and $|\alpha_i| \le 3.0$. Previous results indicate that the influence of viscoelasticity is fully captured by the modified elasticity number, $E = \frac{{(1-\nu)} We}{Re}$, a parameter representing the ratio of the fluid relaxation time to the characteristic time for vorticity diffusion~\citep{Ray2014}. We highlight our instability results versus this parameter. The parameter, $t$, (refer expressions~(\ref{appA:f_and_g}, \ref{appA:f_and_g_derivatives})) representing the time of evolution of the slow manifold is fixed at $t=1.0$. Changes in this parameter within the range from $0.1 \le t \le 10.0$, induces a change of less than $1\%$ in the eigenvalues (or the roots of the dispersion relation, equation~\eqref{appB:DRP}) while leaving the eigenvectors ($f_r, f_i$, equation~\eqref{appA:f_and_g}) qualitatively unaltered. The other parameters fixed in this study are the size of the cell, $L=1.0$ and the control parameter, $\sfrac{1}{B}=1000$ (equation~\eqref{eqn:Interface3}) and the velocity scale, $\mathcal{U}_0 = 1.0$.

For a real wavenumber $\alpha$, the locus of neutrally stable points (refer \S \ref{sec:NSC}) are found after selecting $\omega_i = 0$ in the dispersion relation and solving for the unknowns ($\omega_r, \alpha$), at fixed $Re, E$ using a bivariate Newton-Raphson algorithm~\citep{Bansal2021}. Next, the procedure of finding the most unstable mode (which is the largest positive imaginary component of any root of the dispersion relation or the temporal growth rate, $\omega_{\text{Temp}}$, refer \S \ref{sec:tsa}), consists of detecting the admissible saddle points ($\omega \in \mathbb{C}, \alpha \in \mathbb{R}$) satisfying the equations~\citep{Huerre1990},
\bseq \label{eqn:wTemp}
\begin{align}
&D(\alpha, \omega) = 0, \label{eqn:wTemp1} \\
&\dfrac{\partial \omega_i}{\partial \alpha} = \dfrac{\sfrac{\partial D}{\partial \alpha}}{\sfrac{\partial D}{\partial \omega_i}} = 0, \label{eqn:wTemp2}
\end{align}
\eseq
and then (among all the possible roots of equation~\eqref{eqn:wTemp}) identifying those roots with the largest positive imaginary component of the frequency. Equation~\eqref{eqn:wTemp} (refer equations~(\ref{appB:DRP_Real}, \ref{appB:DRP_Imag}, \ref{appB:DRPi_alpha}), in \S \ref{appB} for the detailed expressions) is again solved using a multivariate Newton-Raphson algorithm.

Next, in the spatiotemporal analysis, eigenpairs with complex wavenumbers and frequencies are permitted in the solution of equation~\eqref{eqn:wTemp}. The necessary (but not sufficient) condition for the presence of absolute instability is the vanishing characteristic of the group velocity of the flow, ${\it v}_g$, at the saddle point in the $\alpha$-plane or the branch point in the $\omega$-plane (${\it v}_g = \frac{\partial \omega}{\partial \alpha} = \sfrac{(\frac{\partial D}{\partial \alpha})}{(\frac{\partial D}{\partial \omega})} = 0$, such that $\omega=D(\alpha)$ satisfies the dispersion relation). But the group velocity is zero at every saddle point, in particular where the two $\alpha$-branches meet, independent of whether the branches originate from the same half of the $\alpha$-plane (i.~e., when evanescent modes are detected) or not. To overcome this inadequacy, Briggs~\citep{Briggs1964} devised the idea of analytic continuation in which the Laplace contour ($L$, described in equation~\eqref{eqn:Green}), is deformed towards the $\omega_r$ axis of the complex $\omega$-plane, with the simultaneous adjustment of the Fourier contour in the $\alpha$-plane to maintain the separation of the $\alpha$-branches; those which originate from the top half (the upstream modes with $\alpha_i > 0$) from those which originate from the bottom half of the $\alpha$-plane (or the downstream modes). The deformation of the Fourier contour (while preserving causality) is inhibited, however, when the paths of the two $\alpha$-branches originating from the opposite halves of the $\alpha$-plane intersect each other, leading to the appearance of saddle points which are the {\it pinch point}, $\alpha^{\text pinch}$. The concurrent branch point appearance in the $\omega$-plane is the {\it cusp point}, $\omega^{\text cusp}$ (i. e., $D(\alpha^{\text pinch}, \omega^{\text cusp})\!=\!\frac{\partial D(\alpha^{\text pinch}, \omega^{\text cusp})}{\partial \alpha}\!=\!0$ but $\frac{\partial^2 D(\alpha^{\text pinch}, \omega^{\text cusp})}{\partial \alpha^2}\!\ne \!0$). \citet{Kupfer1987} employed a local mapping procedure to conceptualize the stability characteristics of this branch point. Near a `reasonably close' neighborhood of the pinch point, a local Taylor series expansion yields a dispersion relation which has a second-order algebraic form in the $\omega$-plane (and which is a first-order saddle point in the $\alpha$-plane), i.~e., $(\omega - \omega^{\text cusp}) \sim (\alpha - \alpha^{\text pinch})^2$. This period-doubling characteristic of the map causes the $\alpha_i$-contours to `rotate' around $\omega^{\text cusp}$, forming a cusp. In the $\omega$-plane, we draw a ray parallel to the $\omega_i$-axis from the cusp point such that it intersects the image of the Fourier contour (or $\alpha_i = 0$ curve) and count the number of intersections (consequently, count the number of times both $\alpha$-branches cross the $\alpha_r$-axis before forming a pinch point in the $\alpha$-plane. If the ray drawn from the cusp point intersects the image of the Fourier contour in the $\omega$-plane (or if either one or both the $\alpha$-branches cross the $\alpha_r$-axis) even number of times, then the flow dynamics correspond to an evanescent mode. Otherwise, in the case of odd intersections the observed cusp point is genuine, leading to either absolutely unstable system (in the upper half of the $\omega$-plane) or convectively unstable system (in the lower half of the $\omega$-plane); provided the system is temporally unstable.

Under the assumption that dispersion relation is a complex analytic function satisfying Cauchy-Riemann relations, other equivalent expressions are chosen preferentially to numerically evaluate the derivatives listed in \S \ref{appB}. The numerical continuation of the points on the neutral stability curves was achieved within the range $Re \in [0.1, 250]$ and at fixed values of $E$ and two specific values of $\nu$: $\nu=0.3$ (or the elastic stress dominated case) and $\nu=0.9$ (or the viscous stress dominated case, refer \S \ref{sec:NSC}), while the continuation of the temporal growth rate curves (\S \ref{sec:tsa}) and absolute growth rate curves (\S \ref{sec:stsa}) were realized within the range $Re \in [0.1, 40]$, $E \in [10^{-3}, 0.05]$ and $\nu \in [0.01, 0.99]$ using a discrete step-size of $\triangle Re = 10^{-3}$, $\triangle E = 10^{-3}$ and $\triangle \nu = 10^{-3}$, respectively. While the (non-dimensional) physical domain spans within the range, $(x, y) \in [-1, 1]$, the neutral stability, the temporal growth rate and the absolute growth rate of the perturbations at the interface, $x = U t$, are reported at four discrete, spatial locations in the transverse direction: $y=\pm 0.99, 0.5, 0.0$, the former (latter) two values chosen to quantitatively probe the near-wall (far away from the wall) effects.

\section{Model validation: comparison with DNS and experiments for Newtonian flows}\label{sec:MV}
First, the numerical method outlined in \S \ref{subsec:NM} is validated by reproducing the results for the relative finger width, normalized with respect to the cell width (i.~e., $\Lambda$, refer figure~\ref{fig1}) for Newtonian Hele-Shaw flows within (a) the Stokes regime and using Darcy's approximation, investigated by~\citet{Saffman1958, McLean1981} and later by~\citet{Tabeling1985} and (b) the inertial regime probed by~\citet{Lindner2006}. First note that the generalized dispersion relation (valid either in the Stokes or the inertial limit and obtained after substituting $\nu=1$ in equation~\eqref{appB:DRP}) for the rectangular Hele-Shaw flows when the driven fluid is Newtonian, is given by 
\beq
\omega = i \dfrac{\left[U_yy - U + 2i\alpha U_y + B\alpha^2\right]\left[ 2 t U_y (f_r + i f_i) \right]}{e^{-\alpha U t} -2(g_{r_x}+i g_{i_x}) -2t^2 U^2_y(-f_{i_x}+i f_{r_x}) + 2tU_y(-g_{i_x} + i g_{r_x} + f_{r_x} + i f_{i_x}) }.
\label{eqn:DRP_Newtonian}
\eeq
Next, based on the physical explanation by~\citet{Chuoke1959} we conjecture that the relative finger width of the advancing interface, $\Lambda$, is the wavelength of the instability found at the maximum temporal growth rate, or
\beq
\Lambda = \dfrac{2\pi}{\alpha}(\omega = \omega_{\text{Temp}}).
\eeq
Thus, $\Lambda$ is found by solving for the dispersion relation~\eqref{eqn:DRP_Newtonian} at the most unstable mode, satisfying the constraints~\eqref{eqn:wTemp}. Figure~\ref{fig2}a presents a comparison of our temporal stability analysis (equation~\eqref{eqn:wTemp} with the experimental results by~\citet{Saffman1958} (`$\color[rgb]{0.5,1,0.83} \Diamond$') and numerical simulation results by~\citet{McLean1981} (`$\color[rgb]{0.8,0.33,0} \Box$') inside a Hele-Shaw cell with air pushing oil (of viscosity $\eta = 4.5$ Pa s, density $\rho = 950$ kg m$^{-3}$ and surface tension $20$ mN m$^{-1}$) inside a cell with aspect ratio, $\dfrac{1}{\kappa} = 31.8$. The other two data sets highlighted in figure~\ref{fig2}a, are from the experimental outcome of~\citet{Tabeling1985} for the case of air pushing a V10 silicone oil (of viscosity $\eta = 0.093$ Pa s, density $\rho = 930$ kg m$^{-3}$ and surface tension $20.1$ mN m$^{-1}$, denoted with `$\color[rgb]{0,0,1} \bullet$') and a V500 silicone oil (of viscosity $\eta = 4.87$ Pa s, density $\rho = 973$ kg m$^{-3}$ and surface tension $21.1$ mN m$^{-1}$, denoted with `$\color[rgb]{0.91,0.45,0.32} \times$') for a cell with aspect ratio, $66.5$. In all the data sets, the cell gap was fixed at $b=7.95 \times 10^{-4}$m and the velocity of the advancing interface was secured at a constant value, $\mathcal{U}_0 = 10^{-3}$m s$^{-1}$, implying that the maximum Reynolds number in these studies was found at $Re = 2.03 \times 10^{-5}$, establishing the nature of these flows well within the Stokes regime.   

The parameter $\sfrac{1}{B}$ represents the ratio of the viscous forces over the capillary forces, and it is predicted to be an exclusive parameter controlling the finger width via a `master curve' in the Stokes limit~\citep{Saffman1958,Lindner2002,Lindner2006}. However, the presence of a thin wetting film left on the plates creates a discrepancy in the stacking of the data sets at different aspect ratios on a universal curve (e.g., refer figure 1 in~\citep{Tabeling1985}). A superposition of these in vitro studies on a universal curve (at small values of $Ca$) is achieved once the surface tension is corrected after accounting for the pressure drop across the interface, as follows,
\beq
\gamma^* = \gamma \left[ \dfrac{\pi}{4} + 1.7 \left(\dfrac{\Lambda}{\kappa}\right) \left(\dfrac{\eta \mathcal{U}_0}{\gamma}\right)^{2/3} \right],
\eeq
ensuing in a renormalized parameter, $\dfrac{1}{B^*} = \dfrac{12\eta}{\kappa^2} \left(\dfrac{\mathcal{U}_0}{\gamma^*}\right)$, thereby rescaling the data onto a single curve, and which is found to be in excellent agreement with our temporal stability analysis ($\nu=1$, equation~\eqref{eqn:wTemp}, denoted with $\rule{15pt}{2pt}$).

Figure~\ref{fig2}b proffers a comparison of our temporal stability analysis (highlighted with solid line, $\rule{15pt}{2pt}$) with the in vitro studies by~\citet{Lindner2006} for the case of air pushing silicone oils, V02 ($\eta = 0.028$ Pa s, $\color[rgb]{0,0,1} \bullet$), V05 ($\eta = 0.05$ Pa s, $\color[rgb]{0.91,0.45,0.32} \times$), V10 ($\eta = 0.1$ Pa s, $\color[rgb]{1,0.65,0} \Box$) and V20 ($\eta = 0.2$ Pa s, $\color[rgb]{0.5,0,0.5} +$) with other material properties of these oils, namely, $\gamma =$ $19.5$ mN m$^{-1}$ and $\rho = 950$ kg m$^{-3}$ held fixed. The Hele-Shaw cell geometry is held constant at $b=7.5 \times 10^{-4}$m and $L=4 \times 10^{-2}$m, resulting in an aspect ratio of $53.3$. The velocity of the advancing interface is presumed to vary with the imposed pressure gradient with a maximum at $\mathcal{U}_0 = 0.15$m s$^{-1}$, which results in a maximum value of the control parameter, $\sfrac{1}{B} \approx 353$ (refer figure 4a in~\citep{Lindner2006}).

Figure~\ref{fig2}b shows a characteristic $Re$-dependent dual regime dictating the dynamics of the finger width. Within the range, $Re \le 0.05$, the relevant forces are the viscous forces (resulting in finger narrowing) and the capillary forces (resulting in finger widening) with the viscous forces becoming dominant at higher velocity leading to a decrease in the finger width. In the range, $Re > 0.05$, the main forces are the viscous forces and inertia. However, since inertia tends to slow down the finger evolution at a given flow rate and the impact of inertia increases at higher velocities, wider finger are observed at higher $Re$. A precise match of these data sets with our model (see \S \ref{subsec:NM}) corroborate the reasoning outlined above.
\bef
\centering
\includegraphics[width=0.495\linewidth, height=0.35\linewidth]{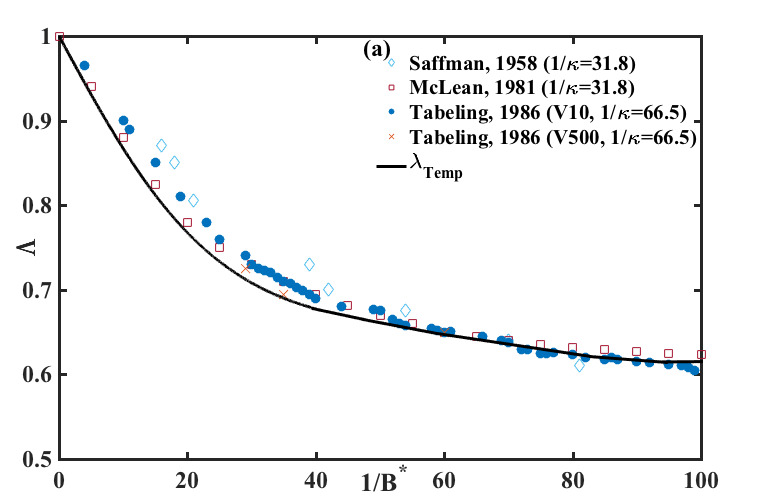}
\includegraphics[width=0.495\linewidth, height=0.35\linewidth]{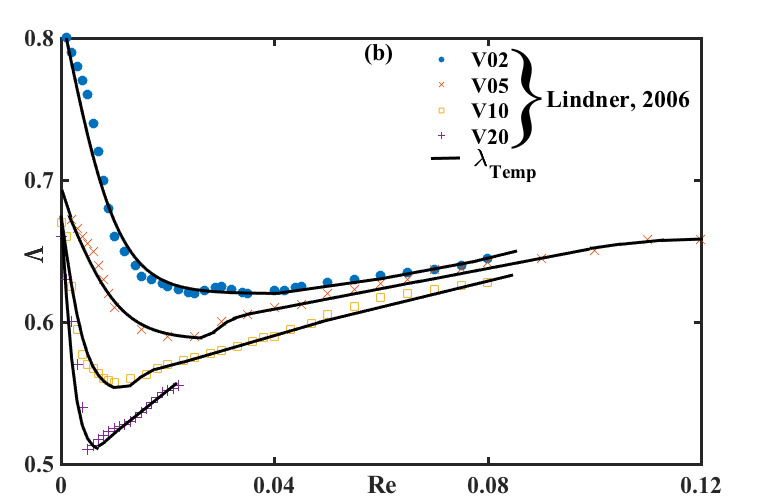}
\caption{Relative finger width, $\Lambda$ (equation~({\bf equation number})) versus (a) rescaled control parameter, $\frac{1}{B^*} = \frac{12\eta}{\kappa^2} \frac{\mathcal{U}_0}{\gamma^*}$, compared with the experimental outcome of~\citet{Saffman1958} (shown with $\color[rgb]{0.5,1,0.83} \Diamond$); \citet{McLean1981} (shown with $\color[rgb]{0.8,0.33,0} \Box$); \citet{Tabeling1985} for V10 oil (shown with $\color[rgb]{0,0,1} \bullet$); \citet{Tabeling1985} for V500 oil (shown with $\color[rgb]{0.91,0.45,0.32} \times$) and (b) $Re = \frac{1}{12}\frac{\kappa \rho \mathcal{U}_0 b}{\eta}$, compared with the in vitro results of~\citet{Lindner2006} for the low viscosity silicone oils: V02 (shown with $\color[rgb]{0,0,1} \bullet$), V05 (shown with $\color[rgb]{0.91,0.45,0.32} \times$), V10 (shown with $\color[rgb]{1,0.65,0} \Box$) and V20 (shown with $\color[rgb]{0.5,0,0.5} +$).}
\label{fig2}
\eef

\section{Neutral stability curves}\label{sec:NSC}
Figures~\ref{fig3}(a) and \ref{fig3}(b) depict the neutral stability curves in the $Re-\alpha$ for fixed values of $\nu$ and $E$. The region enclosed by the curve and bounded below by the coordinate axes,   is temporally stable. Notice that for fixed values of $E$, the onset of instability (denoted by the region outside the one enclosed by the neutral curve) in the elastic stress dominated fluid ($\nu=0.3$ case, figure~\ref{fig3}(a)), occurs at larger values of $Re$ and for shorter wavelengths (or larger $\alpha$). However, there exists a critical Reynolds number ($Re_c$), or the largest Reynolds number beyond which all wavenumbers are temporally unstable. This feature, of the existence of $Re_c$, is reminiscent of a similar in silico study, for miscible, plane Oldroyd-B Hele-Shaw flows by~\citet{Shokri2017}, who concluded (contrarily) that elasticity induces stabilization. Similarly, the viscous stress dominated fluid ($\nu=0.9$ case, figure~\ref{fig3}(b)) indicates a narrow strip of temporally stable wavenumbers for all $Re \le Re_c$. In general, $Re_c$ decreases with increasing $E$. Further, the transition to instability transpires earlier in the elastic stress dominated case (i.~e., the values of $Re_c$ is lower at $\nu=0.3$ than the values for the corresponding curves at $\nu=0.9$). The last two observations suggest a mechanism of elastic destabilization, reported earlier in plane shear flows~\citep{Sircar2019}, and lately observed in plane, viscoelastic, immiscible, Hele-Shaw flows, which we elaborate further in \S \ref{sec:tsa}.
\bef
\centering
\includegraphics[width=0.495\linewidth, height=0.35\linewidth]{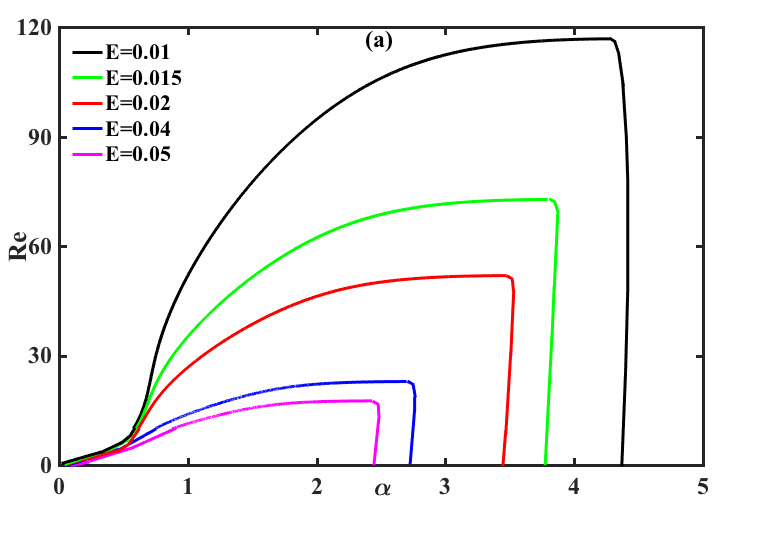}
\includegraphics[width=0.495\linewidth, height=0.35\linewidth]{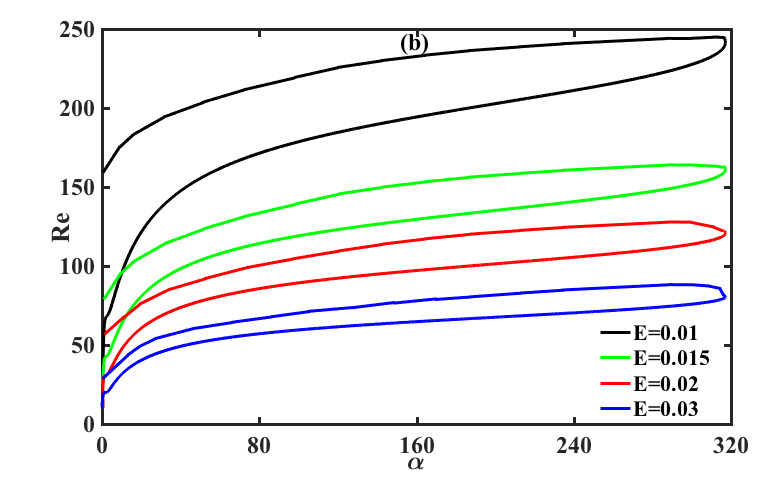}
\caption{Neutral stability curves in the $Re-\alpha$ plane versus elasticity number $E$ at (a) $\nu = 0.3$ and (b) $\nu = 0.9$.}
\label{fig3}
\eef

A qualitatively similar feature of the neutral curves at different values of $E$ in figure~\ref{fig3} is firmly suggestive of a `coincidence' of these curves upon suitable rescaling of both $Re$ and $\alpha$ with elasticity number $E$. Figure~\ref{fig4} highlights that such a coincidence is indeed possible for sufficiently small values of $E$, when $Re$ is rescaled as $Re E^{5/3}$ and $\alpha$ as $\alpha E^{2/3}$. These scaling laws are found to be valid for fixed $\nu$, and the shape of the coinciding curves does depend on $\nu$ (as evident from figures~\ref{fig4}a versus figures~\ref{fig4}b).
\bef
\centering
\includegraphics[width=0.495\linewidth, height=0.35\linewidth]{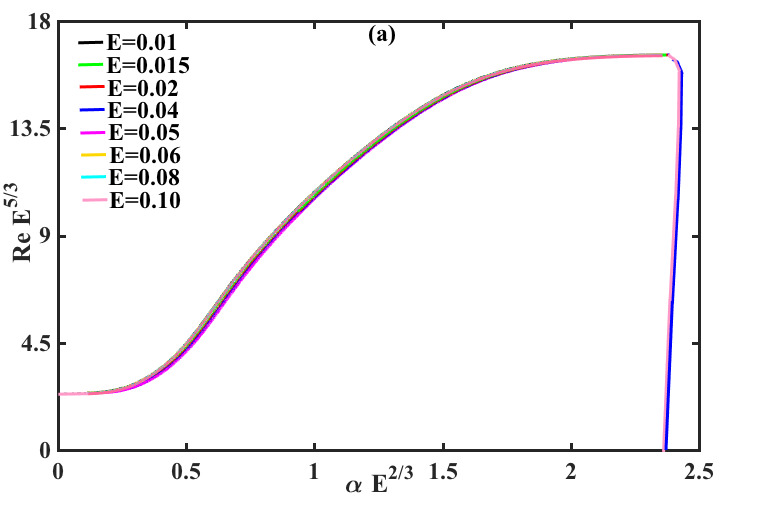}
\includegraphics[width=0.495\linewidth, height=0.35\linewidth]{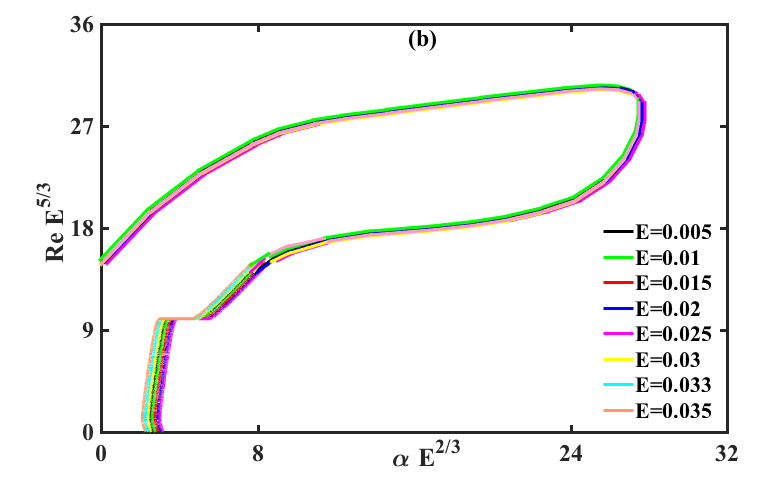}
\caption{Coincidence of neutral curves for various values of $E$ in the $Re-\alpha$ plane for two different values of $\nu$: (a) rescaled neutral curves for $\nu = 0.3$ and (a) rescaled neutral curves for $\nu = 0.9$.}
\label{fig4}
\eef

While coincidence obtained above is for a fixed $\nu$ and for $E \ll 1$, further coincidence is obtained in the dual limit $E(1-\nu) \ll 1$, $(1-\nu) \ll 1$, when the neutral curves are plotted in terms of $Re [(1-\nu)E]^{5/3}$ and $\alpha [(1-\nu)E]^{2/3}$, as shown in figure~\ref{fig5}, implying that the threshold values of $Re$ and $\alpha$ scale as $Re \propto [(1-\nu)E]^{-5/3}$ and $\alpha \propto [(1-\nu)E]^{-2/3}$, respectively, in this dual limit. The rescaled neutral curves in figure~\ref{fig5} begin to coincide onto a single curve for $\nu \ge 0.99$, the coincidence being near perfect for the lower branches, but less so for the upper ones. Thus, as far as the lower branches of the neutral curve is  concerned, the role of the solvent viscosity appears to be universal in the determination of the critical parameters: $Re_c$ and the wavenumber on the neutral curve at $Re=Re_c$ (or $\alpha_c$).
\bef
\centering
\includegraphics[width=0.65\linewidth, height=0.5\linewidth]{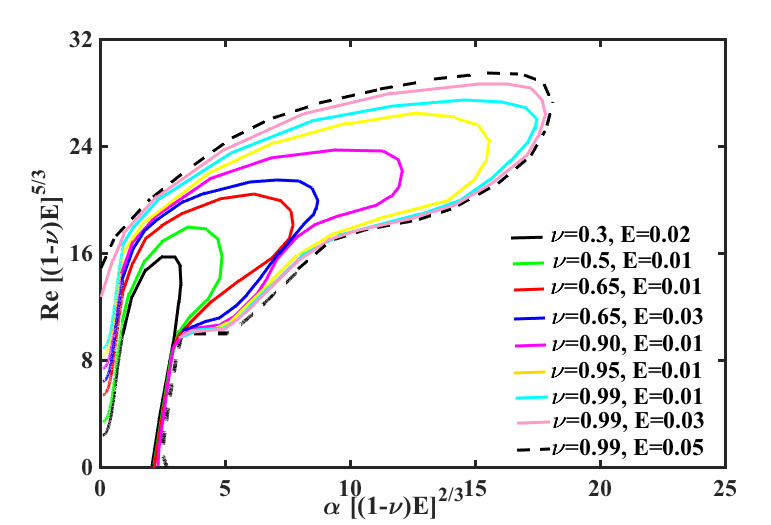}
\caption{Neutral curves for different values of $\nu$ and $E$ shown in terms of $Re\left[ (1-\nu) E\right]^{5/3}$ versus $\alpha \left[ (1-\nu) E\right]^{2/3}$: the rescaled curves coincide in the limit $\nu \rightarrow 1$.}
\label{fig5}
\eef

The variation of these critical parameters, $Re_c$ and $\alpha_c$ with $E(1-\nu)$ for different values of $\nu$ are shown in figures~\ref{fig6}a and~\ref{fig6}b, respectively. Irrespective of $\nu$, the critical parameters conform to the scaling laws for small values of $E(1-\nu)$; thus $Re \sim [(1-\nu)E]^{-5/3}$ and $\alpha \sim [(1-\nu)E]^{-2/3}$. For a fixed $\nu$, as $E(1-\nu)$ increases, $Re_c$ (and $\alpha_c$) decrease as per the scaling law outlined above until it reaches a minimum at a threshold value of $E(1-\nu)$. Beyond this threshold value, $Re_c$ (and $\alpha_c$) deviates from this scaling law and decreases rather sharply indicating the flow to be unstable for all wavenumbers beyond this threshold. However, the threshold $Re_c$ shifts to higher values of $E(1-\nu)$ as $\nu \rightarrow 1$ and further the value of $Re_c$ at the threshold also increases in the limit $\nu \rightarrow 1$, with the highest threshold $Re_c$ found being as large as 6 (albeit for $E \sim 50$). The last observation suggests that the viscoelastic Saffman-Taylor flows for strongly elastic dilute polymer solutions can become unstable for the entire wavenumber spectrum, at a critical value of Reynolds number which is much lower than that of their Newtonian counterpart.
\bef
\centering
\includegraphics[width=0.495\linewidth, height=0.35\linewidth]{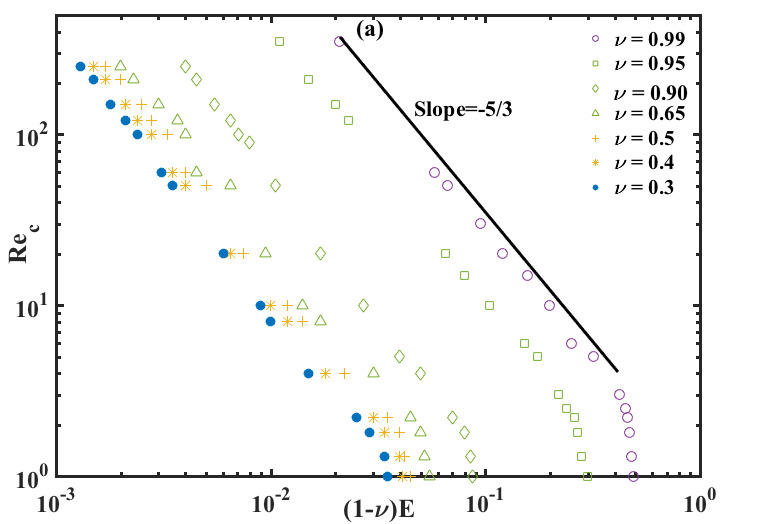}
\includegraphics[width=0.495\linewidth, height=0.35\linewidth]{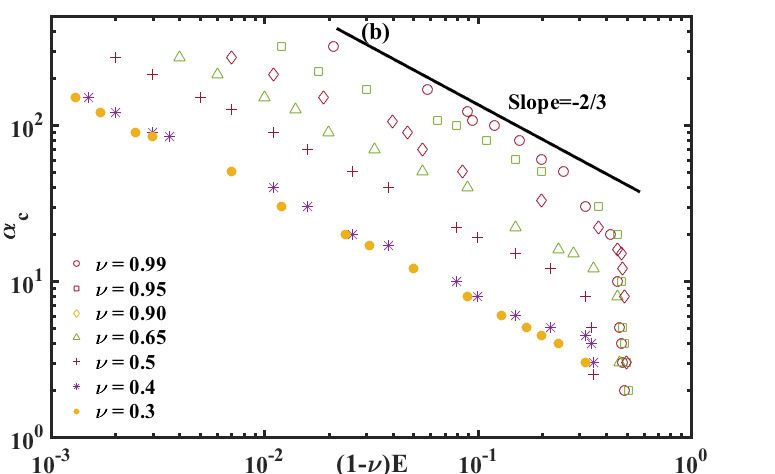}
\caption{Variation of the critical parameters with $(1-\nu)E$ for $\nu$ ranging from $0.3$ to $0.99$, where $Re_c$ and $\alpha_c$ follow the scaling laws $Re_c \sim [(1-\nu)E]^{-3/2}$ and $\alpha \sim [(1-\nu)E]^{-2/3}$, respectively, below a threshold value of $(1-\nu)E$. Plot of (a) $Re_c$ versus $(1-\nu)E$ and (b) $\alpha_c$ versus $(1-\nu)E$.}
\label{fig6}
\eef

\section{Temporal stability analysis}\label{sec:tsa}
A positive sign of the temporal growth rate indicates whether absolute instability is possible. Figure~\ref{fig7}, \ref{fig8} and \ref{fig9} presents the most unstable mode, $\omega_i^{\text{Temp}}$ (which are the eigenvalues of the saddle point problem~\eqref{eqn:wTemp} for purely real wavenumbers, $\alpha_r$) versus $Re$ (at fixed values of viscosity ratio, $\nu=0.3, 0.9$ and elasticity, $E=0.01$ (\protect\blackline), $E=0.015$ (\protect\blueline), $E=0.02$ (\protect\greenline), $E=0.03$ (\protect\cyanline), refer figure~\ref{fig7}), versus $E$ (at fixed values of viscosity ratio, $\nu=0.3, 0.9$ and Reynolds number, $Re=7.0$ (\protect\blackline), $Re=10.0$ (\protect\blueline), $Re=20.0$ (\protect\greenline), $Re=30.0$ (\protect\cyanline), refer figure~\ref{fig8}), versus $\nu$ (at fixed values of Reynolds number, $Re=10.0, 40.0$ and elasticity, $E=0.01$ (\protect\blackline), $E=0.015$ (\protect\blueline), $E=0.02$ (\protect\greenline), $E=0.03$ (\protect\cyanline), refer figure~\ref{fig9}), respectively. Figure~\ref{fig10}, \ref{fig11} and \ref{fig12} highlights the amplitude of the eigenvectors, $|f_r + i f_i|$ (equation~\eqref{appA:f_and_g}), corresponding to the unstable modes. The evolution of the temporal growth rates (and the corresponding eigenvectors) in Hele-Shaw flows in a bounded region is an outcome of an intricate tug-of-war between the inertial destabilization, elastic stabilization and the boundary effects, which we elaborate next.

\subsection{Temporal growth rate}\label{subsec:tgr}
In earlier studies on viscoelastic flows in unbounded domains, elasticity was found to have a destabilizing effect in the dilute regime~\citep{Sircar2019} but a stabilizing effect in the nonaffine regime~\citep{Bansal2021}, for low to moderate $Re$ and $E$. In this study, we partially extend some of these ideas for a non-Newtonian, horizontally aligned, Hele-Shaw flow in the dilute regime, in a bounded domain and within a selected range of parameters, $Re, E, \nu$. Figure~\ref{fig7} presents the variation of the most unstable mode versus $Re$, and at fixed $E$ and $\nu$. Comparing figures~\ref{fig7}a,b (shown at spatial location, $y = 0.99$) with figure~\ref{fig7}g,h (at $y = -0.99$), we find that the eigenvalues are not symmetric about the centerline (i. e.~$y = 0$). This observation is unsurprising since the anisotropic elastic stress ($A_{12} \ne A_{21}$, equation~\eqref{eqn:EStress}) breaks the solution symmetry about the centerline. Next, notice that for larger values of $Re$, elasticity is predominantly destabilizing (i.~e., notice that in figure~\ref{fig7}~a,b,c,f,g,h; $\omega_i^{\text{Temp}}$ is the smallest at $E=0.01$ and increases at higher $E$ values, beyond the range, $Re > 37$). This destabilization is the result of a complex interaction between the inertial forces (operative at larger Reynolds number) and the normal stress anisotropy through elasticity (proportional to $E$) and can be explained via an energy formalism: the stretching of the polymers with increasing elasticity brings about a normal stress anisotropy, leading to an elastically loaded fluid, that is, when the polymers stretch, elastic energy is stored in the sheared fluid. This energy is transferred and released after the fluid element has been adverted to other regions where the shear-induced stretching forces are smaller~\citep{Sircar2019}. 

Further note that in the elastic stress dominated case, at low Reynolds number, elasticity is shown to have a destabilizing influence near the wall (i.~e., $\nu = 0.3$ and $y = \pm 0.99$ case and inside the range $Re \le 5$, figure~\ref{fig7}~a,g) and a stabilizing impact farther away from the wall (i.~e., the $\nu = 0.3$ and $y = 0.5, 0.0$ case, figure~\ref{fig7} c,e). While the former observation can be explained via the boundary effects inducing a local perturbation at the advancing interface coupled with elasticity (a mechanism proposed by \citet{Rabaud1988} in Newtonian Saffman-Taylor flows), the latter observation is attributed to the elasticity-induced stabilization, well documented by Hinch in an appendix to~\citet{Azaiez1994} (albeit in flows without boundaries) and is akin to the action of a `surface tension' effect which can be explained as follows: the stretched polymers at low Reynolds number contribute to an effective tension and this tension damps the local perturbations. This important analogy (of surface tension) helps to provide a physical explanation of the influence of viscoelasticity in bounded, two-dimensional flows.
\bef
\centering
\includegraphics[width=0.495\linewidth, height=0.35\linewidth]{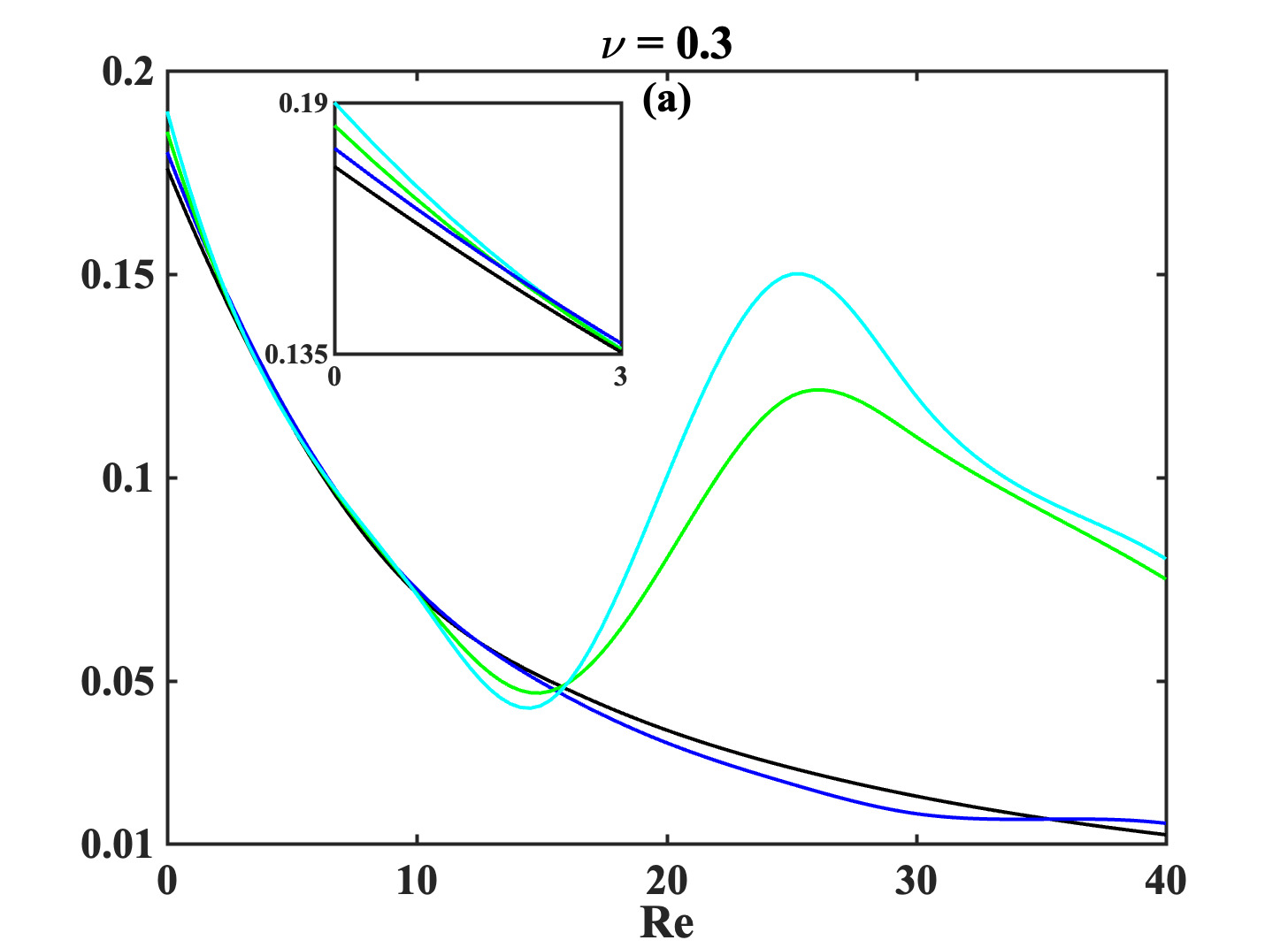}
\includegraphics[width=0.495\linewidth, height=0.35\linewidth]{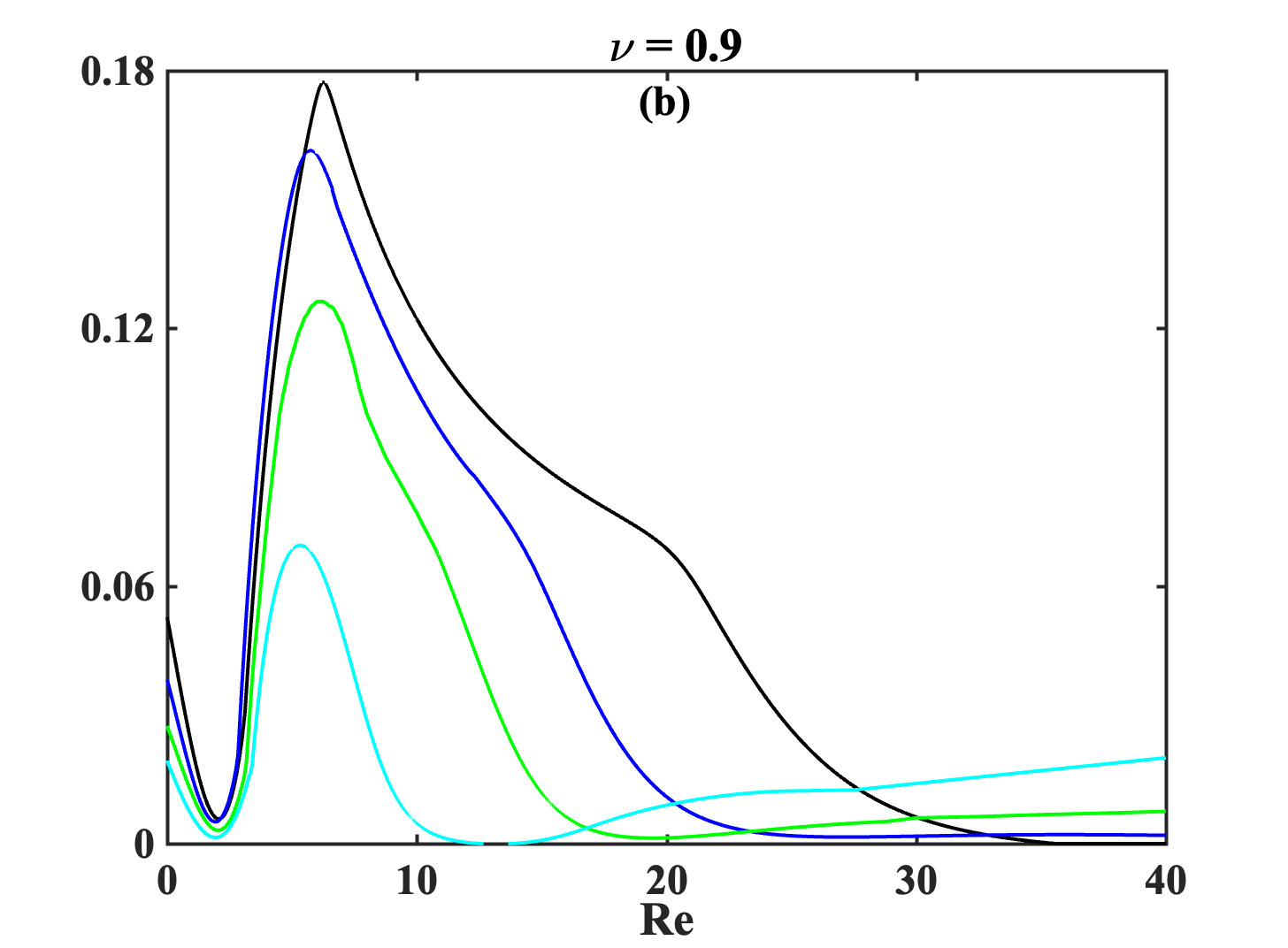}
\vskip 1pt
\includegraphics[width=0.495\linewidth, height=0.35\linewidth]{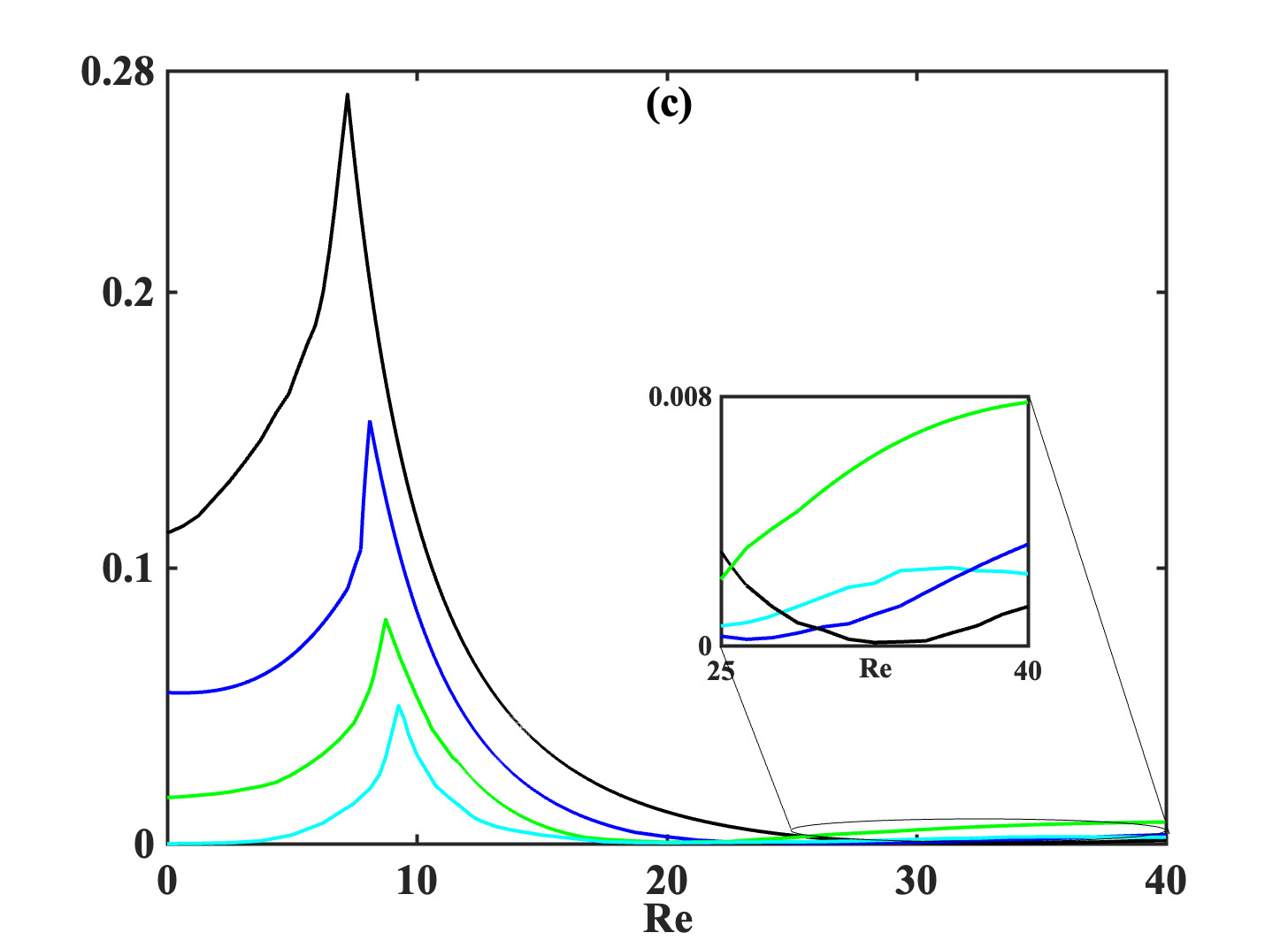}
\includegraphics[width=0.495\linewidth, height=0.35\linewidth]{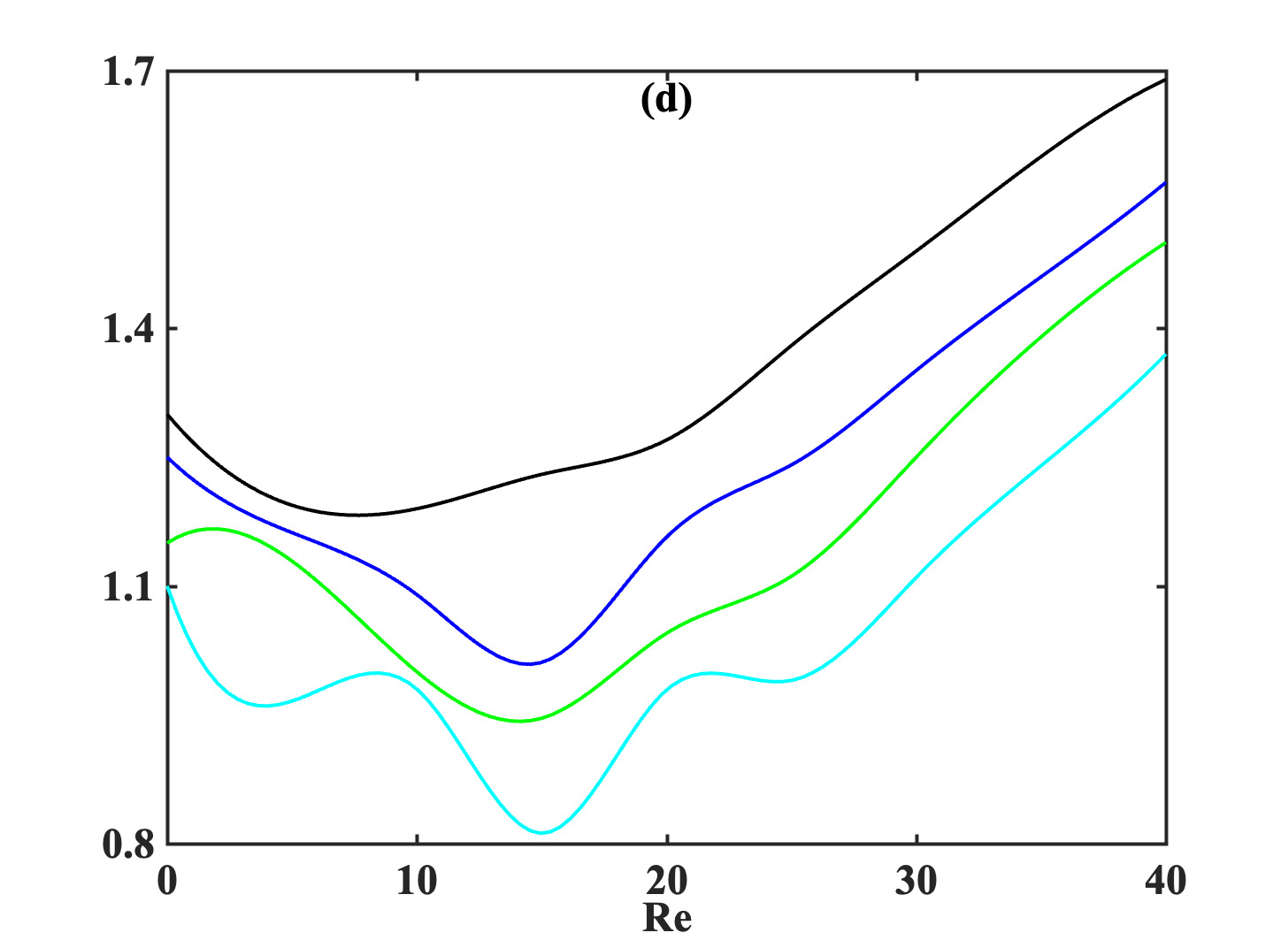}
\vskip 1pt
\includegraphics[width=0.495\linewidth, height=0.35\linewidth]{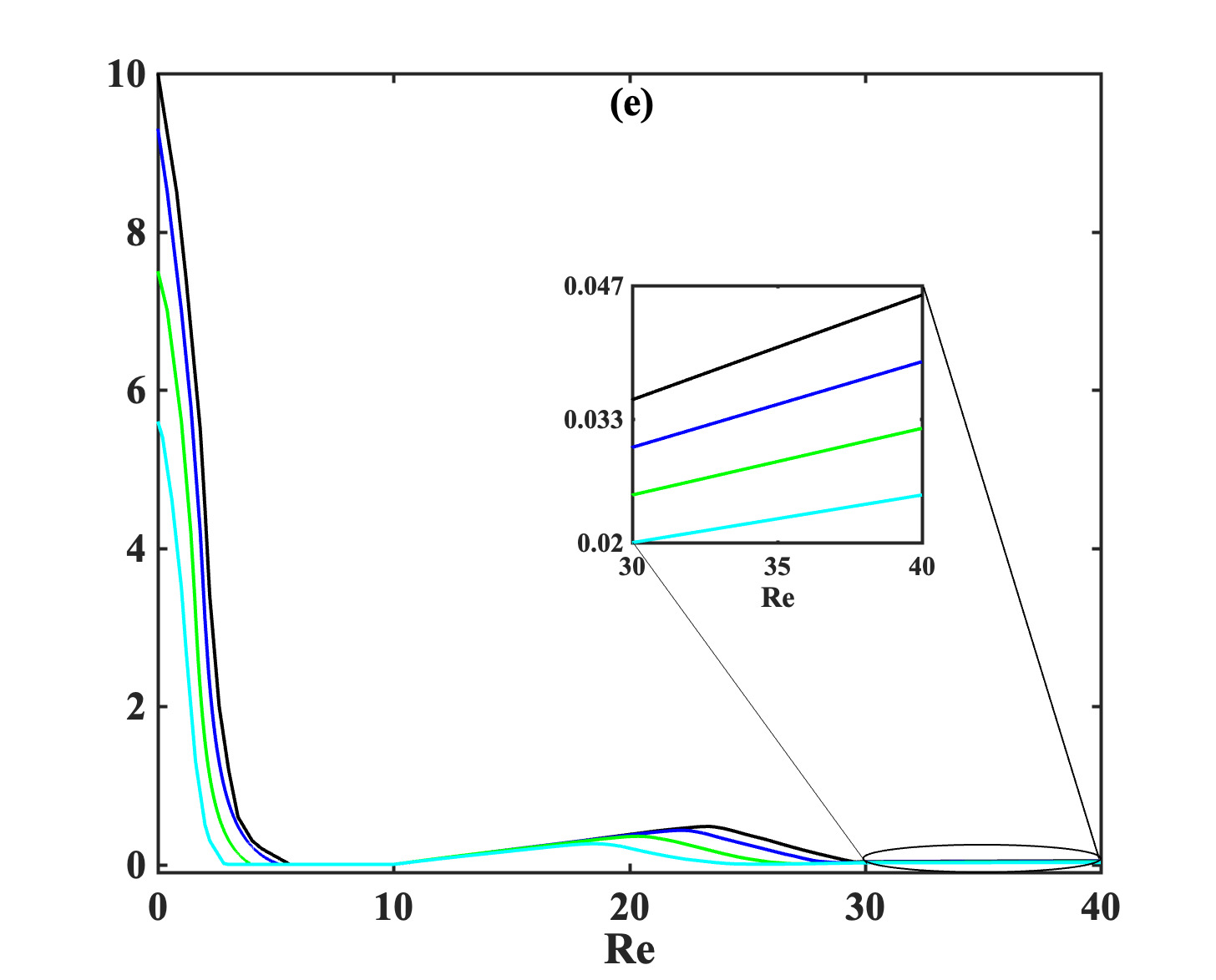}
\includegraphics[width=0.495\linewidth, height=0.35\linewidth]{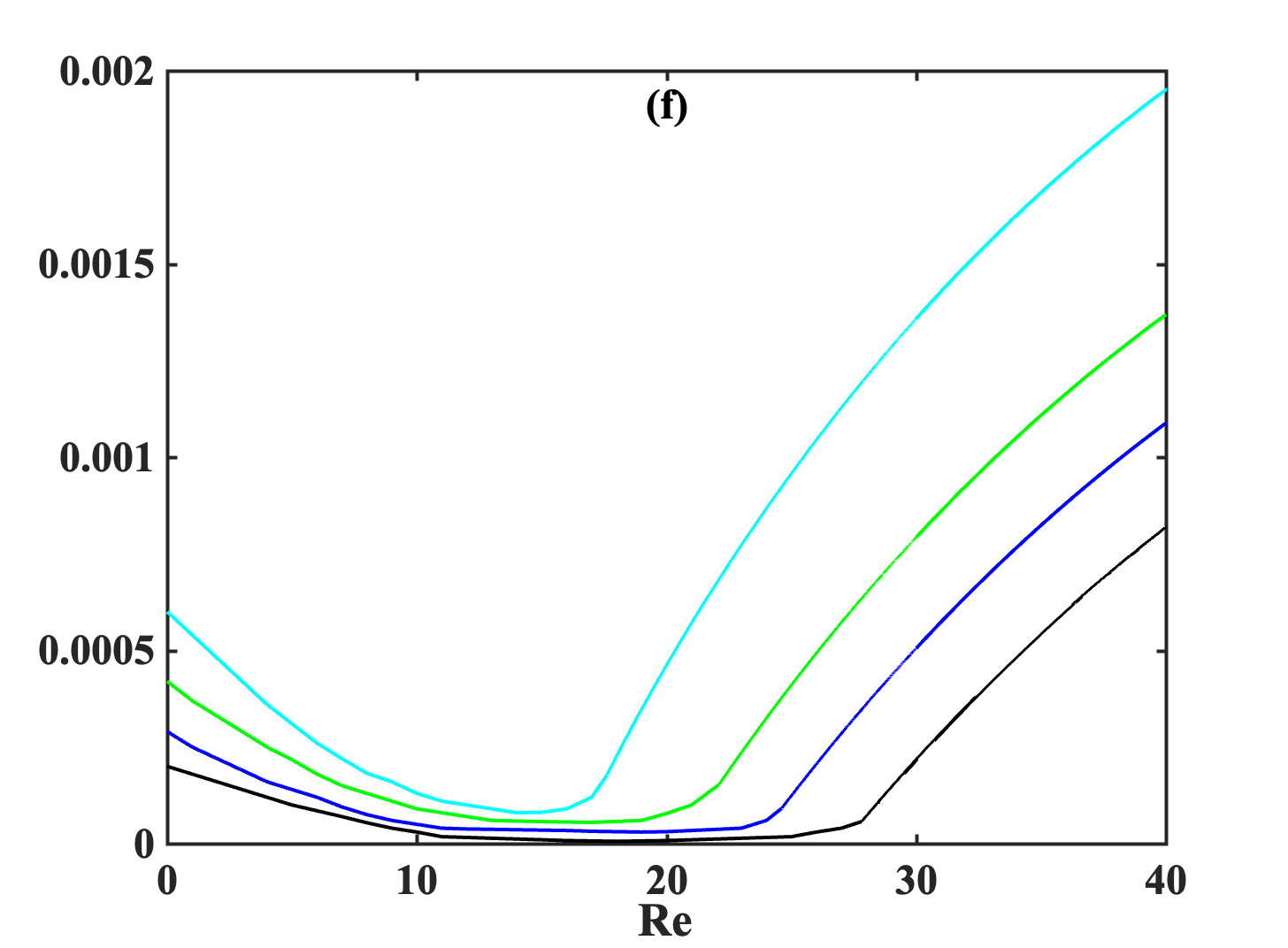}
\vskip 1pt
\includegraphics[width=0.495\linewidth, height=0.35\linewidth]{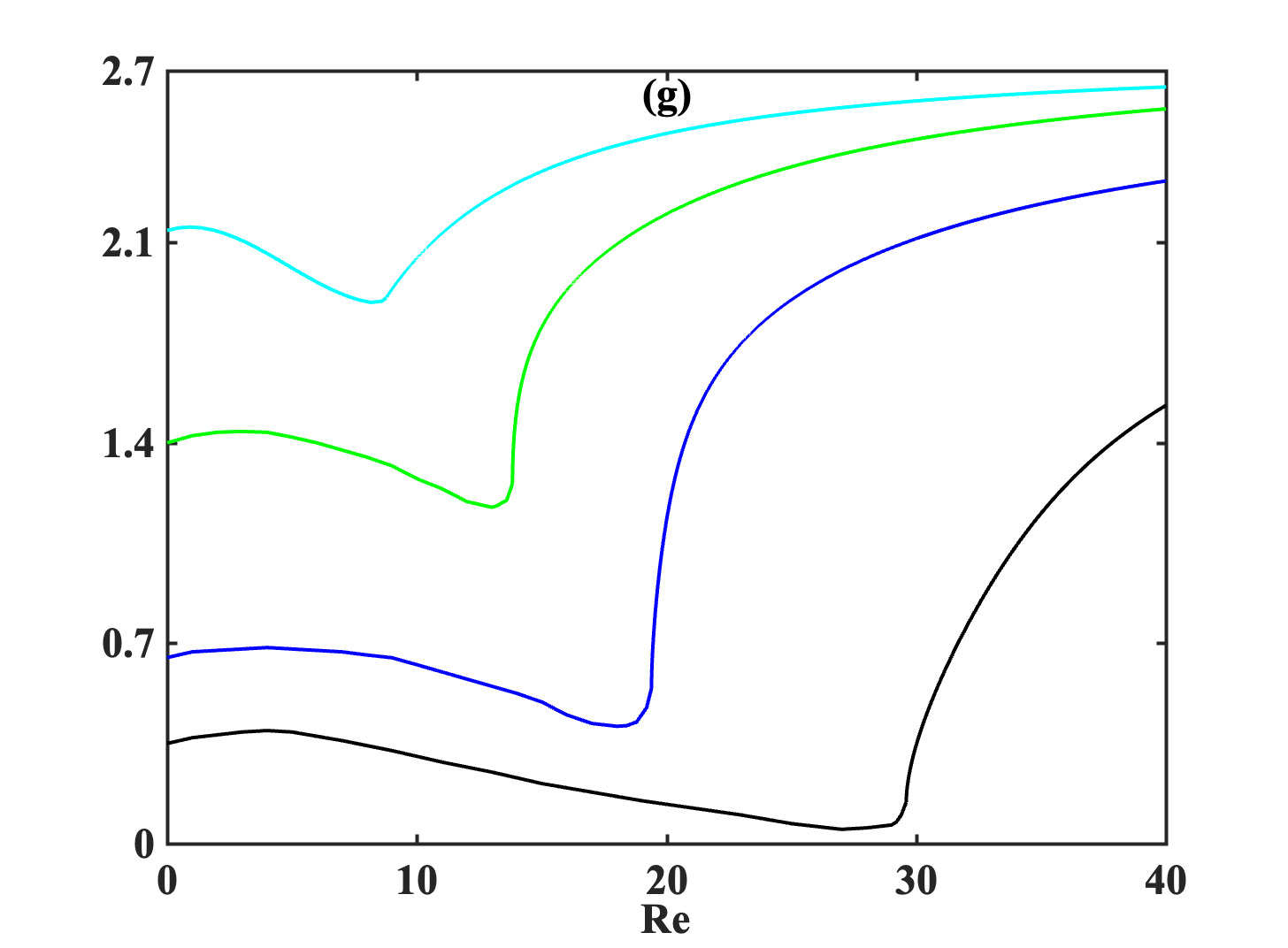}
\includegraphics[width=0.495\linewidth, height=0.35\linewidth]{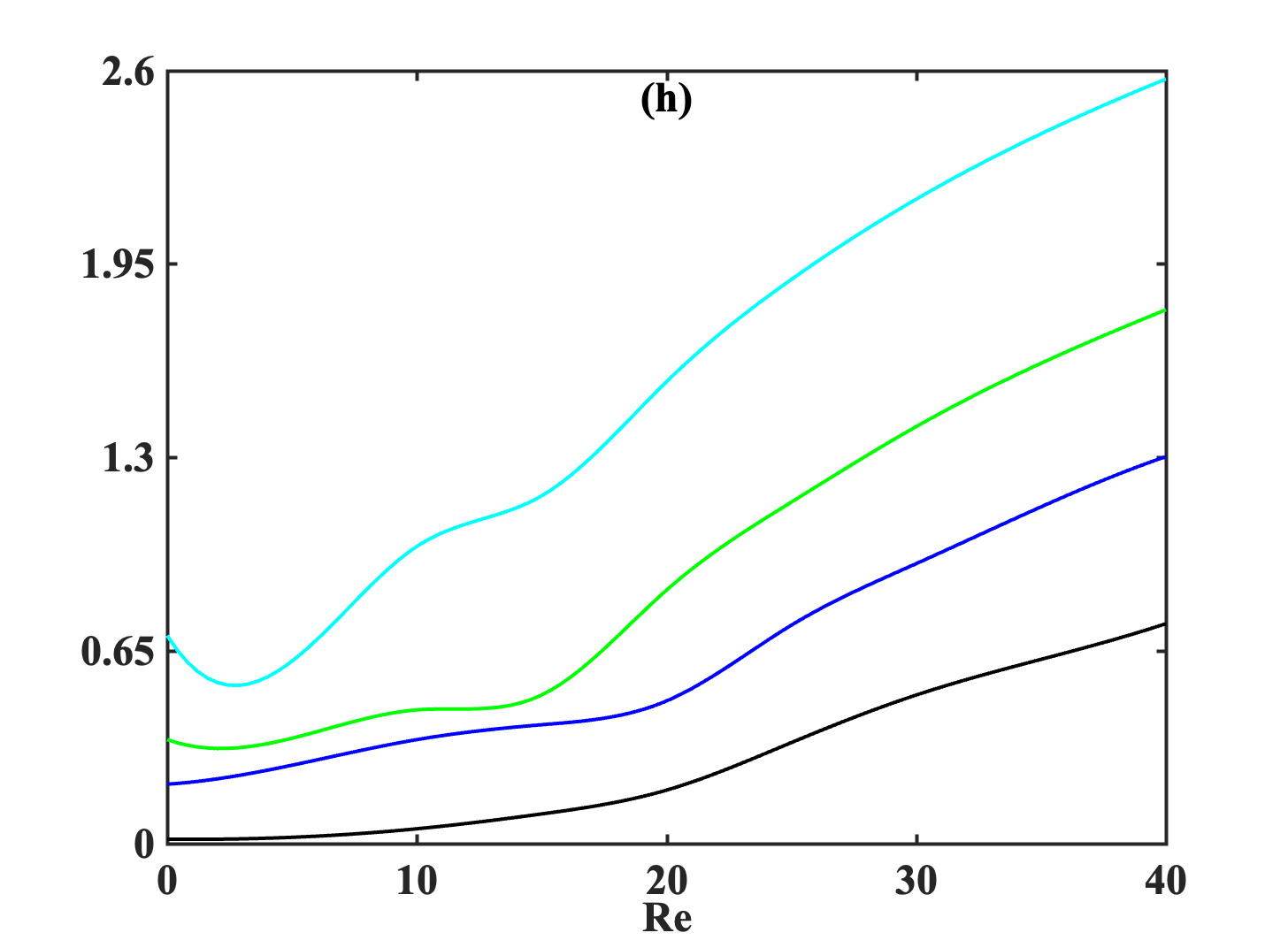}
\vskip 1pt
\caption{Most unstable mode, $\omega_i^{\text{Temp}}$ vs $Re$ at $E=0.01$ (\protect\blackline), $E=0.015$ (\protect\blueline), $E=0.02$ (\protect\greenline) and $E=0.03$ (\protect\cyanline) and evaluated at spatial locations: (a, b) $y=0.99$, (c, d) $y=0.5$, (e, f) $y=0.0$, (g, h) $y=-0.99$ and at fixed values of viscosity ratio, $\nu=0.3$ (left column) and $\nu=0.9$ (right column).}
\label{fig7}
\eef

A refined understanding of the temporal instability emerges via the evolution of the most unstable mode versus the Elasticity number, and at fixed $Re$ and $\nu$ (see figure~\ref{fig8}). Observe that at low $E$ values (or $E \le 0.005$), increasing the Reynolds number has a stabilizing influence (i.~e., note that in figure~\ref{fig8} a,b,c,f,g; $\omega_i^{\text{Temp}}$ is the largest at $Re=7.0$ and decreases at higher $Re$ values). We couple this observation with the one found in figure~\ref{fig7} to conclude that elasticity is stabilizing up until a critical Reynolds number (in our case, $Re \le 30.0$), beyond which the inertial effects dominate and reverses this trend. This particular finding is synonymous with the phenomena of elastic stabilization due to the addition of small amounts of dissolved polymers in free-surface flows such as the temporal stability of the viscoelastic jets by~\citet{Rallison1995} and~\citet{Miller2005} as well as the global stability analysis of viscoelastic cylindrical wakes by~\citet{Sahin2004}, the Floquet analysis of~\citet{Richter2011} and the experimental studies of~\citet{Cadot2000}.
\bef
\centering
\includegraphics[width=0.495\linewidth, height=0.35\linewidth]{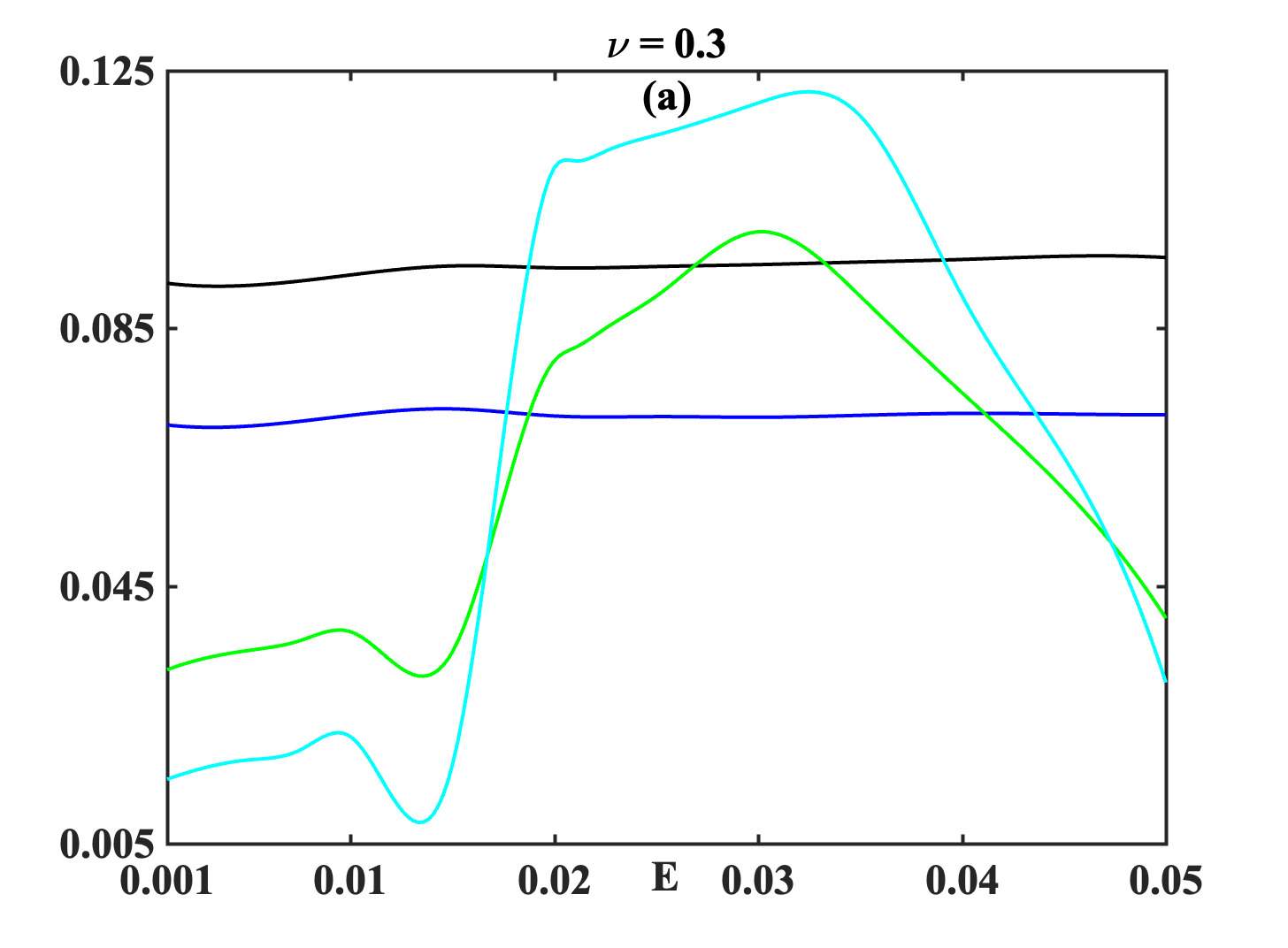}
\includegraphics[width=0.495\linewidth, height=0.35\linewidth]{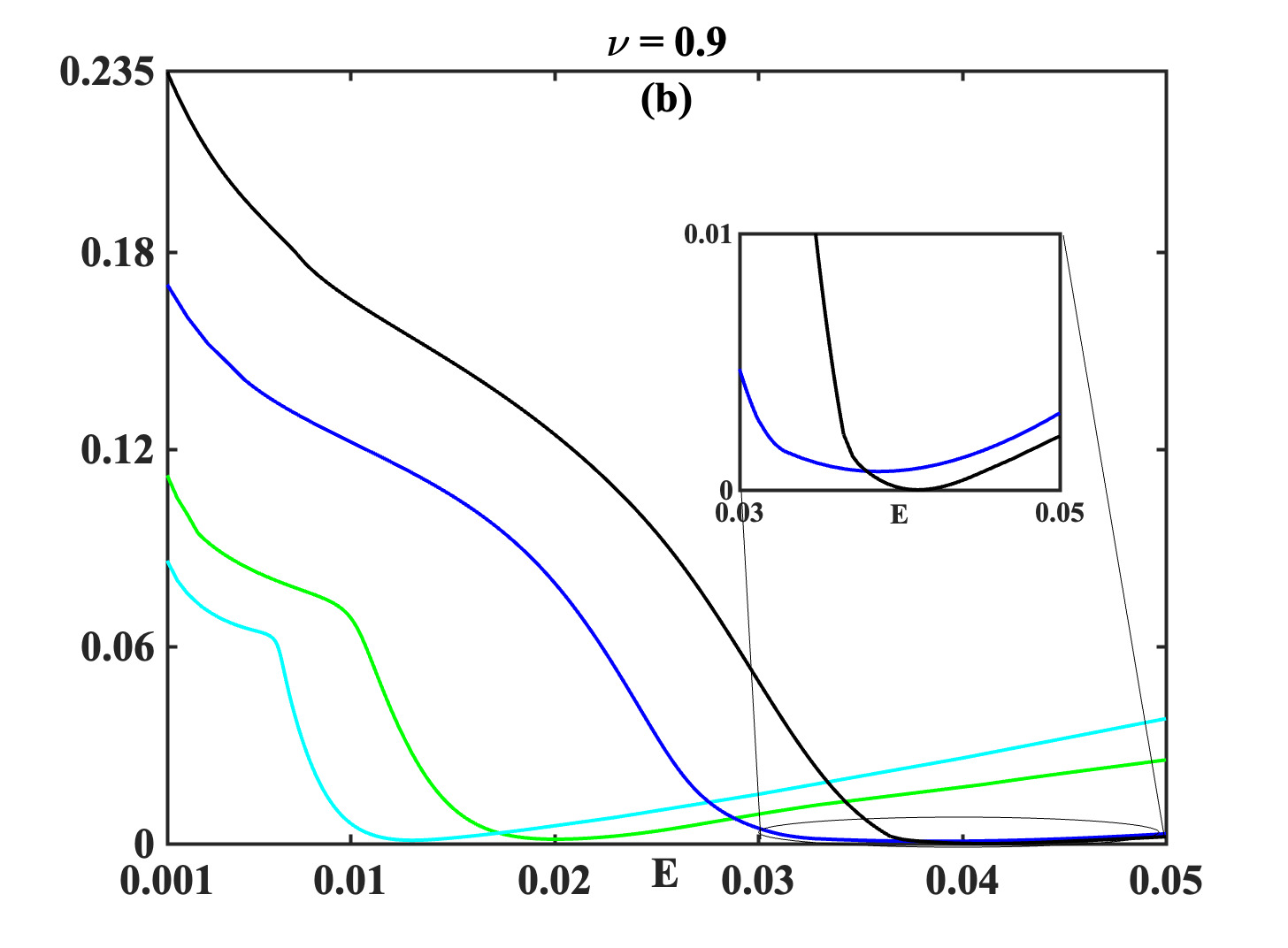}
\vskip 1pt
\includegraphics[width=0.495\linewidth, height=0.35\linewidth]{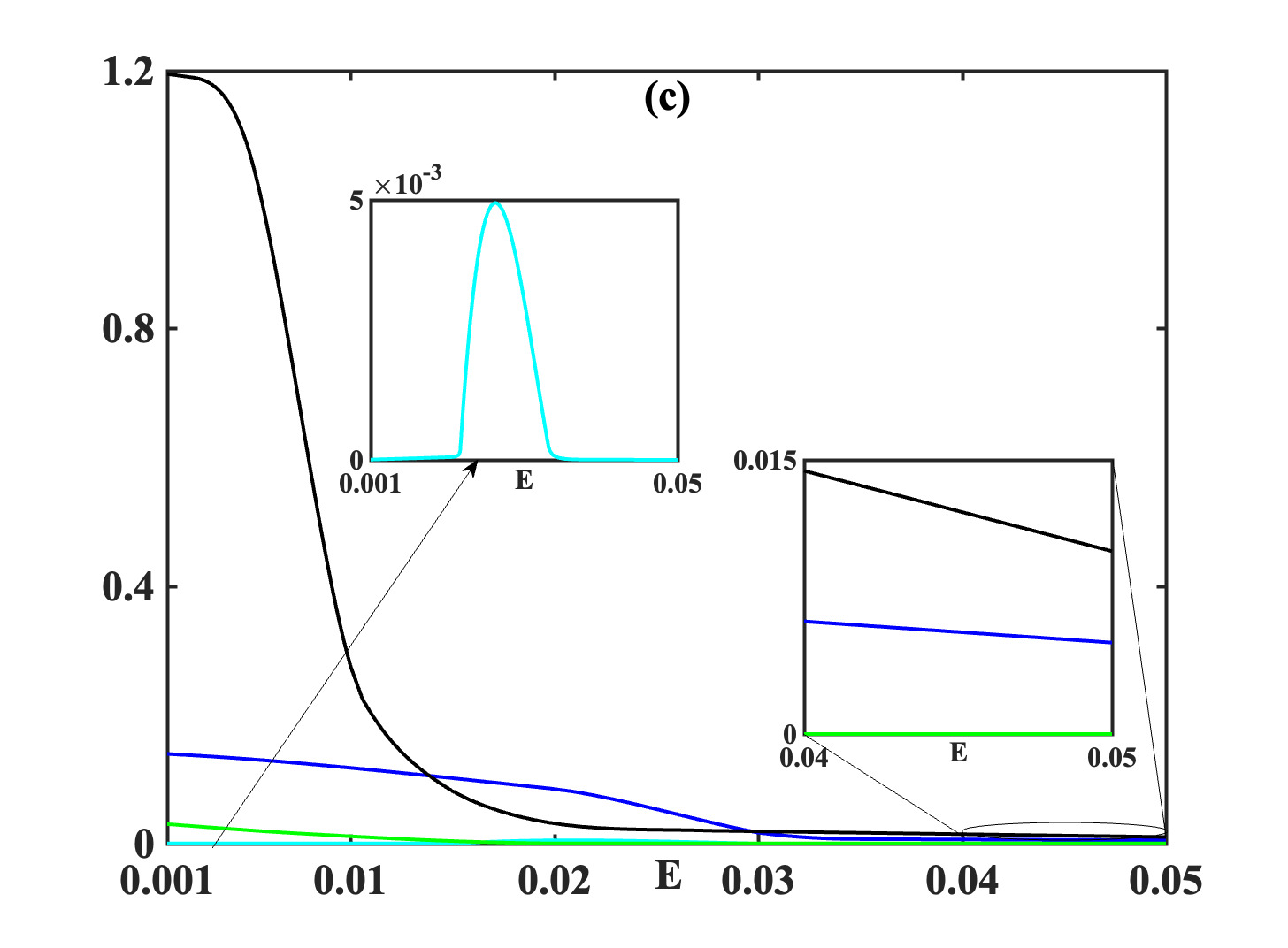}
\includegraphics[width=0.495\linewidth, height=0.35\linewidth]{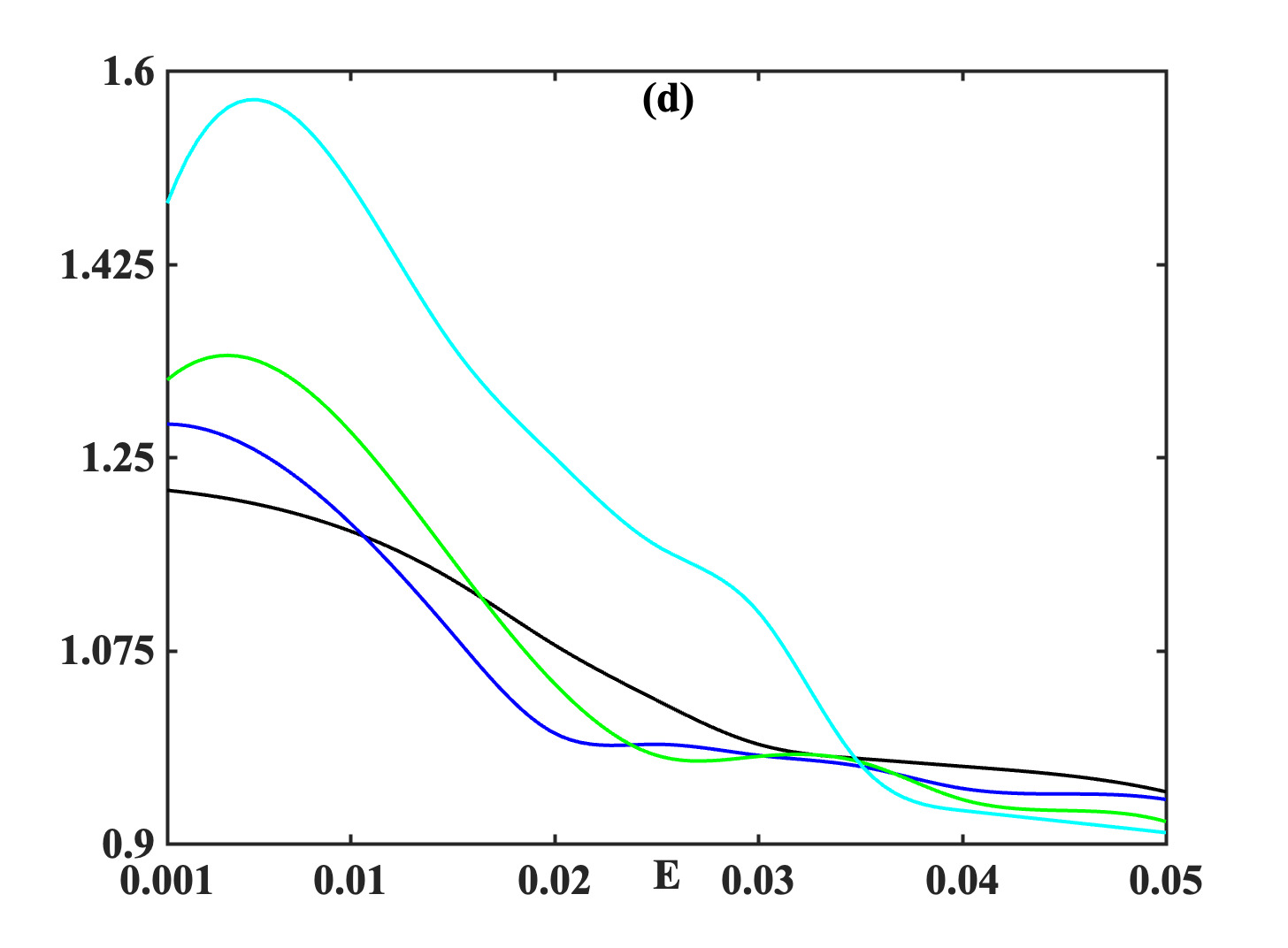}
\vskip 1pt
\includegraphics[width=0.495\linewidth, height=0.35\linewidth]{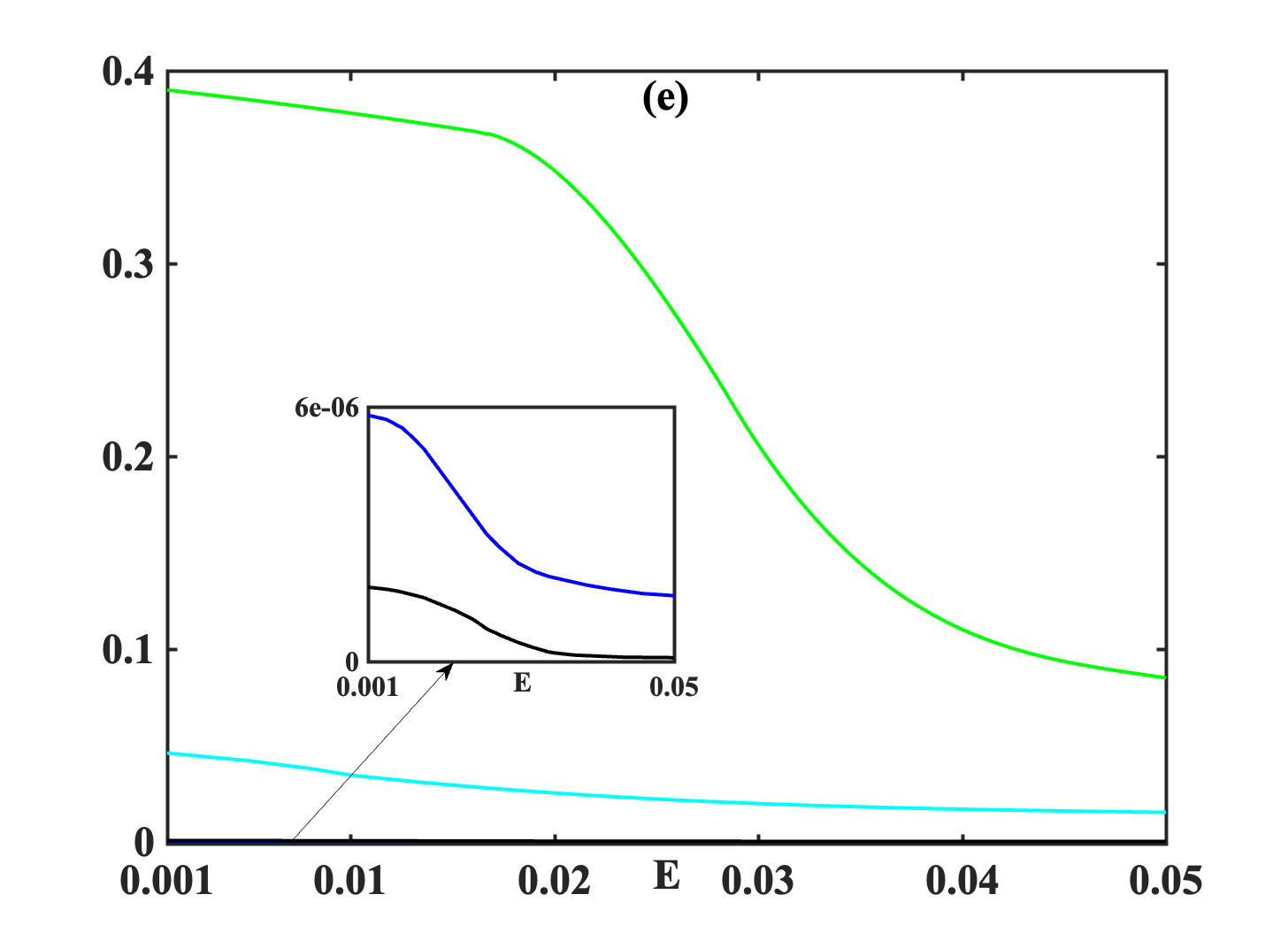}
\includegraphics[width=0.495\linewidth, height=0.35\linewidth]{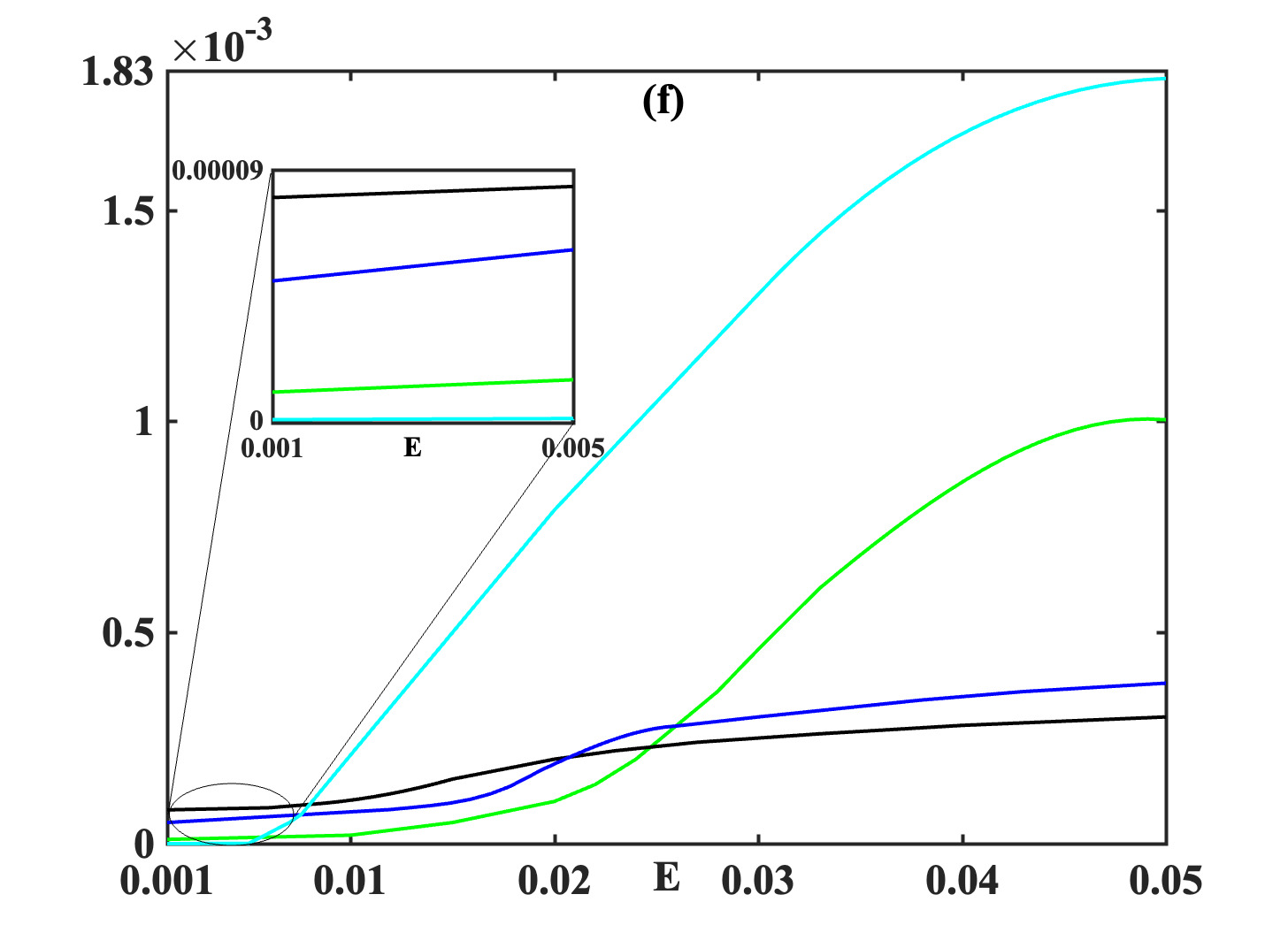}
\vskip 1pt
\includegraphics[width=0.495\linewidth, height=0.35\linewidth]{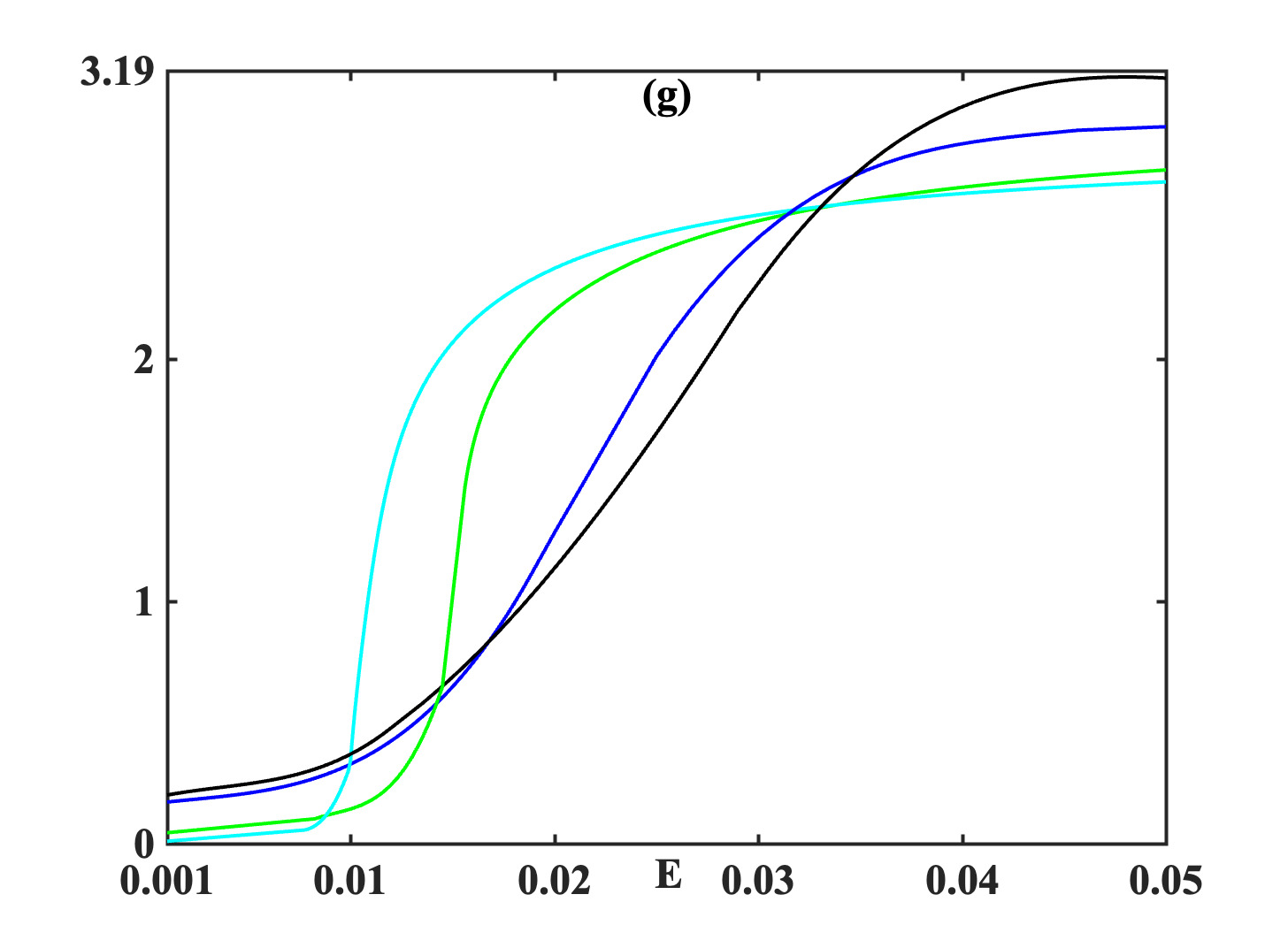}
\includegraphics[width=0.495\linewidth, height=0.35\linewidth]{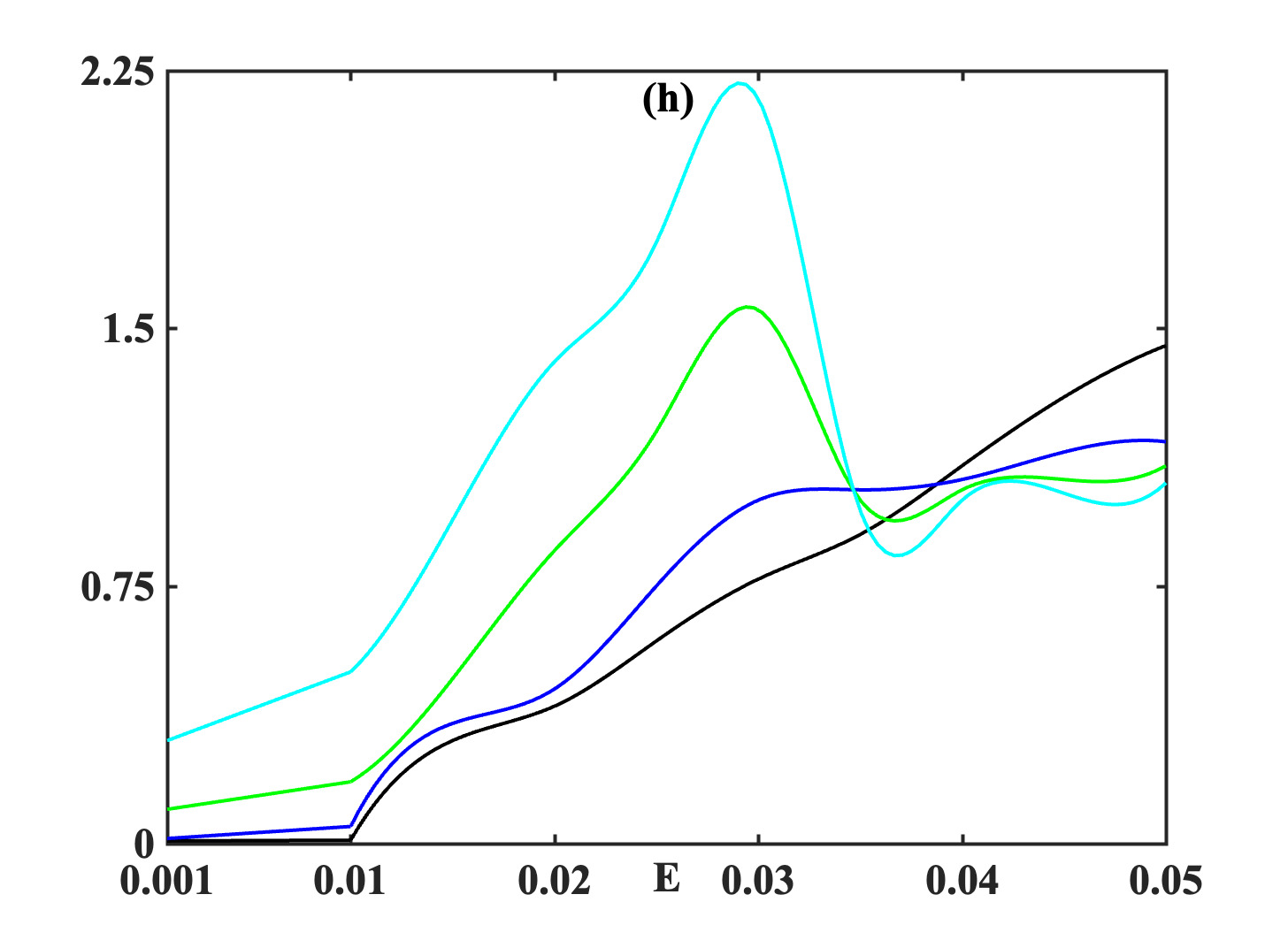}
\vskip 1pt
\caption{Most unstable mode, $\omega_i^{\text{Temp}}$ vs $E$ at $Re=7.0$ (\protect\blackline), $Re=10.0$ (\protect\blueline), $Re=20.0$ (\protect\greenline) and $Re=30.0$ (\protect\cyanline) and evaluated at spatial locations: (a, b) $y=0.99$, (c, d) $y=0.5$, (e, f) $y=0.0$, (g, h) $y=-0.99$ and at fixed values of viscosity ratio, $\nu=0.3$ (left column) and $\nu=0.9$ (right column).}
\label{fig8}
\eef

The plots of the most unstable mode versus the viscosity ratio, $\nu$, at fixed $E$ and $Re$ (figure~\ref{fig9}) further augment our understanding of the temporal instability of viscoelastic Saffman-Taylor flows. First, notice that at lower Reynolds number and in the Newtonian limit (i.~e., at $Re = 10.0$ and $\nu \rightarrow 1$, see left column in figure~\ref{fig9}), elasticity is stabilizing above the centerline (figure~\ref{fig9} a,c) and destabilizing on and below the centerline (figure~\ref{fig9} e,g). Second, notice that at higher Reynolds number (i.~e., at $Re = 40.0$) and in the Newtonian limit, elasticity is largely destabilizing. In the Newtonian limit, we attribute both these observations arising out of the destabilizing effect of the fluid inertia. In contrast, note that elasticity is destabilizing in the strongly elastic limit ($\nu \rightarrow 0$) and deduce that in this limit, viscoelasticity exacerbates the flow instabilities, arising from a combination of normal stress anisotropy and elasticity~\citep{Bansal2021}.

We summarize the interplay of the inertial forces (characterized by the parameter, $Re$) and the elastic forces (represented by the parameter, $E$) as well as the boundary effects on the progression of the temporal instability (exemplified by the most unstable mode) of the viscoelastic Saffman-Taylor flows as follows: (except for some cases in the Newtonian limit as well as in the strongly elastic limit) the inertial forces have a universally destabilizing impact on the evolving flow front, while elasticity combined with low (high) fluid inertia is stabilizing (destabilizing). Finally, the finite boundary is shown to have a destabilizing influence near the wall.
\bef
\centering
\includegraphics[width=0.495\linewidth, height=0.35\linewidth]{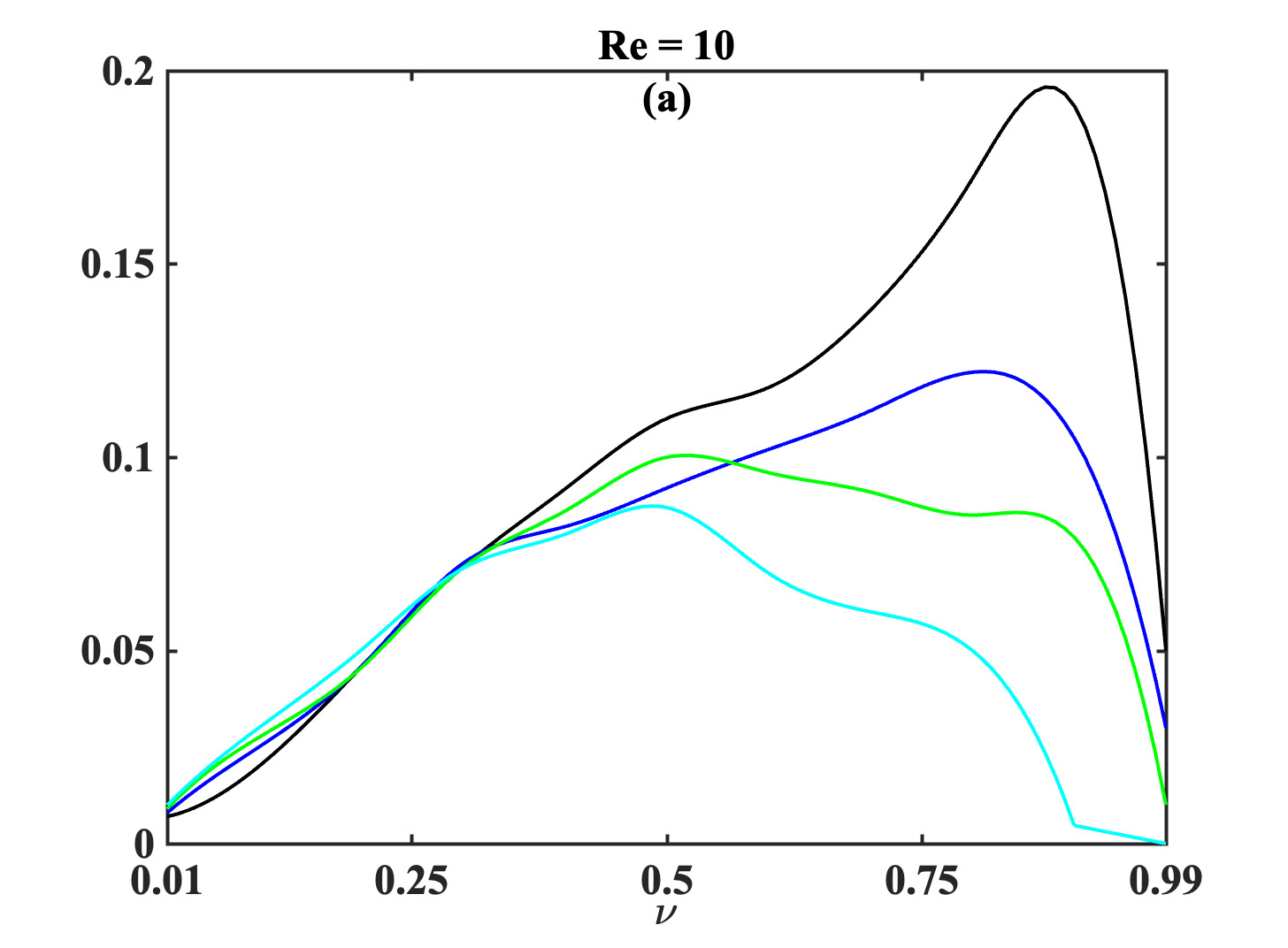}
\includegraphics[width=0.495\linewidth, height=0.35\linewidth]{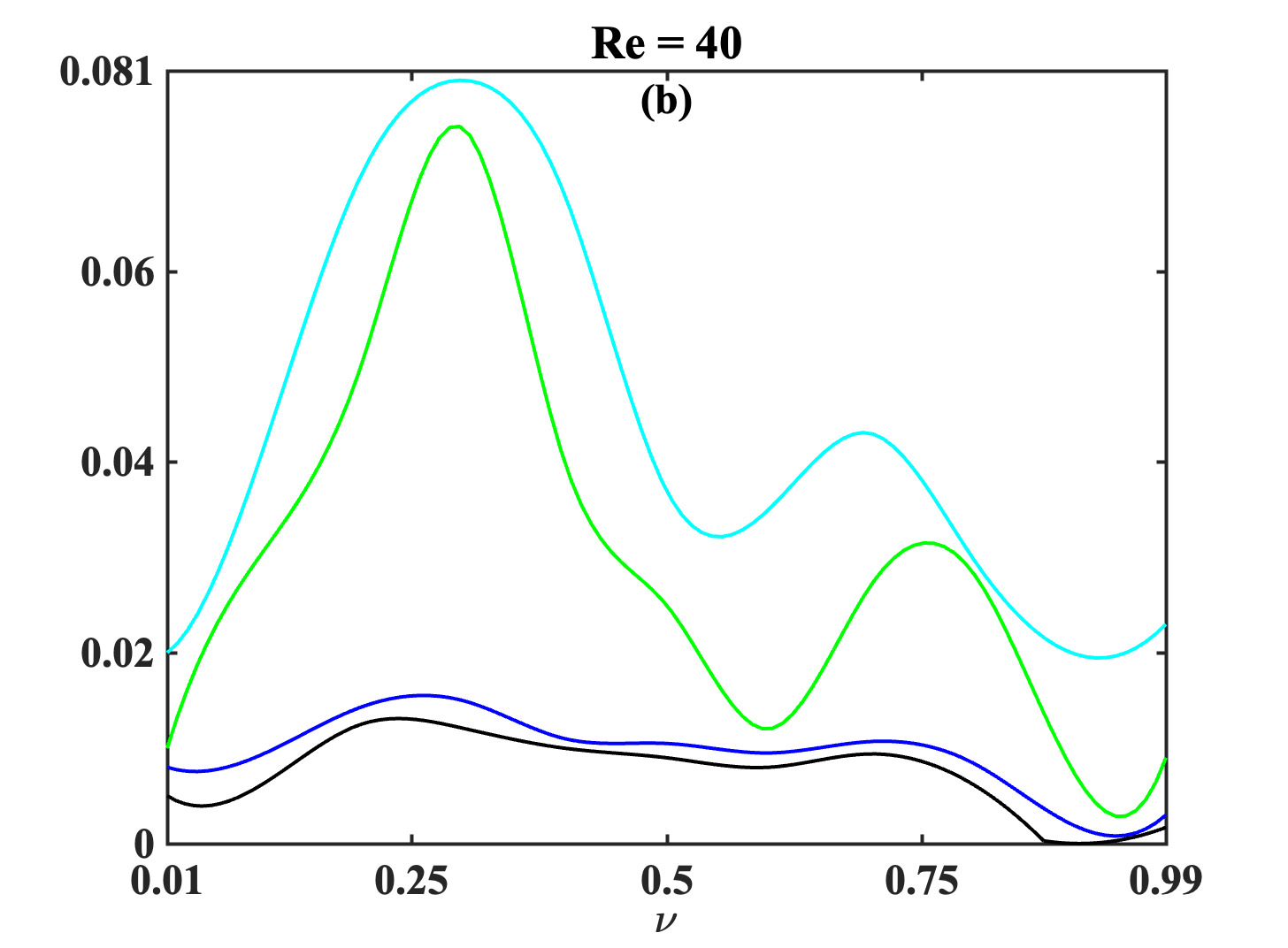}
\vskip 1pt
\includegraphics[width=0.495\linewidth, height=0.35\linewidth]{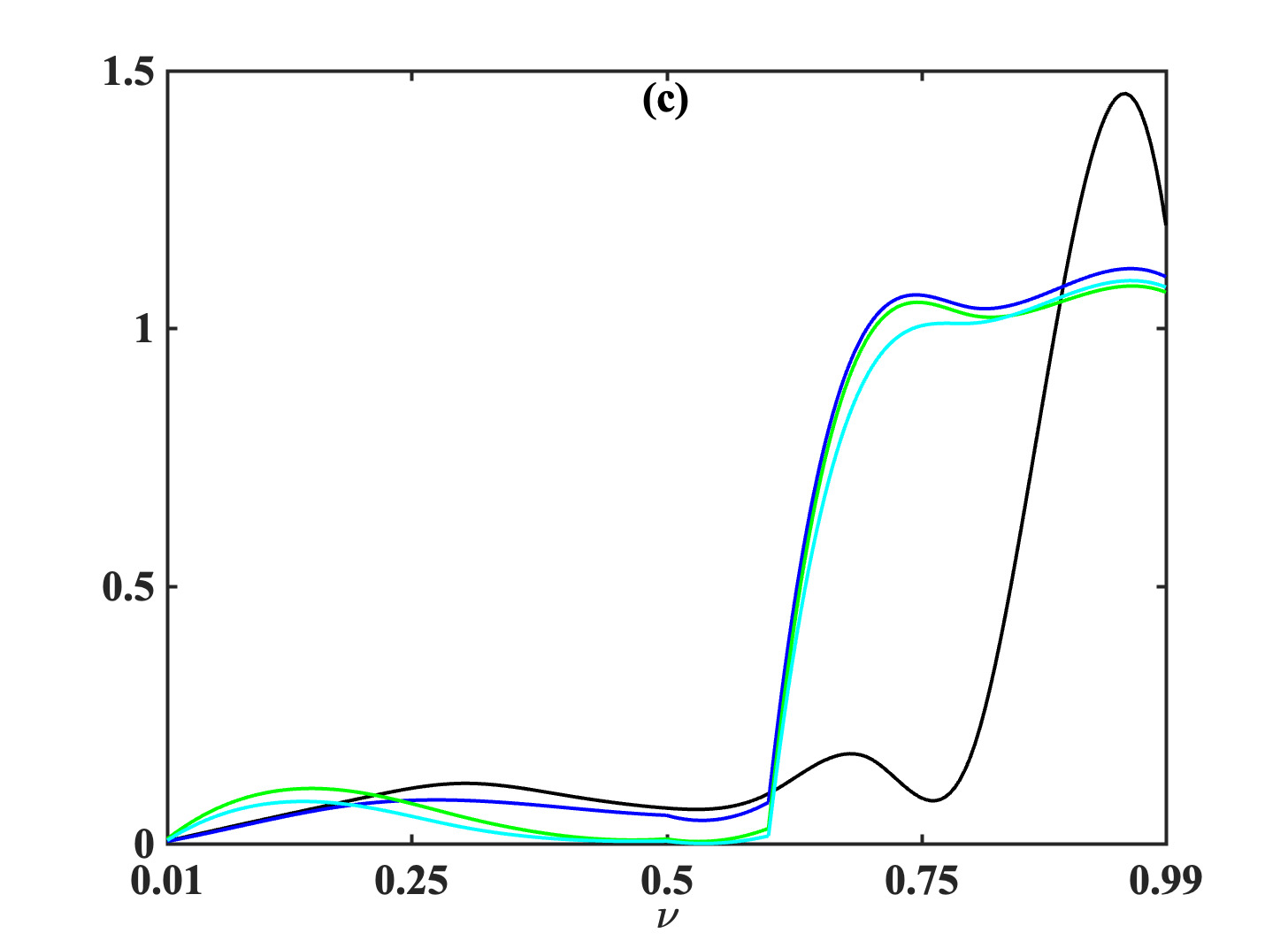}
\includegraphics[width=0.495\linewidth, height=0.35\linewidth]{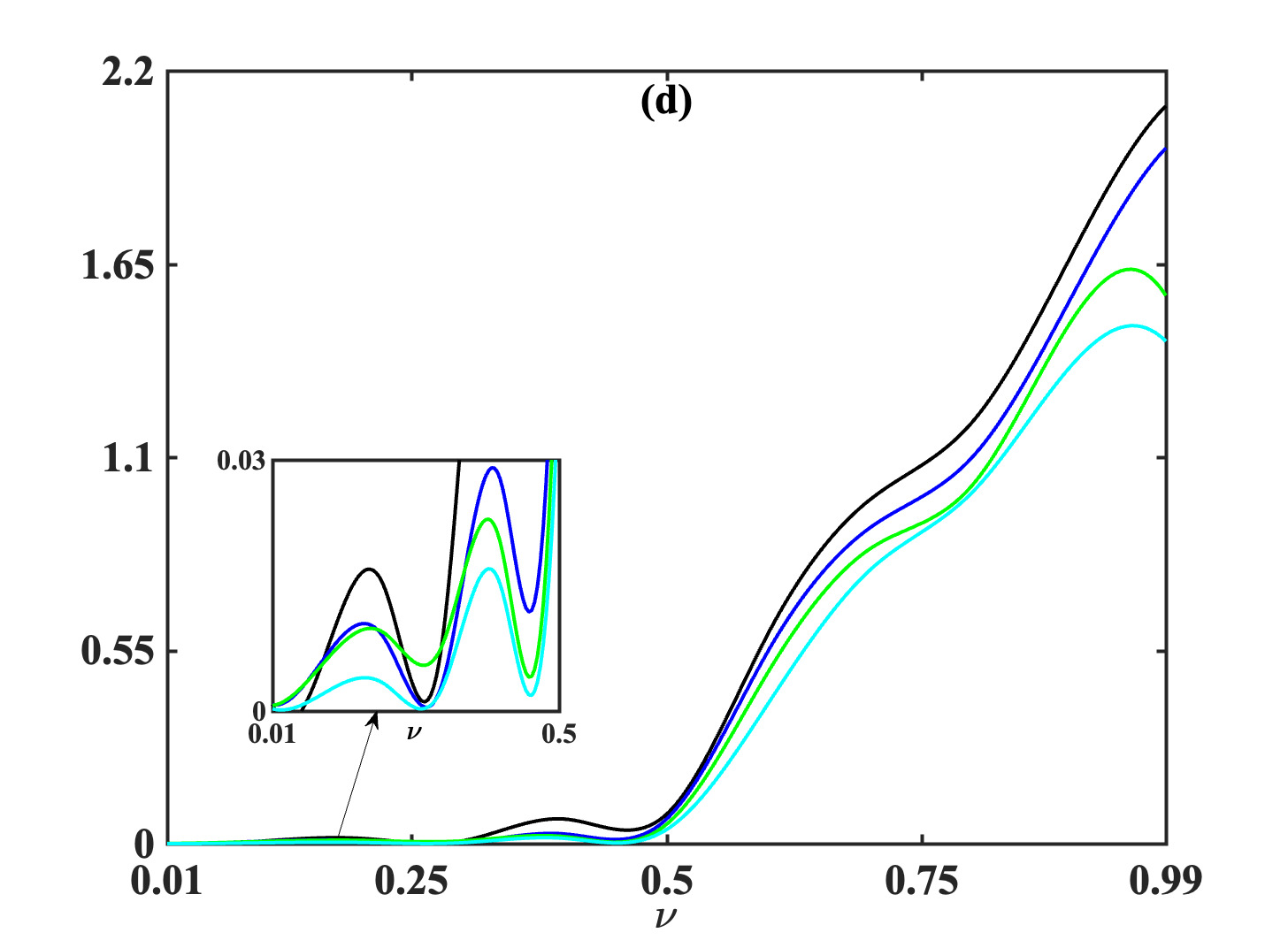}
\vskip 1pt
\includegraphics[width=0.495\linewidth, height=0.35\linewidth]{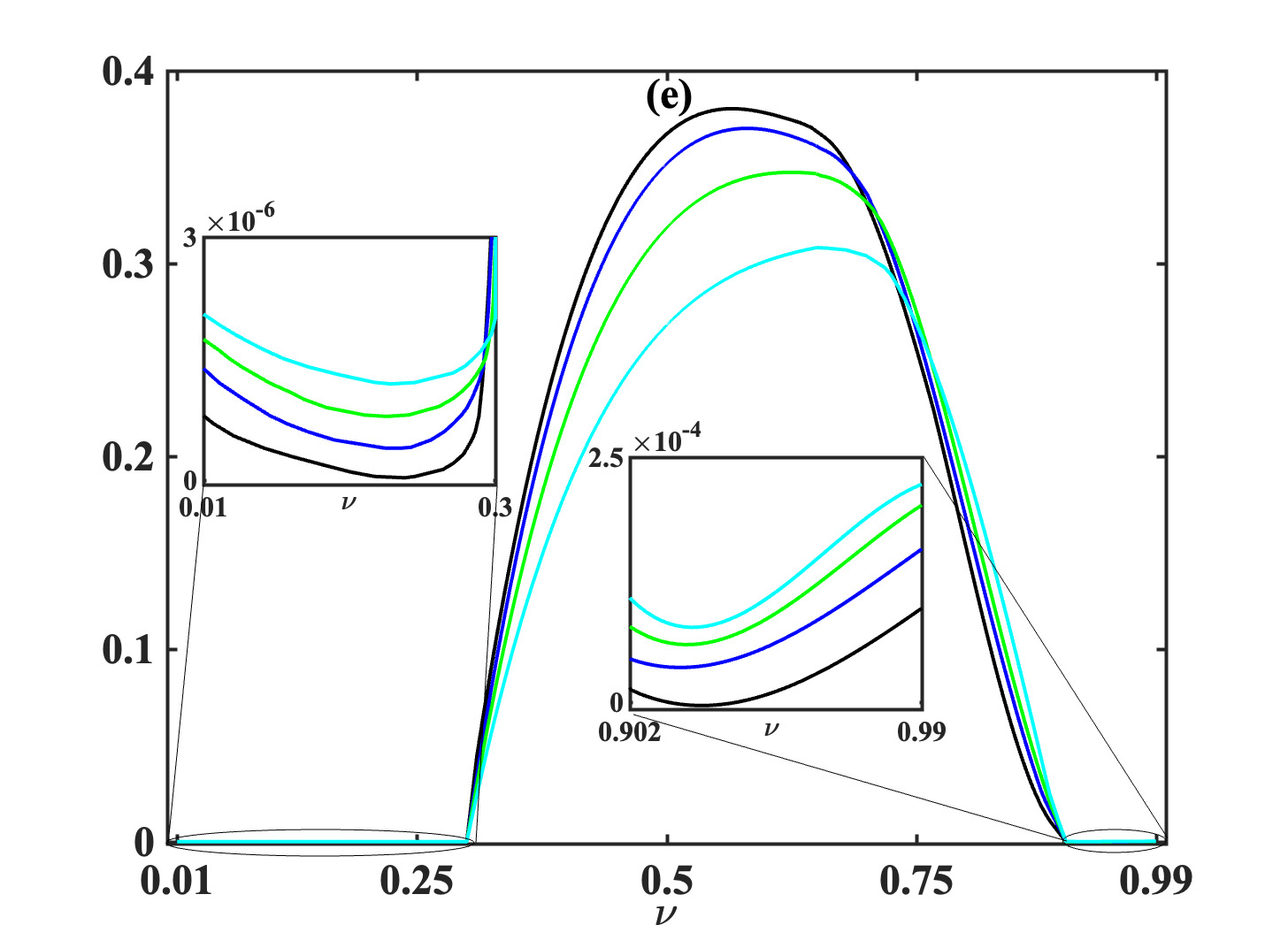}
\includegraphics[width=0.495\linewidth, height=0.35\linewidth]{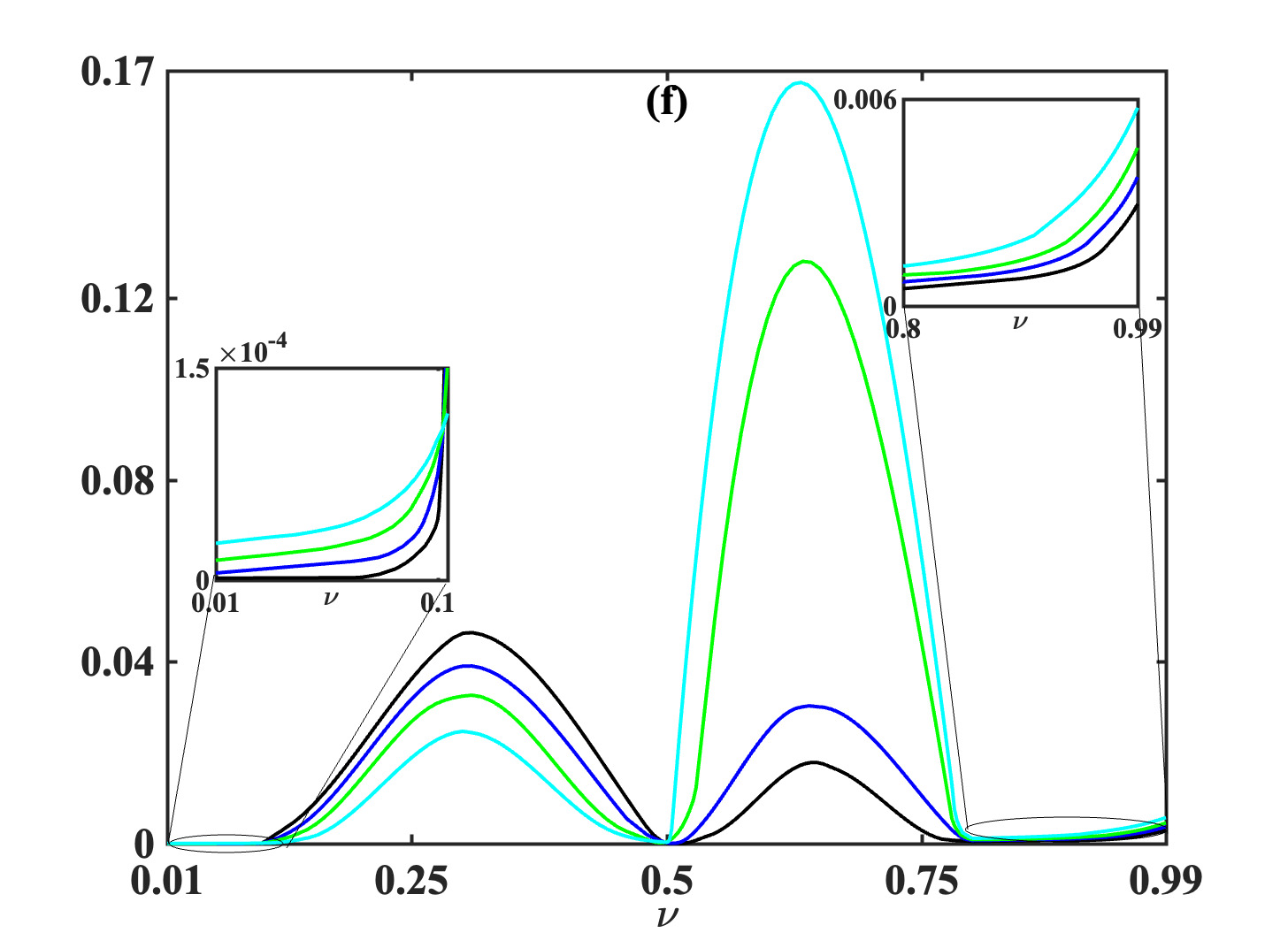}
\vskip 1pt
\includegraphics[width=0.495\linewidth, height=0.35\linewidth]{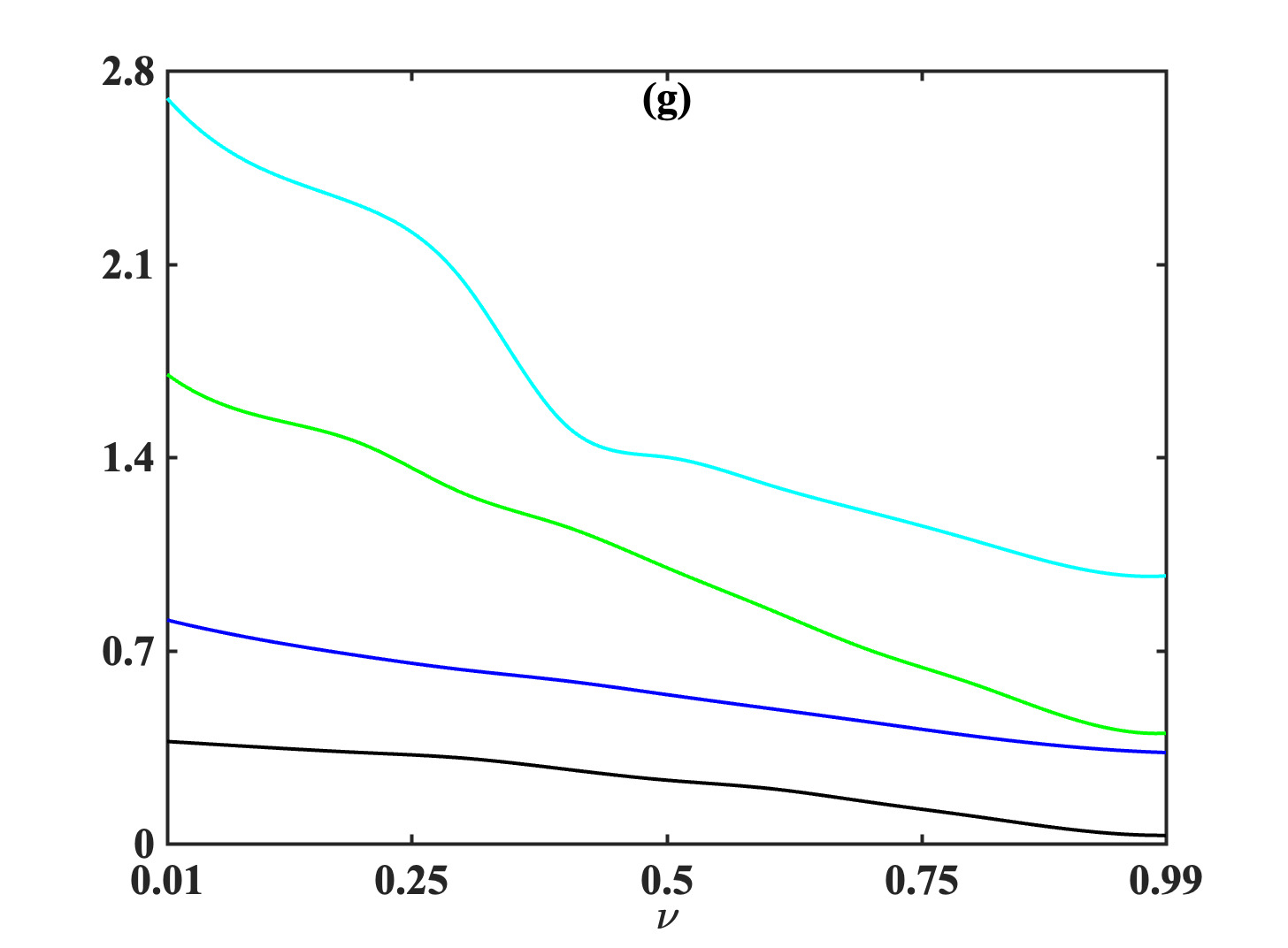}
\includegraphics[width=0.495\linewidth, height=0.35\linewidth]{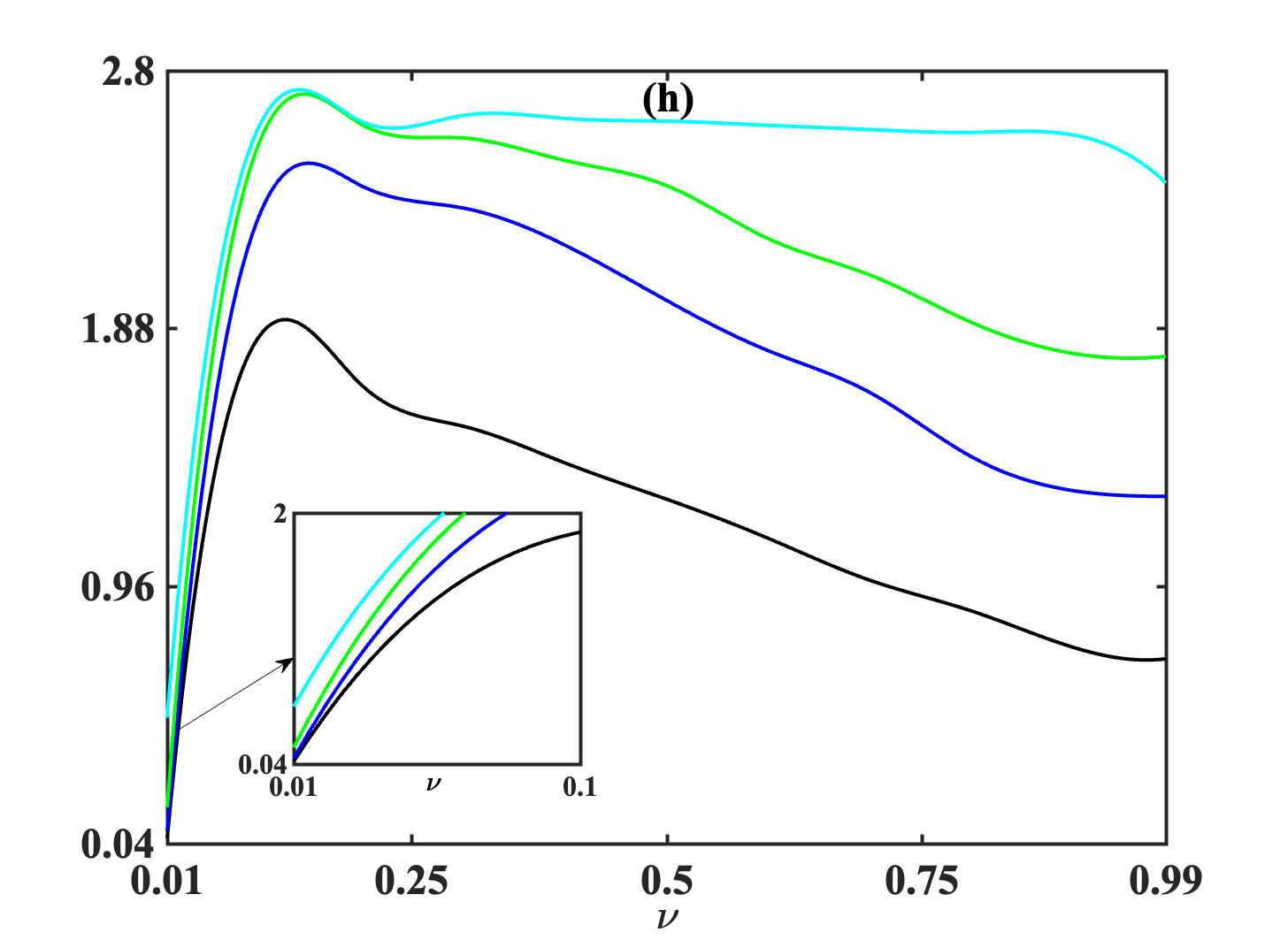}
\vskip 1pt
\caption{Most unstable mode, $\omega_i^{\text{Temp}}$ vs $\nu$ at $E=0.01$ (\protect\blackline), $E=0.015$ (\protect\blueline), $E=0.02$ (\protect\greenline) and $E=0.03$ (\protect\cyanline) and evaluated at spatial locations: (a, b) $y=0.99$, (c, d) $y=0.5$, (e, f) $y=0.0$, (g, h) $y=-0.99$ and at fixed values of Reynolds number, $Re=10.0$ (left column) and $Re=40.0$ (right column).}
\label{fig9}
\eef

\subsection{Velocity perturbation amplitude at the most unstable mode}\label{subsec:EigenV}
Figure~\ref{fig10} exhibits the amplitude of the velocity perturbation, $|f_r + i f_i|$, versus $Re$ at fixed $E$ and $\nu$ (refer equation~\eqref{appA:f_and_g} for details). These eigenfunctions correspond to the most unstable mode showcased in figure~\ref{fig7}. Notice that, in general, the perturbation amplitudes near the wall (figure~\ref{fig10} a,b,g,h) are larger than those farther away from the wall (figure~\ref{fig10} c,d,e,f) reaffirming the observation (as seen in \S \ref{subsec:tgr}) that the destabilizing impact of elasticity is more pronounced near the wall. Near the wall, the disturbance amplitudes are largely insensitive to variations in $Re$. However, farther away from the wall (or at $y=0.5, 0.0$), these eigenfunctions show a distinct and a sharp peak within the range, $Re \le 20.0$, implying that the velocity perturbation of the fluid interface has a larger amplitude at lower Reynolds number, a finding which is in stark contrast with the earlier outcome of \citet{Shokri2017}.
\bef
\centering
\includegraphics[width=0.495\linewidth, height=0.35\linewidth]{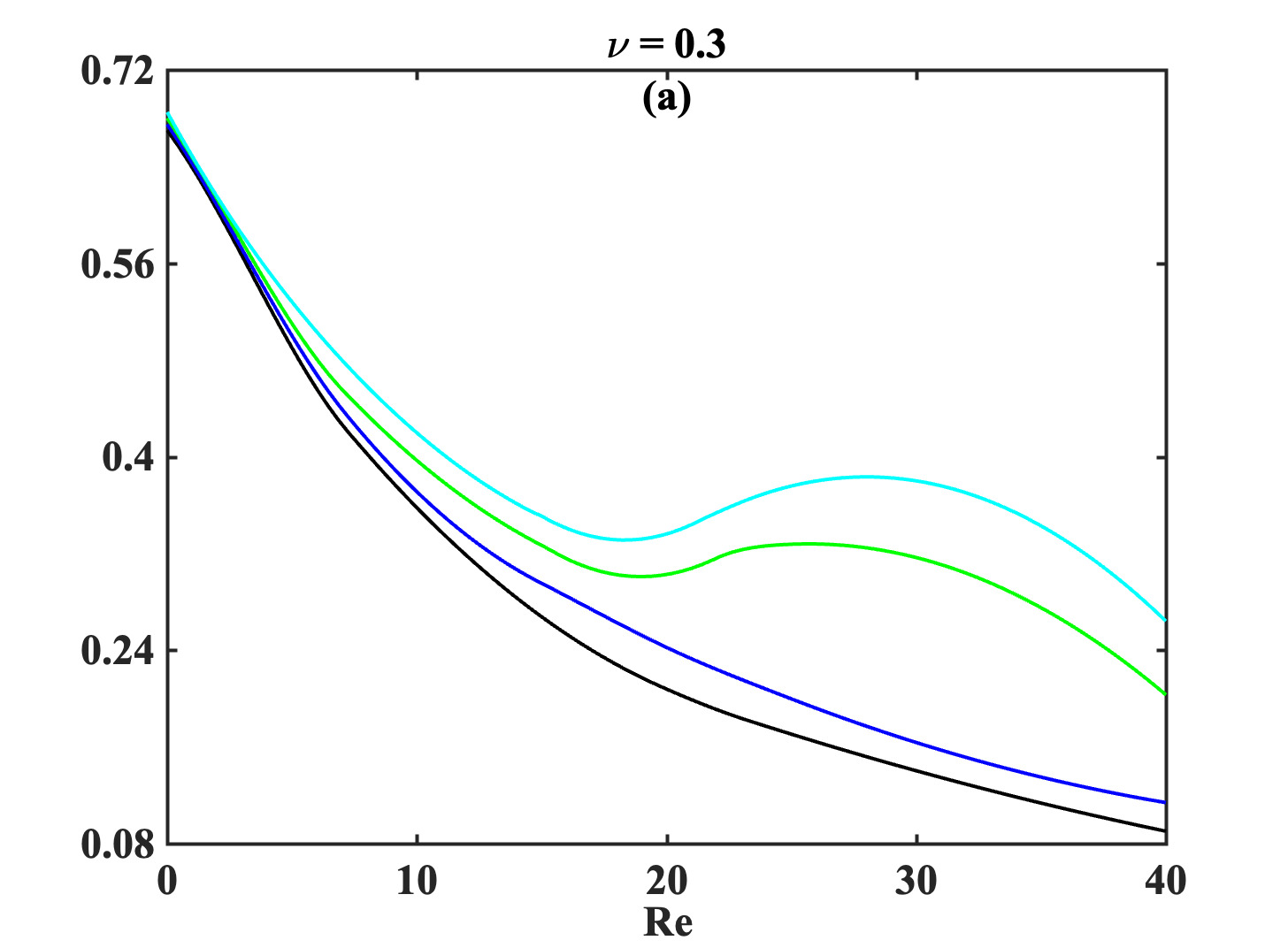}
\includegraphics[width=0.495\linewidth, height=0.35\linewidth]{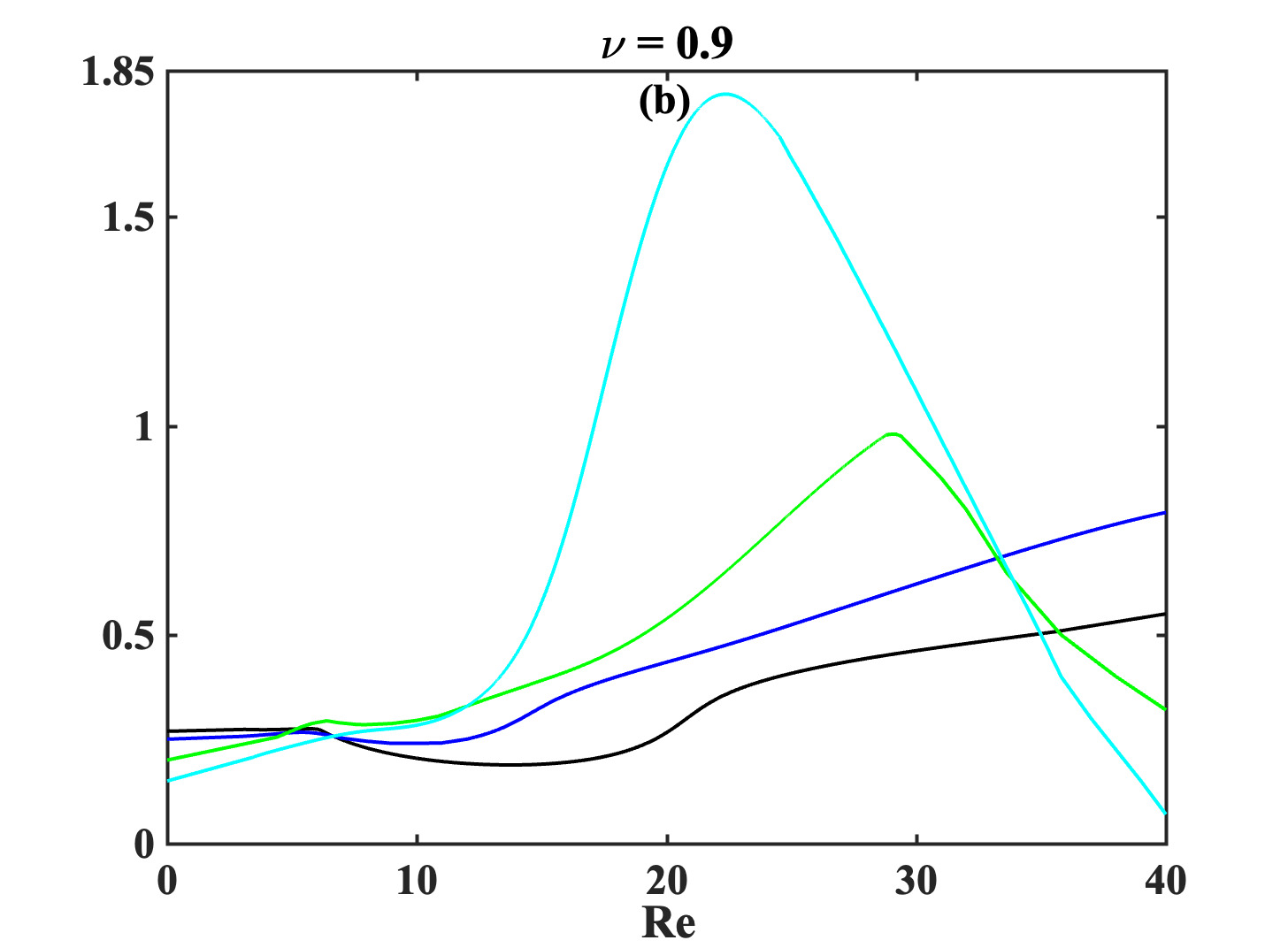}
\vskip 1pt
\includegraphics[width=0.495\linewidth, height=0.35\linewidth]{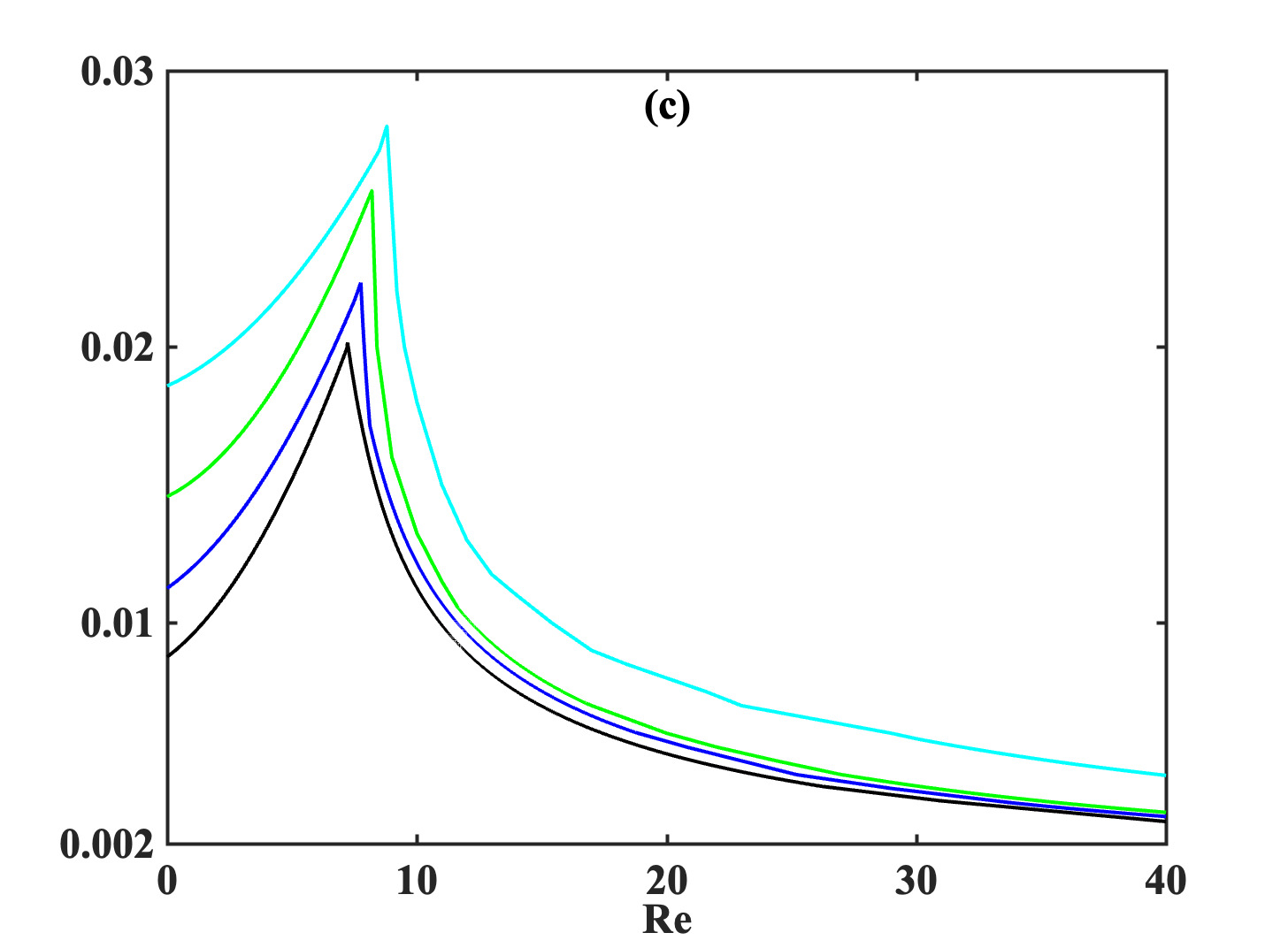}
\includegraphics[width=0.495\linewidth, height=0.35\linewidth]{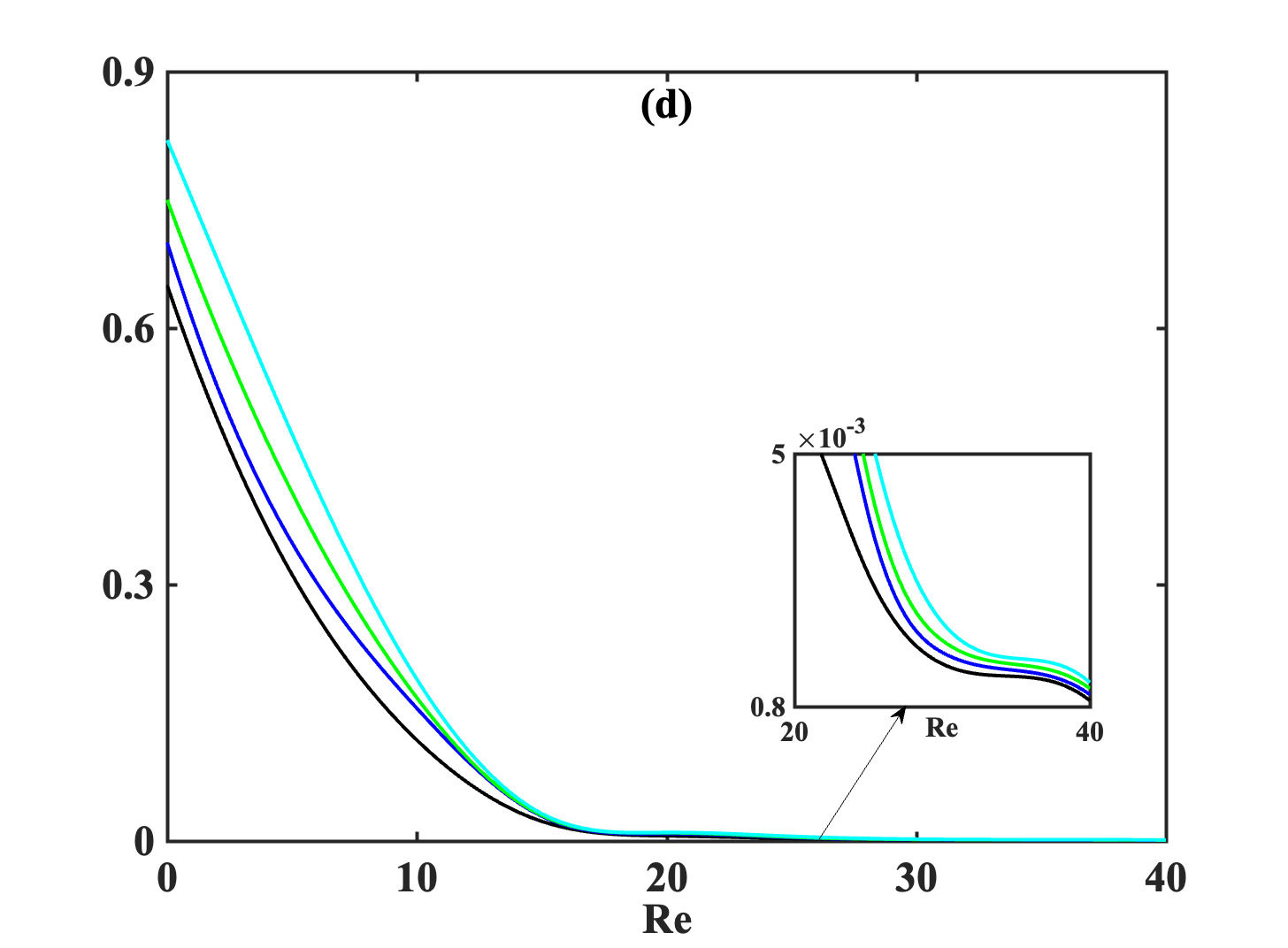}
\vskip 1pt
\includegraphics[width=0.495\linewidth, height=0.35\linewidth]{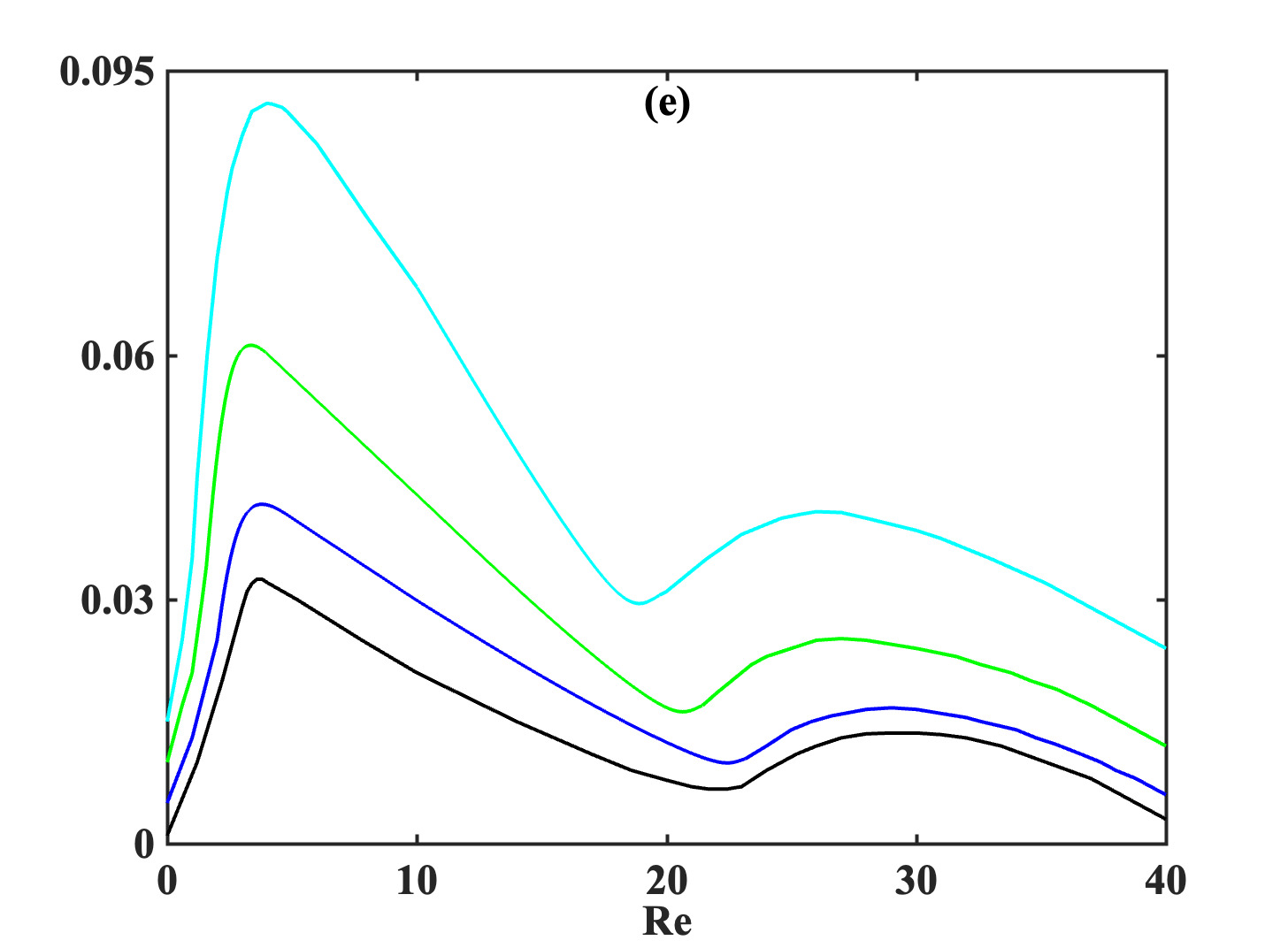}
\includegraphics[width=0.495\linewidth, height=0.35\linewidth]{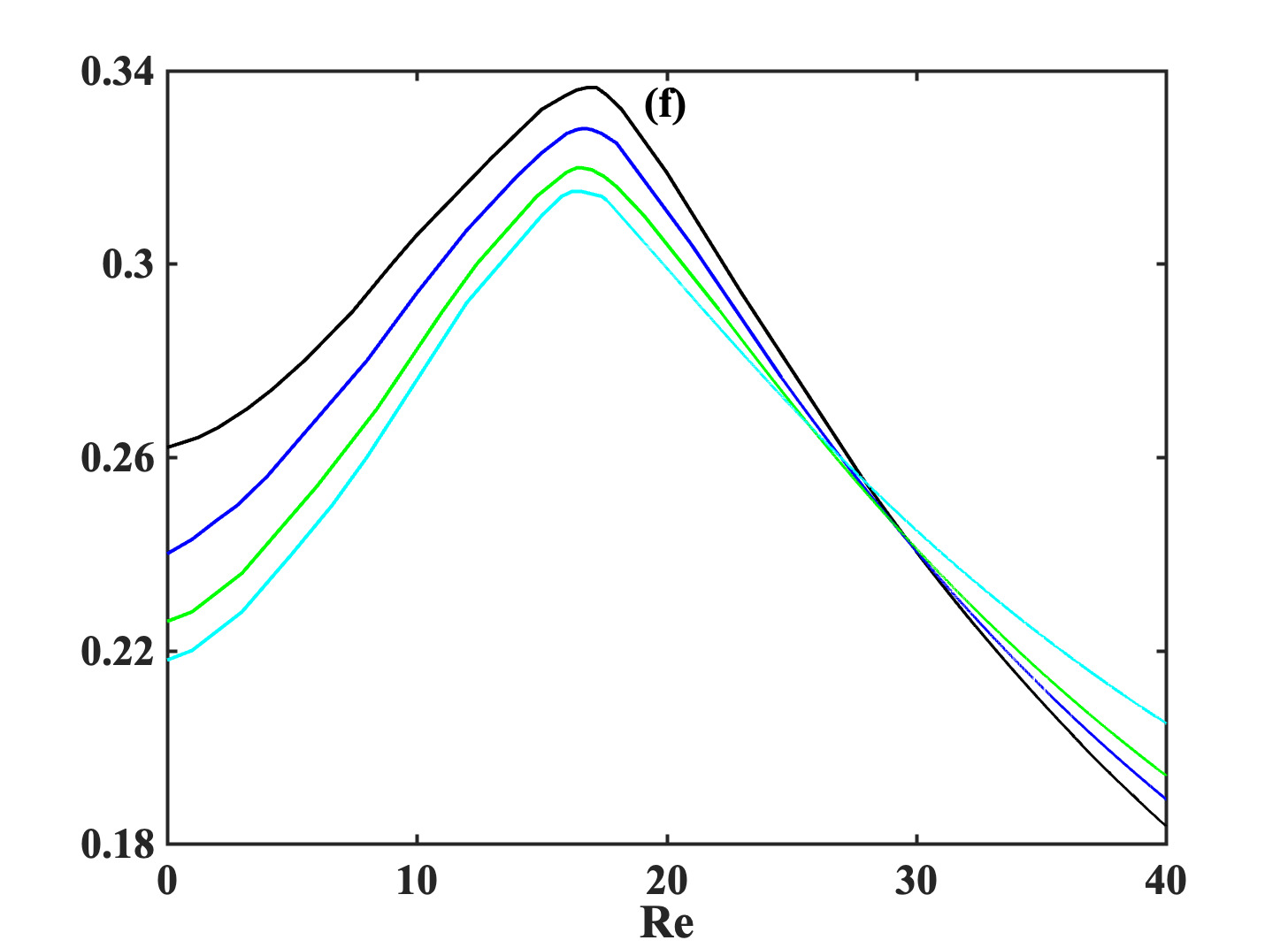}
\vskip 1pt
\includegraphics[width=0.495\linewidth, height=0.35\linewidth]{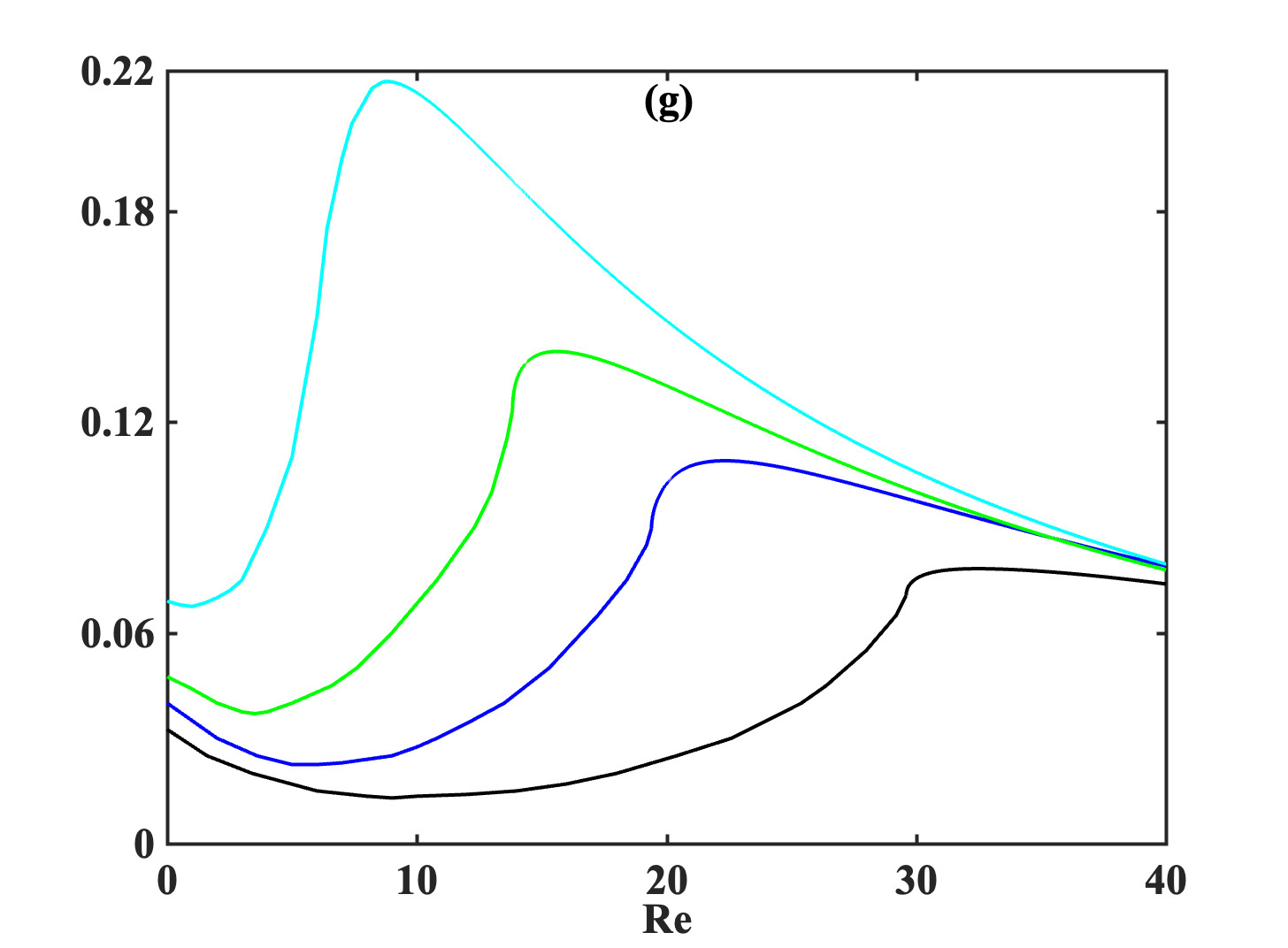}
\includegraphics[width=0.495\linewidth, height=0.35\linewidth]{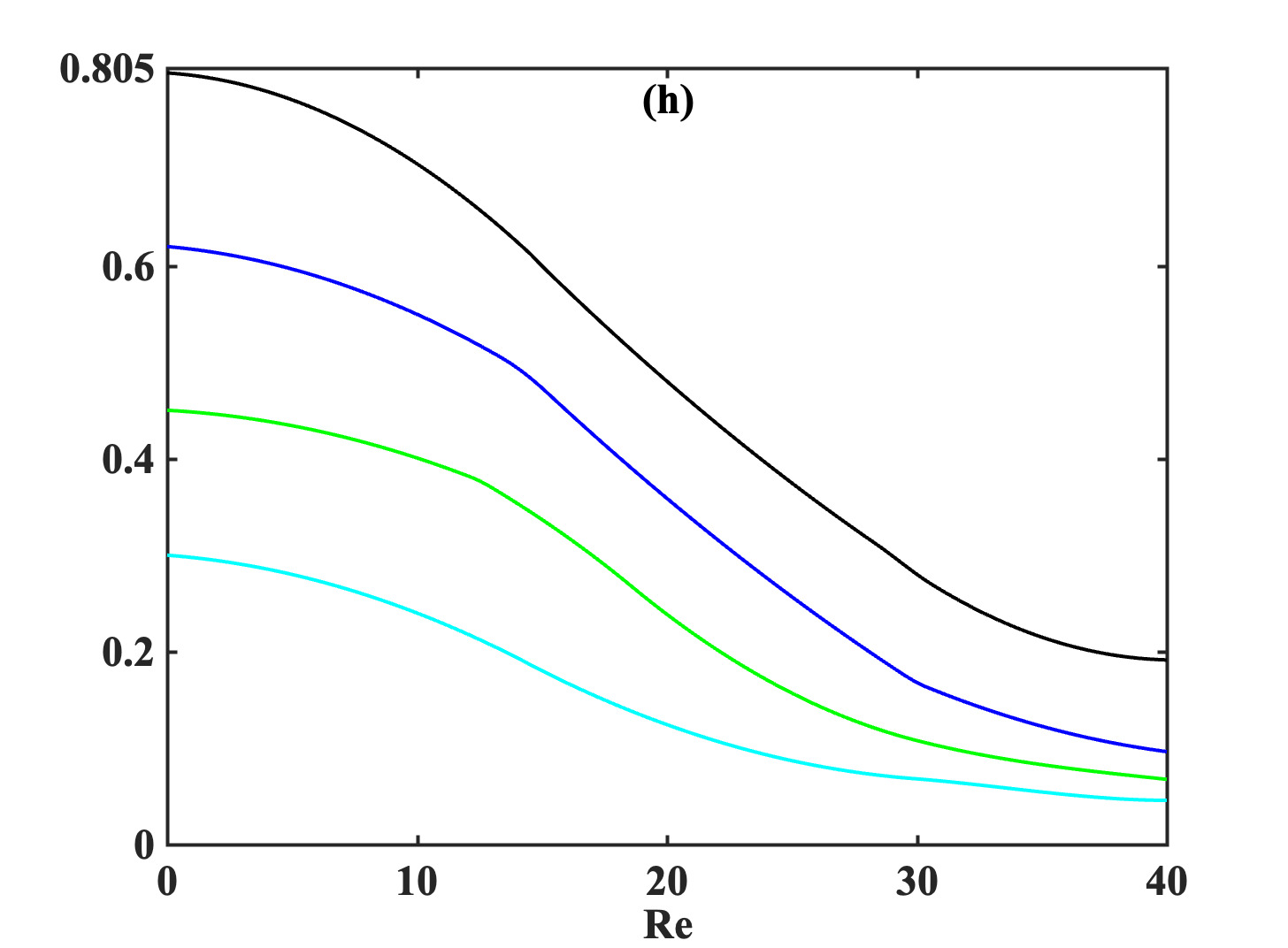}
\vskip 1pt
\caption{Amplitude of the velocity perturbation, $|f_r + i f_i|$ at the most unstable mode (refer figure~\ref{fig7}) vs $Re$ for $E=0.01$ (\protect\blackline), $E=0.015$ (\protect\blueline), $E=0.02$ (\protect\greenline) and $E=0.03$ (\protect\cyanline) evaluated at spatial locations: (a, b) $y=0.99$, (c, d) $y=0.5$, (e, f) $y=0.0$, (g, h) $y=-0.99$ and fixed values of viscosity ratio, $\nu=0.3$ (left column) and $\nu=0.9$ (right column).}
\label{fig10}
\eef

Figure~\ref{fig11} presents the amplitude of the velocity perturbation versus elasticity and at fixed values of the parameters, $Re$, $\nu$, and evaluated at the most unstable mode as shown in figure~\ref{fig8}. Notice, for the $Re$ values considered, the disturbance field is spread across the entire cross-section of the Hele-Shaw cell. Nevertheless, we find that (a) at low values of $E$ (i.~e., $E \le 0.005$), the disturbance amplitudes are suppressed with increasing $Re$ and (b) a similar observation prevails in the elastic stress dominated case, at high values of $E$ (i.~e., $E \ge 0.035$). Both the findings vindicate the stabilizing impact of elasticity within this Reynolds number regime.
\bef
\centering
\includegraphics[width=0.495\linewidth, height=0.35\linewidth]{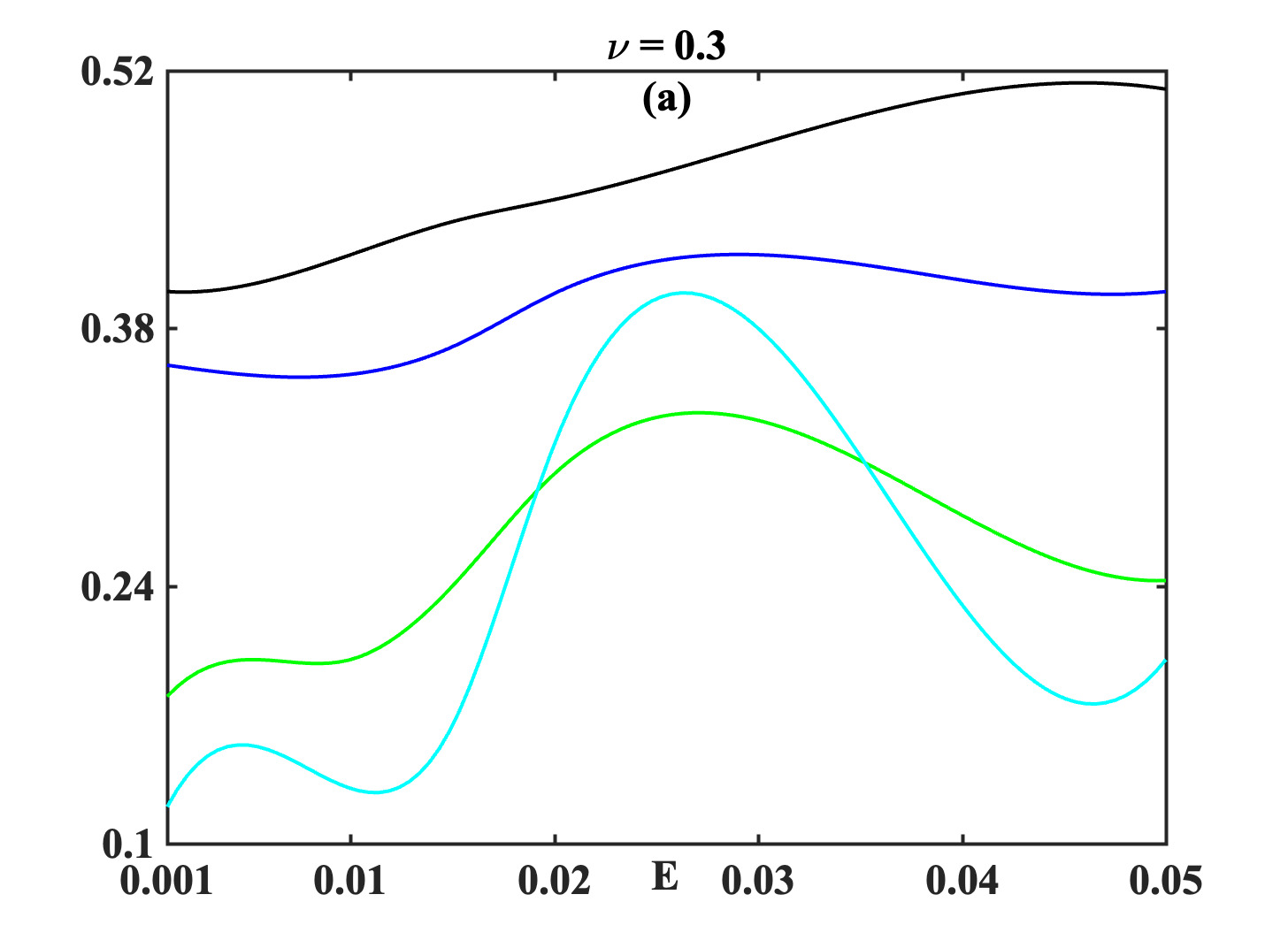}
\includegraphics[width=0.495\linewidth, height=0.35\linewidth]{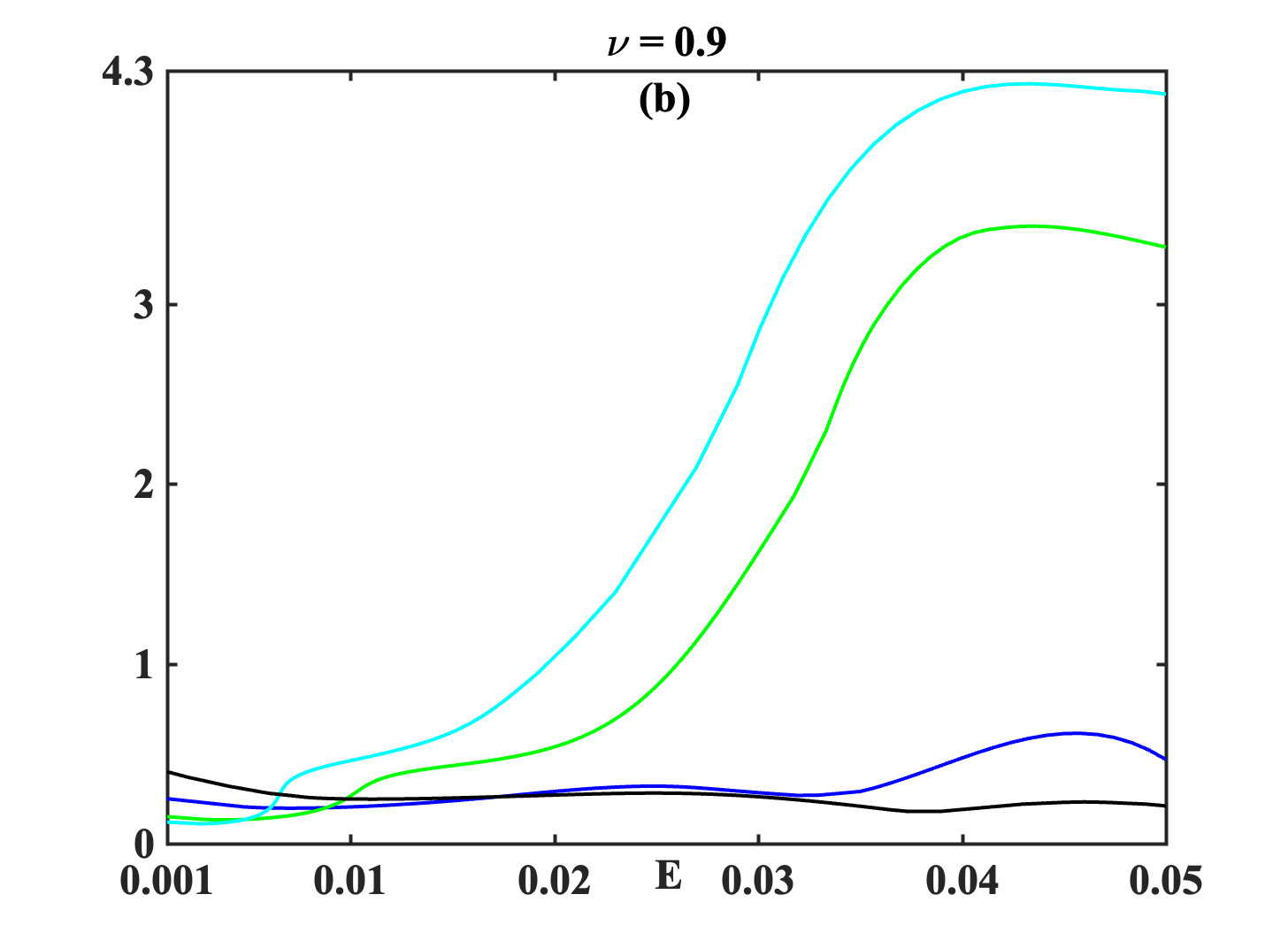}
\vskip 1pt
\includegraphics[width=0.495\linewidth, height=0.35\linewidth]{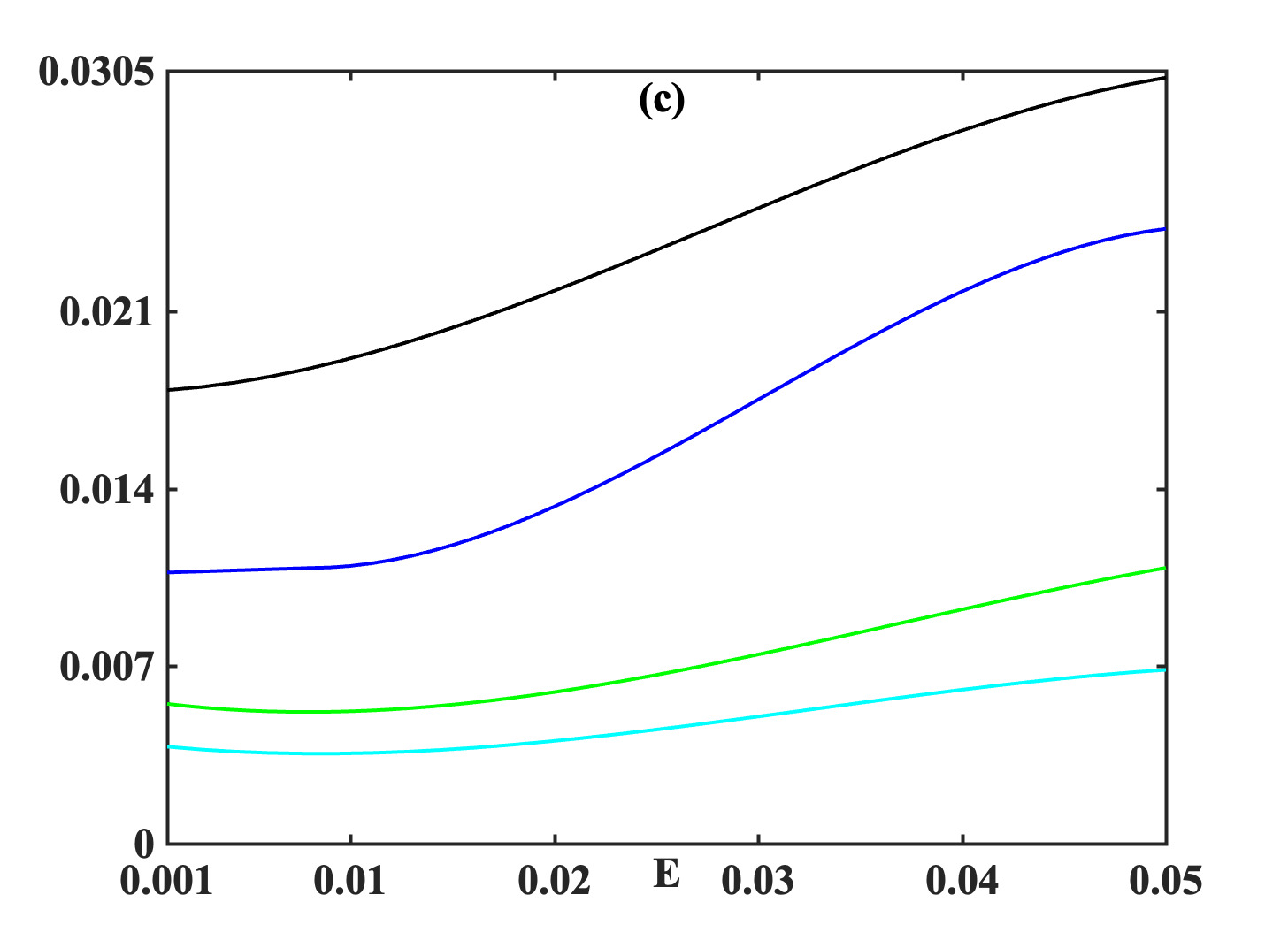}
\includegraphics[width=0.495\linewidth, height=0.35\linewidth]{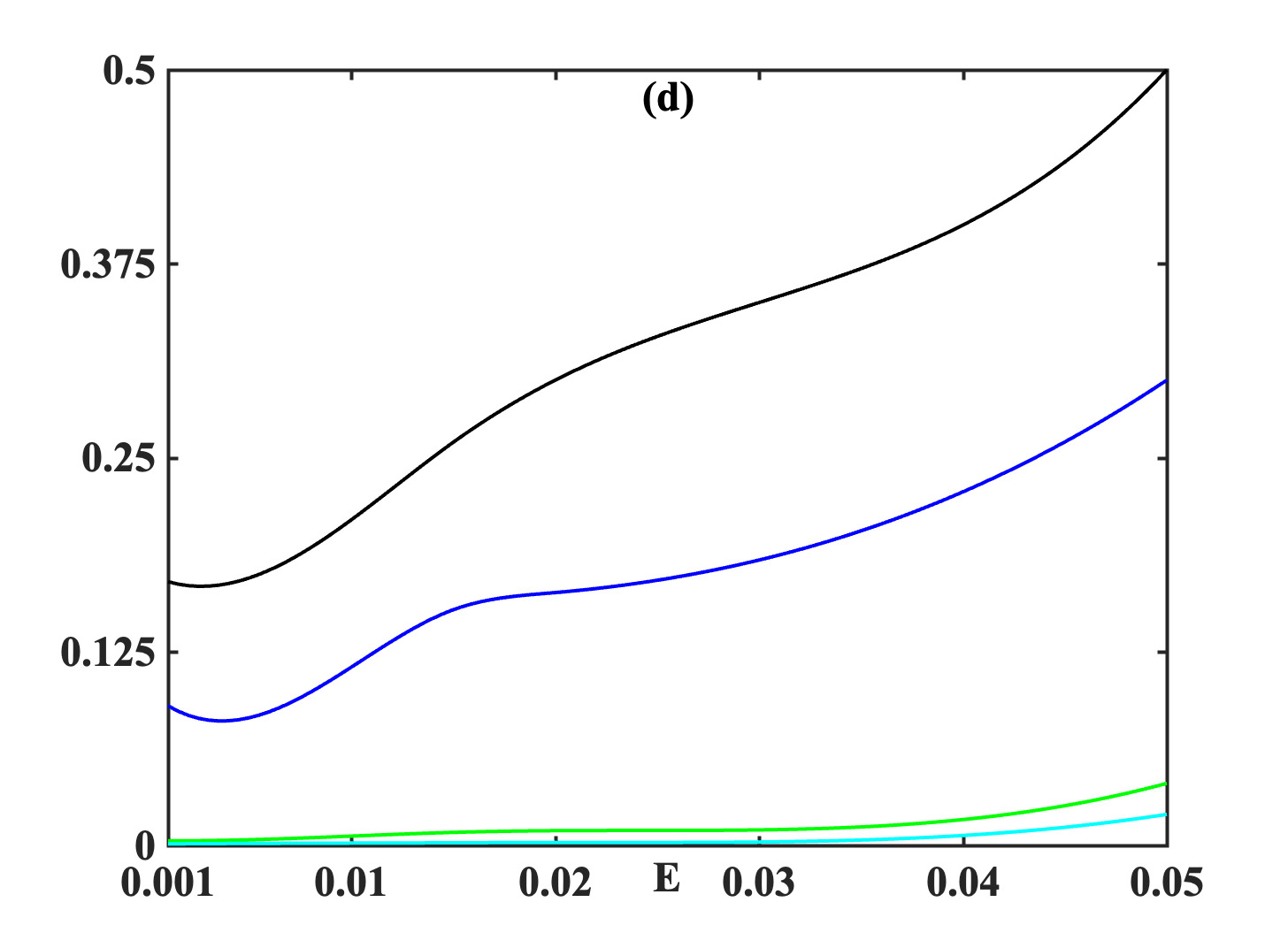}
\vskip 1pt
\includegraphics[width=0.495\linewidth, height=0.35\linewidth]{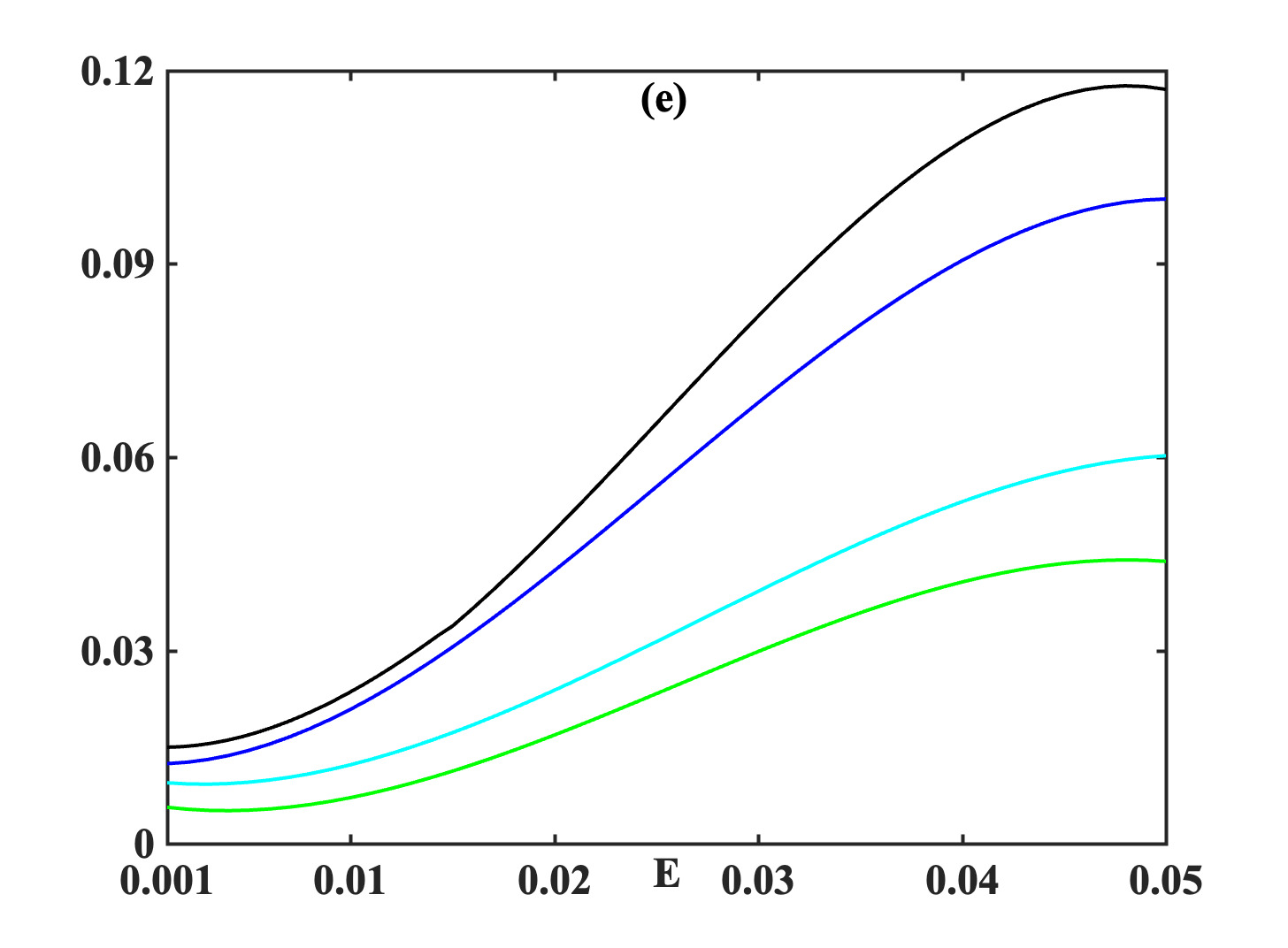}
\includegraphics[width=0.495\linewidth, height=0.35\linewidth]{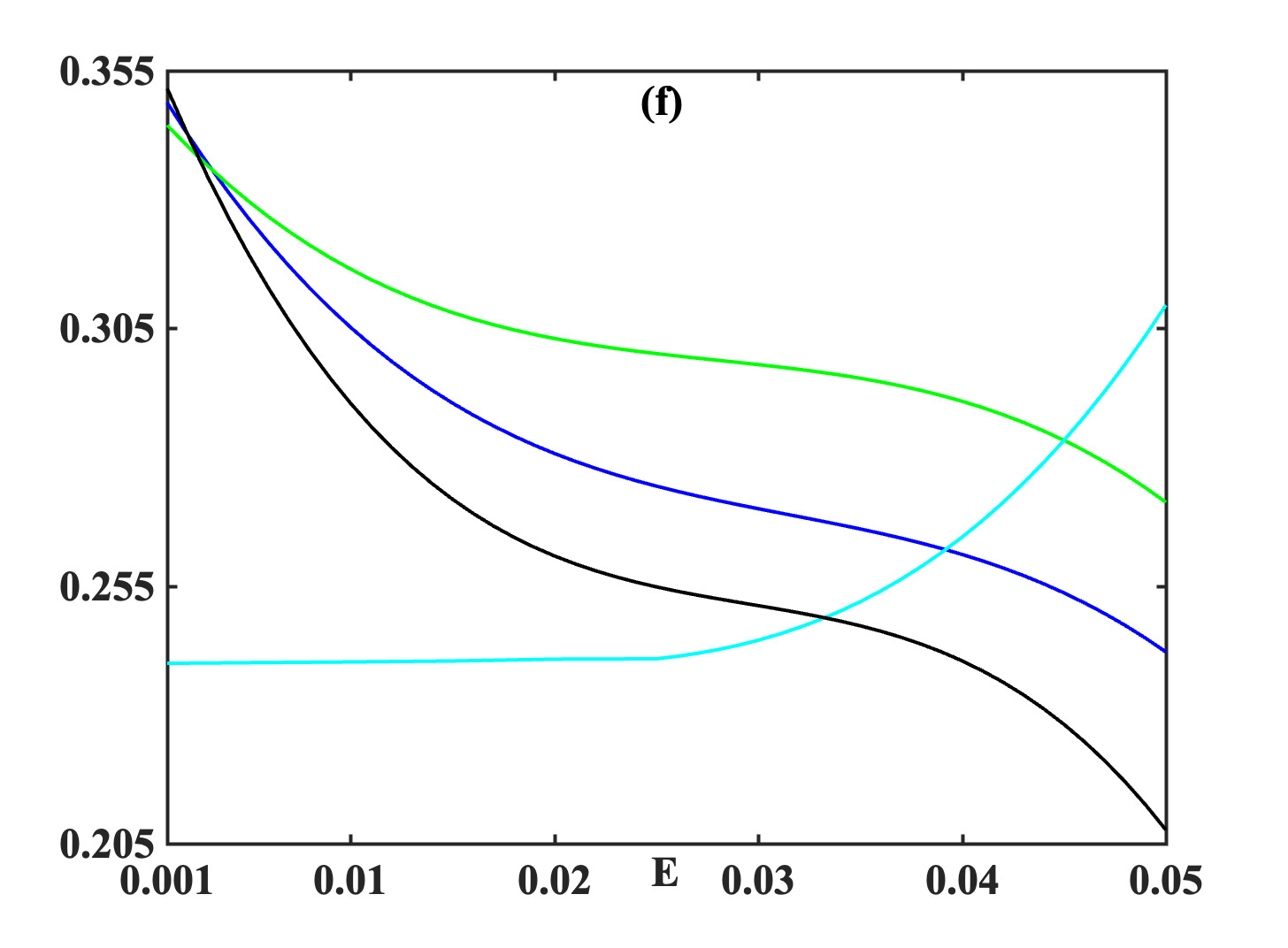}
\vskip 1pt
\includegraphics[width=0.495\linewidth, height=0.35\linewidth]{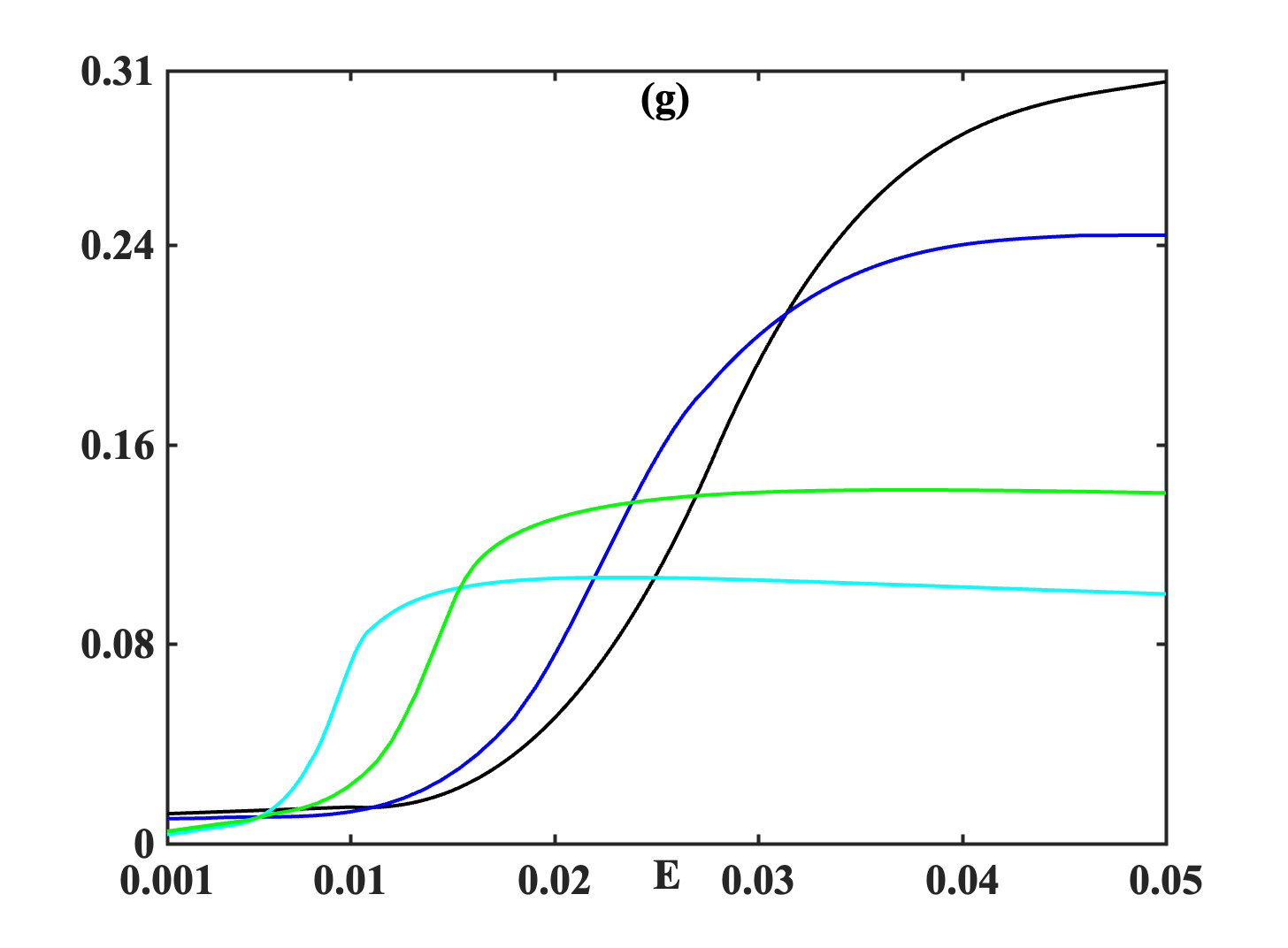}
\includegraphics[width=0.495\linewidth, height=0.35\linewidth]{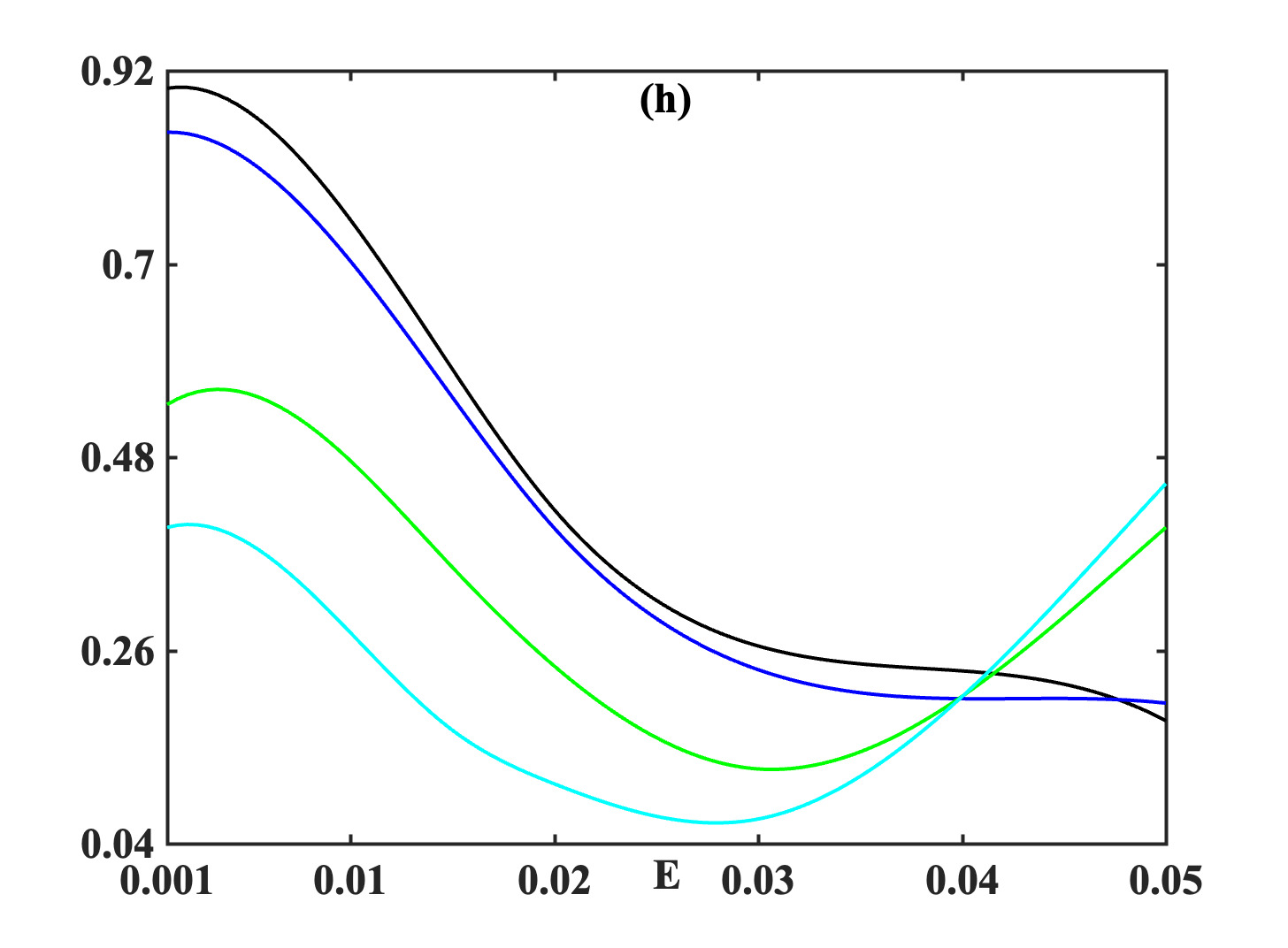}
\vskip 1pt
\caption{Amplitude of the velocity perturbation, $|f_r + i f_i|$ at the most unstable mode (refer figure~\ref{fig8}) vs $E$ for $Re=6.0$ (\protect\blackline), $Re=10.0$ (\protect\blueline), $Re=20.0$ (\protect\greenline) and $Re=30.0$ (\protect\cyanline) evaluated at spatial locations: (a, b) $y=0.99$, (c, d) $y=0.5$, (e, f) $y=0.0$, (g, h) $y=-0.99$ and at fixed values of viscosity ratio, $\nu=0.3$ (left column) and $\nu=0.9$ (right column).}
\label{fig11}
\eef

Finally, figure~\ref{fig12} highlights the velocity perturbation amplitude versus the viscosity ratio and at fixed values of $E, Re$, computed at the most unstable mode as depicted in figure~\ref{fig9}. A notable observation in these disturbance amplitude plots is the appearance of dual maxima near the wall (i.~e., the presence of two peaks at $\nu \approx 0.25, 0.8$ at $y=0.99$ (figure~\ref{fig12} a,b) and at $\nu \approx 0.01, 0.9$ at $y=-0.99$ (figure~\ref{fig12} g,h)) reminding the readers that the anisotropy of the elastic stresses can play a significant role in dictating the dynamics of the instabilities at the flow interface~\citep{Shokri2017}. Although we do not further explore the role of the elastic stresses in instigating the flow instabilities, in the next section, we outline a deeper characterization of these instabilities via the spatiotemporal analysis, to identify the (viable) regions of topological transition of the advancing interface~\citep{Lee2002}.
\bef
\centering
\includegraphics[width=0.495\linewidth, height=0.35\linewidth]{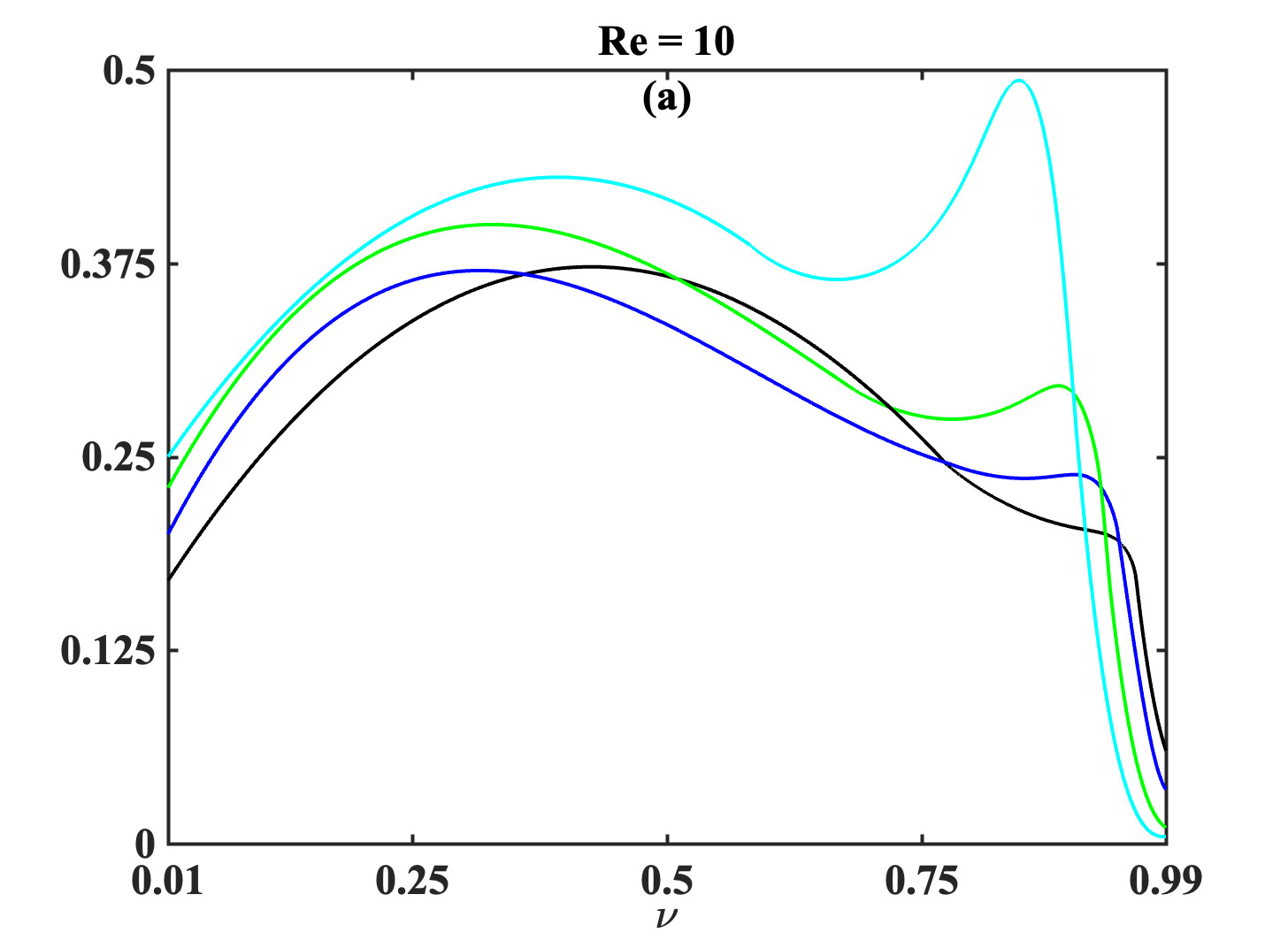}
\includegraphics[width=0.495\linewidth, height=0.35\linewidth]{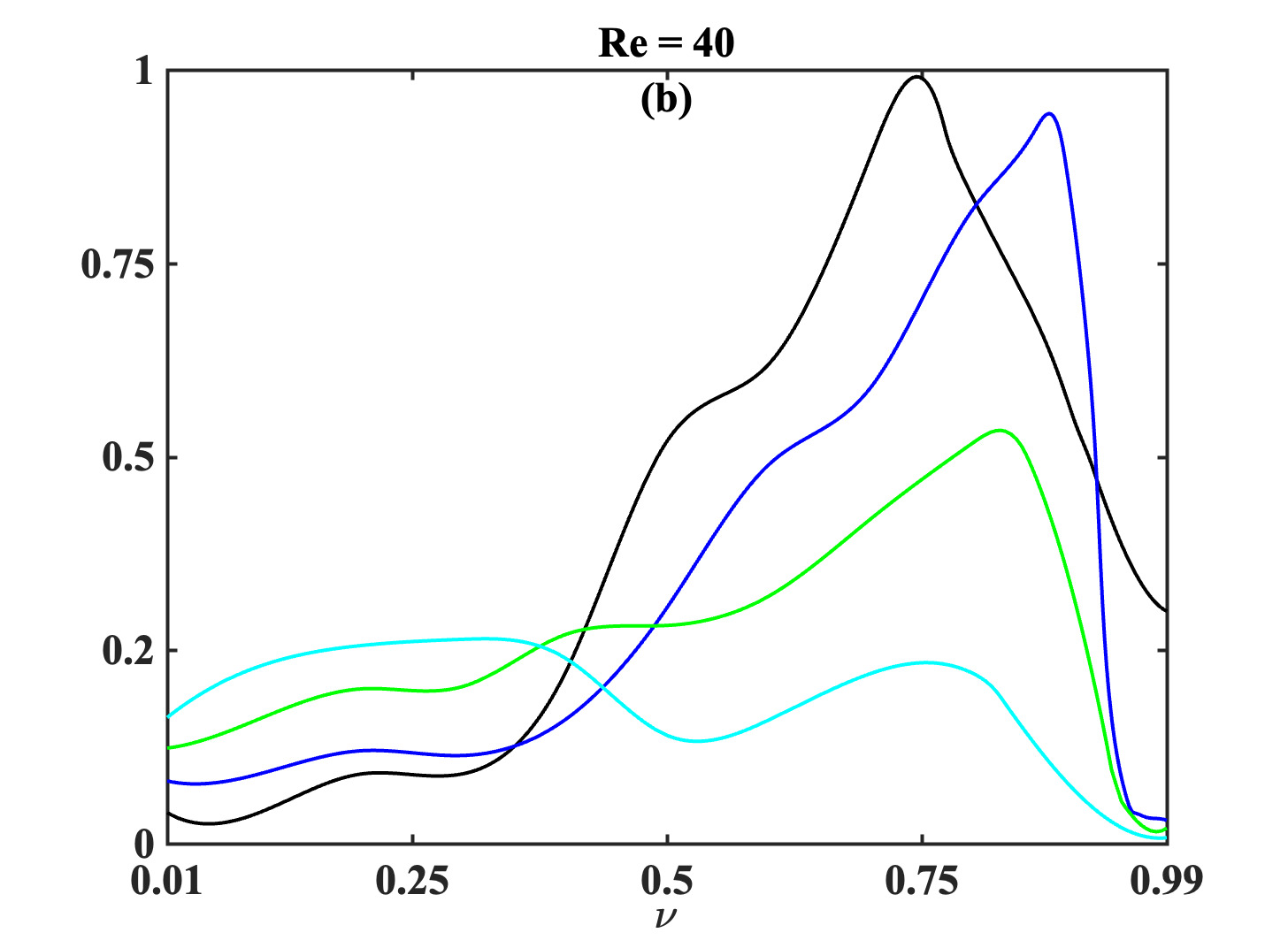}
\vskip 1pt
\includegraphics[width=0.495\linewidth, height=0.35\linewidth]{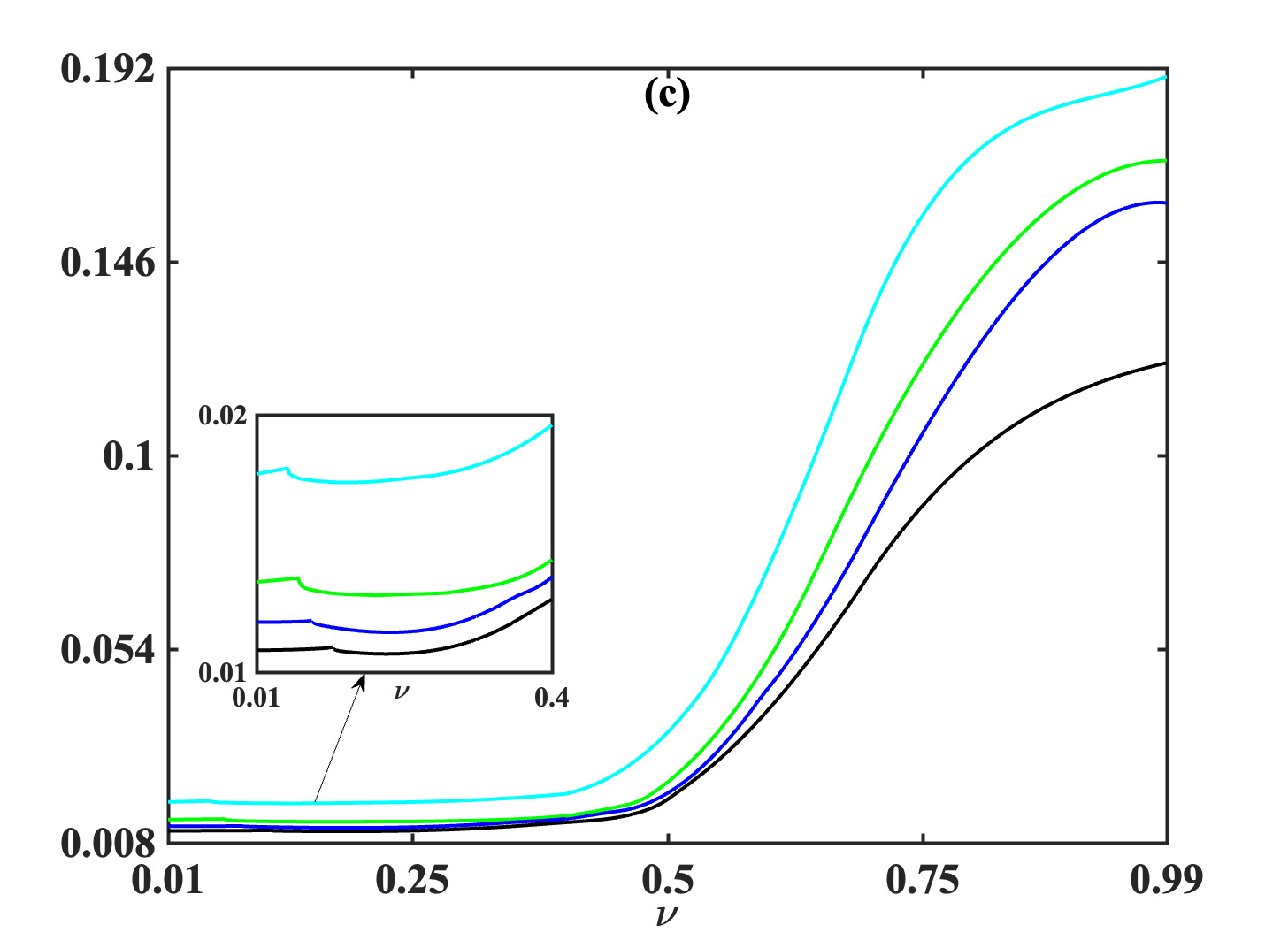}
\includegraphics[width=0.495\linewidth, height=0.35\linewidth]{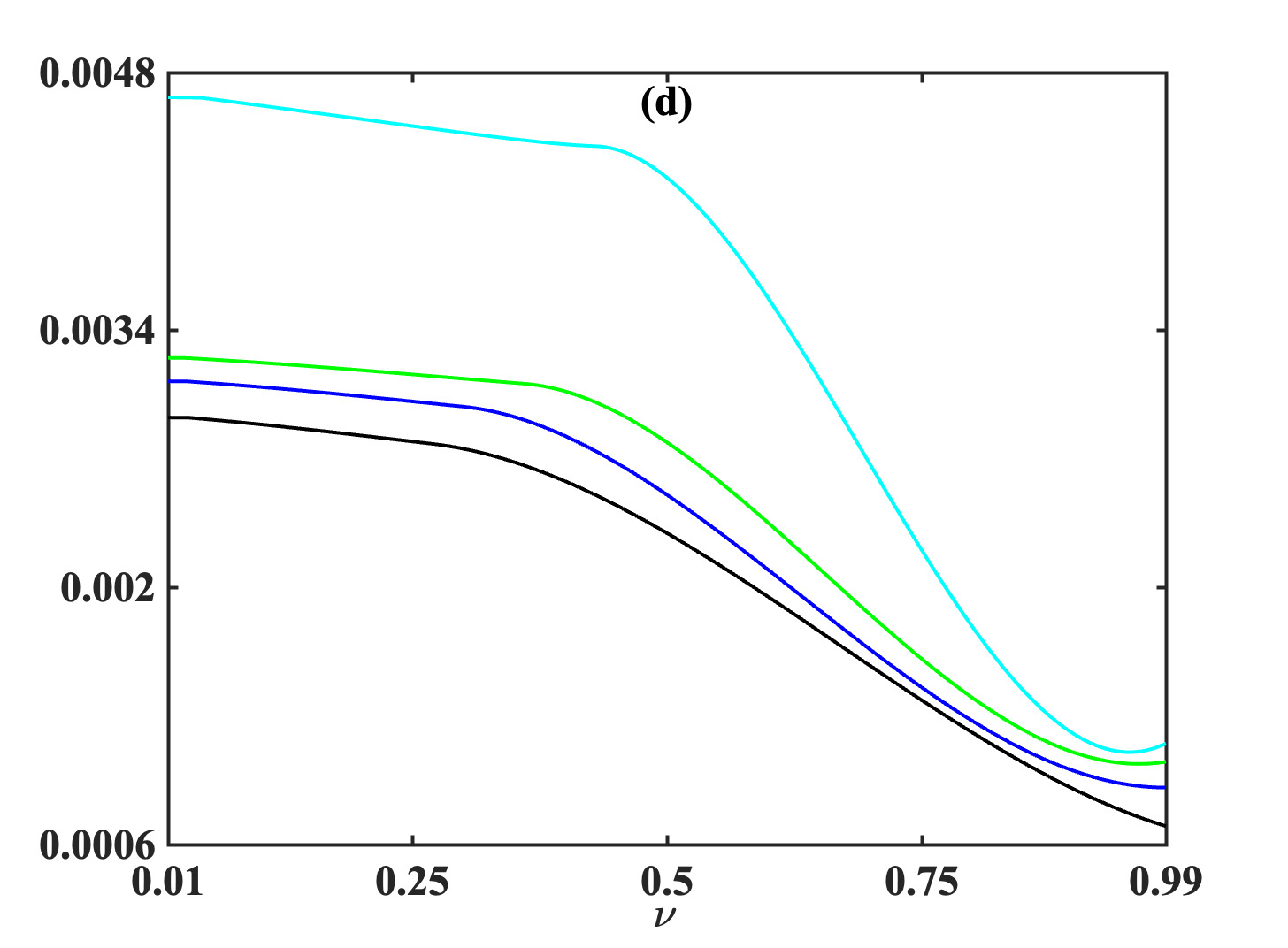}
\vskip 1pt
\includegraphics[width=0.495\linewidth, height=0.35\linewidth]{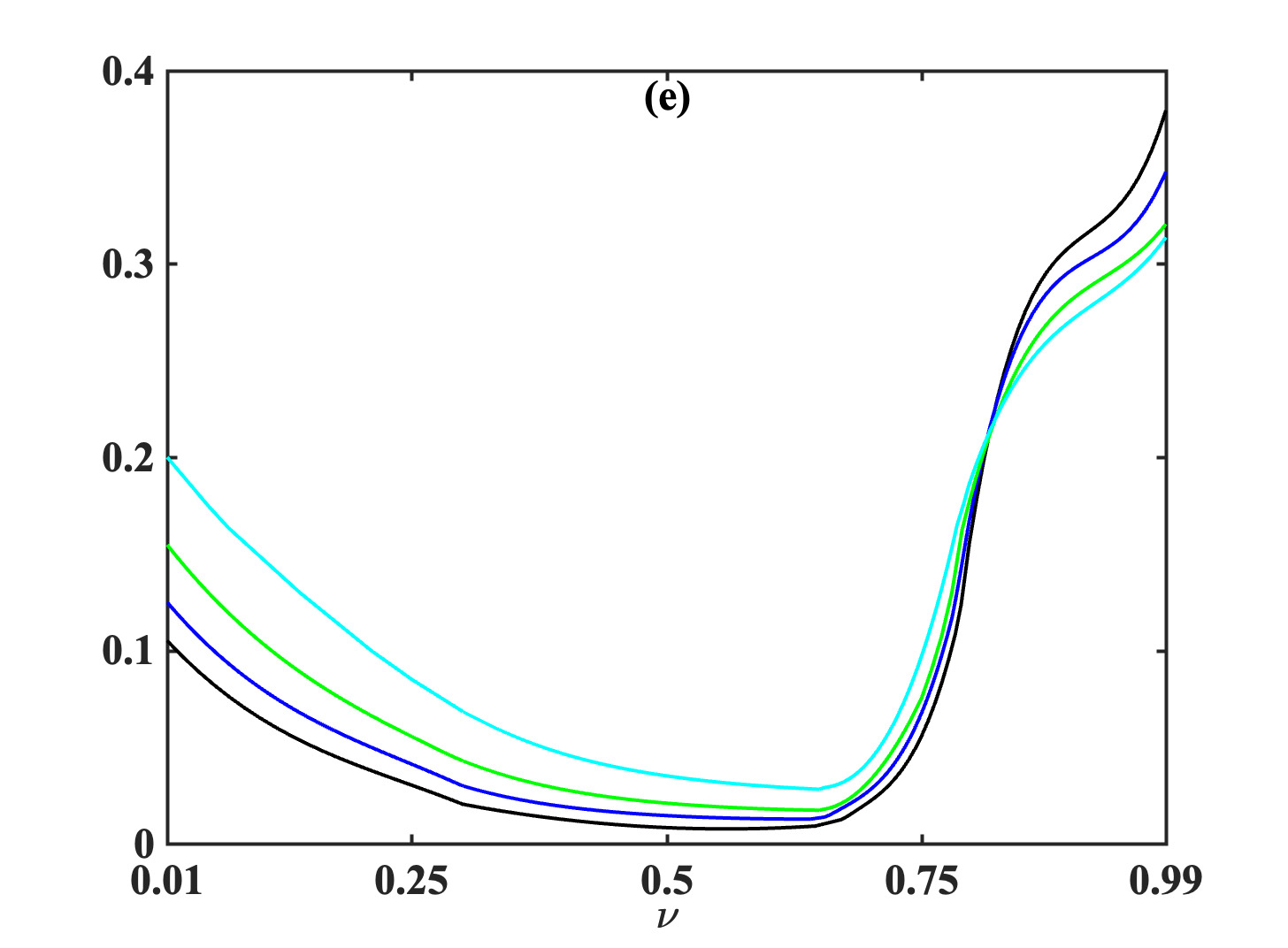}
\includegraphics[width=0.495\linewidth, height=0.35\linewidth]{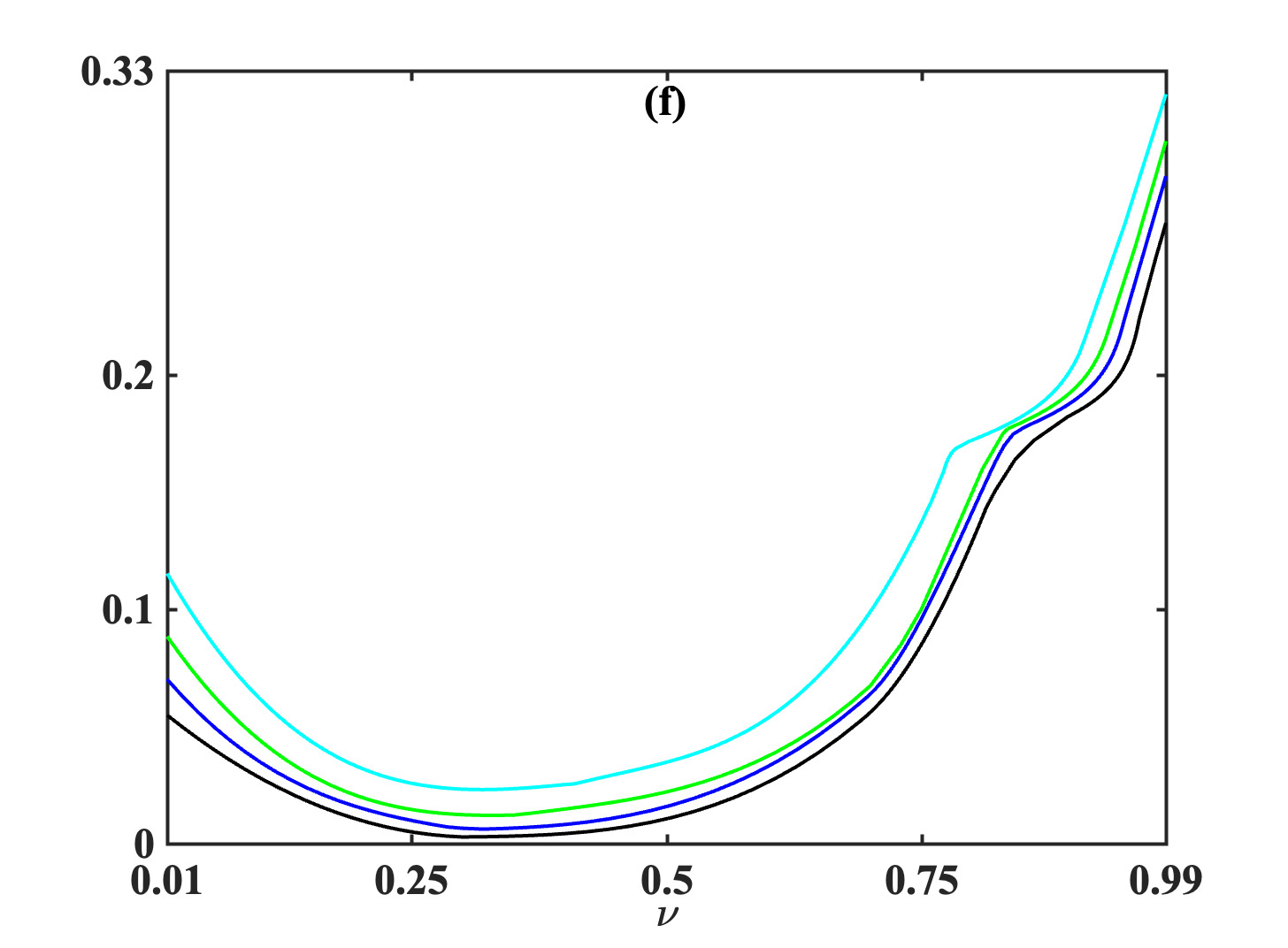}
\vskip 1pt
\includegraphics[width=0.495\linewidth, height=0.35\linewidth]{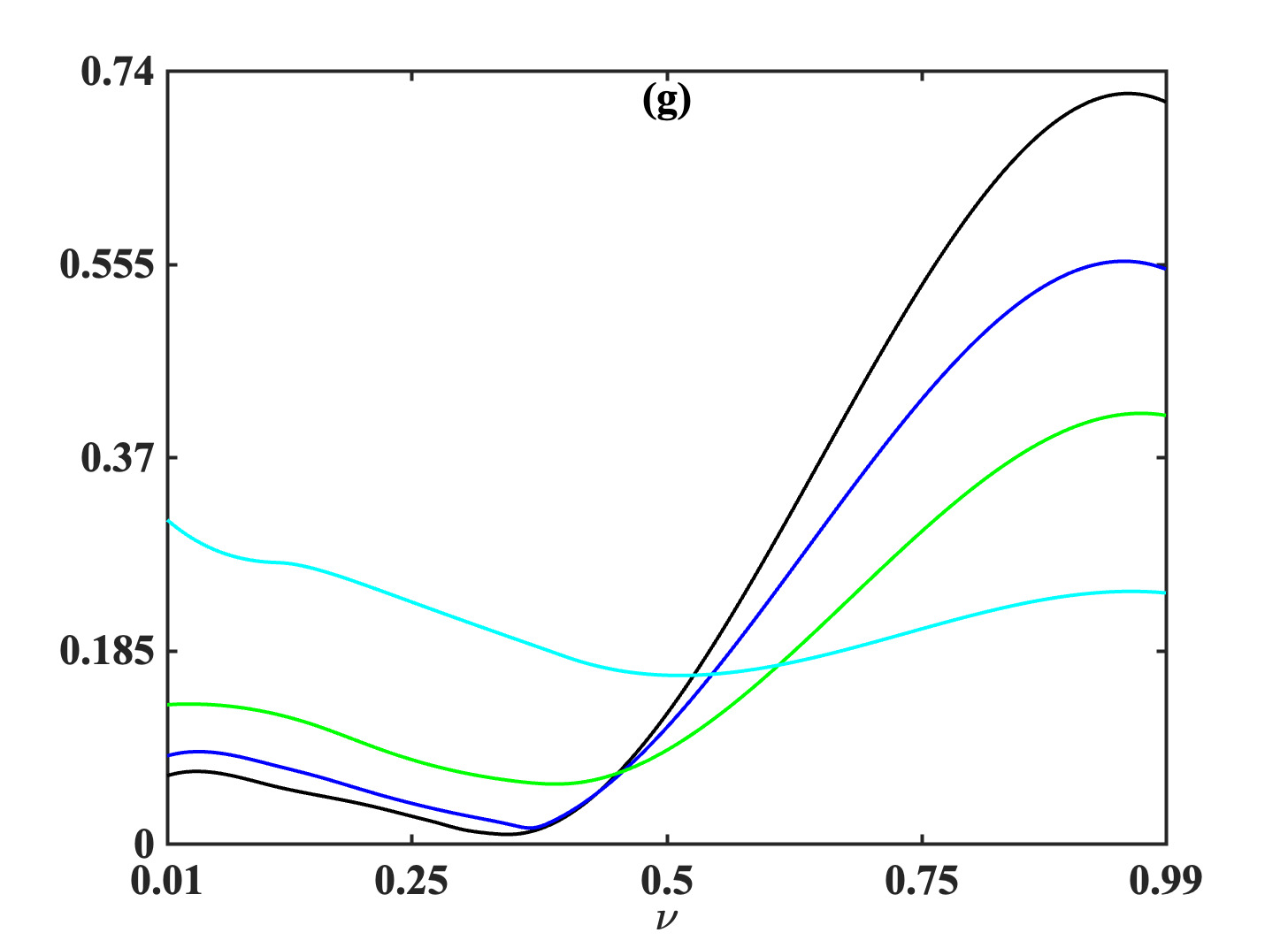}
\includegraphics[width=0.495\linewidth, height=0.35\linewidth]{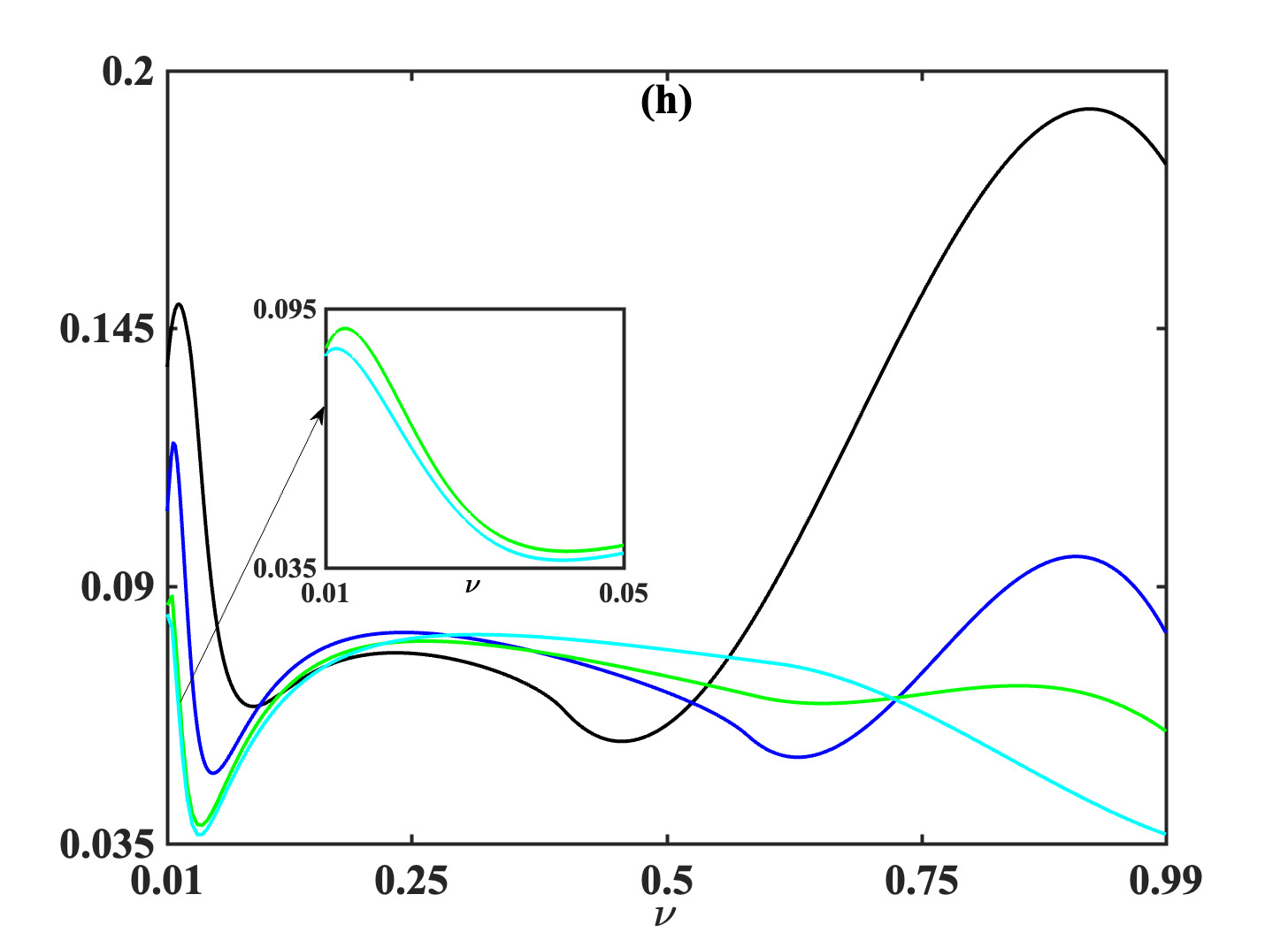}
\vskip 1pt
\caption{Amplitude of the velocity perturbation, $|f_r + i f_i|$ at the most unstable mode (refer figure~\ref{fig9}) vs $\nu$ at $E=0.01$ (\protect\blackline), $E=0.015$ (\protect\blueline), $E=0.02$ (\protect\greenline) and $E=0.03$ (\protect\cyanline) and evaluated at spatial locations: (a, b) $y=0.99$, (c, d) $y=0.5$, (e, f) $y=0.0$, (g, h) $y=-0.99$ and at fixed values of Reynolds number, $Re=10.0$ (left column) and $Re=40.0$ (right column).}
\label{fig12}
\eef

\section{Spatiotemporal stability analysis}\label{sec:stsa}
Spatiotemporal analysis is typically relevant when one introduces an impulse excitation locally in a flow and observes how that disturbance evolves with time~\citep{Huerre1985}. More significantly, we evaluate the absolute growth rate (or $\omega^{\text{cusp}}_i$, details on computing these points are elaborated in \S \ref{subsec:NM}) to identify the region of absolute instability, or the region indicating the topological reconfiguration and subsequent pinch-off of the advancing interface~\citep{Goldstein1993}. However, evanescent modes are also encountered in our analysis~\citep{Sircar2019,Bansal2021}. These modes do not merely depend on the sign of the absolute growth rate and have to be found via the sufficient conditions proposed by~\citet{Briggs1964} (refer \S \ref{subsec:NM}). Detailed discussion of the evanescent modes are provided in the description of the phase diagrams (figures~\ref{fig16}, \ref{fig17}, \ref{fig18}).

As a first step, in figure~\ref{fig13} we present the absolute growth rate in an effort to determine the range of $Re$ (at fixed $E$ and $\nu$) for which the flow regimes are absolutely or convectively unstable. For example, the regions of absolute instability are those demarcated with high Reynolds number (i.~e., $Re \ge 35.0$) and higher values of elasticity number (i.~e., $E \ge 0.015$) as well as the low Reynolds number ($Re \le 2.0$) and lower values of elasticity number (i.~e., $E \le 0.02$) regime. Clearly, while the presence of the former region is the outcome of the fluid inertia, the presence of the latter is the result of the instability generated via polymer elasticity. In contrariety, convectively unstable regions exist at intermediate to high Reynolds number (i.~e., $Re \ge 25.0$) and low elasticity number (i.~e., $E \le 0.01$), near the wall (refer figures~\ref{fig13} a,b,g). At locations which are farther away from the wall, convective instability subsists at intermediate values of $Re$ (i.~e., $5.0 < Re < 15.0$) and for elasticity number values, $E \le 0.02$ (refer figures~\ref{fig13} e,f).
\bef
\centering
\includegraphics[width=0.495\linewidth, height=0.35\linewidth]{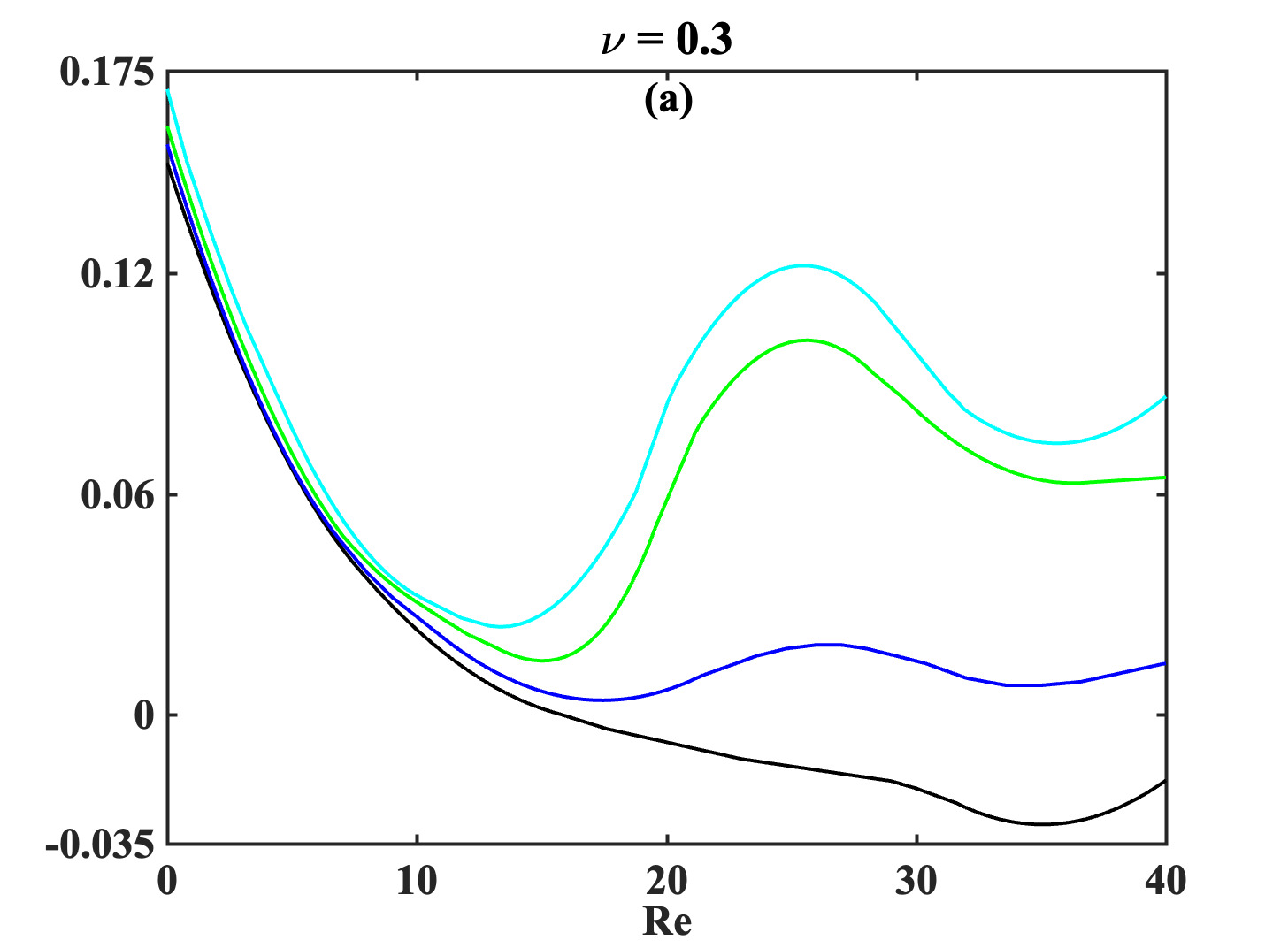}
\includegraphics[width=0.495\linewidth, height=0.35\linewidth]{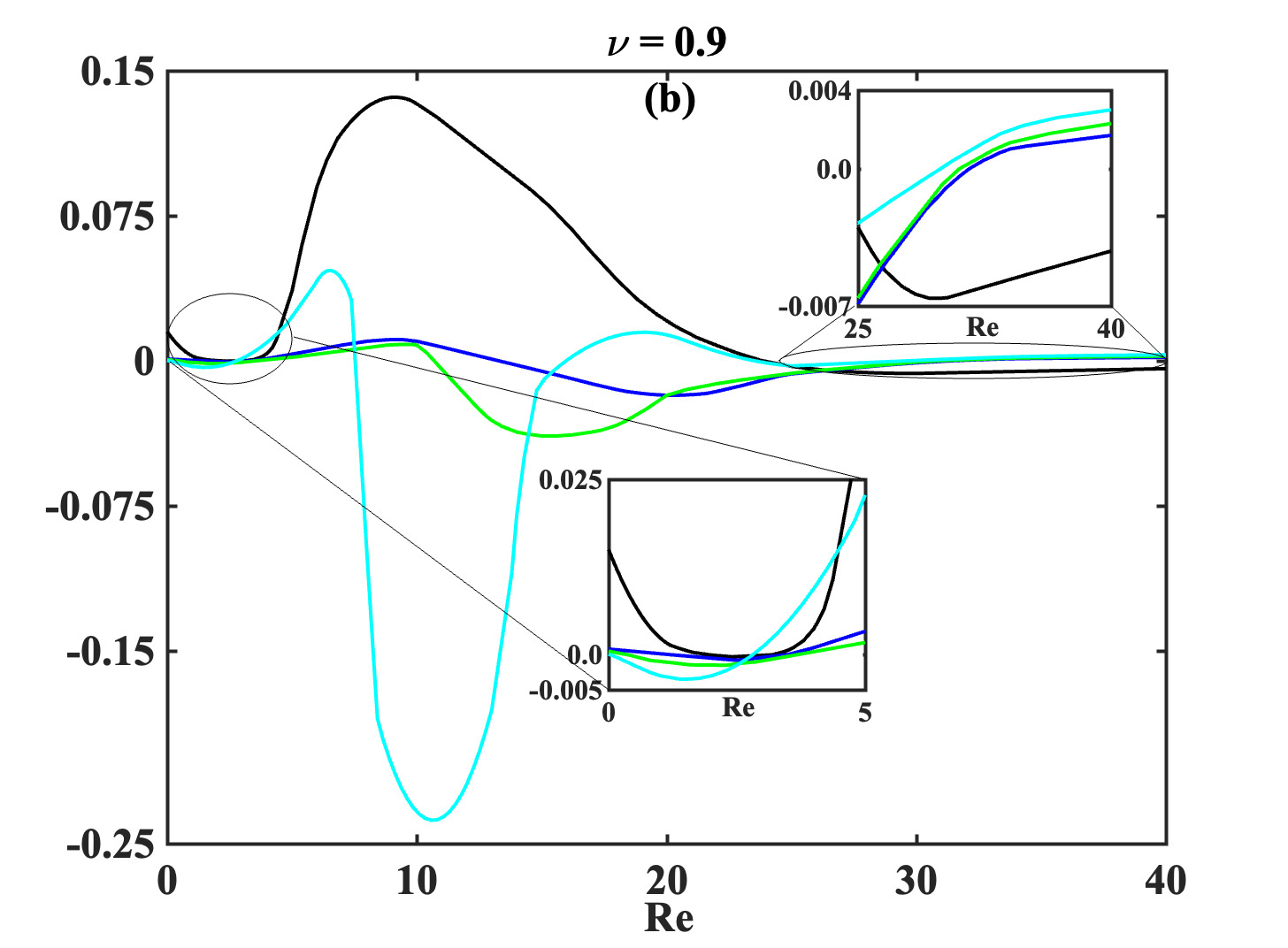}
\vskip 1pt
\includegraphics[width=0.495\linewidth, height=0.35\linewidth]{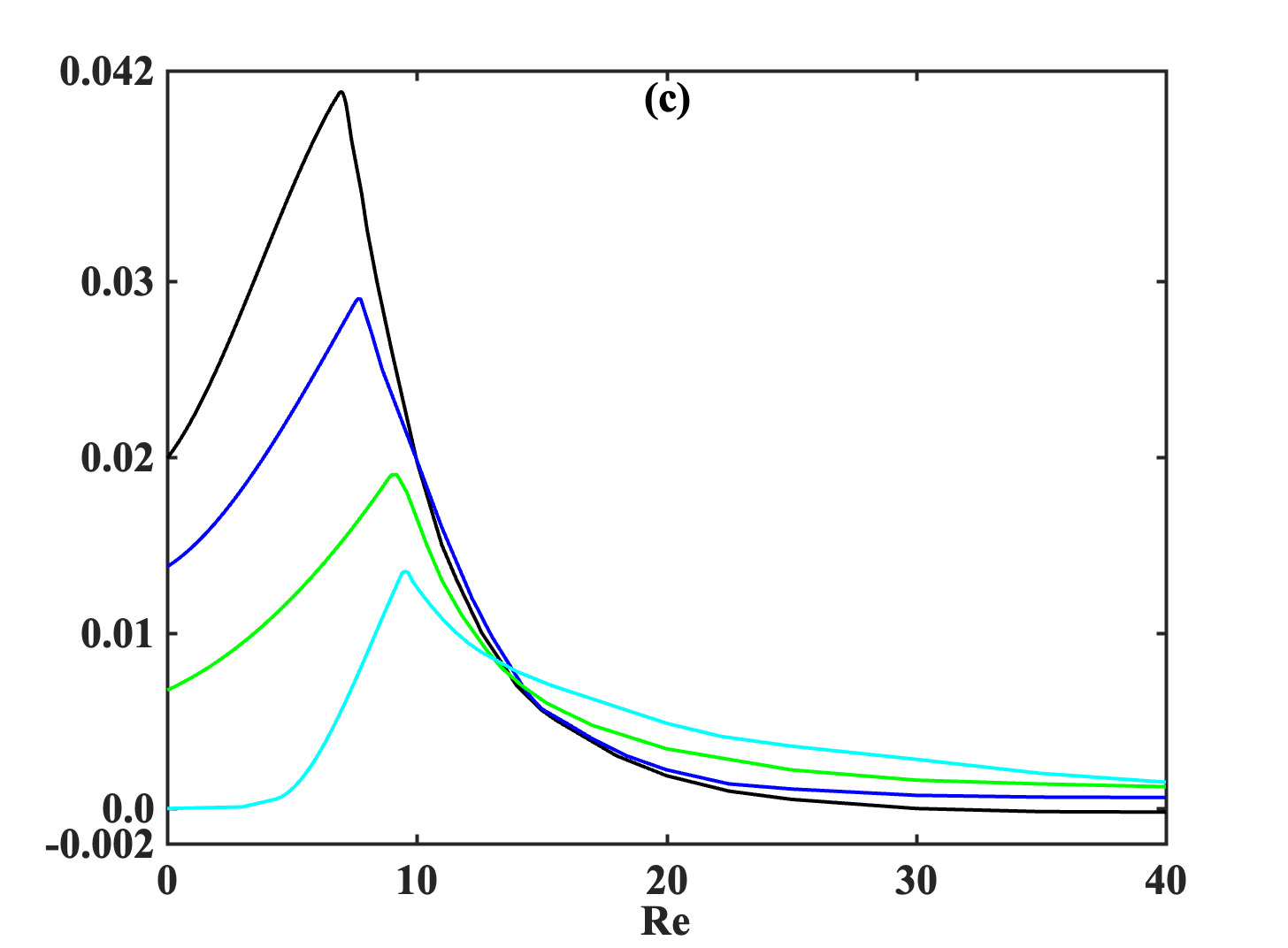}
\includegraphics[width=0.495\linewidth, height=0.35\linewidth]{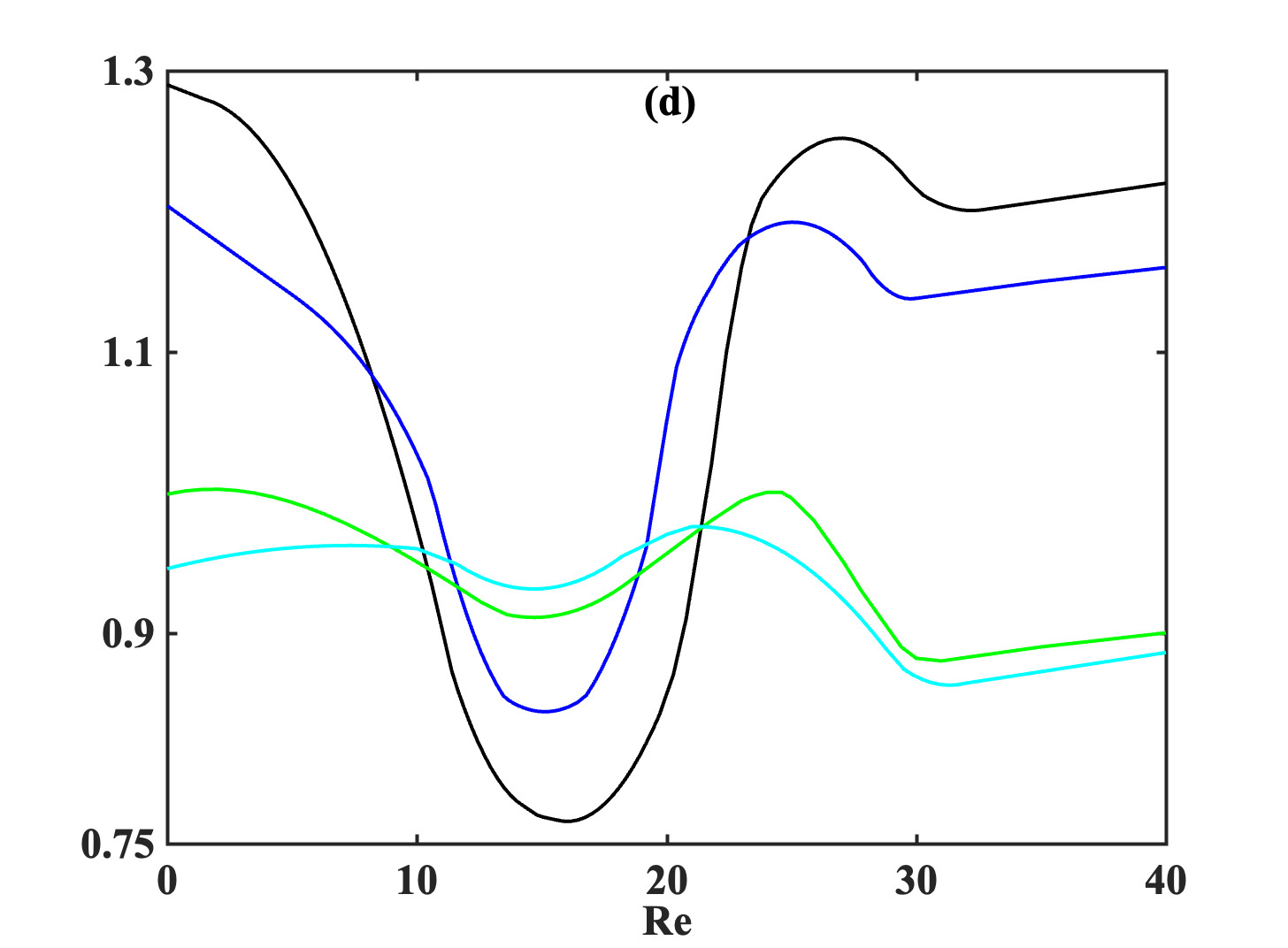}
\vskip 1pt
\includegraphics[width=0.495\linewidth, height=0.35\linewidth]{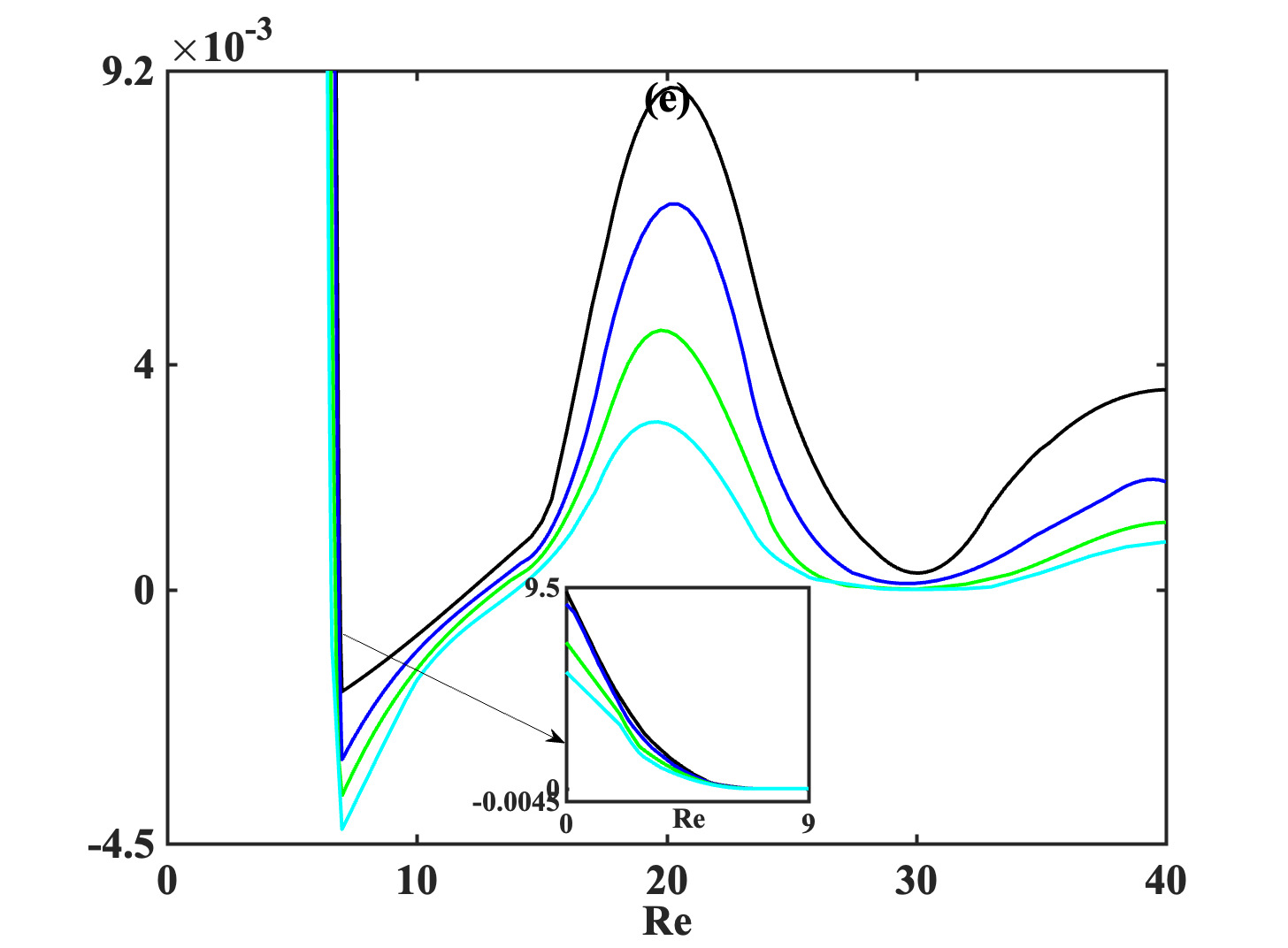}
\includegraphics[width=0.495\linewidth, height=0.35\linewidth]{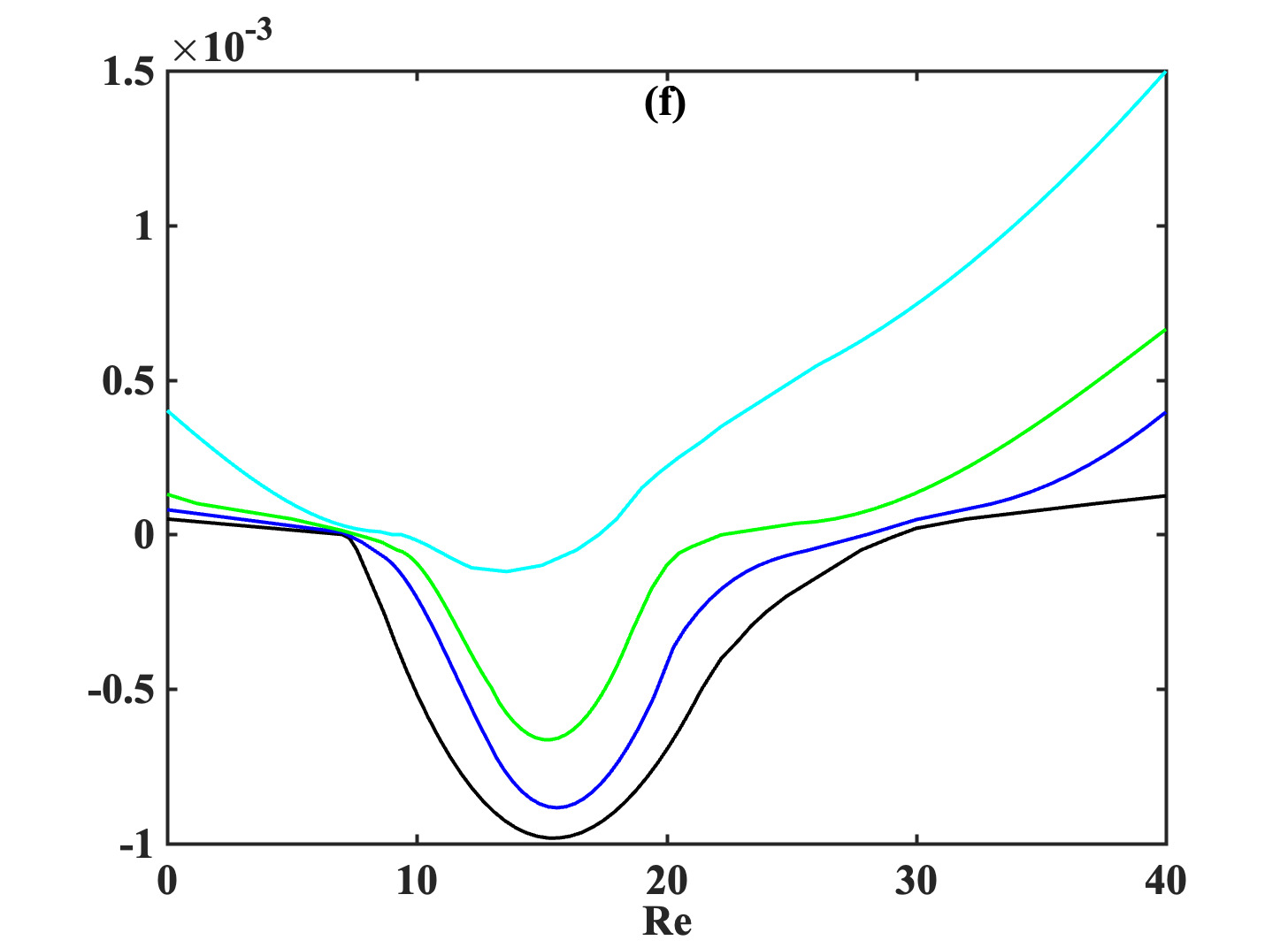}
\vskip 1pt
\includegraphics[width=0.495\linewidth, height=0.35\linewidth]{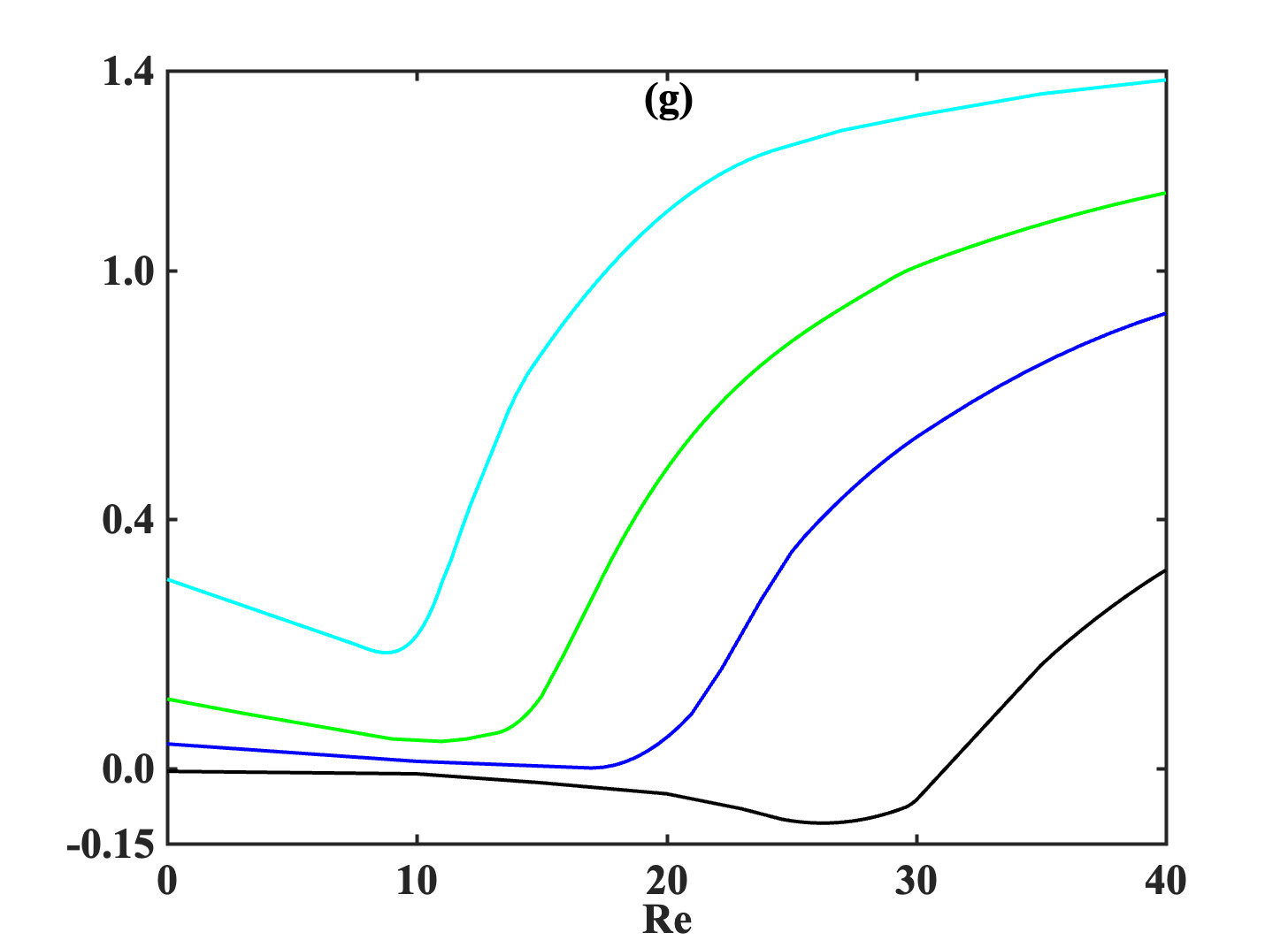}
\includegraphics[width=0.495\linewidth, height=0.35\linewidth]{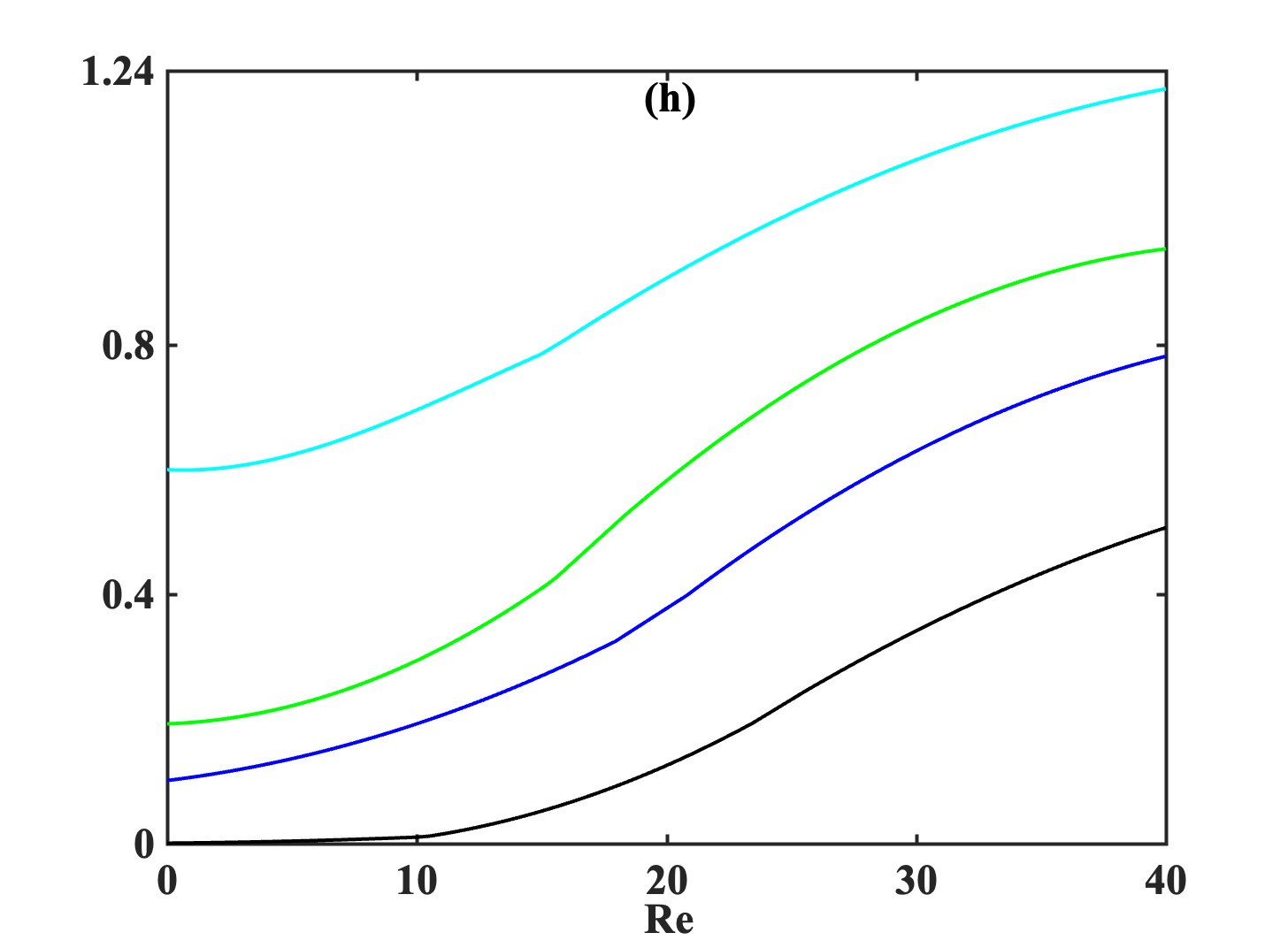}
\vskip 1pt
\caption{Cusp point, $\omega_i^{\text{cusp}}$ vs $Re$ at $E=0.01$ (\protect\blackline), $E=0.015$ (\protect\blueline), $E=0.02$ (\protect\greenline) and $E=0.03$ (\protect\cyanline) and evaluated at spatial locations: (a, b) $y=0.99$, \newline (c, d) $y=0.5$, (e, f) $y=0.0$, (g, h) $y=-0.99$ and at fixed values of viscosity ratio, $\nu=0.3$ (left column) and $\nu=0.9$ (right column).}
\label{fig13}
\eef

Another perspective of the absolute growth rate curves versus $E$, at fixed $Re$ and $\nu$ (figure~\ref{fig14}), provides a polished range of the flow-material parameters where the advancing interface display absolute instability. Notice that above the centerline, the region of absolute instability is found at lower values of Reynolds number (i.~e., $Re \le 10.0$) and low elasticity number (i.~e., $E \le 0.02$, refer figures~\ref{fig14} a-d); while, on and below the centerline, this region shifts at higher values of Reynolds number and intermediate-to-high values of elasticity number (i.~e., $Re \ge 20.0, E \ge 0.02$, refer figures~\ref{fig14} e-h). Conversely, the convectively unstable region appears at higher values of Reynolds number and intermediate-to-high values of elasticity number (i.~e., $Re \ge 20.0, E \ge 0.02$) above the centerline, and low values of Reynolds and elasticity number (i.~e., $Re \le 10.0, E \le 0.02$) below the centerline. We attribute this complete reversal of the nature of these instabilities, across the centerline, due to the anisotropy of the elastic stresses.
\bef
\centering
\includegraphics[width=0.495\linewidth, height=0.35\linewidth]{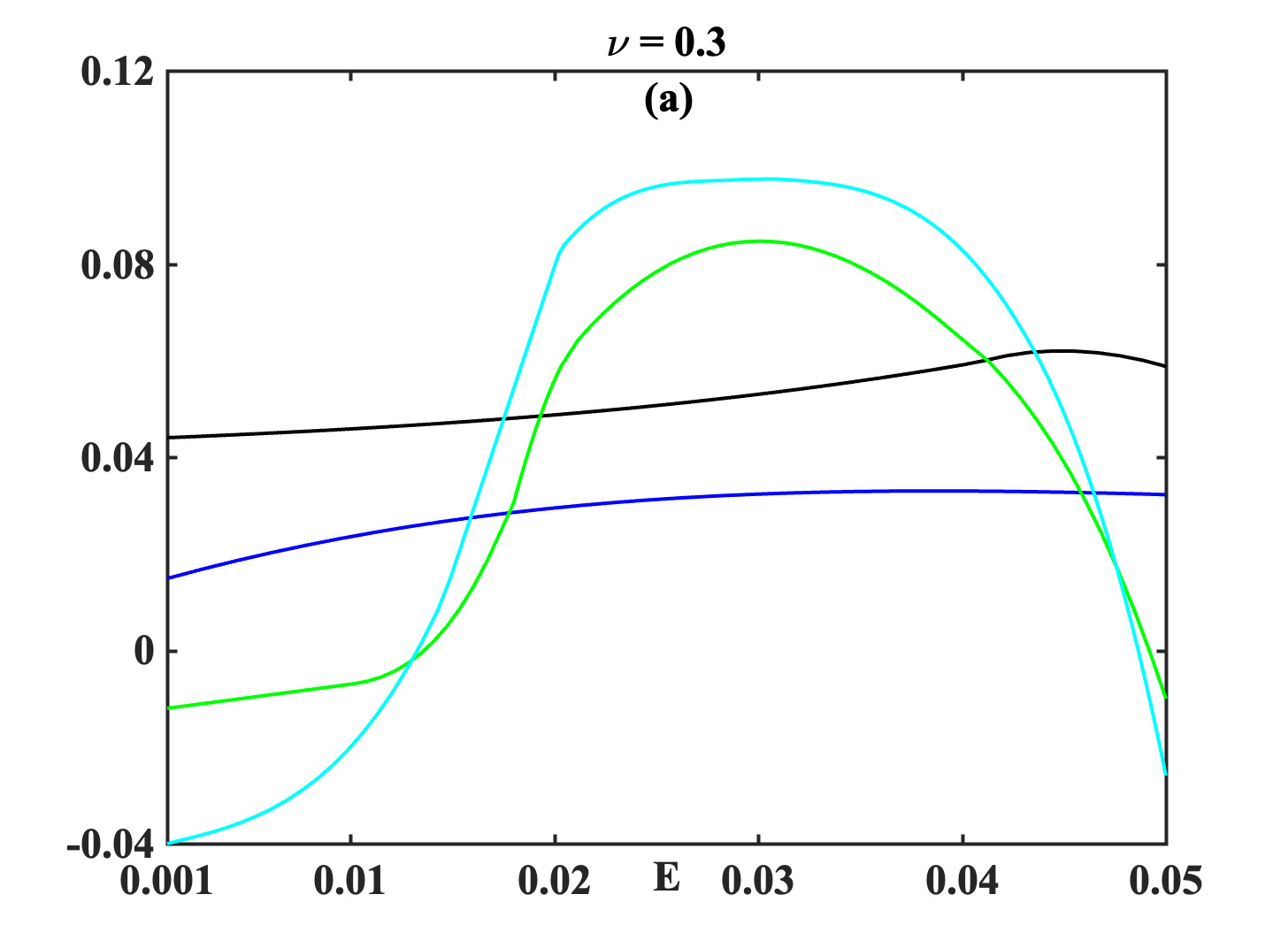}
\includegraphics[width=0.495\linewidth, height=0.35\linewidth]{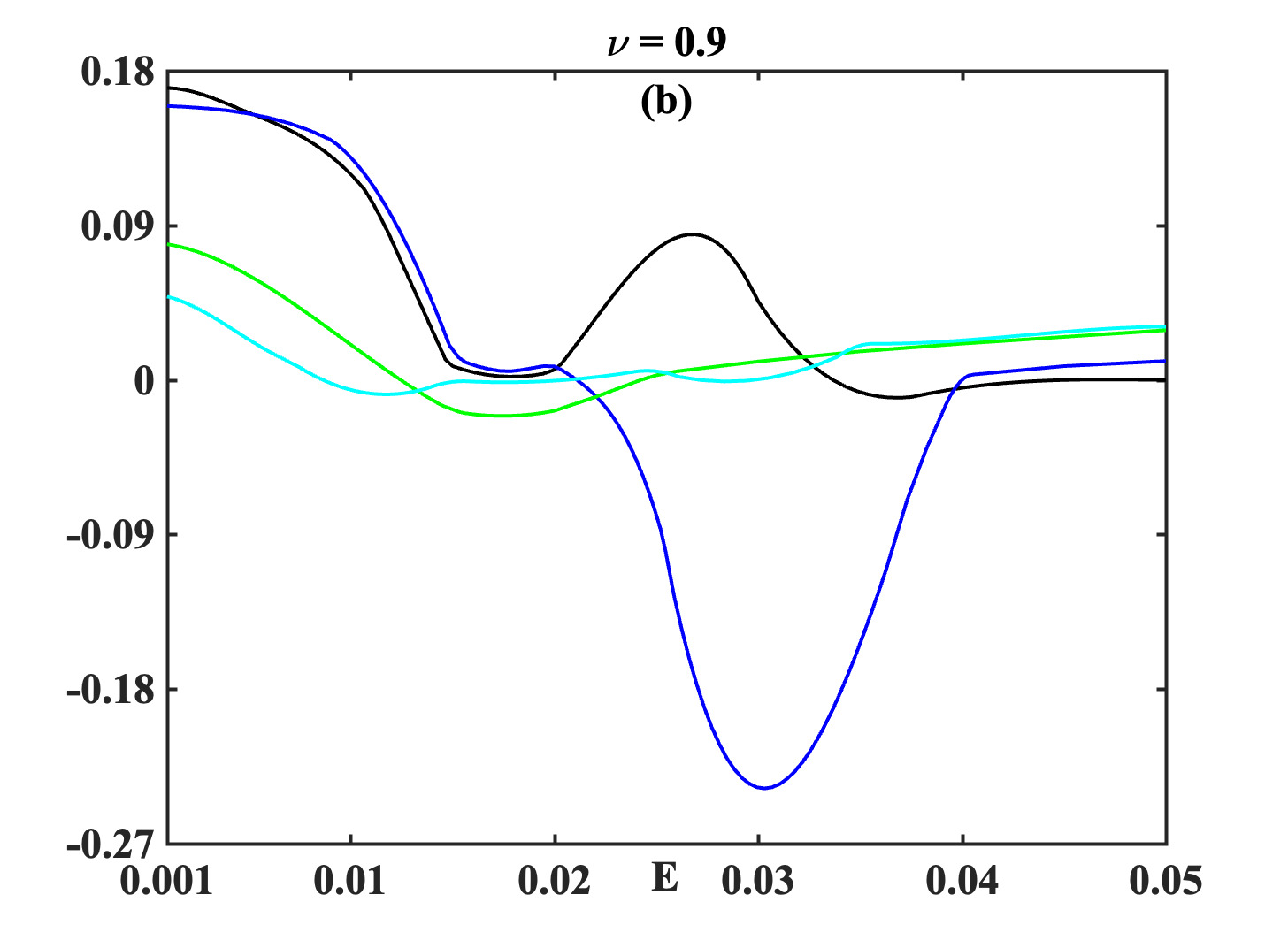}
\vskip 1pt
\includegraphics[width=0.495\linewidth, height=0.35\linewidth]{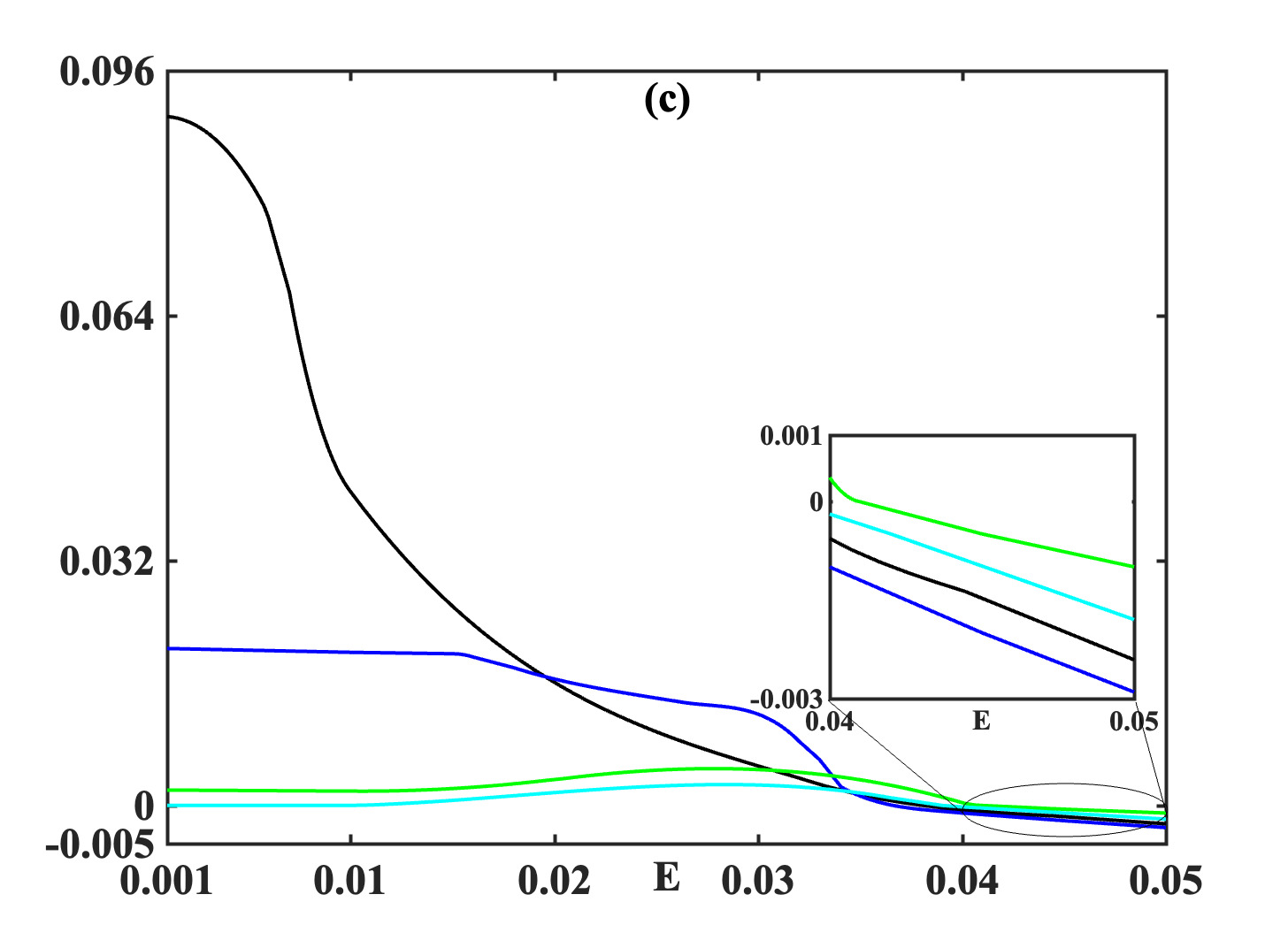}
\includegraphics[width=0.495\linewidth, height=0.35\linewidth]{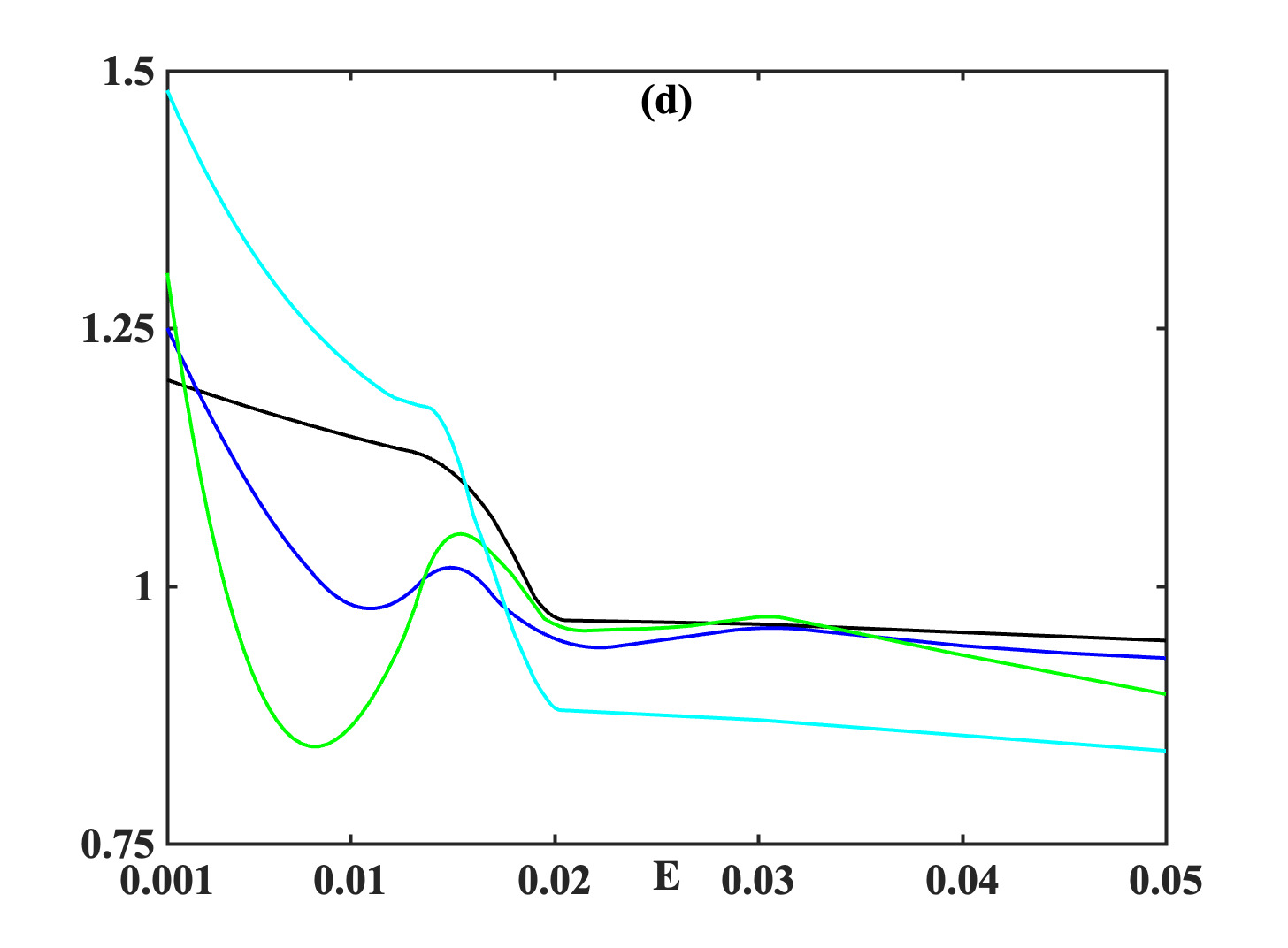}
\vskip 1pt
\includegraphics[width=0.495\linewidth, height=0.35\linewidth]{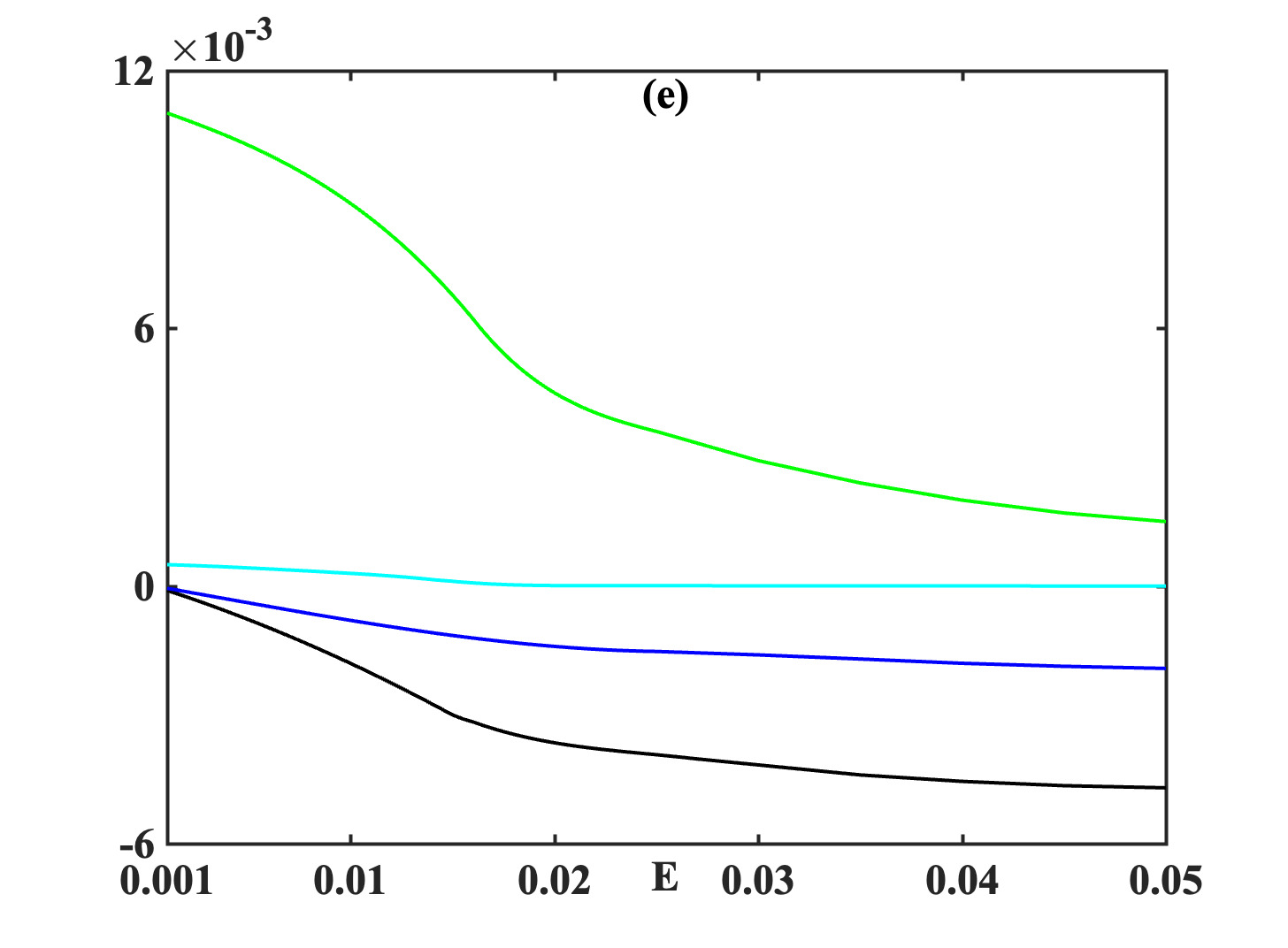}
\includegraphics[width=0.495\linewidth, height=0.35\linewidth]{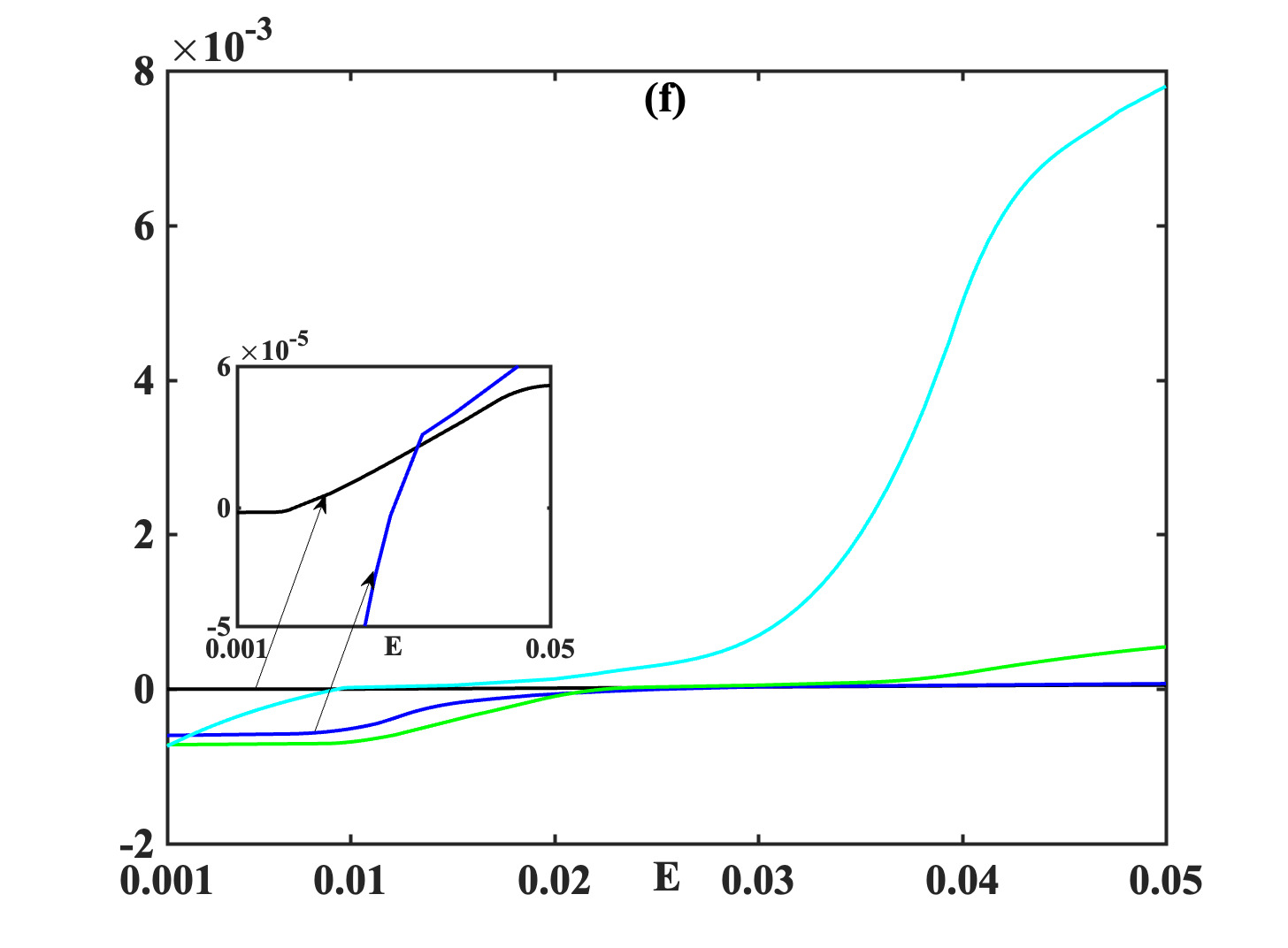}
\vskip 1pt
\includegraphics[width=0.495\linewidth, height=0.35\linewidth]{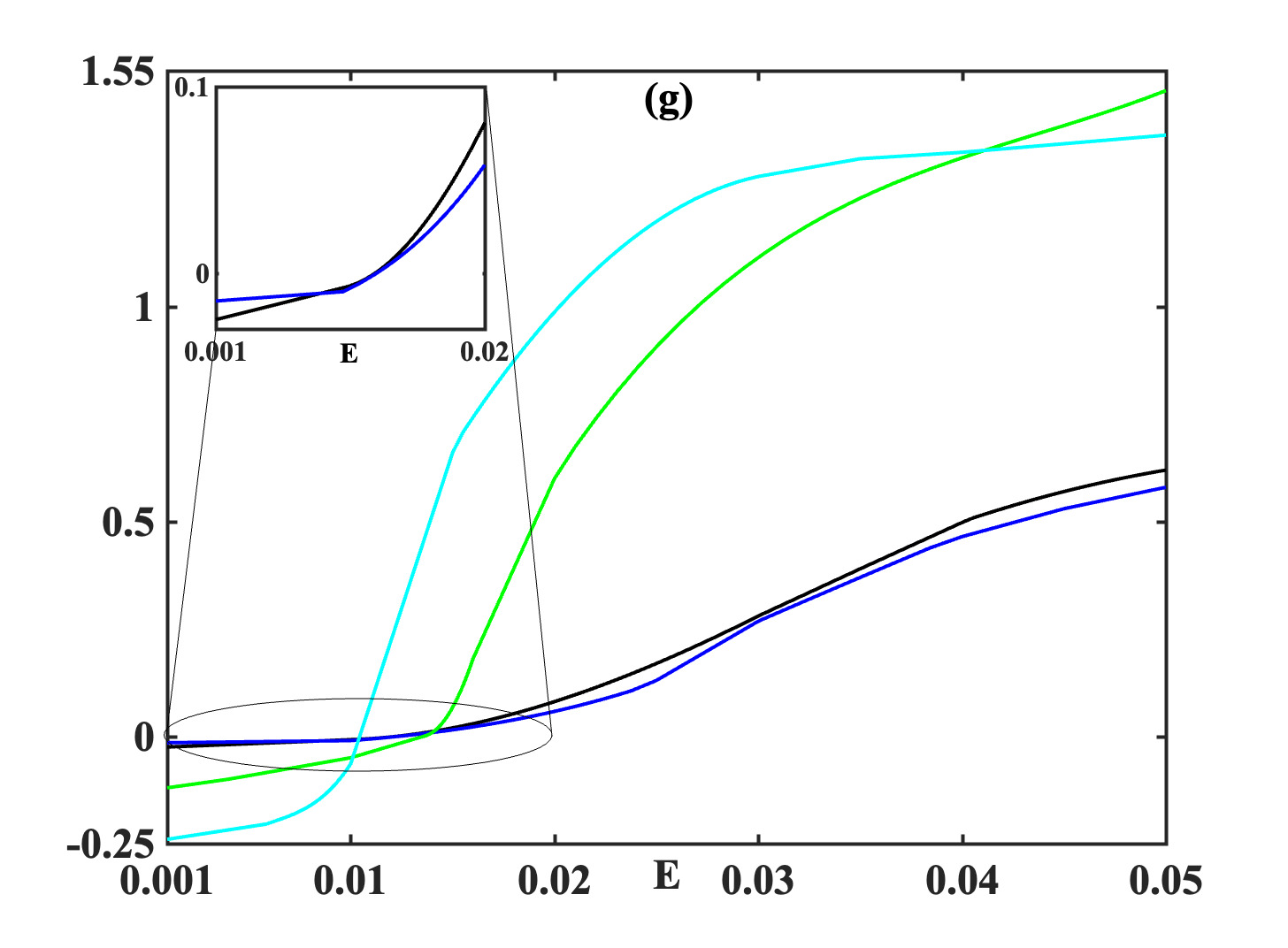}
\includegraphics[width=0.495\linewidth, height=0.35\linewidth]{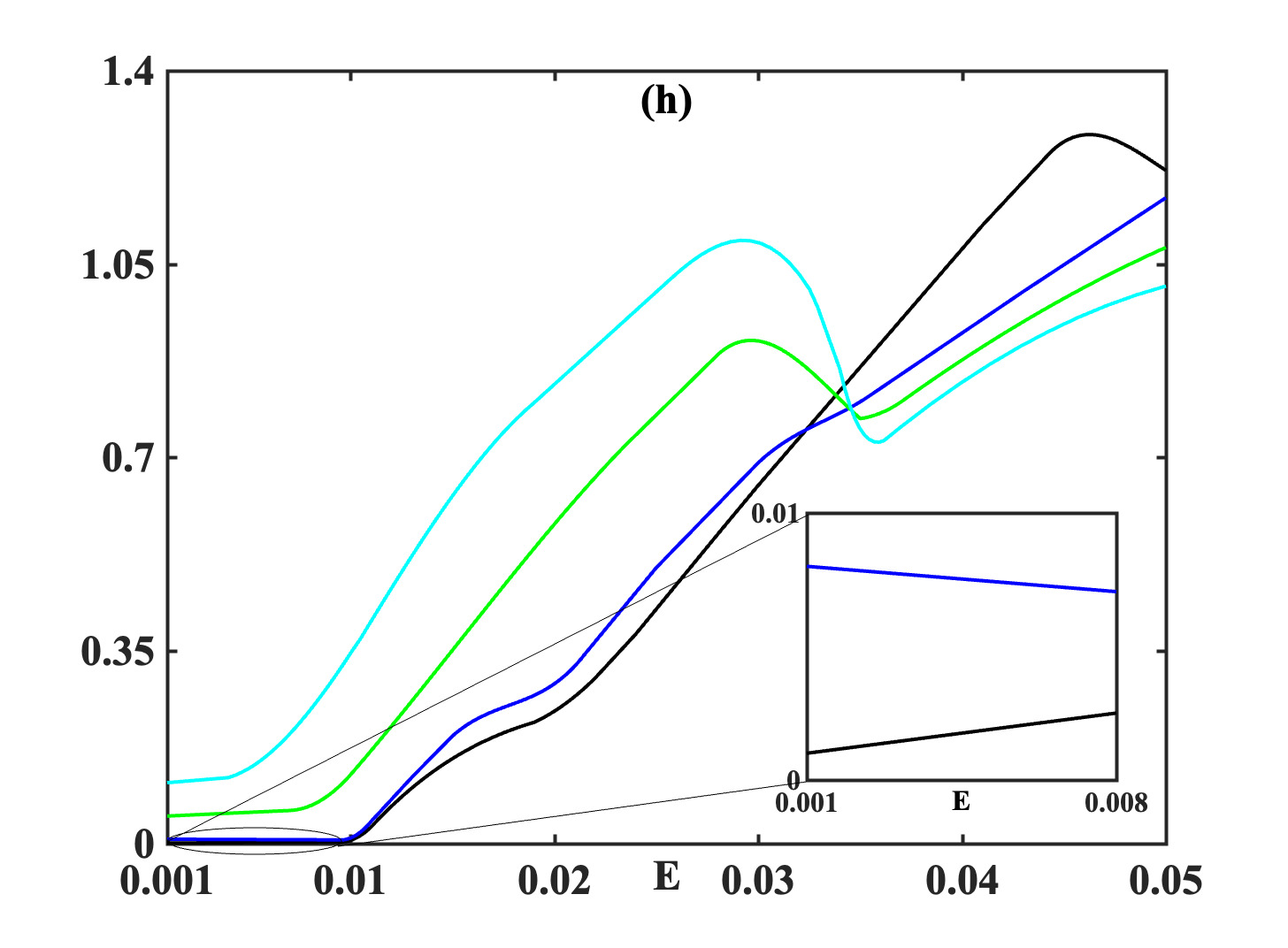}
\vskip 1pt
\caption{Cusp point, $\omega_i^{\text{cusp}}$ vs $E$ at $Re=7.0$ (\protect\blackline), $Re=10.0$ (\protect\blueline), $Re=20.0$ (\protect\greenline) and $Re=30.0$ (\protect\cyanline) and evaluated at spatial locations: (a, b) $y=0.99$, \newline (c, d) $y=0.5$, (e, f) $y=0.0$, (g, h) $y=-0.99$ and at fixed values of viscosity ratio, $\nu=0.3$ (left column) and $\nu=0.9$ (right column).}
\label{fig14}
\eef

Finally, the plots of the absolute growth rate curves versus $\nu$, at fixed $Re$ and $E$ (figure~\ref{fig15}) divulge patches of convective instability at lower values of $E$ (i.~e., $E \le 0.015$) and above the centerline (refer figures~\ref{fig15} a-d). At spatial locations which are on and below the centerline (refer figures~\ref{fig15} e-h), these pockets of convective instability appear in the strongly elastic limit (i.~e., $\nu \rightarrow 0.0$). To recapitulate, our studies disclose the regime of absolute instability in the strongly elastic limit above the centerline as well as at high Reynolds number coupled with high elasticity number, at spatial locations which are on and below the centerline.
\bef
\centering
\includegraphics[width=0.495\linewidth, height=0.35\linewidth]{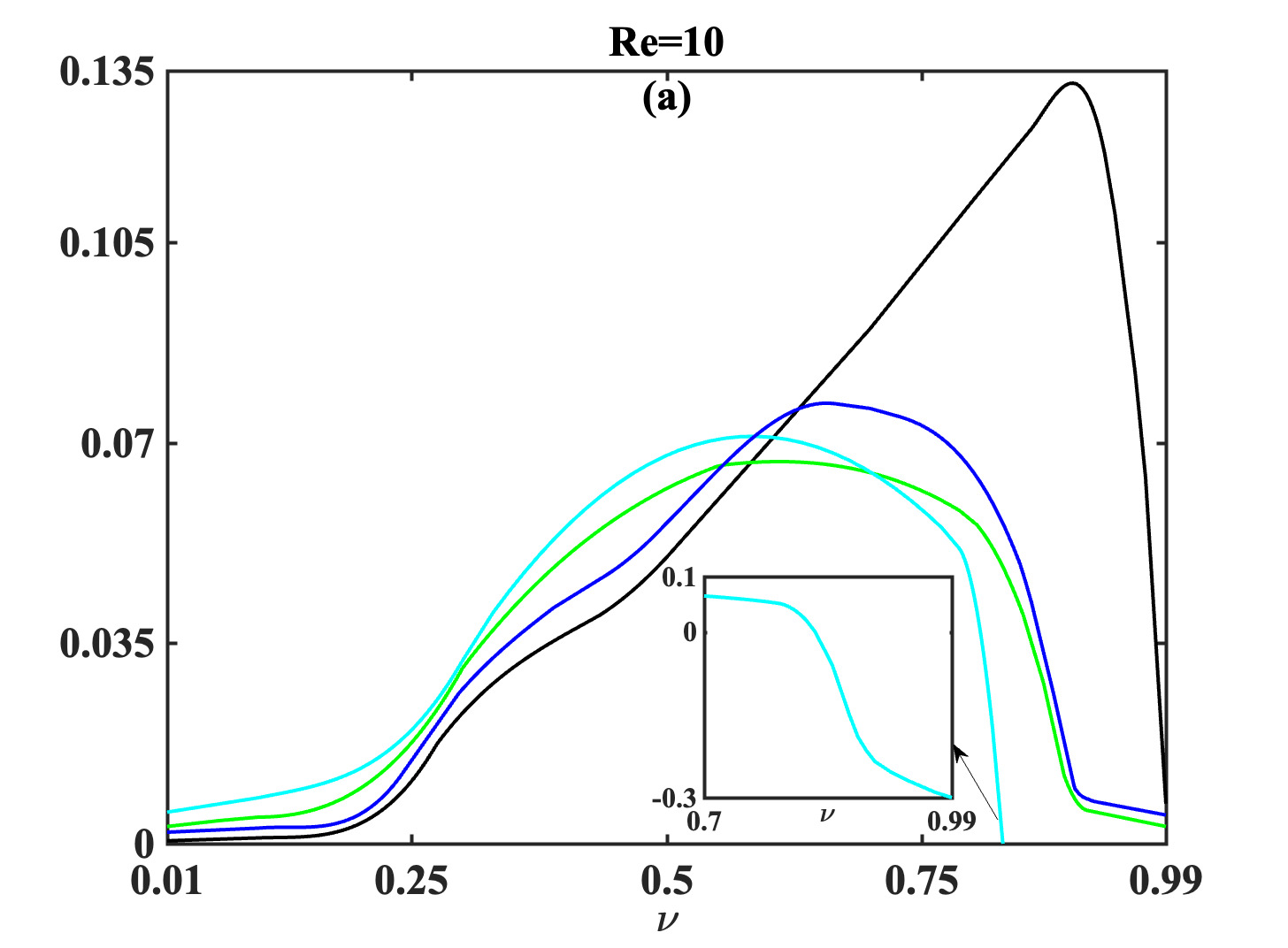}
\includegraphics[width=0.495\linewidth, height=0.35\linewidth]{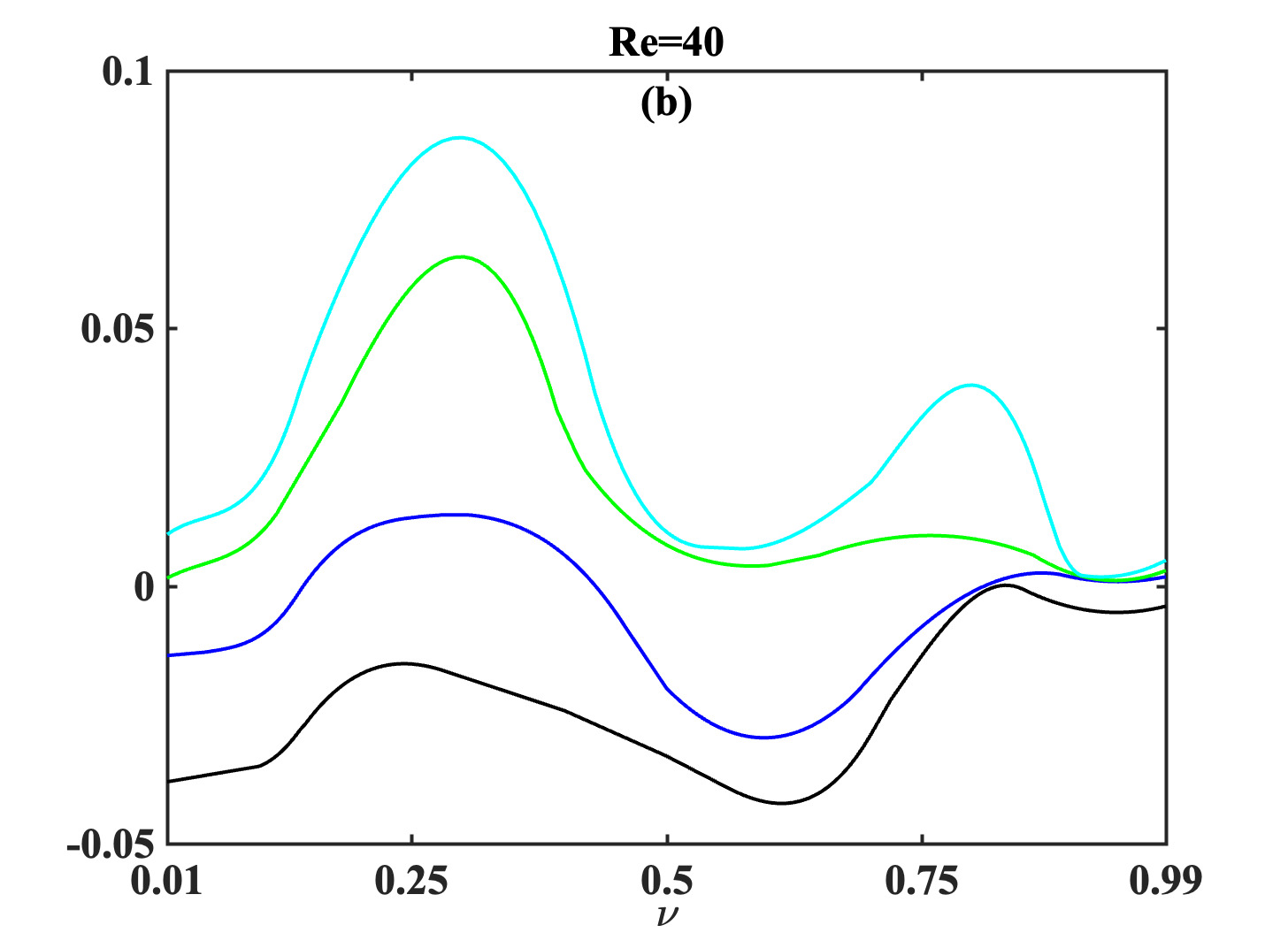}
\vskip 1pt
\includegraphics[width=0.495\linewidth, height=0.35\linewidth]{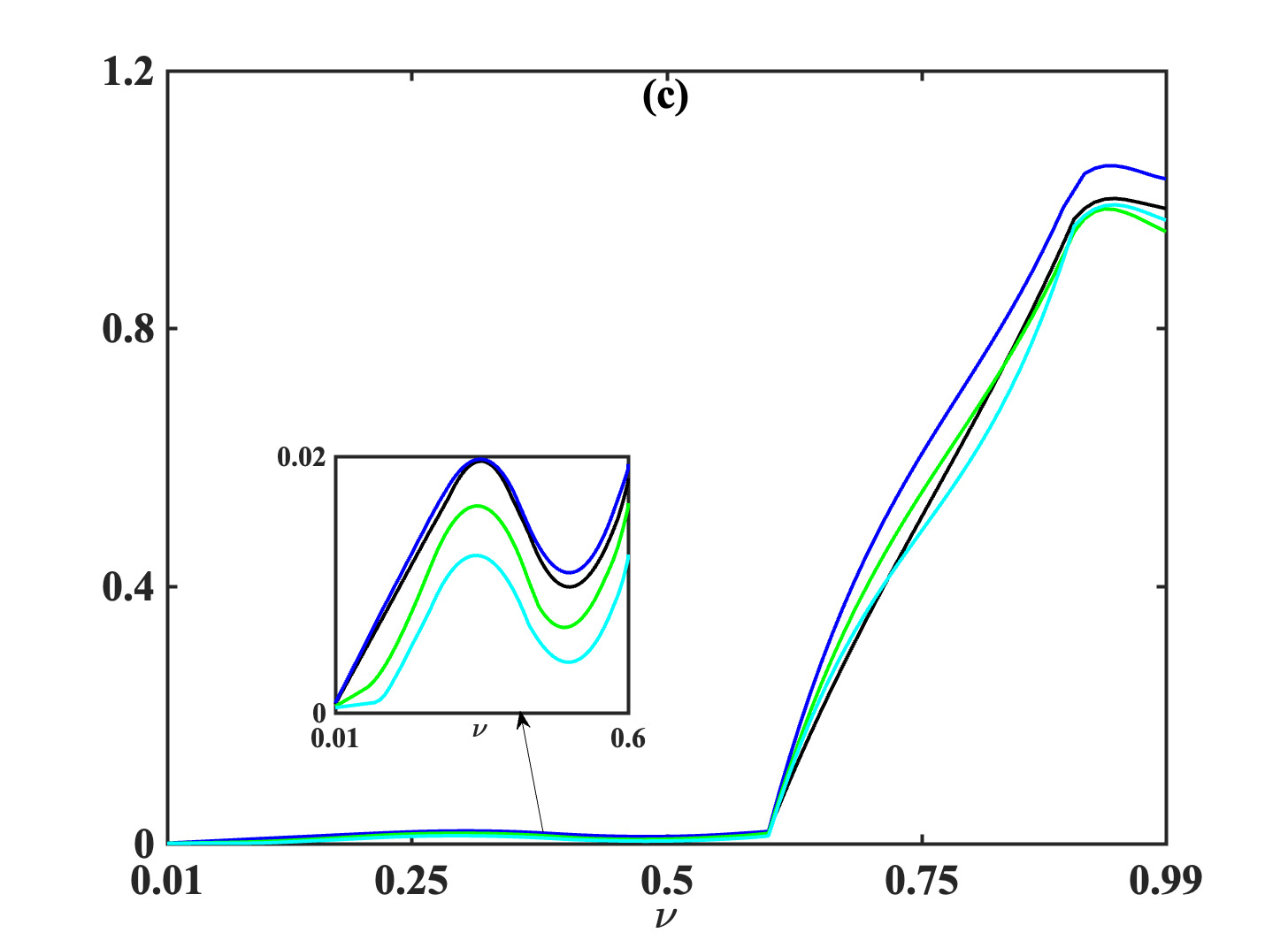}
\includegraphics[width=0.495\linewidth, height=0.35\linewidth]{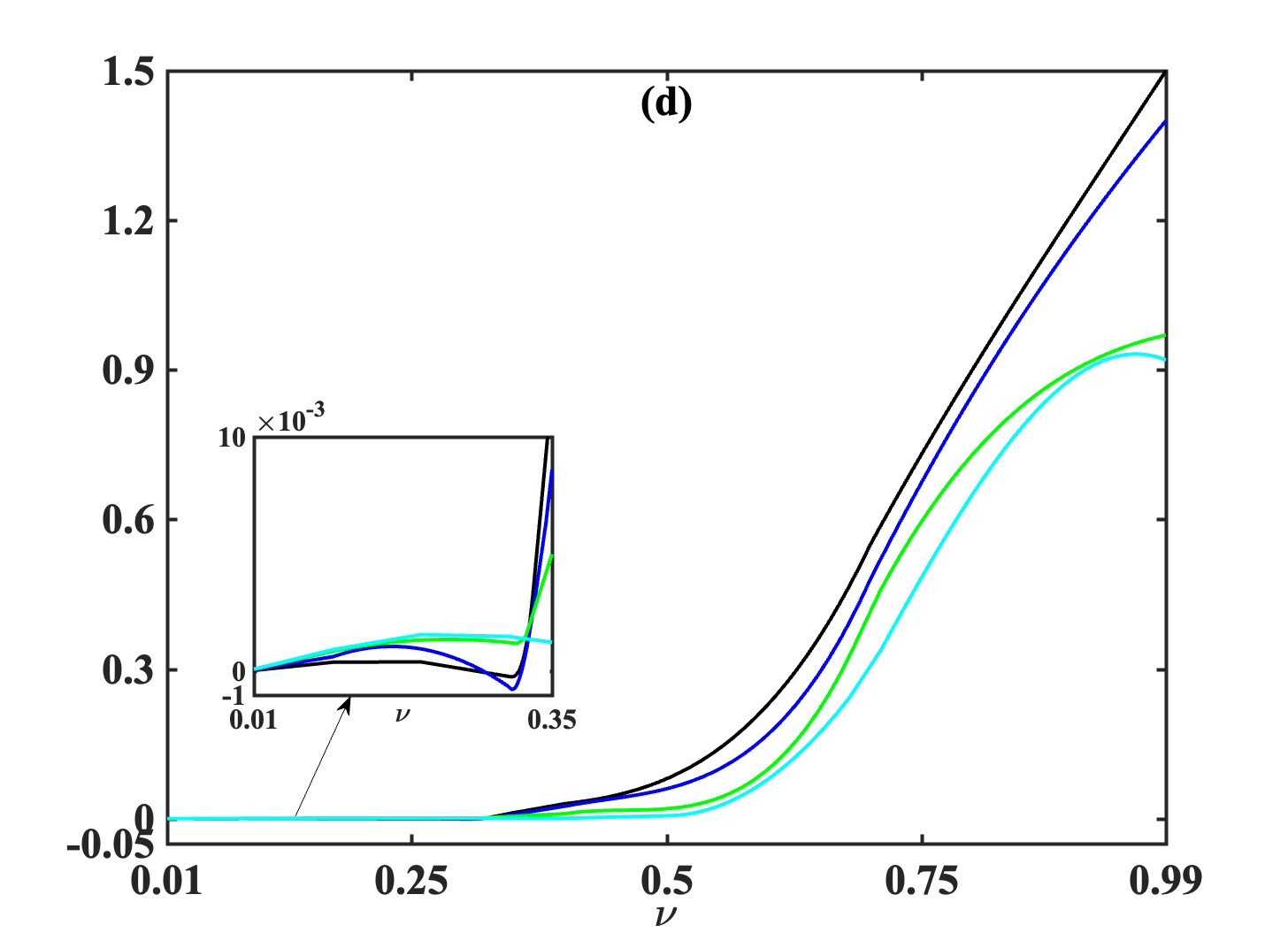}
\vskip 1pt
\includegraphics[width=0.495\linewidth, height=0.35\linewidth]{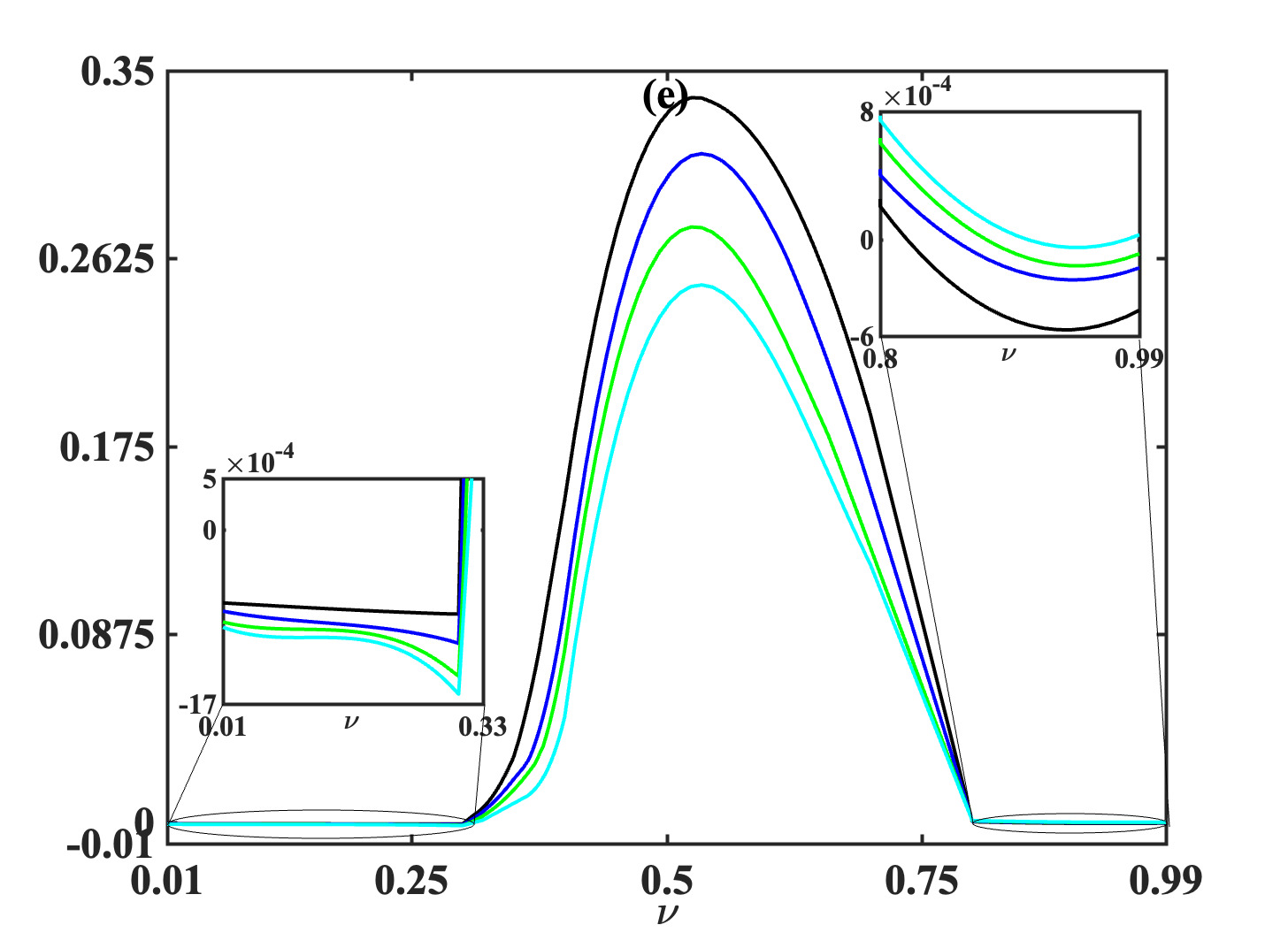}
\includegraphics[width=0.495\linewidth, height=0.35\linewidth]{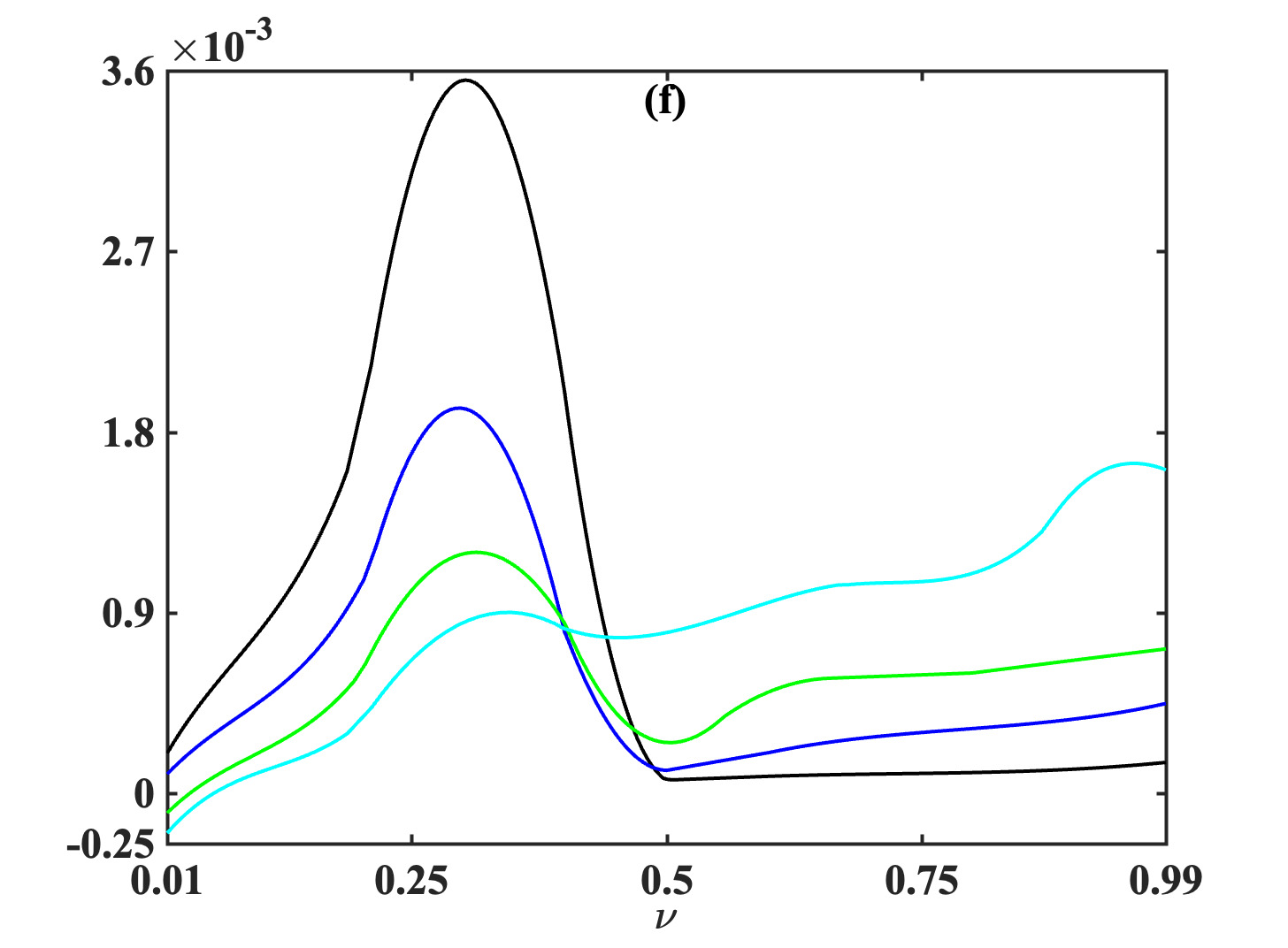}
\vskip 1pt
\includegraphics[width=0.495\linewidth, height=0.35\linewidth]{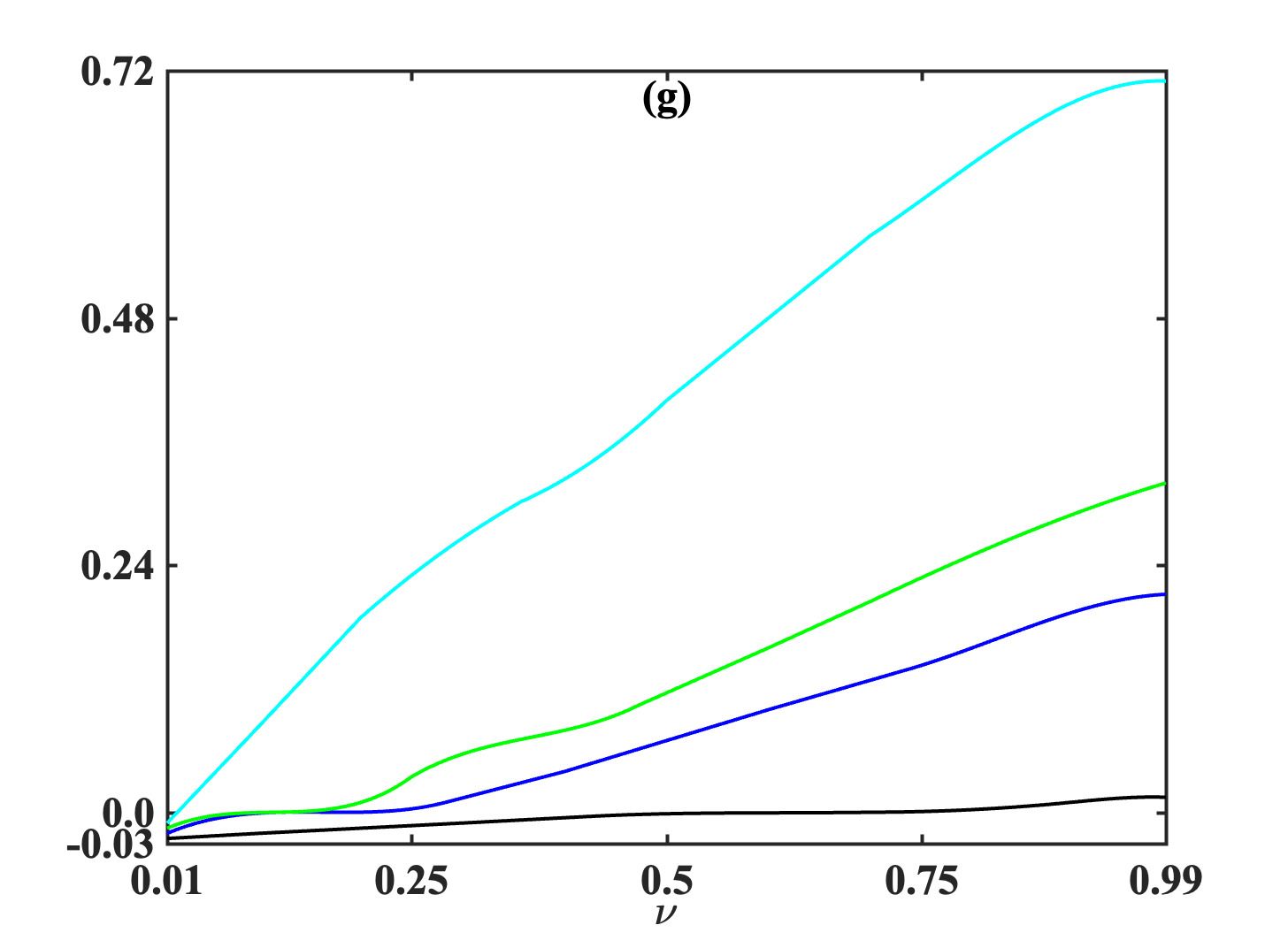}
\includegraphics[width=0.495\linewidth, height=0.35\linewidth]{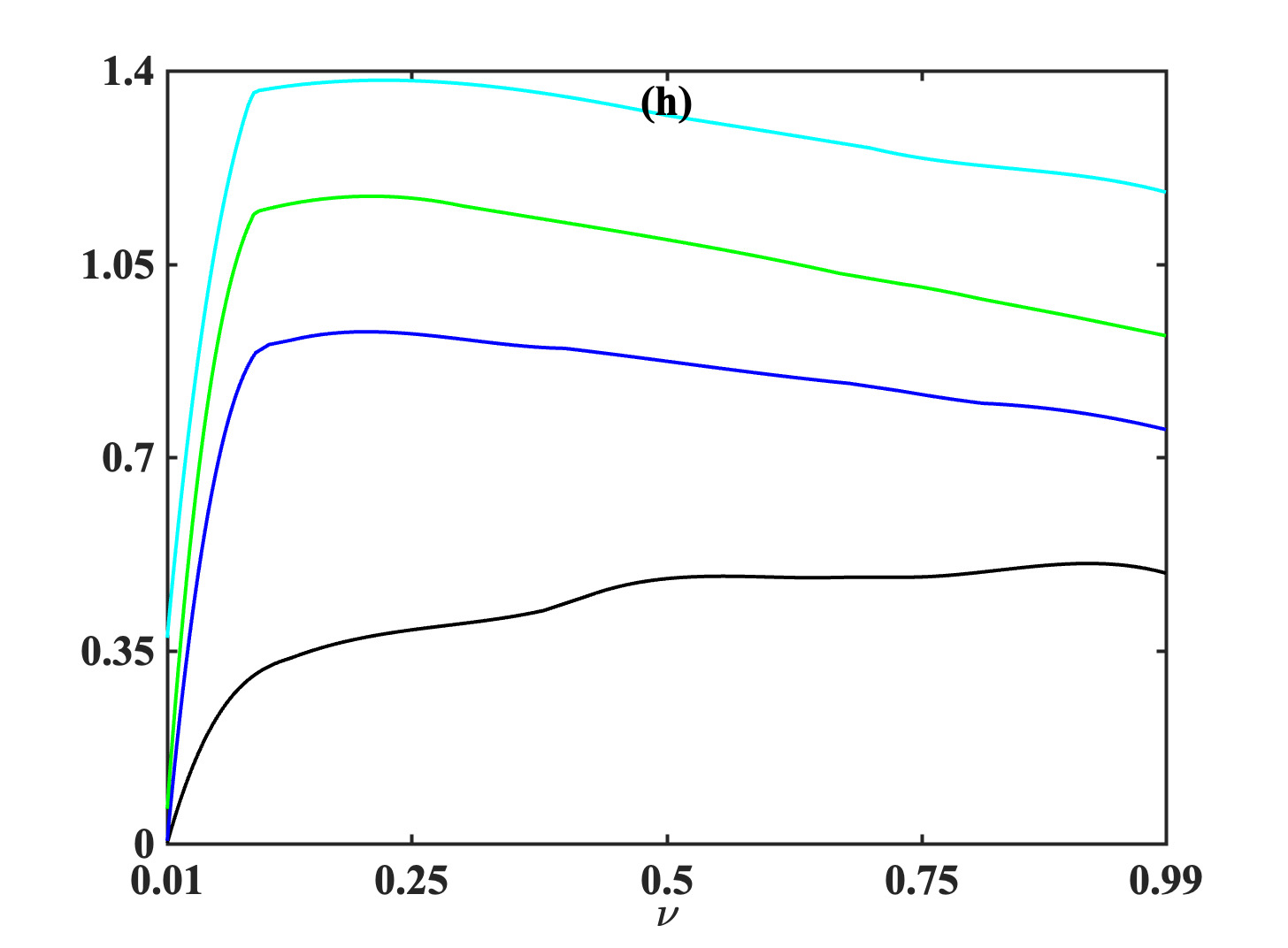}
\vskip 1pt
\caption{Cusp point, $\omega_i^{\text{cusp}}$ vs $\nu$ at $E=0.01$ (\protect\blackline), $E=0.015$ (\protect\blueline), $E=0.02$ (\protect\greenline) and $E=0.03$ (\protect\cyanline) and evaluated at spatial locations: (a, b) $y=0.99$, (c, d) $y=0.5$, (e, f) $y=0.0$, (g, h) $y=-0.99$ and at fixed values of Reynolds number, $Re=10$ (left column) and $Re=40$ (right column).}
\label{fig15}
\eef

Next, we classify the nature of these instabilities by computing the boundaries of the evanescent modes ({\bf E}), the convectively unstable ({\bf C}) and the absolutely unstable regions ({\bf A}) within a selected range of flow-elasticity-viscosity parameter space, i~e., $Re \in [0, 40], E \in [10^{-3}, 0.05]$ and $\nu \in [0.01, 0.99]$. The flow stability phase diagram projected onto the $Re-E$ parameter space (figure~\ref{fig16}) divulge the presence of absolutely unstable region at high values of $Re$ and $E$ ($Re \ge 35.0,\,\, 0.015 \le E \le 0.045$), confirming our presumption that elasticity coupled with high fluid inertia has a destabilizing effect. Additionaly, we report the presence of evanescent modes at intermediate values of Reynolds and elasticity number ($2.0 \le Re \le 35.0,\,\, 0.01 \le E \le 0.04$). 
\bef
\centering
\includegraphics[width=0.495\linewidth, height=0.35\linewidth]{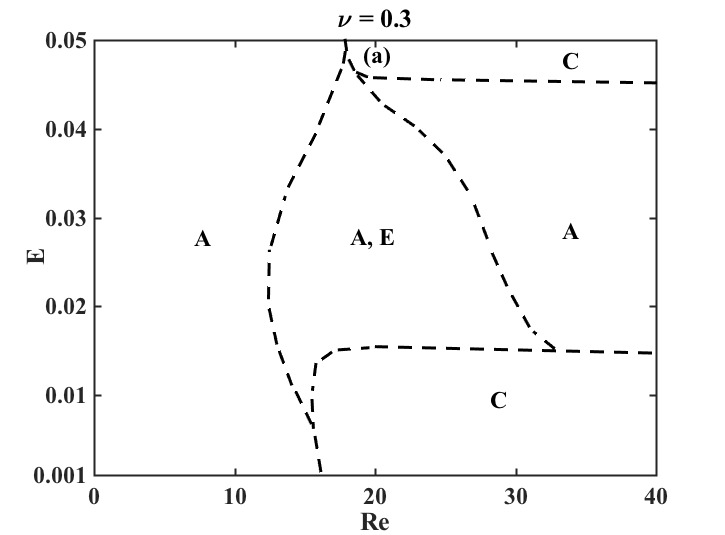}
\includegraphics[width=0.495\linewidth, height=0.35\linewidth]{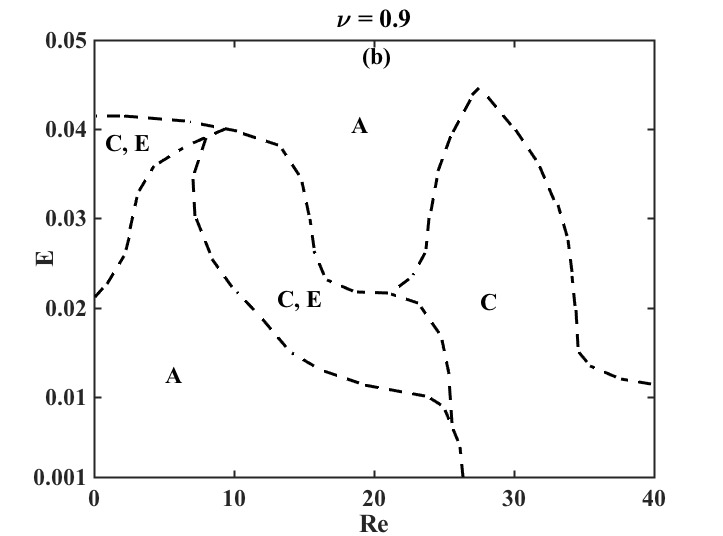}
\vskip 1pt
\includegraphics[width=0.495\linewidth, height=0.35\linewidth]{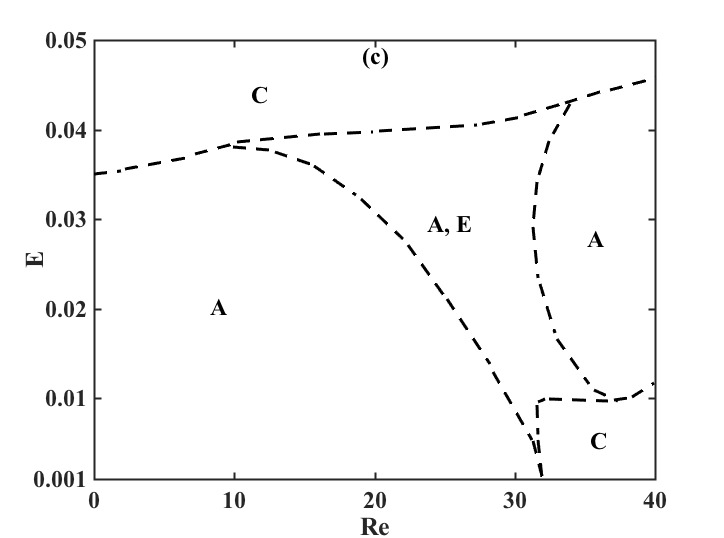}
\includegraphics[width=0.495\linewidth, height=0.35\linewidth]{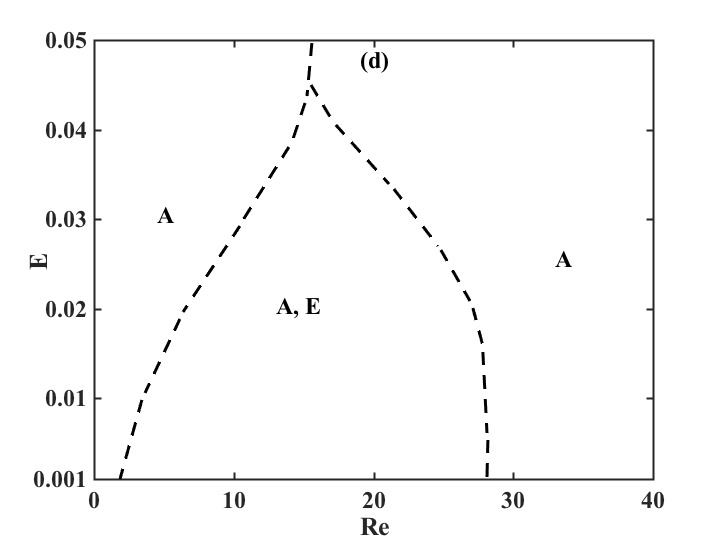}
\vskip 1pt
\includegraphics[width=0.495\linewidth, height=0.35\linewidth]{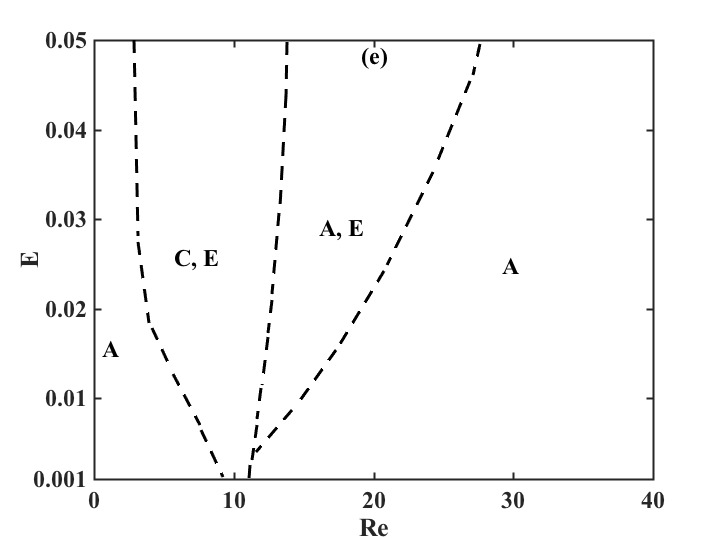}
\includegraphics[width=0.495\linewidth, height=0.35\linewidth]{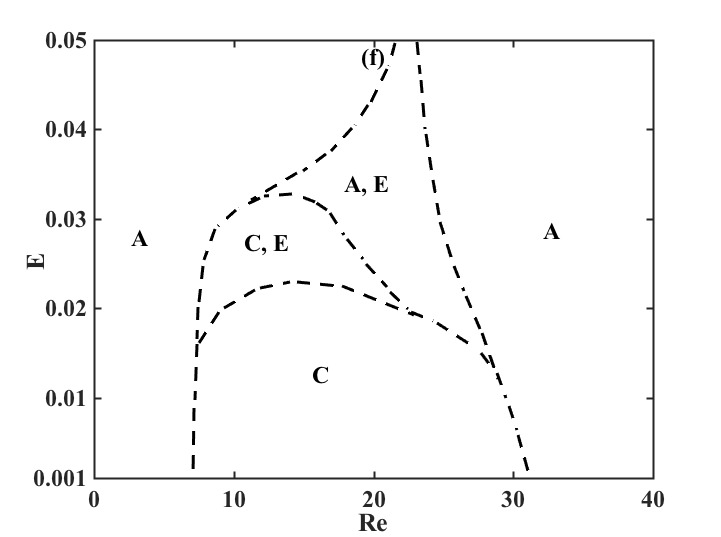}
\vskip 1pt
\includegraphics[width=0.495\linewidth, height=0.35\linewidth]{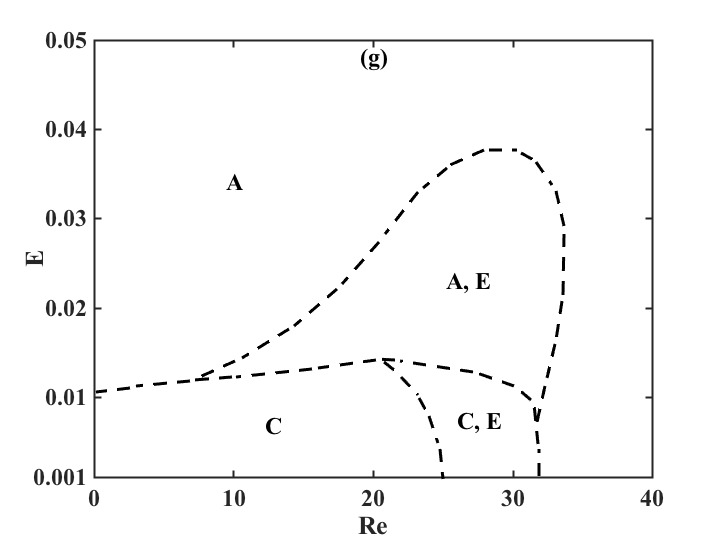}
\includegraphics[width=0.495\linewidth, height=0.35\linewidth]{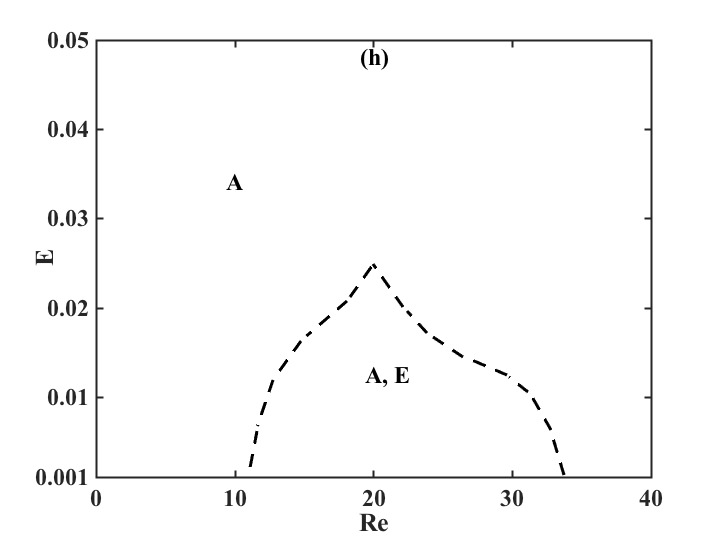}
\vskip 1pt
\caption{Viscoelastic Saffman-Taylor flow stability phase diagram in the $Re-E$ parametric space evaluated at spatial locations: (a, b) $y=0.99$, (c, d) $y=0.5$, \newline (e, f) $y=0.0$, (g, h) $y=-0.99$ and at fixed values of viscosity ratio, $\nu=0.3$ (left column) and $\nu=0.9$ (right column).}
\label{fig16}
\eef

The projection of the flow stability phase diagram onto the $Re-\nu$ parameter space (figure~\ref{fig17}) disclose that the absolutely unstable region is numerically estimated to reside within the limit, $Re \le 3.0$, and for almost the entire range of the viscosity ratio considered. Further, notice that the convectively unstable region appears in succession, after the absolutely unstable region, as the Reynolds number is increased (for example, refer figures~\ref{fig17}b,c,e,f). In particular, note that the convectively unstable region emerges at comparatively lower values of $Re$ (i.~e., $Re \le 32$), at spatial locations on and below the centerline (figures~\ref{fig17}e-h). Thus, at these spatial locations, elasticity blended with low inertia has a stabilizing influence.
\bef
\centering
\includegraphics[width=0.495\linewidth, height=0.35\linewidth]{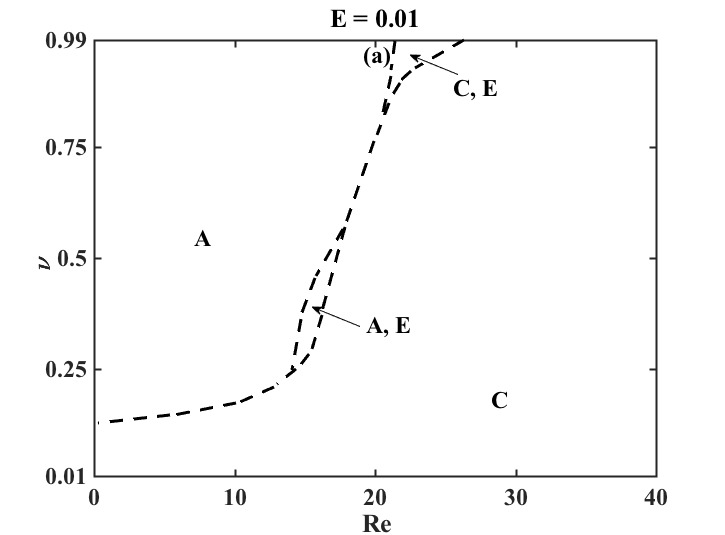}
\includegraphics[width=0.495\linewidth, height=0.35\linewidth]{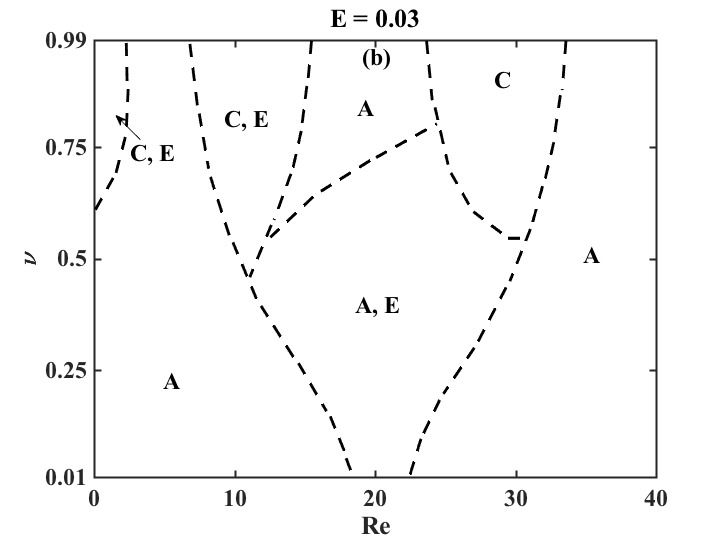}
\vskip 1pt
\includegraphics[width=0.495\linewidth, height=0.35\linewidth]{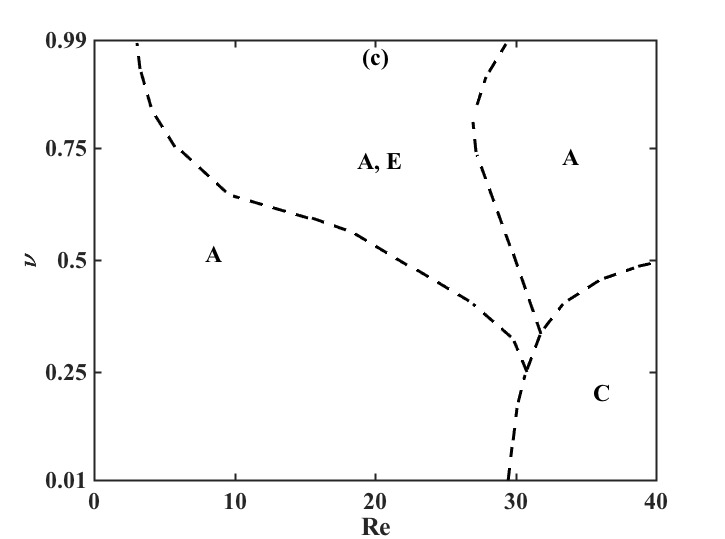}
\includegraphics[width=0.495\linewidth, height=0.35\linewidth]{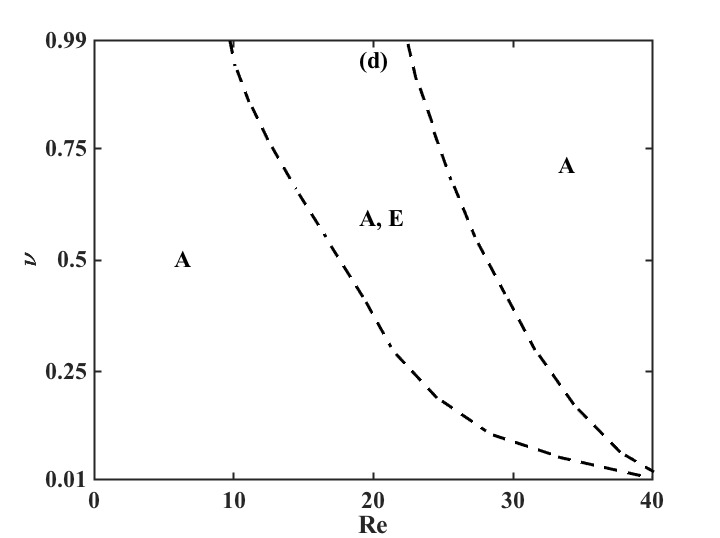}
\vskip 1pt
\includegraphics[width=0.495\linewidth, height=0.35\linewidth]{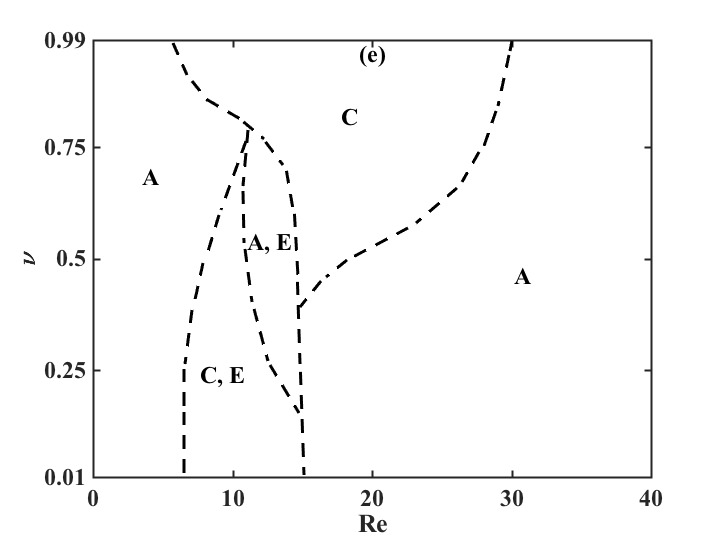}
\includegraphics[width=0.495\linewidth, height=0.35\linewidth]{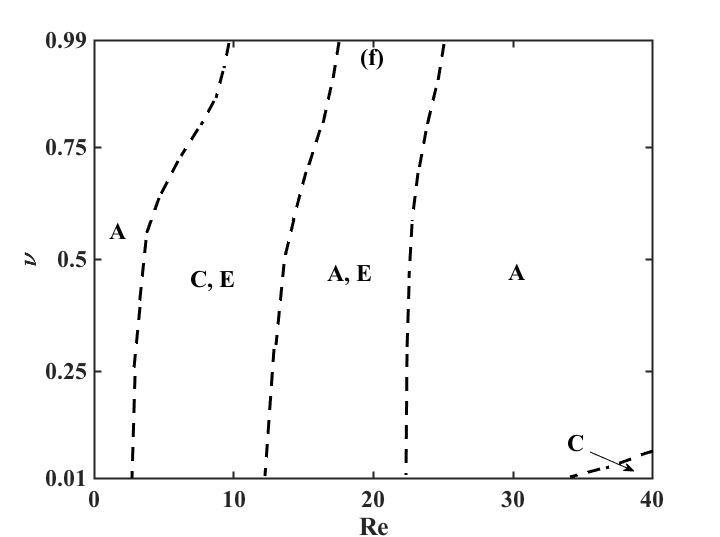}
\vskip 1pt
\includegraphics[width=0.495\linewidth, height=0.35\linewidth]{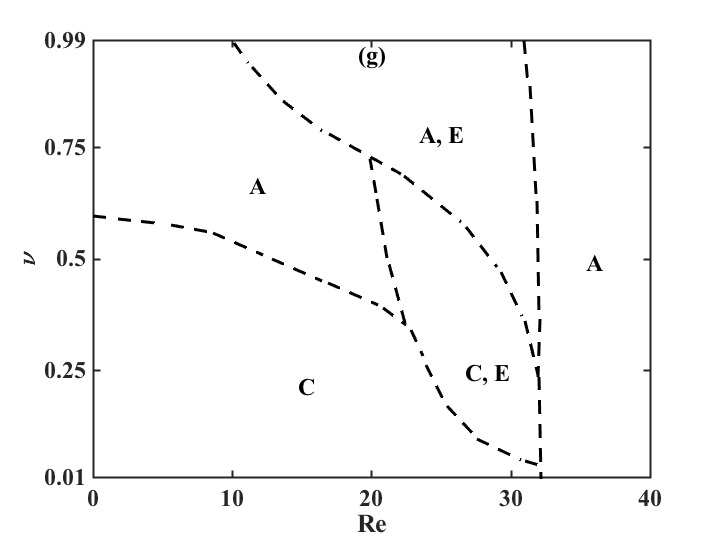}
\includegraphics[width=0.495\linewidth, height=0.35\linewidth]{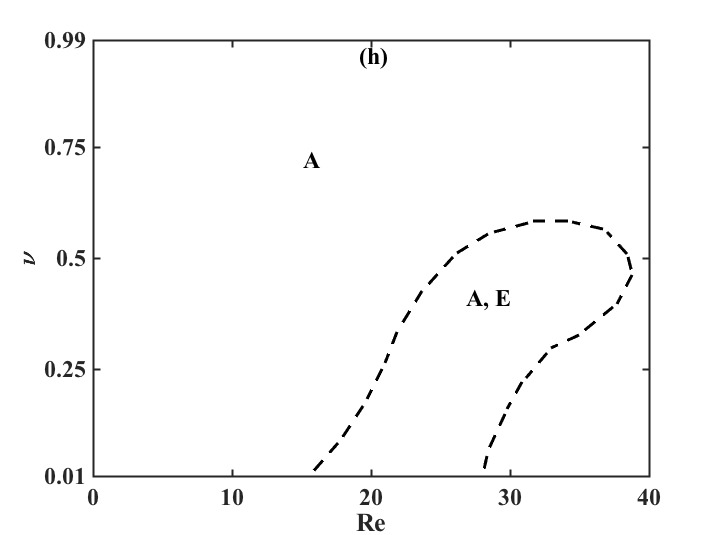}
\vskip 1pt
\caption{Viscoelastic Saffman-Taylor flow stability phase diagram in the $Re-\nu$ parametric space evaluated at spatial locations: (a, b) $y=0.99$, (c, d) $y=0.5$, \newline (e, f) $y=0.0$, (g, h) $y=-0.99$ and at fixed values of elasticity number, $E=0.01$ (left column) and $E=0.03$ (right column).}
\label{fig17}
\eef

Finally, as revealed by the low Reynolds number regime in the flow stability phase diagram in the $E-\nu$ parameter slice (refer the left column in figure~\ref{fig18}), absolutely unstable region appears at high elasticity numbers (i.~e., $E \ge 0.04$) and within the viscous stress dominated case (or $\nu \ge 0.5$ case). 

Building upon an earlier study by~\citet{Constantin1993}, illustrating the interface dynamics in Hele-Shaw flows undergoing topological transitions, we discuss the significance of our temporal stability analysis (figures~\ref{fig7}-\ref{fig12}) as well as the spatiotemporal phase diagram (figures~\ref{fig16}-\ref{fig18}) in relation to the Newtonian as well as viscoelastic surface flows which have experimentally demonstrated these transitions. While convective instability grows in amplitude as it is swept along by the flow, absolute instability occurs at fixed spatial locations, leading to surface transitions (or pinch-off)~\citep{Gallaire2017}, impairment and recoil~\citep{Chang1999}. This type of classification of an unstable spectrum is most pertinent to instabilities which possess a specific sensitive spot - the finger tip in the present example~\citep{Lindner2009}.

Experiments in vertically aligned (or gravity driven) Newtonian Hele-Shaw flows~\citep{Goldstein1993} suggest that the inertial effects are important in triggering a collapse of the moving interface and it's eventual pinch-off, thereby corroborating our numerical outcomes. In the absence of gravity, experiments in Hele-Shaw flows on topological transitions driven by an applied pressure across the finger tip do show shapes at the pinch point much like those predicted in the presence of gravity~\citep{Shelley1993}. Finally, in a related experiment~\citet{Amarouchene2001} studied the role of elasticity on the delayed onset of the interfacial singularity of a (vertically falling) polymeric liquid and observed that the drop detachment and recoil in the strongly elastic limit (and combined with low fluid inertia) occurred at a much larger time-scale. This stabilizing impact of elasticity in their experiment is explained as follows: in a falling viscoelastic droplet, the flow is predominantly elongational, thereby stretching the polymers most efficiently. The growing elastic stresses stabilize the flow interface , since any distortions lead to further stresses.
\bef
\centering
\includegraphics[width=0.495\linewidth, height=0.35\linewidth]{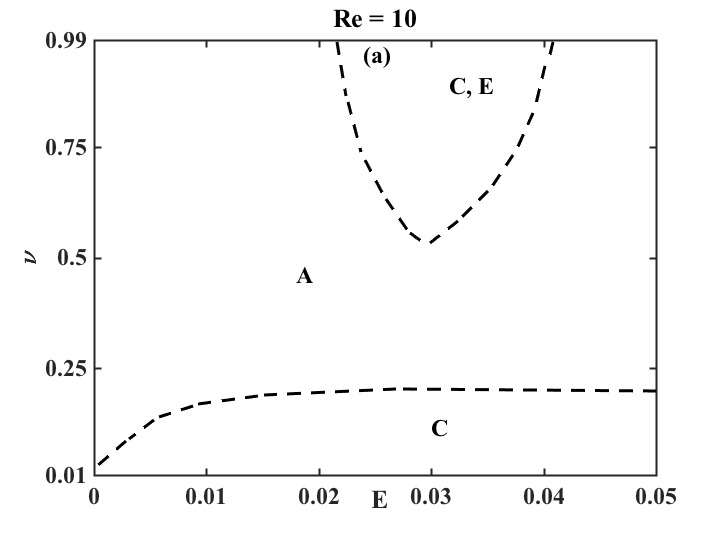}
\includegraphics[width=0.495\linewidth, height=0.35\linewidth]{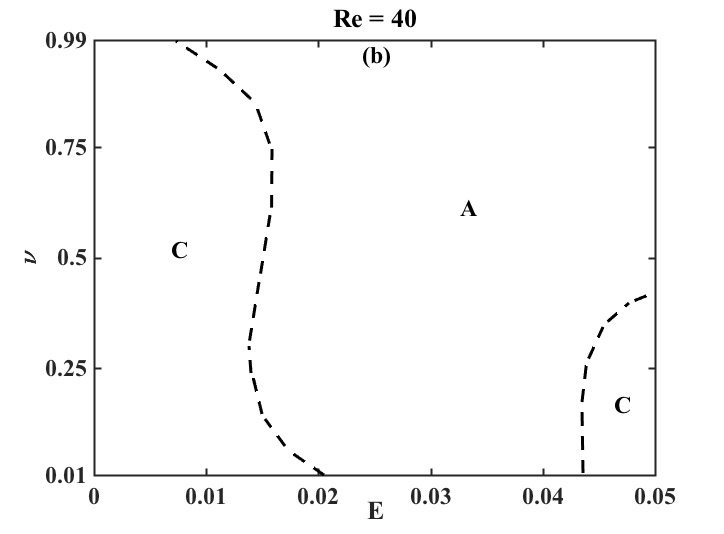}
\vskip 1pt
\includegraphics[width=0.495\linewidth, height=0.35\linewidth]{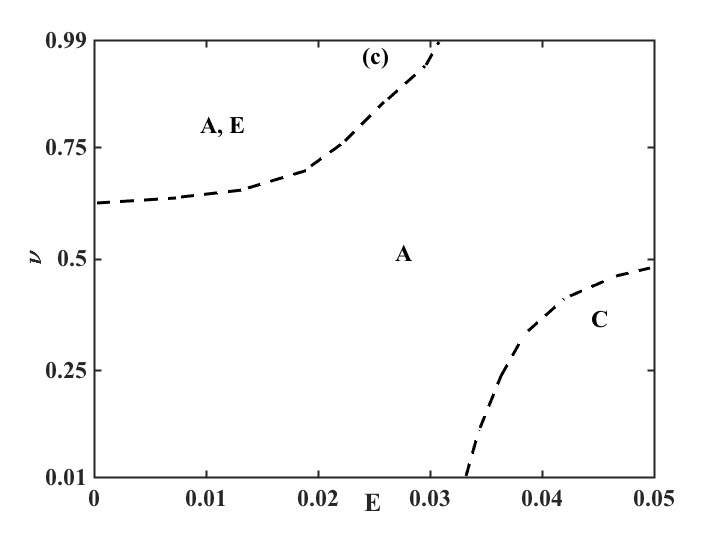}
\includegraphics[width=0.495\linewidth, height=0.35\linewidth]{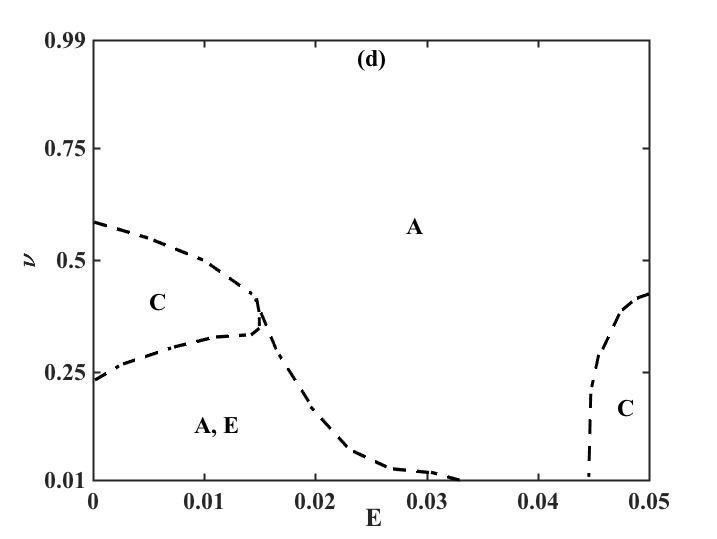}
\vskip 1pt
\includegraphics[width=0.495\linewidth, height=0.35\linewidth]{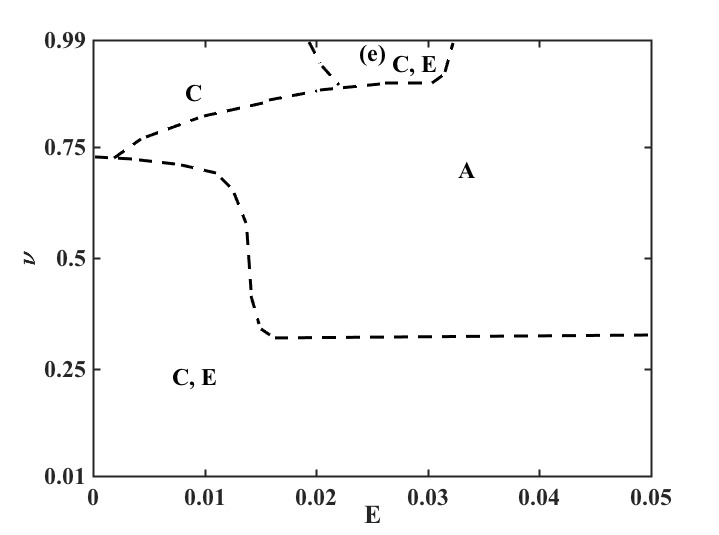}
\includegraphics[width=0.495\linewidth, height=0.35\linewidth]{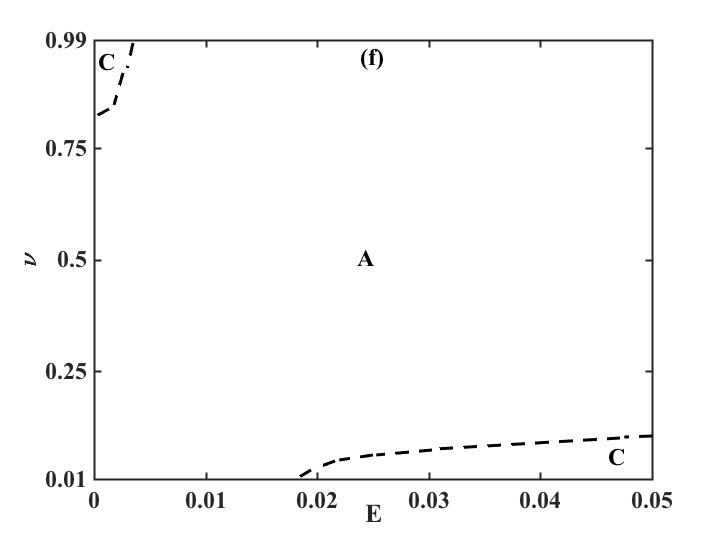}
\vskip 1pt
\includegraphics[width=0.495\linewidth, height=0.35\linewidth]{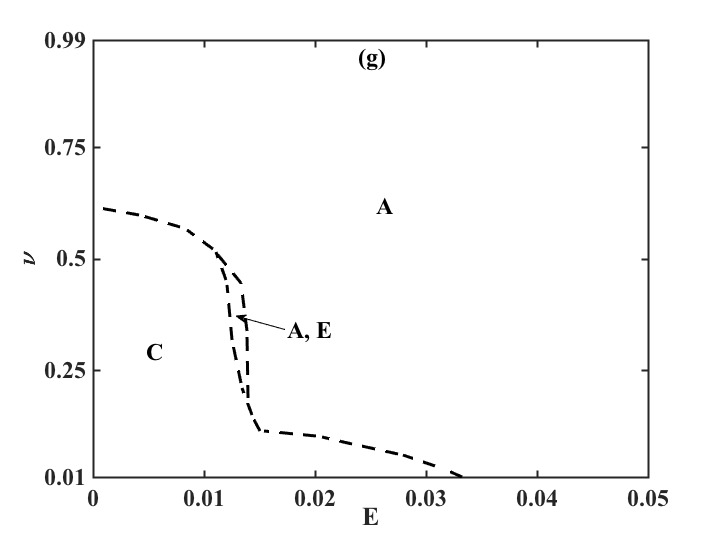}
\includegraphics[width=0.495\linewidth, height=0.35\linewidth]{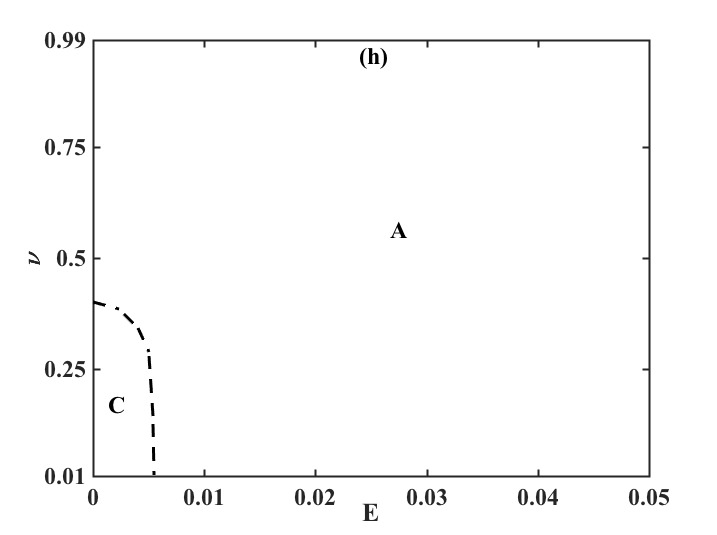}
\vskip 1pt
\caption{Viscoelastic Saffman-Taylor flow stability phase diagram in the $E-\nu$ parametric space evaluated at spatial locations: (a, b) $y=0.99$, (c, d) $y=0.5$, \newline (e, f) $y=0.0$, (g, h) $y=-0.99$ and at fixed values of Reynolds number, $Re=10$ (left column) and $Re=40$ (right column).}
\label{fig18}
\eef
\section{Concluding remarks}\label{sec:conclusion}
This investigation addresses the linear, temporal, and spatiotemporal analyses of the Hele-Shaw flow, by capturing the affine flow response of a Poiseuille base flow profile, for low to moderate Reynolds and elasticity number and when the driven fluid is a dilute polymeric liquid with significantly larger viscosity (compared with the driving, Newtonian fluid). \S \ref{sec:math} presented the viscoelastic Hele-Shaw flow model and the interface conditions, the elements of linear stability analysis as well as the numerical method needed to solve the resulting dispersion relation. \S \ref{sec:MV} validated the model for Newtonian Hele-Shaw flow within the Stokes regime~\citep{Saffman1958, McLean1981, Tabeling1985} as well as within the inertial regime~\citep{Lindner2006}. \S \ref{sec:NSC} identifies the temporally stable region in the Reynolds number-wavenumber plane and determines the specific scaling law on which all the Neutral stability curves coincide. The temporal stability analysis in \S \ref{sec:tsa} indicates that (a) the inertial forces have a predominantly destabilizing impact on the advancing flow front, (b) the finite boundary has a destabilizing influence near the wall, and (c) farther away from the wall, elasticity combined with low (high) fluid inertia has a stabilizing (destabilizing) impact. The spatiotemporal phase diagram in \S \ref{sec:stsa} divulge the presence of absolutely unstable region at high values of Reynolds and elasticity number, evanescent modes at intermediate values of Reynolds and elasticity number and convectively unstable region at comparatively lower values of Reynolds number (irrespective of the elasticity number).

Although this study provides an improved understanding of the linear dynamics of the Saffman-Taylor instability for dilute polymeric liquids, a number of simplifying assumptions were made, including the absence of non-linear terms, the numerical challenges posed in the regime of moderate to high elasticity number, interaction at the interface for miscible as well as chemically reacting fluids and the consideration of the Hele-Shaw flows of polymer melts (or fluids with very large viscosity). The relaxation of these assumptions through the development of an
appropriate stress constitutive relation or otherwise paves a way for further progress. Finally, we note that we have just peripherally considered the role of the finite boundaries on the evolution of the advancing interface. The boundary effects are especially important for cell geometry with small aspect ratio and in the inertial regime. Hence, a deeper consideration of the impact of these boundaries, including elastic boundaries and occlusions, on the evolution of the moving interface may be worthwhile in future.

\section*{Acknowledgements}
D.B., T.C. and S.S. acknowledge the financial support of Grant Nos. CSIR 09/1117 (0004)/2017-EMR-I, CSIR 09/1117(0012)/2020-EMR-I and DST ECR/2017/000632, respectively.

\section*{Declaration of interests}
The authors report no conflict of interest.

\appendix

\section{Functions utilized in the dispersion relation}\label{appA}
The functions ($g(x), f(x)$)|$_{x= U t}$ employed to describe the normal mode perturbation expansion of the velocity, ${\bf u}$ (equation~\eqref{eqn:uvPerturbation}), at the interface, $x= U t$, are given by,
\begin{align}
f_r &= e^{-\alpha U t}\left(W_1 S_1 - W_2 S_2  - \alpha L_2\right) \nonumber \\
f_i &= e^{-\alpha U t}\left(W_1 S_2 + W_2 S_1  + \alpha L_1\right) \nonumber \\
g_r &= g_i = 0,
\label{appA:f_and_g}
\end{align}
and the derivatives of these function with respect to the spatial coordinate, $x$, are given by
\begin{align}
f_{r_x} &= e^{-\alpha U t}\left(W_1 P_1 - W_2 P_2  + \alpha^2 L_2\right) \nonumber \\
f_{i_x} &= e^{-\alpha U t}\left(W_1 P_2 + W_2 P_1  - \alpha^2 L_1\right) \nonumber \\
g_{r_x}&= e^{-\alpha U t}\left(R_1 Y_1 - R_2 Y_2\right) \nonumber \\
g_{i_x} &= e^{-\alpha U t}\left(R_1 Y_2 + R_2 Y_1\right),
\label{appA:f_and_g_derivatives}
\end{align}
where 
\begin{align}
L_1 &= \dfrac{\left(\dfrac{6}{5} Re U \alpha - 1\right)}{\left(\dfrac{6}{5} Re U \alpha - 1\right)^2 + Re^2 \omega^2}, \qquad L_2 =  \dfrac{-Re \omega}{\left(\dfrac{6}{5} Re U \alpha - 1\right)^2 + Re^2 \omega^2}, \nonumber \\
T_1 &=  \sqrt{\dfrac{{\dfrac{9}{25}Re U^2+\nu} + \nu^2 \alpha^2 + \sqrt{\left({\dfrac{9}{25}Re U^2+\nu} + \nu^2 \alpha^2\right)^2 + {\left(\nu Re\right)}^2 \omega^2}}{2}}, \nonumber \\
T_2 &= - \sqrt{\dfrac{- {\dfrac{9}{25}Re U^2+\nu} - \nu^2 \alpha^2 + \sqrt{\left({\dfrac{9}{25}Re U^2+\nu} + \nu^2 \alpha^2\right)^2 + {\left(\nu Re\right)}^2 \omega^2}}{2}}, \nonumber \\
Q_1 &=  \dfrac{1}{\nu} \left(\dfrac{3}{5} Re U_y\right) - \dfrac{1}{\nu} \left(\dfrac{9}{25}\right) {Re}^2 U U_y \left[\dfrac{T_1}{T_1^2 + T_2^2}\right], \qquad Q_2 = \dfrac{1}{\nu} \left(\dfrac{9}{25}\right) {Re}^2 U U_y \left[\dfrac{T_2}{T_1^2 + T_2^2}\right], \nonumber \\
W_1 &=  U t \left[\dfrac{Q_1}{{Q_1}^2 U^2 t^2 + {\left(\alpha+ U t Q_2\right)}^2}\right], \qquad W_2 = -\left[\dfrac{\left(\alpha+U t Q_2\right)}{{Q_1}^2 U^2 t^2 + {\left(\alpha+ U t Q_2\right)}^2}\right], \nonumber \\
R_1 &=  \alpha + \dfrac{3}{5} \dfrac{1}{\nu} Re U - \dfrac{T_1}{\nu}, \qquad R_2 =  - \dfrac{T_2}{\nu}, \nonumber \\
X_1 &=  R_1^2 - \alpha^2 - R_2^2, \qquad X_2 = 2 R_1 R_2, \nonumber \\
Z_1 &=  {\left(R_1-\alpha\right)}^2 - R_2^2, \qquad Z_2 = 2 \left(R_1 - \alpha\right) R_2, \nonumber \\
P_1 &= -\alpha Re  \left(\dfrac{6}{5}\right) U_y \left(X_2 \left(L_1^2 - L_2^2\right) + 2 X_1 L_1 L_2\right) - \alpha \left(L_1 Z_1 - L_2 Z_2\right), \nonumber \\
P_2 &= \alpha Re  \left(\dfrac{6}{5}\right) U_y \left(X_1 \left(L_1^2 - L_2^2\right) - 2 X_2 L_1 L_2\right) - \alpha \left(L_2 Z_1 + L_1 Z_2\right), \nonumber \\
Y_1 &=  \alpha L_1 + 2 \alpha Re  \left(\dfrac{6}{5}\right) U_y L_1 L_2, \qquad Y_2 = \alpha L_2 - \alpha Re  \left(\dfrac{6}{5}\right) U_y \left(L_1^2 - L_2^2\right), \nonumber \\
S_1 &=-2 \alpha Re \left(\dfrac{6}{5}\right) U_y L_1 L_2 \left(R_1 + \alpha\right) - \alpha L_1 \left(R_1 - \alpha\right) - R_2 \left(\alpha Re \left(\dfrac{6}{5}\right) U_y \left(L_1^2 - L_2^2\right) - \alpha L_2\right), \nonumber \\
S_2 &= R_2 \left(-2 \alpha Re \left(\dfrac{6}{5}\right) U_y L_1 L_2 - \alpha L_1\right) + \alpha Re \left(\dfrac{6}{5}\right) U_y \left(L_1^2 - L_2^2\right) \left(R_1+\alpha\right) - \alpha L_2 \left(R_1 - \alpha\right).
\label{appA:f_and_g_exp}
\end{align}
Finally, the expression of the functions utilized in the dispersion relation (refer equation~\eqref{eqn:DRP_matrix}) and evaluated at the interface, are outlined as follows,
\begin{align}
{A} &= 2  \nu {U_y} t \left(g_{r_x}+f_{i_x}+\alpha f_{i}\right)+\left(-g_{i_x}+f_{r_x}\right) \left( \nu-{\left({U_y}\right)}^2 t^2  \nu\right) \nonumber \\
{P} &= 2  \nu {U_y} t \left(g_{i_x}-f_{r_x}-\alpha f_{r}\right)+\left(g_{r_x}+f_{i_x}\right) \left( \nu-{\left({U_y}\right)}^2 t^2  \nu\right) \nonumber  \\
{C} &= \left(1- \nu\right) \left({A_0}^{11}-{A_0}^{22}\right) \alpha-2 \alpha {U_y} t \left( \nu {U_y}+\left(1- \nu\right) {A_0}^{12}\right) \nonumber \\
{D} &= \left(1- \nu\right) {U_y} t \nonumber \\
{E} &= -\left({{U_y}}\right)^2 t^2 \left(1- \nu\right) \nonumber \\
{H} &= -2  \nu g_{r_x}+2  \nu {\left({U_y}\right)}^2 t^2 f_{i_x}+2 {U_y} t  \nu \left(-g_{i_x}+f_{r_x}\right)+e^{-\alpha Ut} \left(1+{{\left(U_y\right)}}^2 t^2\right) \nonumber \\
{J} &= -2  \nu g_{i_x}-2  \nu {\left({U_y}\right)}^2 t^2 f_{r_x}+2 {U_y} t  \nu \left(g_{r_x}+f_{i_x}\right) \nonumber \\
{N} &=  \left(\nu {U_{yy}}-U+\left(1- \nu\right) {{\left({A_0}^{21}\right)}_{y}} \right)\left(1+{{\left(U_y\right)}}^2 t^2\right)+B \alpha^2 \nonumber \\
{M} &= \alpha \left(2  \nu {U_y}+\left(1-\nu\right) \left({A_0}^{12}+{A_0}^{21}\right)-2 {U_y} t \left(1- \nu\right) {A_0}^{22}\right) \nonumber \\
{R} &= {U_y} t f_{r} \nonumber \\
{S} &= {U_y} t f_{i} \nonumber \\
{Q} &= {{\left({A_0}^{11}\right)}_{y}} f_{r}-\left(2 {A_0}^{11}+\dfrac{2}{\text{We}}\right) g_{r_x}-\left({A_0}^{12}+{A_0}^{21}\right) \left(-g_{i_x}\right) \nonumber \\
{W} &= {{\left({A_0}^{11}\right)}_{y}} f_{i}-\left(2 {A_0}^{11}+\dfrac{2}{\text{We}}\right) g_{i_x}-\left({A_0}^{12}+{A_0}^{21}\right) \left(g_{r_x}\right) \nonumber \\
{X} &= -{U_y} + \dfrac{1}{\text{We}} \nonumber \\
{Y} &= {{\left({A_0}^{12}\right)}_{y}} f_{r}-\left({A_0}^{22}+\dfrac{1}{\text{We}}\right) \left(-g_{i_x}\right)-\left({A_0}^{11}+\dfrac{1}{\text{We}}\right) f_{r_x} \nonumber \\
{Z} &= {{\left({A_0}^{12}\right)}_{y}} f_{i}-\left({A_0}^{22}+\dfrac{1}{\text{We}}\right) \left(g_{r_x}\right)-\left({A_0}^{11}+\dfrac{1}{\text{We}}\right) f_{i_x} \nonumber \\
{\delta} &= {{\left({A_0}^{22}\right)}_{y}} f_{r}-\left({A_0}^{12}+{A_0}^{21}\right) f_{rx}-\left(2 {A_0}^{22}+\dfrac{2}{\text{We}}\right) \left(-f_{i_x}-\alpha f_{i}\right) \nonumber \\
{\tau} &= {{\left({A_0}^{22}\right)}_{y}} f_{i}-\left({A_0}^{12}+{A_0}^{21}\right) f_{i_x}-\left(2 {A_0}^{22}+\dfrac{2}{\text{We}}\right) \left(f_{r_x}+\alpha f_{r}\right).
\label{appA:DRP_exp}
\end{align}
%
%
\section{Viscoelastic dispersion relation}\label{appB}
In the expressions below, the derivatives $\dfrac{\partial }{\partial \alpha}, \dfrac{\partial }{\partial \omega}, \dfrac{\partial^2}{\partial \alpha^2}, \dfrac{\partial^2 }{\partial \omega \partial \alpha}$ are denoted by $(\cdot)', \hat{(\cdot)}, (\cdot)'', \hat{(\cdot)}'$, respectively. The dispersion relation for the viscoelastic Hele-Shaw flow, whose derivation is outlined in \S \ref{sec:math}, is given as, 
\begin{align}
&D(\alpha, \omega) = \omega^4 \left[- D (A + i P) + {(1 - \nu)} (H + i J)   \right] + \omega^3 \left[2 \left( \sfrac{1}{\text{We}}\right)(-  i A D +  P D +  i   {(1 - \nu)} H \right. \nonumber \\ 
&- {(1 - \nu)} J) + {(1 - \nu)} ( i X  H - X  J  - i \delta   E  - i  N R +  N S  +  M R  + i  M S - i  D  Y +  D  Z +  E  \tau \nonumber \\ 
&+ i A   {U_y} - P   {U_y}) + {(1 - \nu)}^{2} (- i Q +  W) - i X A D   + X P D - C D R  - i C D S + i D H U_y - \nonumber \\ 
& \left. D J  {U_y} + i E D  Y - E D  Z  + i D^2  Q - D^2  W + D^2  \tau- i \delta  D^2  \right] + 
 \omega^2 \left[\left(\sfrac{1}{\text{We}}\right)^2 (  D (A + i P) - \right. \nonumber \\ 
&{(1 - \nu)} (H + i J)) + \left(\sfrac{1}{\text{We}}\right) ( 2 X A  D + 2 i X P D  - 2 X   {(1 - \nu)} H  - 2 i X   {(1 - \nu)} J  + 2 A   {(\nu -1)} {U_y} \nonumber \\ 
& + 2 i   P {(\nu -1)} {U_y} - 2 i C D R + 2   C D S - 2   D H {U_y} - 2 i   D J {U_y} - 2   D^2 Q - 2 i   D^2 W +    \delta  {(1 - \nu)} E \nonumber \\ 
& + \delta  D^2 -   E D Y - i   E D Z + 2    {(1 - \nu)} N R + 2 i {(1 - \nu)} N S + 2 i {(1 - \nu)} M R - 2 {(1 - \nu)} M S +\nonumber \\ 
& 2   {(1 \!\!-\!\! \nu)}^{2} Q + 2 i   {(1 \!\!-\!\! \nu)}^{2} W+   {(1 \!\!-\!\! \nu)} D Y + i   {(1 \!\!-\!\! \nu)} D Z + i   {(1 \!\!-\!\! \nu)} E \tau + i   D^2 \tau) + (1 \!\!-\!\! \nu)( X \delta E + X N R \nonumber \\ 
& + i X  N S + i X  M R - X  M S + X  D Y + i X  D Z+ i X  E \tau + E  {U_y} Y + i E {U_y} Z + i C  R U_y- C S {U_y} ) +\nonumber \\ 
& {U_y}( D \delta  E + D N R + i D N S + i D M R - D M S + D^2 Y + i D^2  Z  - \delta  E D )- i X C D R + X C D S \nonumber \\ 
& \left. + X \delta D^2 - X E D Y - i X E D Z + i X {D^2} \tau \right] + \omega \!\! \left[\left(\sfrac{1}{\text{We}}\right)^2(i X A D  - X  P D  - i X   {(1 \!\!-\!\! \nu)} H + X  {(1 \!\!-\!\! \nu)} \right. \nonumber \\ 
& J  + i A {(\nu -1)}  {U_y} -  P {(\nu -1)} {U_y}+  C D R + i  C D S - i  D H  {U_y}+  D J  {U_y} - i  D^2  Q +  D^2  W \nonumber \\ 
& + i   {(1\!\! - \!\!\nu)}  N R -  {(1\!\! - \!\!\nu)} N S  -  {(1\!\! - \!\!\nu)} M R - i  {(1\!\! - \!\!\nu)}M S + i  {(1\!\! - \!\!\nu)}^{2}  Q -   {(1\!\! - \!\!\nu)}^{2}  W)+ \left(\sfrac{1}{\text{We}}\right)( 2 X   C D R \nonumber \\ 
& + 2 i X   C D S  + i X   \delta  {(1 - \nu)} E  + i X   \delta  D^2 - i X   E D   Y + X   E D   Z + 2 i X   {(1\!\! - \!\!\nu)} N R  - 2 X   {(1\!\! - \!\!\nu)} N S \nonumber \\ 
& - 2 X   {(1 - \nu)} M R - 2 i X   {(1 - \nu)} M S  + i X   {(1 - \nu)} D   Y - X   {(1 - \nu)}  D   Z - X   {(1 - \nu)} E   \tau - X   D^2  \tau \nonumber \\ 
& + 2   C {(\nu -1)} R  {U_y}- 2 i   C {(1 - \nu )} S {U_y} + i   D \delta  E {U_y} + 2 i   D N R {U_y}- 2    D N S {U_y} - 2   D M R  {U_y}- 2 i  \nonumber \\ 
& D M S  {U_y}+ i   D^2  {U_y} Y - D^2  {U_y} Z-   D E  \tau  {U_y} + i   \delta  {D} {(\nu -1)}  {U_y} - i   E {(\nu -1)}  {U_y}  Y +   E {(\nu -1)}  {U_y} Z \nonumber \\ 
& - {D} {(\nu -1)}  \tau  {U_y} ) + U_y ( X E D  \tau - i X \delta  E D + i X \delta  {(1 - \nu)} D - X {(1 - \nu)} D  \tau )+ {U_y}^2 
(i D^2 \delta - D^2 \tau \nonumber \\ 
& \left. - i \delta  E {(\nu -1)}   + E {(\nu -1)}  \tau  ) \right] + \left[\left(\sfrac{1}{\text{We}}\right)^2( i X  C D R - X  C D S - X  {(1 - \nu)} N R - i X  {(1 - \nu)} N S - \right. \nonumber \\ 
& i X  {(1 - \nu)} M R + X   {(1 - \nu)} M S + i  C {(\nu -1)} R {U_y} + C {(\nu -1)} S {U_y}-  D N R {U_y} - i  D N S {U_y} -  \nonumber \\ 
& \left. i D M R {U_y}+  D M S {U_y}) \right] = 0,
\label{appB:DRP}
\end{align}
with its real part described by,
\begin{align}
&\text{Re}(D(\alpha, \omega)) = \omega^4 \left[ -A D  + {(1 - \nu)} H \right] +
\omega^3 \left[2 \left( \sfrac{1}{\text{We}}\right) (  P D -   {(1 - \nu)} J) + {(1 - \nu)} (- X J  - P  {U_y} \right.\nonumber \\ 
& \left. + N S +  M R  +  D Z  + E \tau) + {(1 - \nu)}^{2} W  + X P D  - C D R - D J {U_y}- D^2 W - E D Z + D^2 \tau \right] + \nonumber \\ 
&  \omega^2 \left[\left( \sfrac{1}{\text{We}}\right)^2(A  D -  {(1 - \nu)} H )  + \left( \sfrac{1}{\text{We}}\right) (2 X A D  - 2 X   {(1 - \nu)} H + 2 A   {(\nu -1)}  {U_y} + 2   C D S \right. \nonumber \\
&\left. - 2 D H {U_y} - 2 D^2  Q + \delta  {(1 - \nu)} E  +   \delta  D^2  + 2    {(1 - \nu)} N R - 2   {(1 - \nu)} M S  + 2   {(1 - \nu)}^{2}  Q +  {(1 - \nu)} \right. \nonumber \\
& \left. D  Y - E D  Y) + ( X C D S  + X \delta  {(1 - \nu)} E + X \delta  D^2 - X E D  Y + X {(1 - \nu)} N R - X {(1 - \nu)} M S  
\right. \nonumber \\
& \left. + X {(1 - \nu)} D  Y) + U_y ( C{(\nu -1)} S + D \delta  E + D N R - D M S + D^2  Y - \delta  E D   - E {(\nu -1)} Y) \right] + \nonumber \\
& \omega \left[ \left( \sfrac{1}{\text{We}}\right)^2 (- X P D  + X  {(1 - \nu)} J  -  P {(\nu -1)}  {U_y}+ C D R + D J  {U_y}+ D^2  W-  {(1 - \nu)} N S
 \right. \nonumber \\
& \left. -  {(1 - \nu)} M R -   {(1 - \nu)}^{2}  W)   + \left( \sfrac{1}{\text{We}}\right) (2 X   C D R + X E D   Z - 2 X   {(1 - \nu)} N S  - 2 X {(1 - \nu)} M R  
 \right. \nonumber \\
 & \left. - X   {(1 - \nu)}  D   Z - X   {(1 - \nu)} E   \tau - X   D^2  \tau + 2   C {(\nu -1)} R  {U_y}- 2    D N S  {U_y} - 2   D M R  {U_y}-   D^2  {U_y} Z  
 \right. \nonumber \\
 & \left. - D E  \tau  {U_y} + E {(\nu -1)}  {U_y} Z -   {D} {(\nu -1)}  \tau  {U_y}) + {U_y}( X E D  \tau  - X {(1 - \nu)}  D  \tau) + {U_y}^2 (- D^2  \tau + \right. \nonumber \\
& \left. E {(\nu \!\!-\!\!1)}  \tau) \right] +  \left[ \left( \sfrac{1}{\text{We}}\right)^2 (- X C D S \!\!-\!\! X {(1 \!\!-\!\! \nu)} N R + X   {(1 \!\!-\!\! \nu)} M S +  C {(1 \!\!-\!\! \nu)} S {U_y} \!\!-\!\! D N R {U_y} \!\!+\!\! D M S {U_y}) \right], 
\label{appB:DRP_Real}
\end{align}
and its imaginary part delineated as,
\begin{align}
& \text{Imag}(D(\alpha, \omega)) =  \omega^4\left[- P D + {(1 - \nu)} J\right] + 
\omega^3 \left[ \left(\sfrac{1}{\text{We}}\right) (- 2 A D + 2 {(1 - \nu)} H ) - {(1 - \nu)}^{2} Q +  
\right. \nonumber \\ 
& \left. (1 - \nu) ( X H + A {U_y} - N R + M S - \delta E -
D Y ) - X A D - C D S + D H {U_y}+ D^2 Q - \delta D^2 + 
\right. \nonumber \\ 
& \left. E D Y
 \right] +
\omega^2 \left[ \left(\sfrac{1}{\text{We}}\right)^2 (P D - {(1 - \nu)} J ) + \left(\sfrac{1}{\text{We}}\right) (2 X P D - 2 X {(1 - \nu)} J + 2 P {(\nu -1)} {U_y} - 2 C D R \right. \nonumber \\ 
& \left. - 2 D J {U_y} - 2 D^2 W- E D Z + 2 {(1 - \nu)} N S + 2 {(1 - \nu)} M R + 2 {(1 - \nu)}^{2} W+ {(1 - \nu)} D Z + 
\right. \nonumber \\  
& \left. {(1 - \nu)} E \tau + D^2 \tau) + U_y (- C {(\nu -1)} R + D N S + D M R + D^2 Z  - E {(\nu -1)} Z ) - X C D R -  
\right. \nonumber \\  
& \left. X E D Z + X {(1 - \nu)} N S + X {(1 - \nu)} M R + X {(1 - \nu)} D Z + X {(1 - \nu)} E \tau + X {D^2} \tau \right] +
\omega \left[ \left(\sfrac{1}{\text{We}}\right)^2 
\right. \nonumber \\ 
& \left. ( X A D
- X {(1 - \nu)} H + A {(\nu -1)} {U_y} + C D S - DH {U_y}- D^2 Q+ {(1 - \nu)} N R - {(1 - \nu)} M S + 
\right. \nonumber \\  
& \left. {(1 - \nu)}^{2} Q) + \left(\sfrac{1}{\text{We}}\right) (2 X C D S + X \delta {(1 - \nu)} E + X \delta D^2 - X E D Y + 2 X{(1 - \nu)} N R - 2 X {(1 - \nu)} 
\right. \nonumber \\  
& \left. M S + X {(1 - \nu)} D Y + 2 C {(\nu -1)} S {U_y} + D \delta E {U_y} + 2 D
N R {U_y}- 2 D M S {U_y} + D^2 {U_y} Y + \delta D \right. \nonumber \\
& \left. {(\nu -1)} {U_y} - E {(\nu -1)} {U_y} Y ) + {U_y}^2 (- \delta E {(\nu -1)} + D^2 \delta ) + U_y(- X \delta E D + X \delta {(1 - \nu)} D ) \right] +  \nonumber \\
& \left[ \left(\sfrac{1}{\text{We}}\right)^2 ( X C D R - X {(1 - \nu)}N S - X {(1 - \nu)} M R- C {(1 - \nu)} R {U_y}- D N S {U_y} - D M R {U_y} )\right].
\label{appB:DRP_Imag}
\end{align}
The various derivatives of the dispersion relation deployed in the derivation of the neutral curves (\S \ref{sec:NSC}), temporal (\S \ref{sec:tsa}) and spatiotemporal analysis (\S \ref{sec:stsa}) are as follows,
\begin{align}
&\dfrac{\partial D_r}{\partial \alpha} = \omega^4 \left[- A' D  + {(1 - \nu)} H'\right] + \omega^3 \left[ \left(\sfrac{1}{\text{We}}\right)( 2  P' D  - 2  {(1 - \nu)} J' ) + (1 - \nu)(- X  J' - P'   {U_y} \right. \nonumber \\
&\left. +  N S' +  N' S +  M R' +  M' R  + D  Z' +  E  \tau') + {(1 - \nu)}^{2}  W' + X P' D - C D R'  - C' D R - D J' \right. \nonumber \\
&\left. {U_y} - D^2  W' - E D  Z' + D^2  \tau'  \right] + \omega^2 \left[  \left(\sfrac{1}{\text{We}}\right)^2 ( A' D  - {(1 - \nu)} H' ) + \left(\sfrac{1}{\text{We}}\right) (2 X A'  D - 2 X \right. \nonumber \\
&\left. {(1 - \nu)} H' + 2 A'  {(\nu -1)}  {U_y} + 2  C D S' + 2   C' D S - 2  D H'  {U_y} - 2  D^2  Q'+  \delta' {(1 - \nu)} E +  \delta' D^2  \right. \nonumber \\
&\left. -  E D  Y' + 2  {(1 - \nu)} N R' + 2  {(1 - \nu)} N' R - 2  {(1 - \nu)} M S' - 2  {(1 - \nu)} M' S + 2  {(1 - \nu)}^{2}  Q'+ \right. \nonumber \\
&\left. {(1 - \nu)} D  Y') + (1 - \nu)( X \delta'  E + X  N R' + X  N' R   - X  M S' - X  M' S  + X  D  Y'- C  S'  {U_y}- C' \right. \nonumber \\
&\left. S  {U_y} + E   {U_y} Y') + X C D S'  + X C' D S + X \delta' D^2  - X E D  Y'+ D \delta' E  {U_y} + D N R'  {U_y}+ D N' R \right. \nonumber \\
&\left. {U_y} - D M S'  {U_y} - D M' S  {U_y} + D^2  {U_y} Y'- \delta' E D  {U_y}  \right] + \omega \left[   \left(\sfrac{1}{\text{We}}\right)^2 (- X   P' D  + X  {(1 - \nu)} J' -  \right. \nonumber \\
&\left.   P' {(\nu -1)}  {U_y}+  C D R'  +  C' D R  +  D J'  {U_y} +  D^2  W' -  {(1 - \nu)} N S'  -  {(1 - \nu)} N' S  -  {(1 - \nu)} M  \right. \nonumber \\
&\left. R' -  {(1 - \nu)} M' R  -  {(1 - \nu)}^{2}  W' ) + \left(\sfrac{1}{\text{We}}\right) ( 2 X  C D R'  + 2 X  C' D R  + X  E D  Z'- 2 X  {(1 - \nu)} N  \right. \nonumber \\
&\left. S'  - 2 X  {(1 - \nu)} N' S  - 2 X  {(1 - \nu)} M R'  - 2 X  {(1 - \nu)} M' R  - X  {(1 - \nu)} D  Z' - X  {(1 - \nu)} E  \tau' - \right. \nonumber \\
&\left. X  D^2  \tau' + 2  C {(\nu -1)} R'  {U_y}+ 2  C' {(\nu -1)} R  {U_y}- 2  D N S'  {U_y}- 2  D N' S  {U_y} -  D E  \tau' {U_y}+  E \right. \nonumber \\
&\left. {(\nu -1)}  {U_y} Z' -  D {(\nu -1)}  \tau'{U_y} - 2  D M R'  {U_y} - 2  D M' R  {U_y} -  D^2  {U_y} Z') + {U_y}^2 (- D^2  \tau' - E \right. \nonumber \\
&\left. {(1 - \nu )}  \tau' ) + {U_y}( X E D  \tau' - X {(1 - \nu)} D  \tau') \right] + \left[\left(\sfrac{1}{\text{We}}\right)^2 (- X {(1 - \nu)}N R' - X  {(1 - \nu)} N' R + \right. \nonumber \\
&\left. X  {(1 - \nu)} M S'+  X  {(1 - \nu)} M' S +  C {(1 - \nu)} S' {U_y} +  C' {(1 - \nu)} S {U_y} - X  C D S' - X  C' D S-  D \right. \nonumber \\
&\left. N R' {U_y}-  D N' R {U_y}+  D M S' {U_y}+  D M' S {U_y}) \right],
\label{appB:DRPr_alpha}
\end{align}

\vskip -20pt
\begin{align}
&\dfrac{\partial D_r}{\partial \omega} = \omega^4 \left[ - \hat{A} D + {(1 - \nu)} \hat{H}    \right]  + \omega^3 \left[ \left(\sfrac{1}{\text{We}}\right)( 2  \hat{P} D  - 2  {(1 - \nu)} \hat{J} ) + {(1 - \nu)}^{2}  \hat{W}  + (1 - \nu)(- X  \right. \nonumber \\
&\left. \hat{J} - \hat{P}   {U_y} + 4  H  +  M \hat{R}+  D  \hat{Z} +  E  \hat{\tau} +  N \hat{S} ) + X \hat{P} D  - 4 A D   - C D \hat{R} - D \hat{J}  {U_y}- D^2  \hat{W} - E D  \hat{Z} \right. \nonumber \\
&\left. + D^2  \hat{\tau}    \right] + \omega^2 \left[ \left(\sfrac{1}{\text{We}}\right)^2 (\hat{A}  D - {(1 - \nu)} \hat{H}) + \left(\sfrac{1}{\text{We}}\right) ( 2 X \hat{A} D - 2 X {(1 - \nu)} \hat{H}  + 2 \hat{A} {(\nu -1)}  {U_y}+ 6 \right. \nonumber \\
&\left. P D  - 2 D \hat{H}  {U_y}- 2 D^2  \hat{Q}+ \hat{\delta} {(1 - \nu)} E  + \hat{\delta} D^2 - E D  \hat{Y} - 6 {(1 - \nu)} J   + 2 {(1 - \nu)} N \hat{R} + 2 {(1 - \nu)}^{2} \right. \nonumber \\
&\left. \hat{Q}+ {(1 - \nu)} D  \hat{Y} + 2 C D \hat{S} - 2 {(1 - \nu)} M \hat{S}) + 3 {(1 - \nu)}^{2}   W + (1 - \nu)(X \hat{\delta}  E - 3 X  J  + X  N \hat{R}  + X \right. \nonumber \\
&\left. D  \hat{Y}- 3 P {U_y} + E   {U_y} \hat{Y} + 3  N S  + 3  M R + 3  D   Z + 3  E   \tau- X  M \hat{S}  - C  \hat{S}  {U_y}) + 3 X P D  + X \hat{\delta} D^2 - \right. \nonumber \\
&\left. X E D  \hat{Y} - 3 C D R  + D \hat{\delta} E  {U_y}- 3 D J   {U_y} + D N \hat{R}  {U_y}- 3 D^2   W+ D^2  {U_y} \hat{Y}- \hat{\delta} E D  {U_y}- 3 E D Z \right. \nonumber \\
&\left. + 3 D^2   \tau+ X C D \hat{S}  - D M \hat{S}  {U_y}   \right] + \omega \left[ \left(\sfrac{1}{\text{We}}\right)^2 ( - X  \hat{P} D + X  {(1 - \nu)} \hat{J}+ 2 A  D   -  \hat{P} {(\nu -1)} {U_y}+  C \right. \nonumber \\
&\left. D \hat{R} +  D \hat{J} {U_y} +  D^2  \hat{W}- 2   {(1 - \nu)} H -  {(1 - \nu)} M \hat{R} -  {(1 - \nu)}^{2}  \hat{W}- {(1 - \nu)}  N \hat{S}) + \left(\sfrac{1}{\text{We}}\right)( 4 X A \right. \nonumber \\
&\left.  D  + 4 A  {(\nu -1)}  {U_y}+ 2 X  C D \hat{R} + X  E D  \hat{Z} - 4 X  {(1 - \nu)} H - 2 X  {(1 - \nu)} M \hat{R}- X  {(1 - \nu)} D  \hat{Z} - \right. \nonumber \\
&\left. X  {(1 - \nu)} E  \hat{\tau} - X  D^2  \hat{\tau}+ 2  C {(\nu -1)} \hat{R} {U_y}+ 4  C D S  - 4  D H  {U_y}- 2  D M \hat{R} {U_y}- 4  D^2  Q-  D^2 {U_y} \hat{Z} \right. \nonumber \\
&\left. -  D E  \hat{\tau} {U_y}+ 2  \delta {(1 - \nu)} E + 2  \delta D^2 +  E {(\nu -1)} {U_y} \hat{Z} - 2  E D  Y + 4  {(1 - \nu)} N R - 4  {(1 - \nu)} M S  + \right. \nonumber \\
&\left.  4  {(1 - \nu)}^{2}  Q + 2  {(1 - \nu)} D  Y -  {D} {(\nu -1)}  \hat{\tau} {U_y} - 2 X {(1 - \nu)} N \hat{S} - 2 D N \hat{S} {U_y} ) + {U_y}^2 (- D^2  \hat{\tau} + E \right. \nonumber \\
&\left. {(\nu -1)}  \hat{\tau} ) + U_y ( X E D  \hat{\tau}  - X {(1 - \nu)} D  \hat{\tau} + 2 C {(\nu -1)} S   + 2 D \delta E  + 2 D N R   - 2 D M S  + 2 D^2   Y  - \right. \nonumber \\
&\left. 2 \delta E D   - 2 E {(\nu -1)}   Y) + 2 X C D S  + 2 X \delta {(1 - \nu)} E + 2 X \delta D^2 - 2 X E D Y+ 2 X {(1 - \nu)} N R - 2 \right. \nonumber \\
&\left. X {(1 - \nu)} M S + 2 X {(1 - \nu)} D  Y  \right] + \left[  \left(\sfrac{1}{\text{We}}\right)^2 (- X  P D + X  {(1 - \nu)} J  - X  {(1 - \nu)} N \hat{R}-  P {(\nu -1)}  {U_y} \right. \nonumber \\
&\left. +  C D R  +  D J  {U_y} -  D N \hat{R} {U_y}+  D^2  W - {(1 - \nu)} N S  -  {(1 - \nu)} M R  -  {(1 - \nu)}^{2}  W - X  C D \hat{S} + X  \right. \nonumber \\
&\left. {(1 - \nu)}   M \hat{S} -  C {(\nu -1)} \hat{S} {U_y} +  D M \hat{S} {U_y}) + \left(\sfrac{1}{\text{We}}\right) (2 C {(\nu -1)} R  {U_y} + 2 X  C D R  + X  E D  Z - 2 X  \right. \nonumber \\
&\left. {(1 - \nu)} N S - 2 X  {(1 - \nu)} M R  - X  {(1 - \nu)} D  Z - X  {(1 - \nu)} E  \tau - X  D^2  \tau - 2  D N S  {U_y} - 2  D M R  {U_y} \right. \nonumber \\
&\left. -  D^2  {U_y} Z \!\!-\!\!  D E  \tau {U_y} \!\!+\!\!  E {(\nu \!\!-\!\!1)}  {U_y} Z \!\!-\!\!  {D} {(\nu \!\!-\!\! 1)}  \tau {U_y}) + {U_y}^2 \!\!(- D^2  \tau \!\!+\!\! E {(\nu \!\!-\!\! 1)}  \tau ) \!\!+\!\! U_y ( X E D  \tau - X {(1 \!\!-\!\! \nu)} D  \tau)  \right], 
\label{appB:DRPr_omega}
\end{align}
\begin{align}
&\dfrac{\partial D_i}{\partial \alpha} = \omega^4 \left[  - P' D + {(1 - \nu)} J'   \right] + \omega^3 \left[   \left(\sfrac{1}{\text{We}}\right) (- 2 A' D  + 2  {(1 - \nu)} H' ) - {(1 - \nu)}^{2}  Q' + (1 - \nu)  \right. \nonumber \\
&\left. (X H' + A'  {U_y}- \delta' E - N' R - N R'  + M' S + M S'  - D  Y') - X A' D - C' D S  - C D S'  + D H' \right. \nonumber \\
&\left. {U_y} + D^2  Q'- \delta' D^2 + E D  Y' \right]  + \omega^2 \left[  \left(\sfrac{1}{\text{We}}\right)^2  (P' D-  {(1 - \nu)} J') + \left(\sfrac{1}{\text{We}}\right)( 2 X  P' D  - 2 X  {(1 - \nu)} \right. \nonumber \\
&\left. J'  + 2  P' {(\nu -1)}  {U_y}- 2  C' D R - 2  C D R' - 2  D J'  {U_y} - 2  D^2  W' -  E D  Z'+ 2  {(1 - \nu)} N' S  + 2 \right. \nonumber \\
&\left. {(1 - \nu)} N S'  + 2  {(1 - \nu)} M' R + 2  {(1 - \nu)} M R' + 2  {(1 - \nu)}^{2}  W'+  {(1 - \nu)} D  Z'+  {(1 - \nu)} E   \tau' + \right. \nonumber \\
&\left.  D^2  \tau') + (1 - \nu)( X  N' S  + X  N S' + X  M' R + X  M R' + X  D  Z' + X  E  \tau'+ C'  R  {U_y}+ C  R'  {U_y}  +  \right. \nonumber \\
&\left. E   {U_y} Z') + U_y (D N' S  + D N S'  + D M' R  + D M R'  + D^2  Z'  ) - X C' D R - X C D R' - X E D  Z' \right. \nonumber \\
&\left. + X D^2  \tau'   \right] + \omega \left[ \left(\sfrac{1}{\text{We}}\right)^2 ( X A'  D - X  {(1 - \nu)} H'  + A'  {(\nu -1)}  {U_y} +  C' D S + C D S' -  D H'  {U_y} - \right. \nonumber \\
&\left.  D^2  Q'+  {(1 - \nu)} N' R +  {(1 - \nu)} N R'  -  {(1 - \nu)}  M' S  -  {(1 - \nu)} M S'  +  {(1 - \nu)}^{2}  Q') +\left(\sfrac{1}{\text{We}}\right)  (2 X \right. \nonumber \\
&\left.  C' D S  + 2 X  C D S' + X  \delta' {(1 - \nu)} E  + X  \delta' D^2  - X  E D  Y' + 2 X   {(1 - \nu)} N' R + 2 X  {(1 - \nu)} N R' -  \right. \nonumber \\
&\left.  2 X  {(1 - \nu)} M' S  - 2 X  {(1 - \nu)} M S'  + X  {(1 - \nu)} D  Y' + 2  C' {(\nu -1)} S  {U_y}+ 2  C {(\nu -1)} S'  {U_y}+  D \delta' \right. \nonumber \\
&\left. E  {U_y}+ 2  D N' R  {U_y} + 2  D N R'  {U_y}- 2  D M' S  {U_y}- 2  D M S' {U_y}+  D^2  {U_y} Y'+  \delta' {D} {(\nu -1)} {U_y} -  E \right. \nonumber \\
&\left. {(\nu -1)}  {U_y} Y' ) +{U_y}^2 (D^2 \delta'   - \delta' E {(\nu -1)}  ) + U_y (- X \delta' E D + X \delta' {(1 - \nu)} D)  \right] +\left[\left(\sfrac{1}{\text{We}}\right)^2 (X  C' \right. \nonumber \\
&\left. D R + X  C D R' - X  {(1 - \nu)} N' S - X  {(1 - \nu)} N S'- X  {(1 - \nu)} M' R - X  {(1 - \nu)} M R' +  C' {(\nu -1)} \right. \nonumber \\
&\left. R {U_y}+  C {(\nu -1)} R' {U_y}-  D N' S {U_y} -  D N S' {U_y} -  D M' R {U_y} -  D M R' {U_y})\right],
\label{appB:DRPi_alpha}
\end{align} 
\begin{align}
&\dfrac{\partial D_i}{\partial \omega} = \omega^4 \left[ - \hat{P} D + {(1 - \nu)} \hat{J}   \right] + \omega^3 \left[  \left(\sfrac{1}{\text{We}}\right) (- 2 \hat{A}  D + 2  {(1 - \nu)} \hat{H} ) - {(1 - \nu)}^{2}  \hat{Q} + (1 - \nu)(X \hat{H} \right. \nonumber \\
&\left.  + \hat{A}   {U_y}- \hat{\delta}  E + 4  J  -  N \hat{R}  + M \hat{S}  -  D  \hat{Y}) - X \hat{A} D  - 4 P D   - C D \hat{S} + D \hat{H}  {U_y}+ D^2  \hat{Q}- \hat{\delta} D^2  + E D  \hat{Y} \right]  \nonumber \\
& + \omega^2 \left[ \left(\sfrac{1}{\text{We}}\right)^2 (\hat{P} D -  {(1 - \nu)} \hat{J}) + \left(\sfrac{1}{\text{We}}\right)(2 X  \hat{P} D  - 2 X  {(1 - \nu)} \hat{J} - 6 A  D+ 2  \hat{P} {(\nu -1)}  {U_y} - 2  C D \hat{R} \right. \nonumber \\
&\left. - 2  D \hat{J}  {U_y}- 2  D^2  \hat{W}-  E D  \hat{Z}+ 6  {(1 - \nu)} H  + 2  {(1 - \nu)} N \hat{S}  + 2  {(1 - \nu)} M \hat{R}  + 2  {(1 - \nu)}^{2} \hat{W} +  {(1 - \nu)} \right. \nonumber \\
&\left.  D  \hat{Z} +  {(1 - \nu)} E  \hat{\tau} +  D^2  \hat{\tau} ) - 3 {(1 - \nu)}^{2}   Q + (1 - \nu)( 3 X  H + 3 A U_y  + X  N \hat{S}  + X  M \hat{R} + X  D \hat{Z} +     \right. \nonumber \\
&\left. X E \hat{\tau}- 3 \delta  E - 3  N R  + 3  M S   - 3  D   Y + C  \hat{R}  {U_y}+ E   {U_y} \hat{Z} )- 3 X A D  - X C D \hat{R} - X E D  \hat{Z} + X D^2  \hat{\tau} \right. \nonumber \\
&\left. - 3 C D S  + 3 D H   {U_y} + D N \hat{S}  {U_y}+ D M \hat{R}  {U_y} + 3 D^2   Q+ D^2  {U_y} \hat{Z} - 3 \delta D^2  + 3 E D   Y    \right] + \omega \left[ \left(\sfrac{1}{\text{We}}\right)^2 \right. \nonumber \\
&\left. (X \hat{A}  D  - X  {(1 - \nu)} \hat{H} + \hat{A}  {(\nu -1)}  {U_y}+ 2  P    D +  C D \hat{S} -  D \hat{H}  {U_y}-  D^2  \hat{Q}- 2  {(1 - \nu)} J   +  {(1 - \nu)} \right. \nonumber \\
&\left. N \hat{R} -  {(1 - \nu)} M \hat{S} +  {(1 - \nu)}^{2}  \hat{Q}) + \left(\sfrac{1}{\text{We}}\right) (4 X  P D  + 2 X  C D \hat{S} + X  \hat{\delta} {(1 - \nu)} E  + X  \hat{\delta} D^2 - X  E D  \hat{Y} \right. \nonumber \\
&\left. - 4 X  {(1 - \nu)} J   + 2 X  {(1 - \nu)} N \hat{R}  - 2 X  {(1 - \nu)} M \hat{S}  + X  {(1 - \nu)} D  \hat{Y} + 4  P {(\nu -1)}   {U_y}+ 2  C {(\nu -1)} \right. \nonumber \\
&\left. \hat{S} {U_y}- 4  C D R  +  D \hat{\delta} E  {U_y} - 4  D J   {U_y}+ 2  D N \hat{R}  {U_y}- 2  D M \hat{S}  {U_y}- 4  D^2   W +  D^2  {U_y} \hat{Y} +  \hat{\delta} {D} {(\nu -1)} \right. \nonumber \\
&\left. {U_y}-  E {(\nu -1)}  {U_y}  \hat{Y}- 2  E D   Z + 4  {(1 - \nu)} N S  + 4  {(1 - \nu)} M R  + 4  {(1 - \nu)}^{2}   W + 2  {(1 - \nu)} D   Z+ \right. \nonumber \\
&\left. 2  {(1 - \nu)} E   \tau + 2  D^2   \tau) + (1 - \nu)( X \hat{\delta}  D  {U_y} + 2  X  N S + 2 X  M R   + 2 X  D   Z + 2 X  E   \tau + \hat{\delta} E   {U_y}^2 +  \right. \nonumber \\
&\left.2 C  R   {U_y}+ 2 E    {U_y} Z )- 2 X C D R  - X \hat{\delta} E D  {U_y} - 2 X E D   Z+ 2 X D^2   \tau + D^2 \hat{\delta} {U_y}^2 + 2 D N S   {U_y}+ \right. \nonumber \\
&\left. 2 D M R   {U_y} + 2 D^2   {U_y} Z   \right] + \left[ \left(\sfrac{1}{\text{We}}\right)^2 ( X A  D + X  C D \hat{R}- X  {(1 - \nu)} H - X  {(1 - \nu)} N \hat{S} - X  {(1 - \nu)} \right. \nonumber \\
&\left.  M \hat{R} + A {(\nu -1)}  {U_y}+  C {(\nu -1)} \hat{R} {U_y}+  C D S -  D H  {U_y} -  D N \hat{S} {U_y}-  D M \hat{R} {U_y}-  D^2  Q +  {(1 - \nu)}\right. \nonumber \\
&\left.  N R -  {(1 - \nu)} M S  +  {(1 - \nu)}^{2}  Q) + \left(\sfrac{1}{\text{We}}\right) ( 2 X   C D S  + X  \delta {(1 - \nu)} E + X  \delta D^2 - X  E D  Y+ 2 X \right. \nonumber \\
&\left. {(1 - \nu)} N R - 2 X  {(1 - \nu)} M S  + X  {(1 - \nu)} D  Y+ 2  C  {(\nu -1)} S  {U_y} +  D \delta E  {U_y} + 2  D N R  {U_y} - 2  D M \right. \nonumber \\
&\left. S  {U_y} \!\!+\!\!  D^2  {U_y} Y\!\!+\!\!  \delta {D} {(\nu \!\!-\!\!1)}  {U_y} \!\!-\!\!  E {(\nu \!\!-\!\! 1)}  {U_y} Y ) \!\!+\!\! {U_y}^2 (D^2 \delta \!\!-\!\! \delta E {(\nu -1)}  ) \!\!+\!\! U_y (- X \delta E D  + X \delta {(1 \!\!-\!\! \nu)} D ) \right], 
%
\label{appB:DRPi_omega}
\end{align}
\begin{align}
&\dfrac{\partial^2 D_i}{\partial \alpha^2} = \omega^4 \left[- P'' D + {(1 - \nu)} J'' \right] + \omega^3 \left[  \left(\sfrac{1}{\text{We}}\right) (- 2 A'' D + 2  {(1 - \nu)} H'' ) - {(1 - \nu)}^{2}  Q'' + \right. \nonumber \\
&\left.(1 - \nu) ( X  H''  + A''   {U_y} - \delta''  E  -  N'' R  - 2  N' R' -  N R''  + 2  M' S' +  M S''  -  D  Y'') - X A'' D \right. \nonumber \\
&\left. - 2 C' D S' - C D S'' + D H''  {U_y} + D^2  Q''- \delta'' D^2  + E D  Y''  \right] + \omega^2 \left[ \left(\sfrac{1}{\text{We}}\right)^2 (P'' D -  {(1 - \nu)} J'') \right. \nonumber \\
&\left. + \left(\sfrac{1}{\text{We}}\right) (2 X P'' D - 2 X  {(1 - \nu)} J'' + 2  P'' {(\nu -1)} {U_y}- 4  C' D R' - 2  C D R''  - 2  D J''  {U_y} - 2  D^2 \right. \nonumber \\
&\left.  W''-  E D  Z'' + 2  {(1 - \nu)} N'' S + 4  {(1 - \nu)} N' S'  + 2  {(1 - \nu)} N S''  + 4  {(1 - \nu)} M' R'   + 2  {(1 - \nu)} \right. \nonumber \\
&\left. M R''  + 2  {(1 - \nu)}^{2}  W''+  {(1 - \nu)} D  Z''+  {(1 - \nu)} E  \tau'' +  D^2  \tau'') + (1 - \nu)( X  N'' S + 2 X  N' S' \right. \nonumber \\
&\left. + X  N S''  + 2 X  M' R' + X  M R'' + X  D Z'' + X  E \tau''+ 2 C' {(1 - \nu )} R'  {U_y}+ C  R''  {U_y} + E   {U_y} Z'') \right. \nonumber \\
&\left. + U_y ( D N'' S  + 2 D N' S'   + D N S''   + 2 D M' R'  + D M R''  + D^2   Z'' ) - 2 X C' D R' - X C D R'' \right. \nonumber \\
&\left. - X E D Z''+ X D^2 \tau''   \right] + \omega \left[  \left(\sfrac{1}{\text{We}}\right)^2 (X A''  D  - X   {(1 - \nu)}  H'' + A''    {(\nu -1)}  {U_y} + 2  C' D S'   +  C  \right. \nonumber \\
&\left. D S'' -  D H''  {U_y} -  D^2  Q''+  {(1 - \nu)} N'' R  + 2 {(1 - \nu)} N' R' +   {(1 - \nu)} N R'' - 2  {(1 - \nu)} M' S'  - \right. \nonumber \\
&\left. {(1 - \nu)} M S'' +  {(1 - \nu)}^{2}  Q'') + \left(\sfrac{1}{\text{We}}\right) ( 4 X  C' D S'  + 2 X  C D S'' + X  \delta'' {(1 - \nu)} E + X  \delta'' D^2 - \right. \nonumber \\
&\left. X  E D  Y''+ 2 X  {(1 - \nu)} N'' R + 4 X  {(1 - \nu)} N' R' + 2 X   {(1 - \nu)} N   R'' - 4 X  {(1 - \nu)} M' S' - 2 X \right. \nonumber \\
&\left.  {(1 - \nu)} M S''  + X  {(1 - \nu)} D  Y''+ 4  C' {(\nu -1)} S'  {U_y} + 2  C {(\nu -1)} S''  {U_y}+ D \delta'' E  {U_y} + 2  D N'' R \right. \nonumber \\
&\left. {U_y}+ 4  D N' R'  {U_y} + 2  D N R''  {U_y} - 4  D M' S'    {U_y} - 2  D M  S''  {U_y}+  D^2  {U_y} Y''+  \delta'' {D} {(\nu -1)}  {U_y} - \right. \nonumber \\
&\left. E {(\nu -1)}  {U_y} Y'') +  {U_y}^2 ( D^2 \delta''   - \delta'' E {(\nu -1)}) + U_y (- X \delta'' E D  + X \delta'' {(1 - \nu)} D  )  \right]  + \left[  \left(\sfrac{1}{\text{We}}\right)^2 \right. \nonumber \\
&\left. ( 2 X  C' D R'+ X  C D R''- X  {(1 - \nu)} N'' S - 2 X  {(1 - \nu)} N' S' - X  {(1 - \nu)} N S'' - 2 X  {(1 - \nu)} M' R' \right. \nonumber \\
&\left. + 2  C' {(\nu \!\!-\!\!1)} R' {U_y} \!\!+\!\!  C {(\nu \!\!-\!\!1)} R'' {U_y} \!\!-\!\!  D N'' S {U_y} \!\!-\!\! 2  D N' S' {U_y} \!\!-\!\! D N S'' {U_y} \!\!-\!\! 2  D M' R' {U_y} - D M R'' {U_y})  \right], \nonumber \\
%
\label{appB:DRPi_alpha2}
\end{align}
\begin{align}
&\dfrac{\partial^2 D_i}{\partial \omega \partial \alpha} = \omega^{4}\left[- \hat{P'} D  + {(1 - \nu)} \hat{J'}\right] + 
\omega^3 \left[ \left(\sfrac{1}{\text{We}}\right)(- 2 \hat{A'}  D + 2  {(1 - \nu)} \hat{H'} ) - {(1 - \nu)}^{2}  \hat{Q'} + (1 - \nu) \right. \nonumber \\
&\left. ( X  \hat{H'} + \hat{A'}   {U_y}-  N' \hat{R} -  N \hat{R'}  +  M' \hat{S}  +  M \hat{S'} -  D  \hat{Y'} + 4  J'  - \hat{\delta'}  E ) - X \hat{A'} D - 4 P' D  - C' D \hat{S}\right. \nonumber \\
&\left. - C D \hat{S'}  + D \hat{H'}  {U_y}+ D^2  \hat{Q'} - \hat{\delta'} D^2  + E D  \hat{Y'}   \right] + \omega^2 \left[  \left(\sfrac{1}{\text{We}}\right)^{2} ( \hat{P'} D - {(1 - \nu)} \hat{J'}) + \left(\sfrac{1}{\text{We}}\right) (2 X \right. \nonumber \\
&\left. \hat{P'} D - 2 X {(1 - \nu)} \hat{J'} - 6 A' D + 2 \hat{P'} {(\nu -1)}  {U_y} - 2 C' D \hat{R} - 2 C D \hat{R'} - 2 D \hat{J'}  {U_y} - 2 D^2  \hat{W'} - \right. \nonumber \\
&\left. E D  \hat{Z'} + 6 {(1 - \nu)} H'   + 2 {(1 - \nu)} N' \hat{S} + 2  {(1 - \nu)} N \hat{S'} + 2 {(1 - \nu)} M' \hat{R}  + 2 {(1 - \nu)} M \hat{R'} + 2 \right. \nonumber \\
&\left. {(1 - \nu)}^{2}  \hat{W'}+ {(1 - \nu)} D  \hat{Z'}+ {(1 - \nu)} E  \hat{\tau'}+ D^2  \hat{\tau'} ) + (1 - \nu)( 3 X  H'  + X  N' \hat{S} + X  N \hat{S'} +  X  M' \right. \nonumber \\
&\left.\hat{R} +  X  M \hat{R'} + X  D  \hat{Z'} + X  E  \hat{\tau'} + 3 A'    {U_y}+ C'  \hat{R}  {U_y}+ C  \hat{R'}  {U_y}- 3 \delta'  E  + E   {U_y} \hat{Z'} - 3  N' R  - 3  N R' \right. \nonumber \\
&\left. + 3  M' S  + 3  M S'   - 3 \omega^{2}   Q' - 3  D   Y' )- 3 X A' D   - X C' D \hat{R} - X C D \hat{R'} -  X E D  \hat{Z'}+ X D^2  \hat{\tau'} - \right. \nonumber \\
&\left. 3 C' D S   - 3 C D S'  + 3 D H'   {U_y}+ D N' \hat{S}  {U_y}+ D N \hat{S'}  {U_y}+ D M' \hat{R}  {U_y} + D M \hat{R'}  {U_y} + 3 D^2   Q'+ \right. \nonumber \\
&\left. D^2  {U_y} \hat{Z'} - 3 \delta' D^2  + 3 E D   Y' \right] + \omega \left[ \left(\sfrac{1}{\text{We}}\right)^{2} (X \hat{A'} D - X {(1 - \nu)} \hat{H'} + \hat{A'} {(\nu -1)}  {U_y}+ 2   P' D + \right. \nonumber \\
&\left. C' D \hat{S} + C D \hat{S'} - D \hat{H'}  {U_y} - D^2  \hat{Q'}- 2   {(1 - \nu)} J'+ {(1 - \nu)} N' \hat{R}  + {(1 - \nu)} N \hat{R'}  - {(1 - \nu)} M' \hat{S} \right. \nonumber \\
&\left.  - {(1 - \nu)} M \hat{S'}  + {(1 - \nu)}^{2}  \hat{Q'} ) + \left(\sfrac{1}{\text{We}}\right) ( 4 X  P' D  + 2 X C' D \hat{S} + 2 X C D \hat{S'}  + X \hat{\delta'} {(1 - \nu)} E  + X \hat{ \delta' } \right. \nonumber \\
&\left. D^2 - X E D  \hat{Y'}- 4X {(1 - \nu)} J'   + 2 X {(1 - \nu)}  N' \hat{R} + 2 X {(1 - \nu)} N \hat{R'} - 2 X {(1 - \nu)} M' \hat{S} - 2 X  \right. \nonumber \\
&\left. {(1 - \nu)} M \hat{S'} + X {(1 - \nu)} D  \hat{Y'}+ 4 P' {(\nu -1)}   {U_y}+ 2 C' {(\nu -1)} \hat{S}  {U_y}+ 2 C {(\nu -1)} \hat{S'}  {U_y} - 4 C' D R \right. \nonumber \\
&\left. - 4 C D R'   + D \hat{\delta'} E  {U_y} - 4 D J'   {U_y} + 2 D N' \hat{R}  {U_y} + 2 D N \hat{R'}  {U_y} - 2 D M' \hat{S}  {U_y}- 2 D M \hat{S'}  {U_y} - \right. \nonumber \\
&\left. 4 D^2   W' + D^2  {U_y} \hat{Y'} + \hat{\delta'} {D} {(\nu -1)} {U_y} - E {(\nu -1)}  {U_y} \hat{Y'}- 2 E D    Z ' + 4  {(1 - \nu)} N' S  + 4 {(1 - \nu)} N S' \right. \nonumber \\
&\left.  + 4 {(1 - \nu)} M' R  + 4 {(1 - \nu)} M R'   + 4 {(1 - \nu)}^{2}   W'+ 2 {(1 - \nu)} D   Z'+ 2 {(1 - \nu)} E   \tau' + 2 D^2   \tau' ) + {U_y}^{2}  \right. \nonumber \\
&\left. (D^2 \hat{\delta'}  - \hat{\delta'} E {(\nu -1)}  ) + U_y (- X \hat{\delta'} E D   + X \hat{\delta'} {(1 - \nu)} D  - 2 C' {(\nu -1)} R   - 2 C {(\nu -1)} R'   + 2 D \right. \nonumber \\
&\left. N' S   + 2 D N S'   + 2 D M' R   + 2 D M R'   + 2 D^2    Z' - 2 E {(\nu -1)}    Z' )- 2 X C' D R  - 2 X C D R'  - \right. \nonumber \\
&\left. 2 X E D   Z' + 2 X {(1 - \nu)} N' S   + 2 X {(1 - \nu)} N S'  + 2 X {(1 - \nu)} M' R  + 2 X {(1 - \nu)} M R'  + 2 X \right. \nonumber \\
&\left. {(1 - \nu)} D   Z'+ 2 X {(1 - \nu)} E   \tau' +  2 X D^2   \tau'   \right] + \left[ \left(\sfrac{1}{\text{We}}\right)^{2}( X A'  D + X  C' D \hat{R}+ X  C D \hat{R'} - X  {(1 - \nu)} \right. \nonumber \\
&\left.  H'  - X  {(1 - \nu)} N' \hat{S}- X  {(1 - \nu)} N \hat{S'}- X  {(1 - \nu)} M' \hat{R}- X  {(1 - \nu)} M \hat{R'} + A'  {(\nu -1)}  {U_y} +  C' \right. \nonumber \\
&\left. {(\nu -1)} \hat{R} {U_y}+  C {(\nu -1)} \hat{R'} {U_y} +   C' D S +  C D S'  -  D H'  {U_y}-  D N' \hat{S} {U_y}-  D N \hat{S'} {U_y} -  D M' \hat{R} {U_y} \right. \nonumber \\
&\left.  - D M \hat{R'} {U_y}-  D^2  Q' +   {(1 - \nu)} N' R +  {(1 - \nu)} N R'  -  {(1 - \nu)} M' S-  {(1 - \nu)} M S'  +  {(1 - \nu)}^{2}  Q')\right. \nonumber \\
&\left.  + \left(\sfrac{1}{\text{We}}\right) (2 X  C' D S + 2 X C D S'  + X  \delta' {(1 - \nu)} E + 2  C' {(\nu -1)} S  {U_y} + 2  C {(\nu -1)} S'  {U_y} +  D \delta' E  {U_y} \right. \nonumber \\
&\left. + 2  D N' R  {U_y}  + 2  D N R'  {U_y} - 2  D M' S  {U_y} - 2  D M S'  {U_y} +  D^2  {U_y} Y' +  \delta' {D} {(\nu -1)}  {U_y} -  E {(\nu -1)} \right. \nonumber \\
&\left.  {U_y} Y'+ X  \delta' D^2 - X  E D  Y' + 2 X  {(1 - \nu)} N' R  + 2 X  {(1 - \nu)} N R' - 2 X  {(1 - \nu)} M' S - 2 X   {(1 - \nu)} \right. \nonumber \\
&\left. M S' + X  {(1 - \nu)} D  Y') + {(U_y)}^{2} ( D^2 \delta'  - \delta' E {(\nu -1)} ) + U_y (- X \delta' E D  + X \delta' {(1 - \nu)} D  )  \right],
\label{appB:DRPi_alpha_omega}
\end{align}
%
%
%
\bibliographystyle{jfm}

\end{document}